\documentclass{mn2e}

\usepackage{amsfonts}
\usepackage{amsmath}
\usepackage{graphicx}

\title[Galaxy distributions per morphological type]
      {Galaxy luminosities, stellar masses, sizes, 
       velocity dispersions as a function of morphological type}

\author[M. Bernardi et. al.]
{M. Bernardi$^1$\thanks{E-mail: bernardm@physics.upenn.edu}, 
 F. Shankar$^2$, J. B. Hyde$^1$, S. Mei$^{3,4}$, F. Marulli$^5$ 
 \& R. K. Sheth$^{1,6}$ \\
 $^1$Department of Physics \& Astronomy, University of Pennsylvania, 
      209 S. 33rd St., Philadelphia, PA 19104, USA\\
 $^2$ Max-Planck-Instit\"{u}t f\"{u}r Astrophysik,
Karl-Schwarzschild-Str. 1, D-85748, Garching, Germany \\
 $^3$ University of Paris Denis Diderot,  75205 Paris Cedex 13, France\\
 $^4$ GEPI, Observatoire de Paris, Section de Meudon, 
      5 Place J. Janssen, 92195 Meudon Cedex, France\\
 $^5$ Dipartimento di Astronomia, Universit\'{a} degli Studi di Bologna, 
      via Ranzani 1, I-40127 Bologna, Italy\\
 $^6$ Center for Particle Cosmology, University of Pennsylvania, 
      209 S. 33rd St., Philadelphia, PA 19104, USA}

\begin{document}

\maketitle

\label{firstpage}

\begin{abstract}
We provide fits to the distribution of galaxy luminosity, size, 
velocity dispersion and stellar mass as a function of concentration 
index $C_r$ and morphological type in the SDSS.  (Our size estimate, 
a simple analog of the SDSS {\tt cmodel} magnitude, is new:  it 
is computed using a combination of seeing-corrected quantities 
in the SDSS database, and is in substantially better agreement 
with results from more detailed bulge/disk decompositions.)  
We also quantify how estimates of the fraction of `early' or `late' type 
galaxies depend on whether the samples were cut in color, concentration 
or light profile shape, and compare with similar estimates based on 
morphology.  
Our fits show that ellipticals account for about 20\% of the 
$r$-band luminosity density, $\rho_{L_r}$, and 25\% of the stellar mass 
density, $\rho_*$; including S0s and Sas increases these numbers to 
33\% and 40\%, and 50\% and 60\%, respectively.  
The values of $\rho_{L_r}$ and $\rho_*$, and the mean sizes, of E, E+S0 and 
E+S0+Sa samples are within 10\% of those in the Hyde \& Bernardi (2009), 
$C_r\ge 2.86$ and $C_r\ge 2.6$ samples, respectively.  
Summed over all galaxy types, we find
 $\rho_* \sim 3\times 10^8 M_\odot$Mpc$^{-3}$ at $z\sim 0$.
This is in good agreement with expectations based on integrating 
the star formation history.
However, compared to most previous work, we find an excess of objects at large
masses, up to a factor of $\sim 10$ at $M_*\sim 5 \times 10^{11}M_\odot$. 
The stellar mass density further increases at large masses 
if we assume different IMFs for 
elliptical and spiral galaxies, as suggested by some recent chemical 
evolution models, and results in a better agreement with the dynamical 
mass function.

We also show that the trend for ellipticity to decrease with luminosity 
is primarily because the E/S0 ratio increases at large $L$.  However, the 
most massive galaxies, $M_*\ge 5\times 10^{11}M_\odot$, are less concentrated 
and not as round as expected if one extrapolates from lower $L$, and they 
are not well-fit by pure deVaucouleur laws.  This suggests formation 
histories with recent radial mergers.  
Finally, we show that the age-size relation is flat for ellipticals of 
fixed dynamical mass, but, at fixed $M_{\rm dyn}$, S0s and Sas with 
large sizes tend to be younger.  Hence, samples selected on the basis of 
color or $C_r$ will yield different scalings.  Explaining this difference 
between E and S0 formation is a new challenge for models of early-type 
galaxy formation.  
\end{abstract}


\section{Introduction}
Each galaxy has its own peculiarities.  
Nevertheless, even to the untrained eye, sufficiently well-resolved 
galaxies can be separated into three morphological types:  
disky spirals, bulgy ellipticals, and others which are neither.  
The morphological classification of galaxies is a field that is 
nearly one hundred years old, and sample sizes of a few thousand 
morphologically classified galaxies are now available 
(e.g. Fukugita et al. 2007; Lintott et al. 2008).  However, such
eyeball classifications are prohibitively expensive in the era 
of large scale sky surveys, which image upwards of a few million 
galaxies.  Moreover, the morphological classification of even 
relatively low redshift objects from ground-based data is difficult.  
Thus, a number of groups have devised automated algorithms for 
discerning morphologies from such data (e.g. Ball et al. 2004 and 
references therein).  

In parallel, it has been recognized that relatively simple 
criteria, using either crude measures of the light profile 
(e.g. Strateva et al. 2001), the colors (e.g. Baldry et al. 2004), 
or some combination of photometric and spectroscopic information 
(Bernardi et al. 2003; Bernardi \& Hyde 2009) allow one to separate 
early-type galaxies from the rest.  Because they are 
so simple, these tend to be more widely used.  
The main goal of this paper is to show how samples based on such 
crude cuts compare with those which are based on the eyeball 
morphological classifications of Fukugita et al. (2007).  
We do so by comparing the luminosity, stellar mass, size and 
velocity dispersion distributions for cuts based on photometric 
parameters with those based on morphology.  These were chosen 
because the luminosity function is standard, although it is 
becoming increasingly common to compare models with $\phi(M_*)$ 
rather than $\phi(L)$ (e.g. Cole et al. 2001; Bell et al. 2003; 
Panter et al. 2007; Li \& White 2009); 
the size distribution $\phi(R)$ has also begun to receive 
considerable attention recently (Shankar et al. 2009b); 
and the distribution of velocity dispersions $\phi(\sigma)$ 
(Sheth et al. 2003) is useful, amongst other things, to reconstruct
the mass distribution of super-massive black holes (e.g.
Shankar et al. 2004; Tundo et al. 2007; Bernardi et al. 2007; 
Shankar, Weinberg \& Miralda-Escud{\'e} 2009; Shankar et al. 2009a) 
and in studies of gravitational lensing (Mitchell et al. 2005).  

Section~\ref{data} describes the dataset, the photometric and 
spectroscopic parameters derived from it, and the subsample 
defined by Fukugita et al. (2007).  
This section shows how we use quantities output from the 
SDSS database to define seeing-corrected half-light radii which 
closely approximate the result of bulge + disk decompositions.  
We describe our stellar mass estimator in this section as well;
a detailed comparison of it with stellar mass estimates computed 
by three different groups is presented in Appendix~\ref{Mscomp}.
The result of classifying objects into two classes, on the 
basis of color, concentration index, or morphology are compared 
in Section~\ref{sec:mcba}.
Luminosity, stellar mass, size and velocity dispersion distributions, 
for the Fukugita et al. morphological types are presented 
in Section~\ref{DFs}, where they are compared with those based on 
the other simpler selection cuts.  This section includes a discussion 
of the functional form, a generalization of the Schechter function, 
which we use to fit our measurements.  
We find more objects with large stellar masses than in previous work 
(e.g. Cole et al. 2001; Bell et al. 2003; Panter et al. 2007; 
Li \& White 2009); this is the subject of Section~\ref{phiMass}, 
where implications for the match with the integrated star formation 
rate, and the question of how the most massive galaxies have evolved 
since $z\sim 2$ are discussed.  

While we believe these distributions to be interesting in their 
own right, we also study a specific example in which correlations 
between quantities, rather than the distributions themselves, 
depend on morphology.  This is the correlation between the 
half-light radius of a galaxy and the luminosity weighted age of 
its stellar population.  Section~\ref{ageRe} shows that the 
morphological dependence of this relation means it is sensitive to 
how the `early-type' sample was selected, potentially resolving a 
discrepancy in the recent literature (Shankar \& Bernardi 2009; 
van der Wel et al. 2009; Shankar et al. 2009c).  
A final section summarizes our results, many of which are 
provided in tabular form in Appendix~\ref{tables}.  

Except when stated otherwise, we assume a spatially flat 
background cosmology dominated by a cosmological constant, 
with parameters $(\Omega_m,\Omega_\Lambda)=(0.3,0.7)$, and 
a Hubble constant at the present time of
 $H_0=70$~km~s$^{-1}$Mpc$^{-1}$.  
When we assume a different value for $H_0$, we write it as 
$H_0=100h$~km~s$^{-1}$Mpc$^{-1}$.

\section{The SDSS dataset}\label{data}

\subsection{The full sample}\label{sample}
In what follows, we will study the luminosities, sizes, velocity 
dispersions and stellar masses of a magnitude limited sample of
$\sim 250,000$ SDSS galaxies with 
$14.5 < m_{\tt Petrosian} < 17.5$ in the $r-$band, selected from 
4681 deg$^2$ of sky.  In this band, the absolute magnitude of the 
Sun is $M_{r,\odot} = 4.67$.

The SDSS provides a variety of measures of the light profile of 
a galaxy.  Of these the Petrosian magnitudes and sizes are the 
most commonly used, because they do not depend on fits to models.  
However, for some of what is to follow, the Petrosian magnitude is 
not ideal, since it captures a type-dependent fraction of the total 
light of a galaxy.  In addition, seeing compromises use of the 
Petrosian sizes for almost all the distant lower luminosity objects, 
leading to systematic biases (see Hyde \& Bernardi 2009 for examples).  

Before we discuss the alternatives, we note that there is one 
Petrosian based quantity which will play an important role in 
what follows.  This is the concentration index $C_r$, which 
is the ratio of the scale which contains 90\% of the Petrosian 
light in the $r$-band, to that which contains 50\%.  
Early-type galaxies, which are more centrally concentrated, are 
expected to have larger values of {\tt $C_r$}, and two values are 
in common use:  a more conservative {\tt $C_r$}$\ge 2.86$ (e.g. 
Nakamura et al. 2003; Shen et al. 2003) and a more cavalier 
{\tt $C_r$}$\ge 2.6$ (e.g. Strateva et al. 2001; Kauffmann et al. 2003; 
Bell et al. 2003).  We show below that from the first approximately 
two-thirds of the sample comes from E+S0 types, whereas the second selects a 
mix in which E+S0+Sa's account for about two-thirds of the objects.

\begin{figure*}
 \centering
 \includegraphics[width=0.95\hsize]{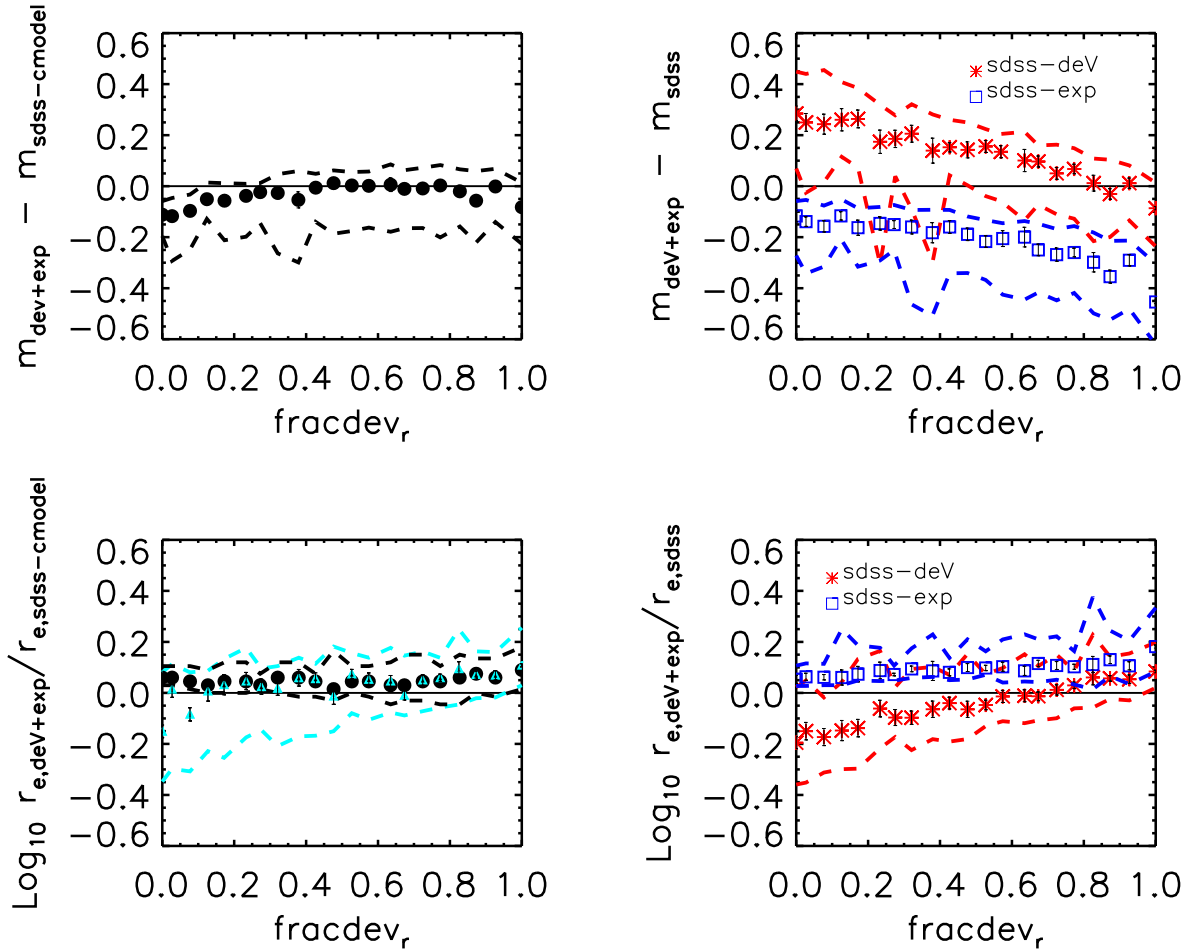}
 \caption{Comparison between apparent magnitudes and sizes obtained 
          from performing full bulge+disk decompositions (denoted 
          {\tt deV+exp}), with those output by or constructed from 
          parameters in the SDSS database.  In all cases, symbols 
          with error bars show the mean relation and the error on 
          the mean, and dashed lines show the actual rms scatter.  
          Filled circles in left panels show results for the SDSS 
          {\tt cmodel} magnitudes and effective radii (see text for details) 
          and right panels are for the SDSS {\tt deV} 
          (red stars; the effective radius is the value in the SDSS database 
          multiplied by $\sqrt{b/a}$, i.e. $\sqrt{ba}$) or 
          {\tt exp} (blue open squares; the effective radius is the value
          in the SDSS database $a$) quantities.  
          Cyan triangles in bottom left panel show the result of 
          picking either the {\tt deV} or {\tt exp} size, 
          based on which of the corresponding magnitudes were 
          closer to the {\tt cmodel} magnitude.  Although these 
          triangles are almost indistinguishable from the filled circles, 
          the rms scatter is substantially larger, particularly at 
          small {\tt fracdev}.  }
 \label{cmodel}
\end{figure*}

\begin{figure*}
 \centering
 \includegraphics[width=0.95\hsize]{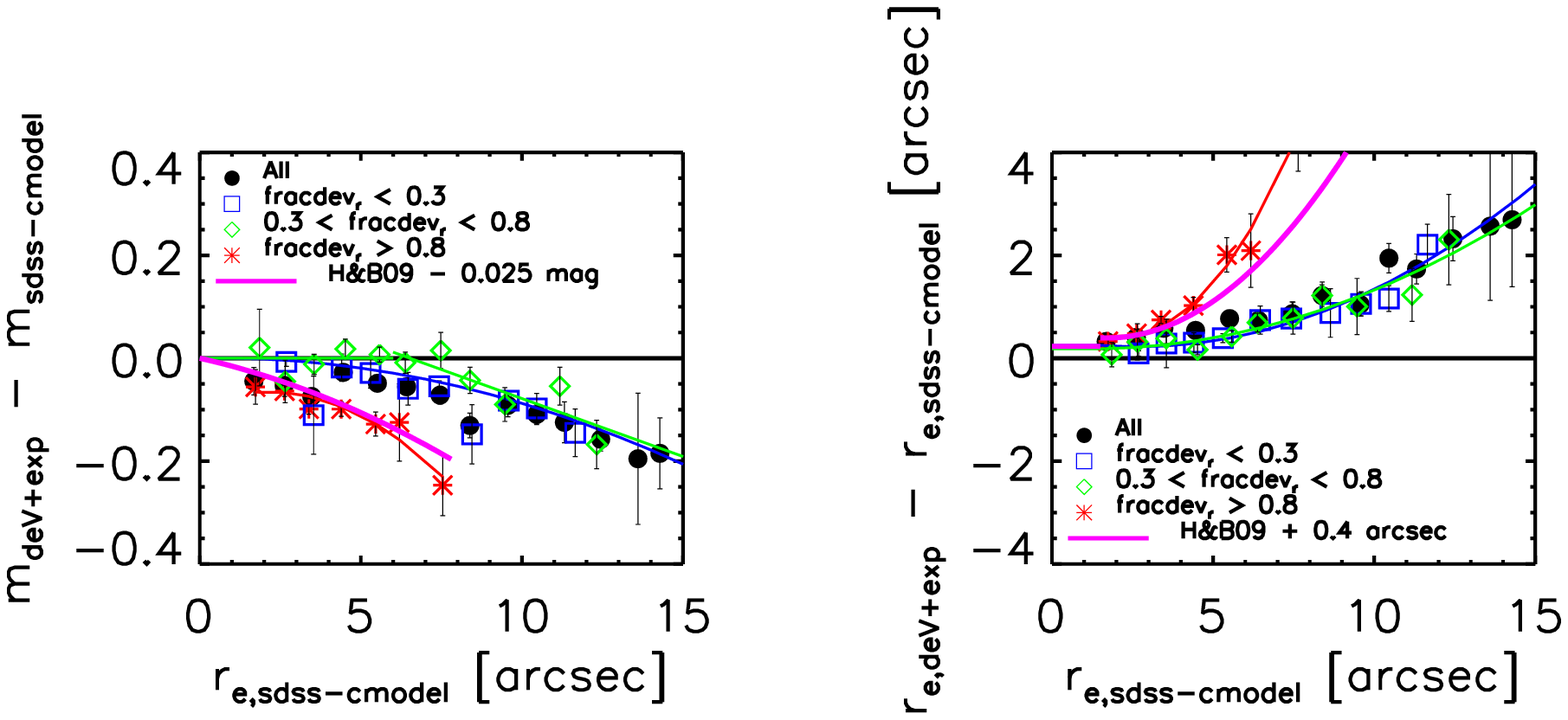}
 \includegraphics[width=0.95\hsize]{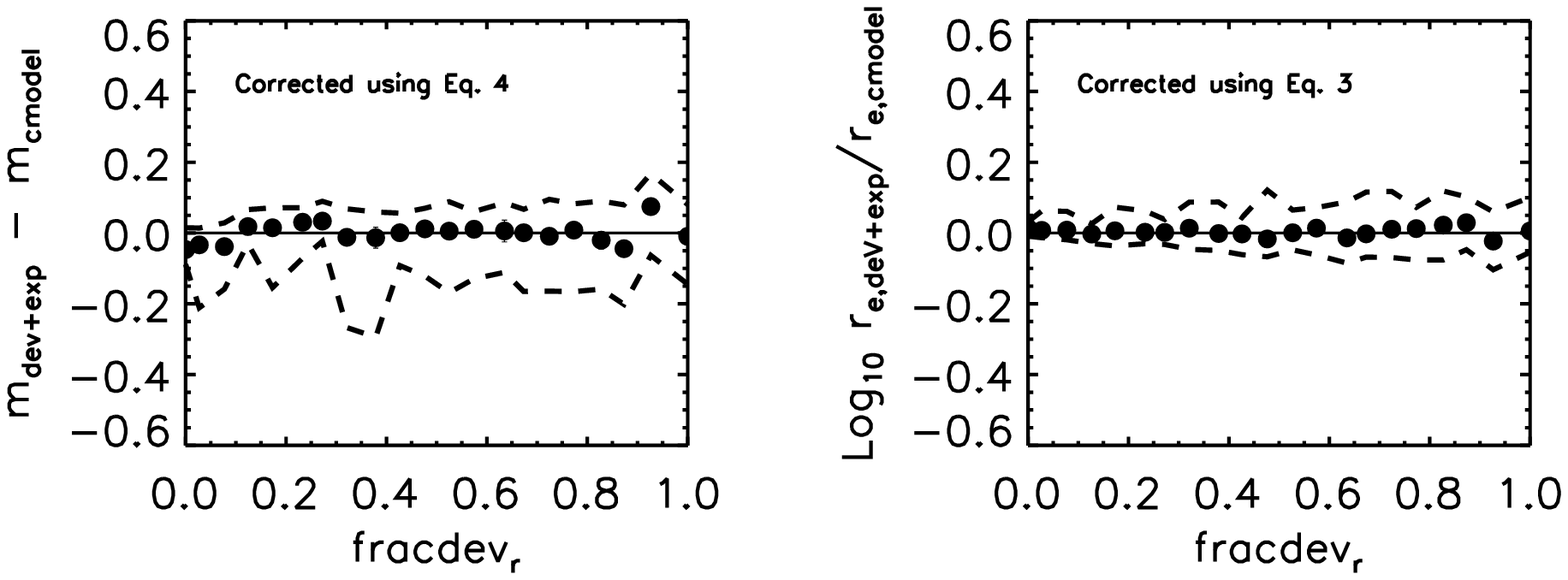}
 \caption{Top panels: Comparison of {\tt cmodel} magnitudes (right) and 
          sizes (left) with those obtained from performing full 
          bulge+disk decompositions as function of {\tt cmodel} sizes.
          The sample was divided in three bins based on the shape of the 
          light profile (as indicated in the panels).  Thin solid curves 
          (blue, green and  red) show fits to equations~(\ref{skysubr}) 
          and~(\ref{skysubm}).  
          Except for the sample with ${\tt fracDev}>0.8$, the coefficients 
          of these fits are given in Table~\ref{tabCorr}.  
          For ${\tt fracDev}>0.8$, the coefficients in Table~\ref{tabCorr} 
          are based on the larger sample of Hyde \& Bernardi (2009) (see
          text for the origin of the small offsets); this results in the 
          thick (magenta) solid curve shown.  
          Bottom panels: Similar to panels on left of Figure~\ref{cmodel}, 
          but with {\tt cmodel} magnitudes and sizes corrected following 
          equations~(\ref{skysubr}) and~(\ref{skysubm}).
          In all panels, symbols show the mean relation, error  
          bars show the error on the mean and dashed lines (bottom) 
          show the rms scatter.}
 \label{cmodel2}
\end{figure*}

The SDSS also outputs {\tt deV} or {\tt exp} magnitudes and sizes 
which result from fitting to a deVaucouleur or exponential profile, 
and {\tt fracDev}, a quantity which takes values between 0 and 1, 
which is a measure of how well the deVaucouleur profile actually 
fit the profile (1 being an excellent fit).  In addition, it 
outputs {\tt cmodel} magnitudes; this is a very crude disk+bulge 
magnitude which has been seeing-corrected.  Rather than resulting 
from the best-fitting linear combination of an exponential disk and a 
deVaucouleur bulge, the {\tt cmodel} magnitude comes from separately 
fitting exponential and deVaucouleur profiles to the image, and 
then combining these fits by finding that linear combination of them 
which best-fits the image.  Thus, if $m_{\rm exp}$ and $m_{\rm deV}$ 
are the magnitudes returned by fitting the two models, then 
\begin{eqnarray}
 10^{-0.4m_{\tt sdss-cmodel}} &=& 
  ({\tt 1} - {\tt fracdeV})\, 10^{-0.4m_{\tt Exp}} \nonumber\\
  &&\qquad         +\quad {\tt fracdeV} \, 10^{-0.4m_{\tt deV}}.
\end{eqnarray}

Later in this paper, we will be interested in seeing-corrected
half-light radii.  We use the {\tt cmodel} fits to define these
sizes by finding that scale $r_{e,{\tt cmodel}}$ where
\begin{eqnarray}
 \frac{10^{-0.4m_{\tt sdss-cmodel}}}{2} &=&
 ({\tt 1} - {\tt fracdeV})\,2\pi\int_0^{r_{{\tt sdss-cmodel}}}
   d\theta\, \theta\, I_{\rm exp}(\theta) \nonumber\\
 && +\ {\tt fracdeV}\,2\pi\int_0^{r_{{\tt sdss-cmodel}}}
        \!\! d\theta\, \theta\, I_{\rm deV}(\theta),
 \label{Rcmodel}
\end{eqnarray}
where $I$ is the surface brightness associated with the two fits.  
Note that the SDSS actually performs a two dimensional fit to the 
image, and it outputs the half-light radius of the long axis of 
the image {\tt a}, and the axis ratio {\tt b/a}.  
The expression above assumes one dimensional profiles, so we use 
the half-light radius {\tt a} of the exponential fit, and 
$\sqrt{\tt b/a}\times {\tt a} = \sqrt{\tt ba}$ for the deVaucouleur 
fit.  We describe some tests of these {\tt cmodel} quantities shortly.  

We would also like to study the velocity dispersions of these objects.
One of the important differences between the SDSS-DR6 and previous 
releases is that the low velocity dispersions 
($\sigma< 150$~km~s$^{-1}$) were biased high; this has been 
corrected in the DR6 release (see DR6 documentation, or discussion 
in Bernardi 2007).
The SDSS-DR6 only reports velocity dispersions if the $S/N$ 
in the spectrum in the restframe $4000-5700$~\AA\ is larger 
than 10 or with the {\tt status} flag equal to 4 (i.e. this 
tends to exclude galaxies with emission lines). 
To avoid introducing a bias from these cuts, we have also 
estimated velocity dispersions for all the remaining objects 
(see Hyde \& Bernardi 2009 for more discussion).
These velocity dispersions are based on spectra measured through 
a fiber of radius 1.5~arcsec; they are then corrected to $r_e/8$, as 
is standard practice. (This is a small correction.) 
The velocity dispersion estimate for emission line galaxies can be 
compromised by rotation.  In addition, the dispersion limit of the 
SDSS spectrograph is $69$~km~s$^{-1}$, so at small $\sigma$ the 
estimated velocity dispersion may both noisy and biased.  We will 
see later that this affects the velocity dispersion function.  
The size and velocity dispersion can be combined to estimate a 
dynamical mass; we do this by setting $M_{\rm dyn} = 5R_e\sigma^2/G$.

\subsection{A morphologically selected subsample}\label{morphs}
Recently, Fukugita et al. (2007) have provided morphological 
classifications (Hubble type T) for a subset of 2253 SDSS 
galaxies brighter than $m_{\tt Pet}=16$ in the $r-$band, 
selected from 230 deg$^2$ of sky.  
Of these, 1866 have spectroscopic information.
Since our goal is to compare these morphological 
selected subsamples with those selected based on relatively simple 
criteria (e.g. concentration index), we group galaxies classified
with half-integer T into the smaller adjoining integer bin 
(except for the E class; see also Huang \& Gu 2009 and 
Oohama et al. 2009).  
In the following, we set E (T = 0 and 0.5), S0 (T = 1), 
Sa (T = 1.5 and 2), Sb (T = 2.5 and 3), and 
Scd (T = 3.5, 4, 4.5, 5, and 5.5).
This gives a fractional morphological mix of 
(E, S0, Sa, Sb, Scd) = (0.269, 0.235, 0.177, 0.19, 0.098).  
Note that this is the mix in a magnitude limited catalog -- meaning 
that brighter galaxies (typically earlier-types) are over-represented.  

\subsection{{\tt cmodel} magnitudes and sizes}\label{cmodels}

As a check of our {\tt cmodel} sizes, we have performed deVaucouleurs 
bulge + exponential disk fits to light profiles of a subset of the 
objects; see Hyde \& Bernardi (2009) for a detailed description and 
tests of the fitting procedure.  
If we view these as the correct answer, then the top left panel 
of Figure~\ref{cmodel} shows that the {\tt cmodel} magnitudes are 
in good agreement with those from the full bulge+disk fit, except 
at {\tt fracDev}$\approx 0$ and {\tt fracDev}$\approx 1$ where 
{\tt cmodel} is fainter by 0.05~mags (top left).  This is 
precisely the regime where the agreement should have been best.
As discussed shortly, the discrepancy arises mainly because the SDSS 
reductions suffer from sky subtraction problems (see, e.g., SDSS DR7 
documentation), whereas our bulge-disk fits do not (see 
Hyde \& Bernardi 2009 for details).
Comparison with the top right panel shows that {\tt cmodel} is a 
significant improvement on either the {\tt deV} or the {\tt exp} 
magnitudes alone.  

\begin{table}
\caption[]{Coefficients used in equations~(\ref{skysubr}) and~(\ref{skysubm}) to correct sizes and magnitudes for sky subtraction problems.\\}
\begin{tabular}{lccc}
 \hline 
  Sample & $Cr_0$ & $Cr_1$ & $Cr_2$\\ 
 \hline
  {\tt fracDev} $> 0.8$ & & & \\
   \qquad \& $r_{e,{\tt sdss-cmodel}} > 1.5$~arcsec & 0.582 & $-0.221$ & 0.065 \\
   \qquad \& $r_{e,{\tt sdss-cmodel}} < 1.5$~arcsec & 0.249  & 0 & 0 \\
  $0.3 <$ {\tt fracDev} $< 0.8$ & & & \\
   \qquad \& $r_{e,{\tt sdss-cmodel}} > 1.5$~arcsec & 0.201 & $-0.034$ & 0.015 \\
   \qquad \& $r_{e,{\tt sdss-cmodel}} < 1.5$~arcsec & 0.182 & 0 & 0 \\
  {\tt fracDev} $< 0.3$ & & & \\
   \qquad \& $r_{e,{\tt sdss-cmodel}} > 1.5$~arcsec & 0.368  & $-0.110$ & 0.021 \\
   \qquad \& $r_{e,{\tt sdss-cmodel}} < 1.5$~arcsec & 0.231  & 0 & 0 \\
 \hline
  Sample & $Cm_0$ & $Cm_1$ & $Cm_2$\\ 
 \hline
  {\tt fracDev} $> 0.8$ & 0 & $-0.014$ & $-0.001$ \\
  $0.3 <$ {\tt fracDev} $< 0.8$  &   &  &  \\
   \qquad \& $r_{e,{\tt sdss-cmodel}} > 6$~arcsec &  0.147 & $-0.023$ & 0 \\
   \qquad \& $r_{e,{\tt sdss-cmodel}} < 6$~arcsec & 0 & 0 & 0 \\
  {\tt fracDev} $< 0.3$ & 0 & 0.001 & $-0.001 $\\
 \hline 
\end{tabular}
\label{tabCorr} 
\end{table}

The bottom panels show a similar comparison of the sizes.  
At intermediate values of {\tt fracDev}, neither the pure 
deVaucouleur nor the pure exponential fits are a good description 
of the light profile, so the sizes are also biased (bottom right).  
However, at {\tt fracDev}=1, where the deVaucouleurs model should 
be a good fit, the {\tt deV} sizes returned by the SDSS are about 
0.07~dex smaller than those from the bulge+disk decomposition.
There is a similar discrepancy of about 0.05~dex with the SDSS 
Exponential sizes at {\tt fracDev}=0.  We argue below that these 
offsets are related to those in the magnitudes, and are primarily 
due to sky subtraction problems.  

The filled circles in the bottom left panel show that our 
{\tt cmodel} sizes (from equation~\ref{Rcmodel}) are in 
substantially better agreement with those from the bulge+disk 
decomposition over the entire range of {\tt fracDev}, with a typical 
scatter of about 0.05~dex (inner set of dashed lines).  
For comparison, the triangles show the result of picking 
either the deVaucouleur or exponential size, based on which of 
these magnitudes were closer to the {\tt cmodel} magnitude (this 
is essentially the scale that the SDSS uses to compute {\tt model} 
colors).  Note that while this too removes most of the bias 
(except at small {\tt fracDev}), it is a substantially noisier 
estimate of the true size (outer set of dashed lines).  
This suggests that our {\tt cmodel} sizes, which are seeing 
corrected, represent a significant improvement on what has been 
used in the past. 

The SDSS reductions are known to suffer from sky subtraction problems 
which are most dramatic for large objects or objects in crowded 
fields (see DR7 documentation).  
The top panels in Figure~\ref{cmodel2} show this explicitly:  
while there is little effect at small size, the SDSS underestimates 
the brightnesses and sizes when the half-light radius is larger than 
about 5~arcsec.  Note that this is actually a small fraction of the 
objects:  6\% of the objects have {\tt sdss-cmodel} sizes larger than 
5~arcsec; 13\% are larger than 4~arcsec.  Whereas previous work has 
concentrated on mean trends for the full sample, Figure~\ref{cmodel2} 
shows that, in fact, the difference depends on the type of light 
profile -- galaxies with {\tt fracDev}~$> 0.8$ (i.e. close to 
deVaucouleur profiles) are more sensitive to sky-subtraction problems 
than later-type galaxies.  
Some of this is due to the fact that such galaxies tend to populate 
more crowded fields.  

To correct for this effect, we have fit low order polynomials to the 
trends; the solid curves in the top panels of Figure~\ref{cmodel2} show 
these fits.  Except for the sample with {\tt fracDev}~$> 0.8$, we use 
these fits to define our final corrected {\tt cmodel} sizes by:
\begin{eqnarray}
  r_{e,{\tt cmodel}} &=& r_{e,{\tt sdss-cmodel}}
                + Cr_0 + Cr_1\, r_{e,{\tt sdss-cmodel}}  \nonumber \\
           && \qquad\qquad\quad +\ Cr_2\, r_{e,{\tt sdss- cmodel}}^2 
 \label{skysubr}
\end{eqnarray}  
and 
\begin{eqnarray}
  m_{\tt cmodel} &=& m_{\tt sdss-cmodel} 
                         + Cm_0 + Cm_1\, r_{e,{\tt sdss-cmodel}} \nonumber \\
           && \qquad\quad  +\ Cm_2\, r_{e,{\tt sdss-cmodel}}^2,
\label{skysubm}
\end{eqnarray}  
where the coefficients $Cm_0$, $Cm_1$, $Cm_2$, $Cr_0$, $Cr_1$ and 
$Cr_2$ are reported in Table~\ref{tabCorr}.  

For objects with {\tt fracDev}~$>0.8$, the trends we see are similar 
to those shown in Fig.~5 of Hyde \& Bernardi (2009), which were based 
on a (larger) sample of about $6000$ early-type galaxies.  The thick 
solid (magenta) line in the top panels of Figure~\ref{cmodel2} show the 
Hyde-Bernardi trends, with a small offset to account for the fact that  
they did not integrate the fitted profiles to infinity (because the 
SDSS, to which they were comparing, does not), whereas we do.  The 
thick solid curve differs from the thin one at sizes larger than about 
5~arcsec.  Since the thick curve is based on a larger sample, we use it 
to define our final corrected {\tt cmodel} sizes.  The corrections are 
again described by equations~(\ref{skysubr}) and~(\ref{skysubm}), with 
coefficients that are reported in Table~\ref{tabCorr}.  

The bottom panels of Figure~\ref{cmodel2} show that these corrected 
quantities agree quite well with those from the full bulge+disk fit, 
even at small and large {\tt fracDev}.

\subsection{Stellar Masses}\label{mstars}
Stellar masses $M_*$ are typically obtained by estimating $M_*/L$ 
(in solar units), and then multiplying by the restframe $L$ (which is 
not evolution corrected).  In the following we compare three different 
estimates of $M_*$.  All these estimates depend on the assumed IMF:  
Table~\ref{tabIMF} shows how we transform between different IMFs.  

The first comes from Bell et al. (2003), who report that, 
at $z=0$, 
 $\log_{10}(M_*/L_r)_0 = 1.097\,(g - r)_0 + zp$, where the zero-point $zp$ 
depends on the IMF (see their Appendix~2 and Table~7).  Their standard 
diet-Salpeter IMF has $zp=-0.306$, which they state has 70\% smaller 
$M_*/L_r$ at a given color than a Salpeter IMF.  In turn, a 
Salpeter IMF has 0.25~dex more $M_*/L_r$ at a given color than the 
Chabrier (2003) IMF used by the other two groups whose mass estimates 
we use (see Table~\ref{tabIMF} for conversion of different IMFs).  
Therefore, we set $zp=-0.306 + 0.15 - 0.25 = -0.406$, making 
\begin{equation}
 \log_{10}(M_{*{\rm Bell}}/L_r)_0 = 1.097\,(g - r)_0 - 0.406.
 \label{MbellKroupa}
\end{equation} 
We then obtain $M_{*{\rm Bell}}$ by multiplying by the SDSS 
$r-$band luminosity.  When comparing with previous work, we 
usually use {\tt Petrosian} magnitudes, although our final 
results are based on the {\tt cmodel} magnitudes which we 
believe are superior.  

Note that this expression requires 
luminosities and colors that have been $k$- and evolution-corrected 
to $z=0$ (E. Bell, private communication).  Unfortunately, these 
corrections are not available on an object-by-object basis.  
Bell et al. (2003) report that the absolute magnitudes brighten as 
$1.3z$ and $g-r$ color becomes bluer as $0.3z$.  Although these 
estimates differ slightly from independent measurements of evolution 
by Bernardi et al. (2003) and Blanton et al. (2003), and more 
significantly from more recent determinations (Roche et al. 2009a), 
we use them, because they are the ones from which 
equation~(\ref{MbellKroupa}) was derived.  Thus, in terms of 
restframe quantities, 
\begin{equation}
 \log_{10}\left(\frac{M_{*{\rm Bell}}}{M_\odot}\right) 
  = 1.097\,(g - r) - 0.406 - 0.4 (M_r - 4.67) - 0.19z.
 \label{gmrBell}
\end{equation} 
If we use the restframe $r-i$ color and $L_r$ luminosity instead, then 
\begin{equation}
 \log_{10}\left(\frac{M_{*{\rm Bell}}}{M_\odot}\right) 
  = 1.431\,(r - i) - 0.122 - 0.4 (M_r - 4.67) - 0.23z.
 \label{rmiBell}
\end{equation} 

\begin{table}
\caption[]{$M_*/L$ IMF offsets used in this work. 
  Offset is with respect to the Salpeter (1955) IMF: 
  log$_{10} M_*/L$ (IMF Salpeter) = log$_{10} M_*/L$ (IMF Reference) + Offset\\}
\begin{tabular}{lcl}
 \hline 
  IMF & Offset [dex] & Refecence \\
 \hline
  Kennicut & 0.30 & Kennicut (1983)\\
  Scalo & 0.25 & Scalo (1986) \\
  diet-Salpeter & 0.15 & Bell \& de Jong (2001)\\
  pseudo-Kroupa & 0.20 & Kroupa (2001) \\
  Kroupa & 0.30 & Kroupa (2002)\\
  Chabrier & 0.25 & Chabrier (2003)\\
  Baldry \& Glazbrook & 0.305 & Baldry \& Glazbrook (2003)\\
 \hline
\end{tabular}
\label{tabIMF} 
\end{table}

These two estimates of $M_*$ will differ because there is scatter 
in the $(g-r)-(r-i)$ color plane.  Unfortunately, Bell et al. do 
not provide a prescription which combines different colors.  
Although we could perform a straight average of these two estimates, 
this is less than ideal because the value of $r-i$ at fixed $g-r$ 
may provide additional information about $M_*/L$.  In practice, 
we will use the $g-r$ estimate as our standard, and $r-i$ to 
illustrate and quantify intrinsic uncertainties with the current 
approach.  



The second estimate of $M_*$ is from Gallazzi et al. (2005).  
This is based on a likelihood analysis of the spectra, 
assumes the Chabrier (2003) IMF, and returns $M_*/L_z$.  
The stellar mass $M_{*{\rm Gallazzi}}$ is then computed using SDSS 
{\tt petrosian} $z$-band restframe magnitudes (i.e., they were 
$k$-corrected, but no evolution correction was applied).  In this 
respect, they differ in philosophy from $M_{*,{\rm Bell}}$, in that 
the $M_*$ estimate is not corrected to $z=0$.  
In practice, since we are mainly interested in small lookback times 
from $z=0$, for which the expected mass loss to the IGM is almost 
negligible, this almost certainly makes little difference for the 
most massive galaxies.  
These estimates are only available for $205,510$ of the 
objects in our sample ($\sim 82\%$).  The objects for which 
Gallazzi et al. do not provide stellar mass estimates are lower 
luminosity, typically lower mass objects; we show this explicitly 
in Figure~\ref{MsFpet}. 

A final estimate of $M_*/L_r$ comes from Blanton \& Roweis (2007), 
and is based on fitting the observed colors in all the SDSS bands 
to templates of a variety of star formation histories and 
metallicities, assuming the Chabrier (2003) IMF. 
In the following we use the Blanton \& Roweis stellar masses 
computed by applying the SDSS {\tt petrosian} and {\tt model} 
(restframe) $r$-magnitudes to these mass-to-light estimates:  
$M_{*{\rm Petro}}$ and $M_{*{\rm Model}}$.  
Blanton \& Roweis also provide mass estimates from a template 
which is designed to match luminous red galaxies; we call these 
$M_{*{\rm LRG}}$.  Note that in this case $M_*/L$ is converted 
to $M_*$ using (restframe) {\tt model} magnitudes only, since 
the Petrosian magnitude is well-known to underestimate (by about 0.1~mags) 
the magnitudes of LRG-like objects (Blanton et al. 2001, DR7 documentation).  
To appreciate how different the LRG template is from the others, note 
that it allows ages of upto 10~Gyrs, whereas that for the others, the 
age is more like 7~Gyrs.  

A detailed comparison of the different mass estimates is presented
in Appendix~\ref{Mscomp}.  This shows that to use the 
Blanton \& Roweis (2007) masses, one {\em must} devise an algorithm 
for choosing between $M_{*{\rm Model}}$ and $M_{*{\rm LRG}}$.  
In principle, some of the results to follow allow one to do this, 
but exploring this further is beyond the scope of the present paper.  
On the other hand, to use the Gallazzi et al. (2005) estimates, one 
must be wary of aperture effects.  Finally, stellar masses based on 
the $k+e$ corrected $r-i$ color show stronger systematics than do 
the $g-r$ based estimates of $M_*$.  
Therefore, in what follows, our prefered mass estimate will be 
that based on $k+e$ corrected {\tt cmodel} $r-$magnitudes and 
$g-r$ colors (i.e., equation~\ref{gmrBell} with restframe and 
evolution corrected magnitudes 
and colors).  Note that we believe the {\tt cmodel} magnitudes to 
be far superior to the {\tt Petrosian} ones.  Of course, when 
we compare our results with previous work which used Petrosian 
magnitudes (Section~\ref{phiMass}), we do so too.


\begin{figure*}
 \centering
 \includegraphics[width=0.45\hsize]{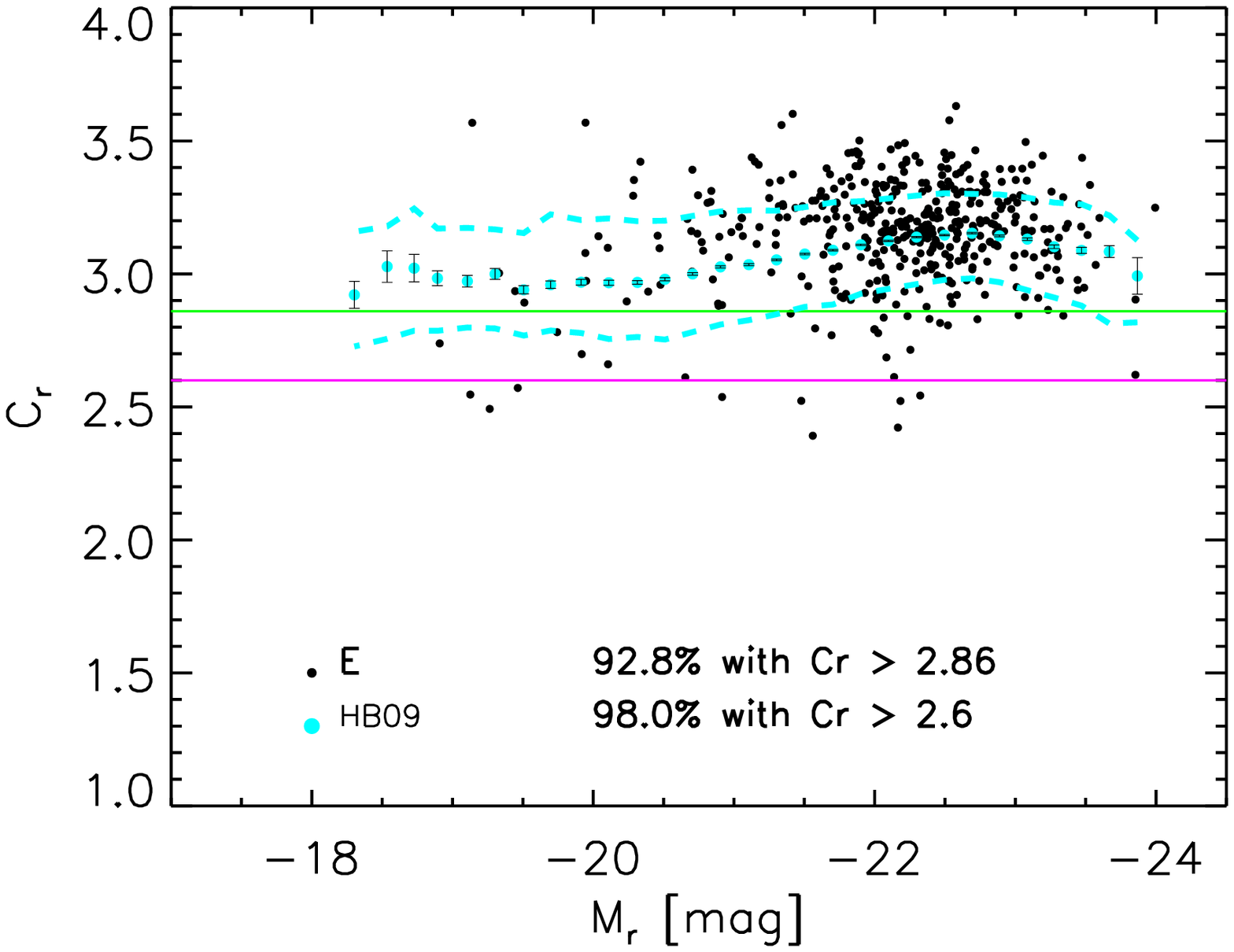}
 \includegraphics[width=0.45\hsize]{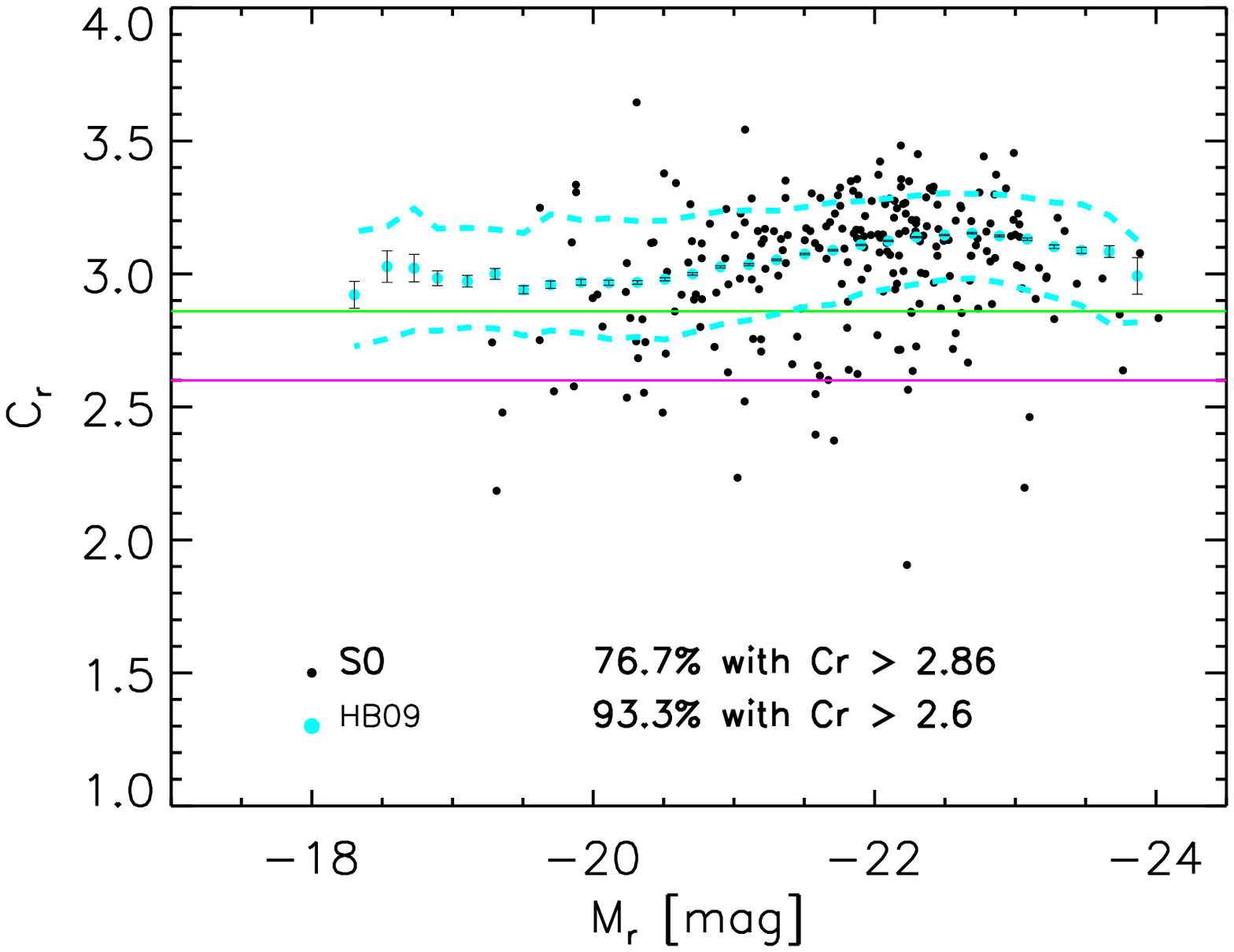}
 \includegraphics[width=0.45\hsize]{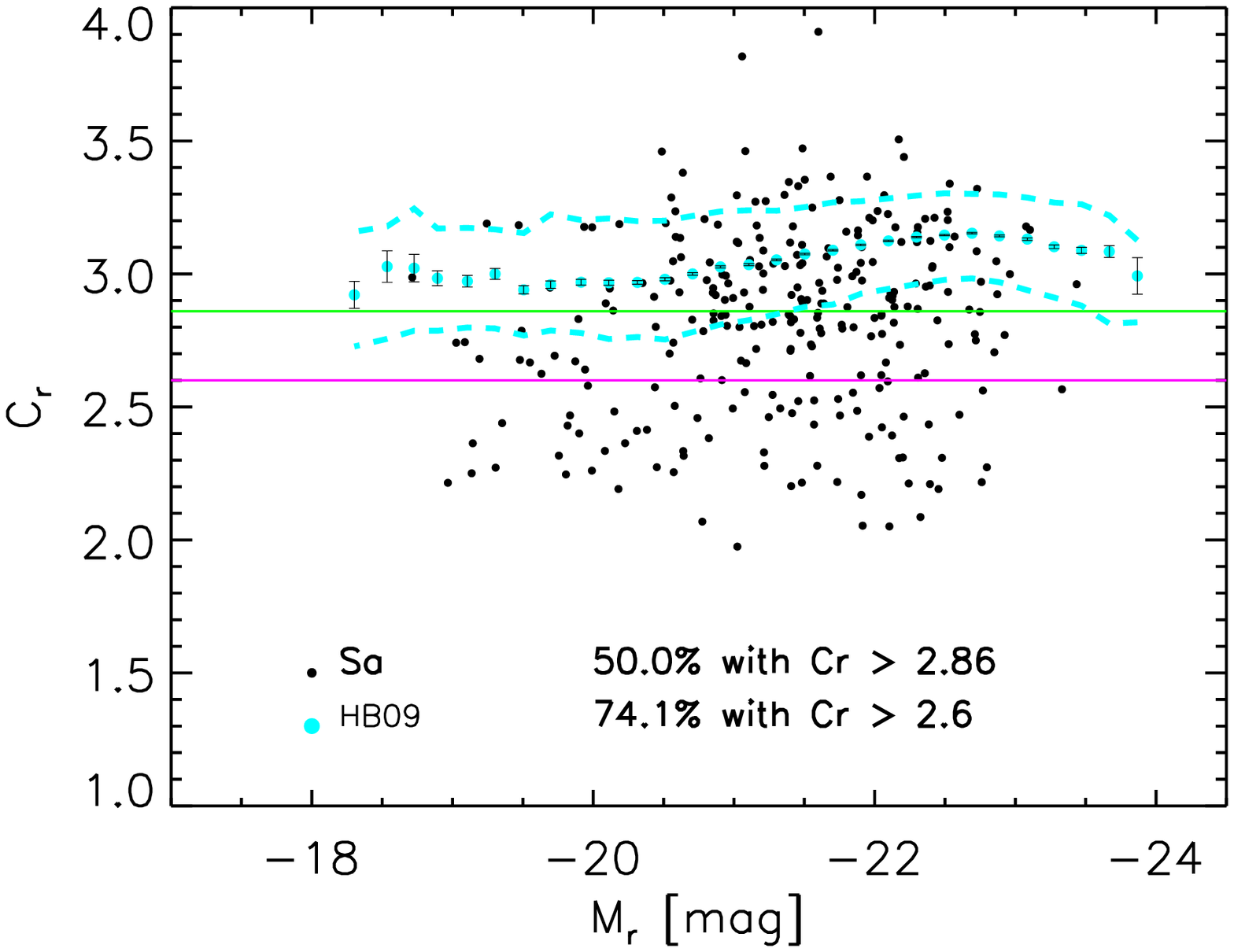}
 \includegraphics[width=0.45\hsize]{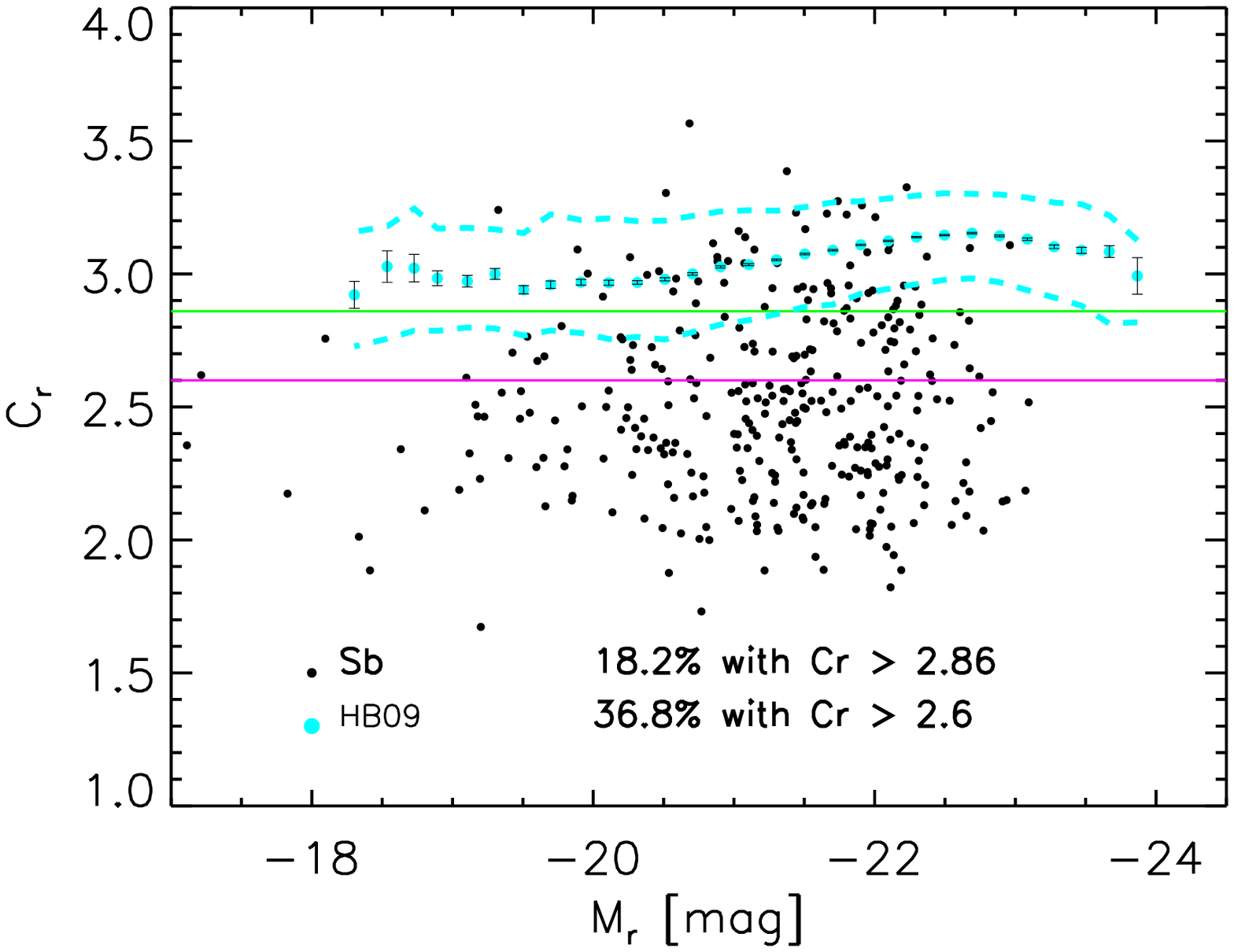}
 \includegraphics[width=0.45\hsize]{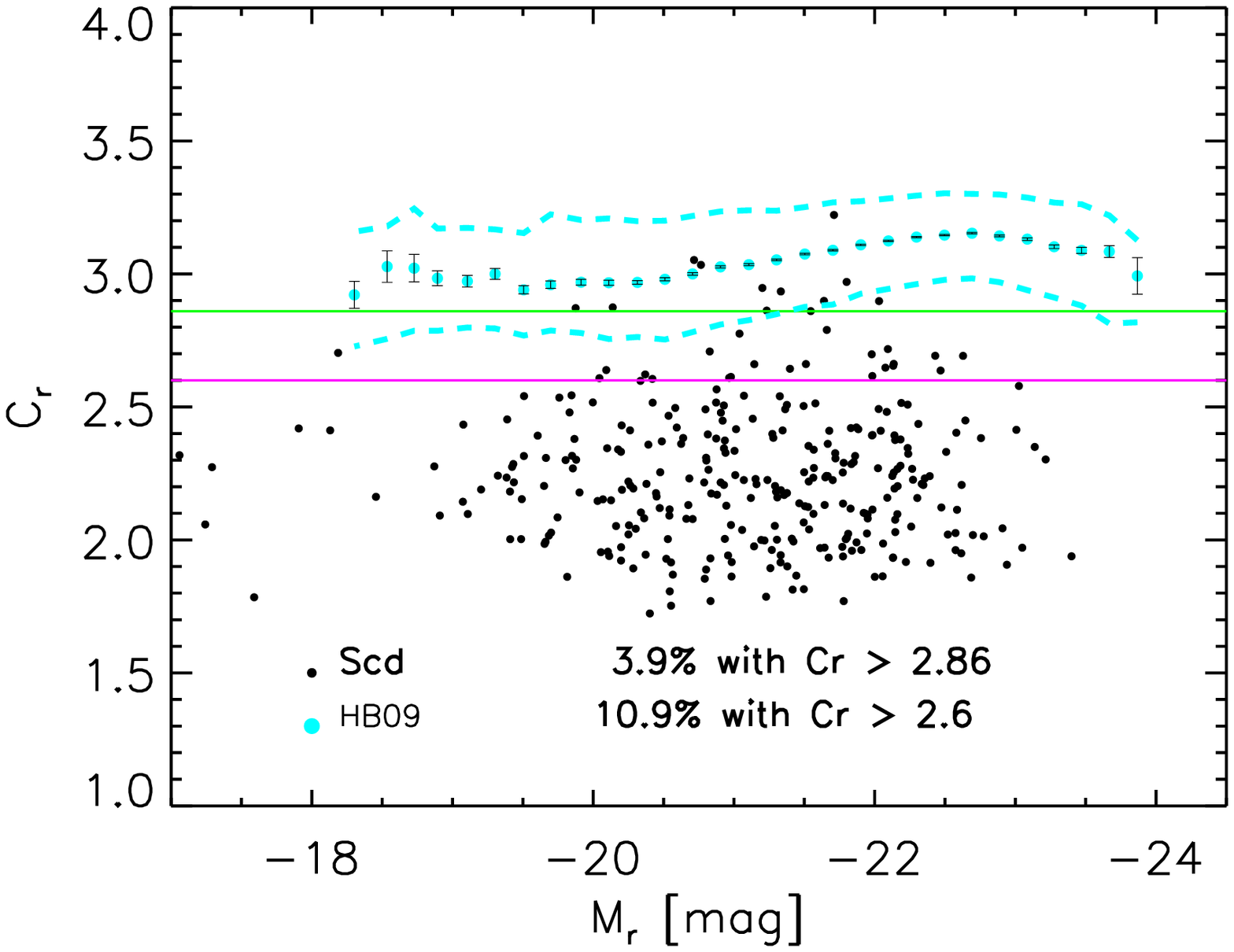}
 \includegraphics[width=0.45\hsize]{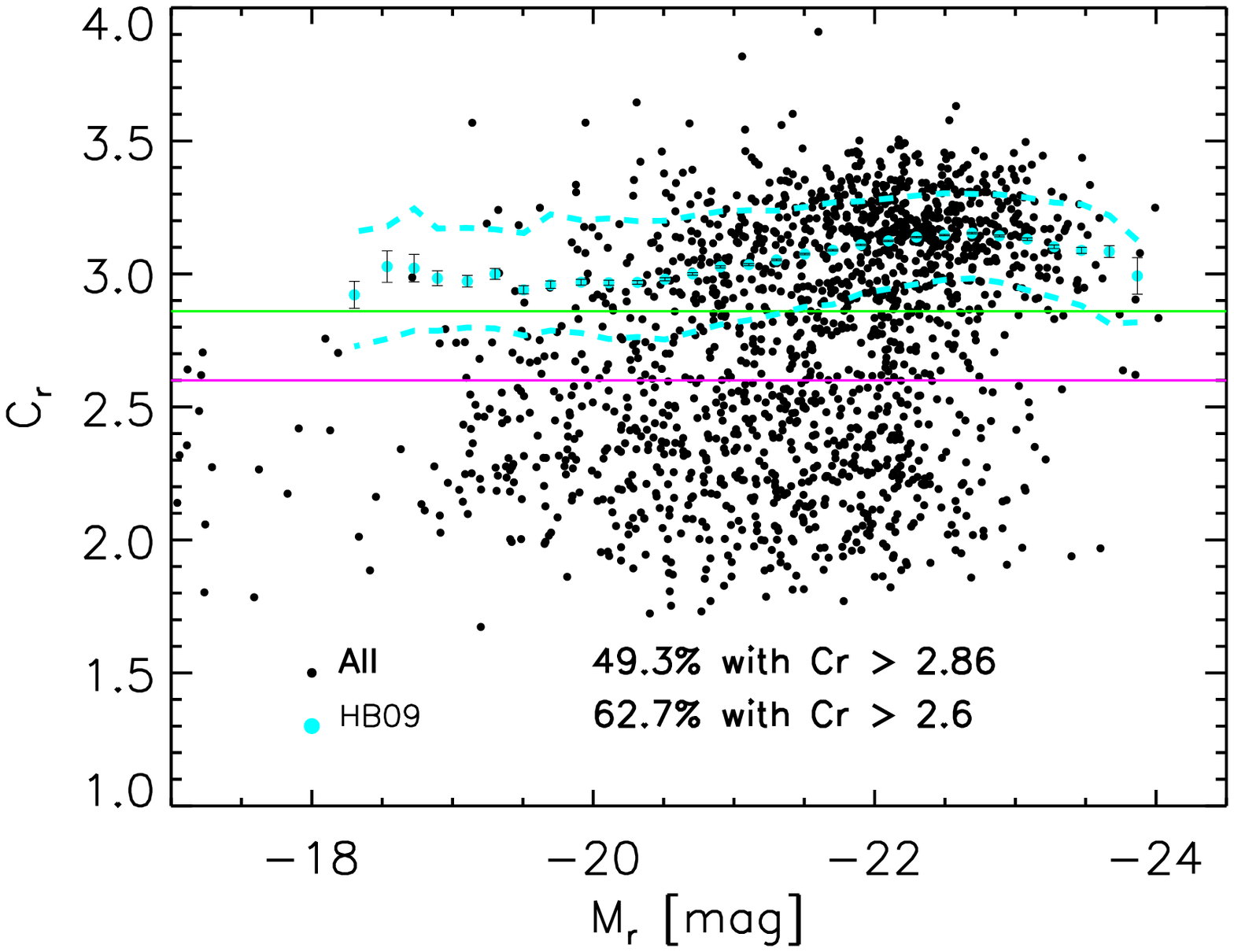}
 \caption{Distribution of morphological types from Fukugita et al. (2007)
          in the space of $C_r$ vs luminosity.  Bottom right panel shows 
          the full sample.  Horizontal lines (same in all panels) show two  
          popular cuts for selecting early-types.  Text in each panel 
          shows the fraction of objects in the panel which lie above 
          these lines.  Cyan symbols with error bars and flanked by dashed 
          curves (same in each panel) show the median relation and the rms 
          scatter defined by a sample selected following Hyde \& Bernardi (2009).  }
 \label{cmorph}
\end{figure*}

\begin{figure*}
 \centering
 \includegraphics[width=0.45\hsize]{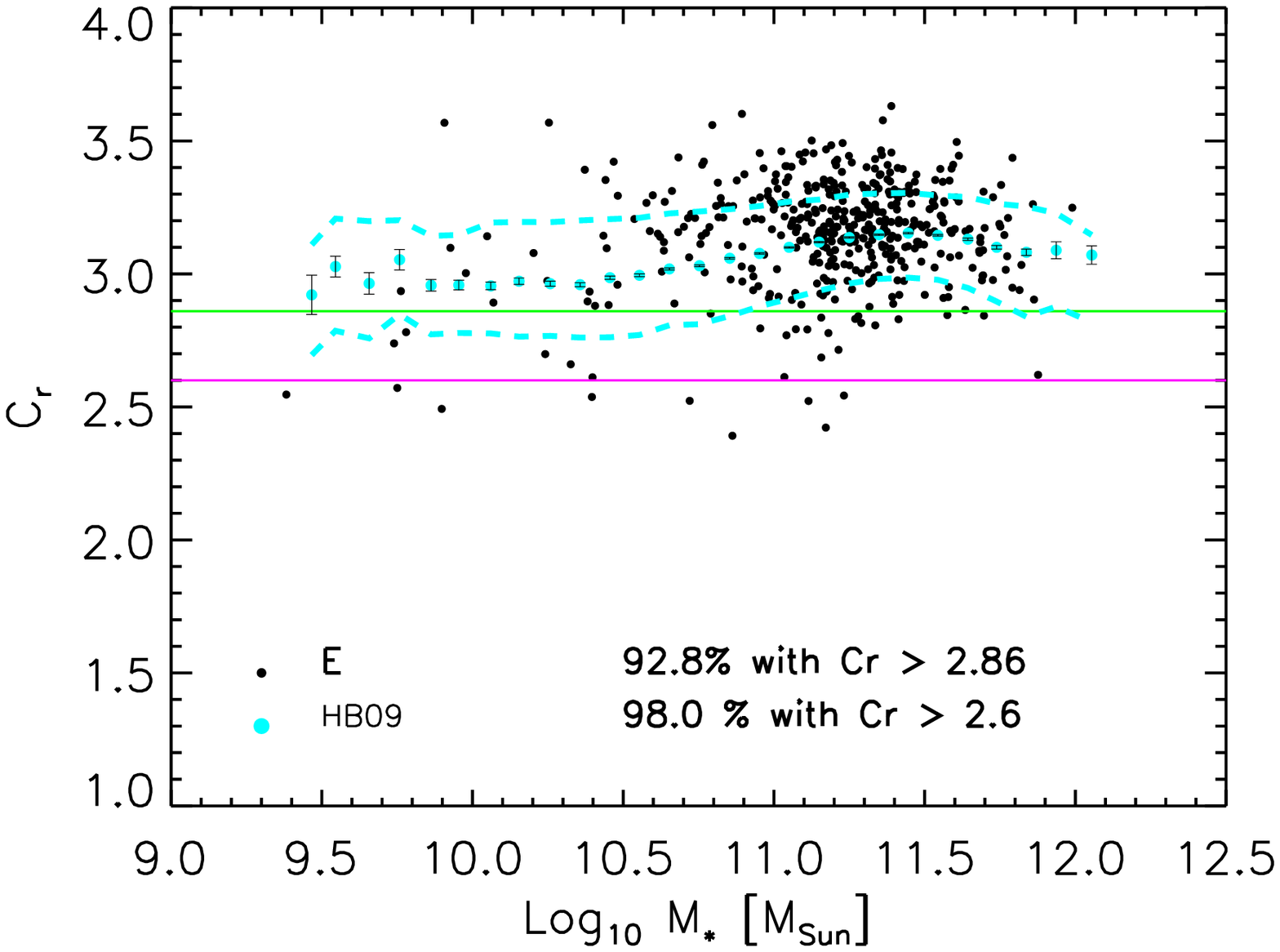}
 \includegraphics[width=0.45\hsize]{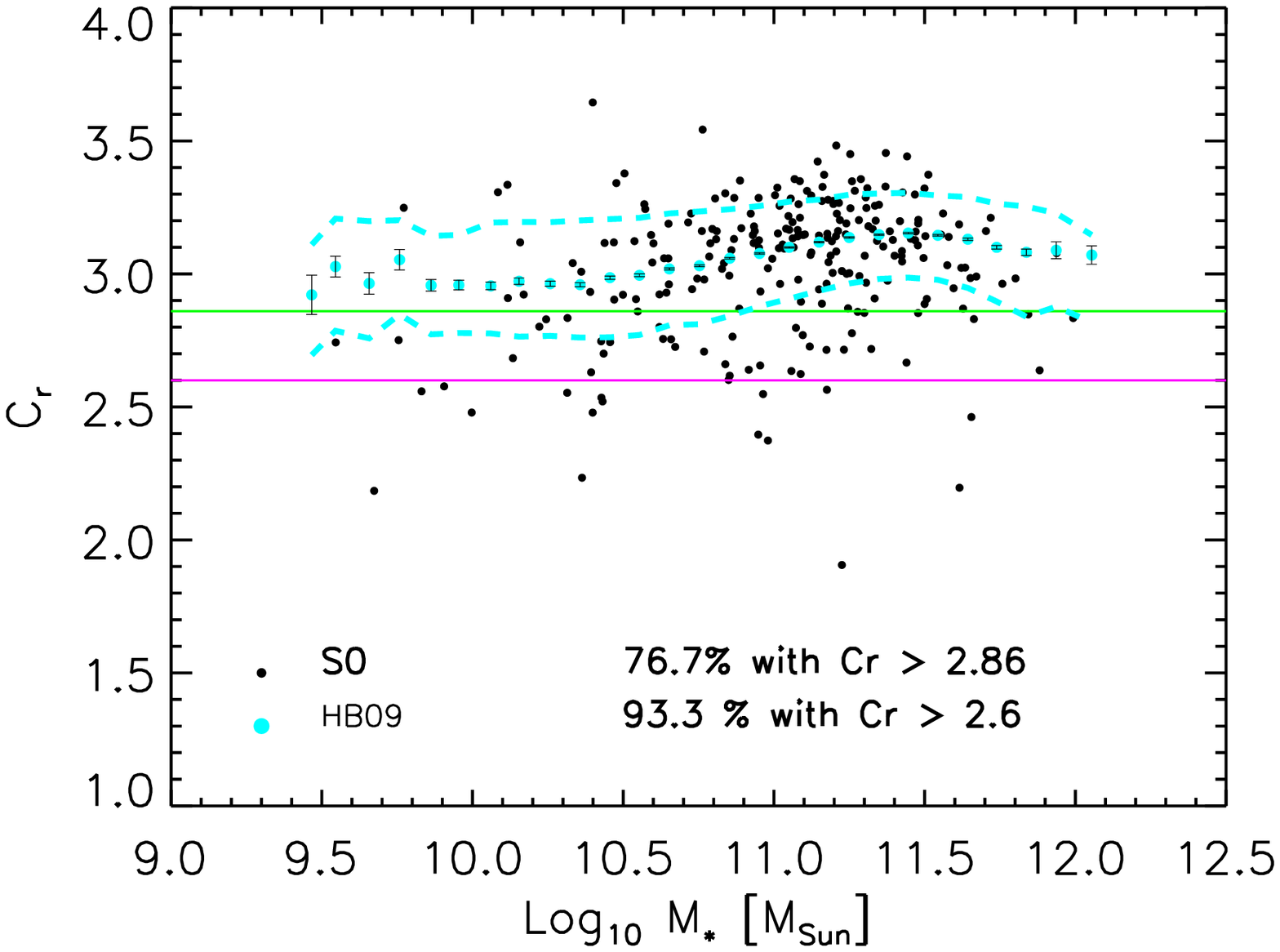}
 \includegraphics[width=0.45\hsize]{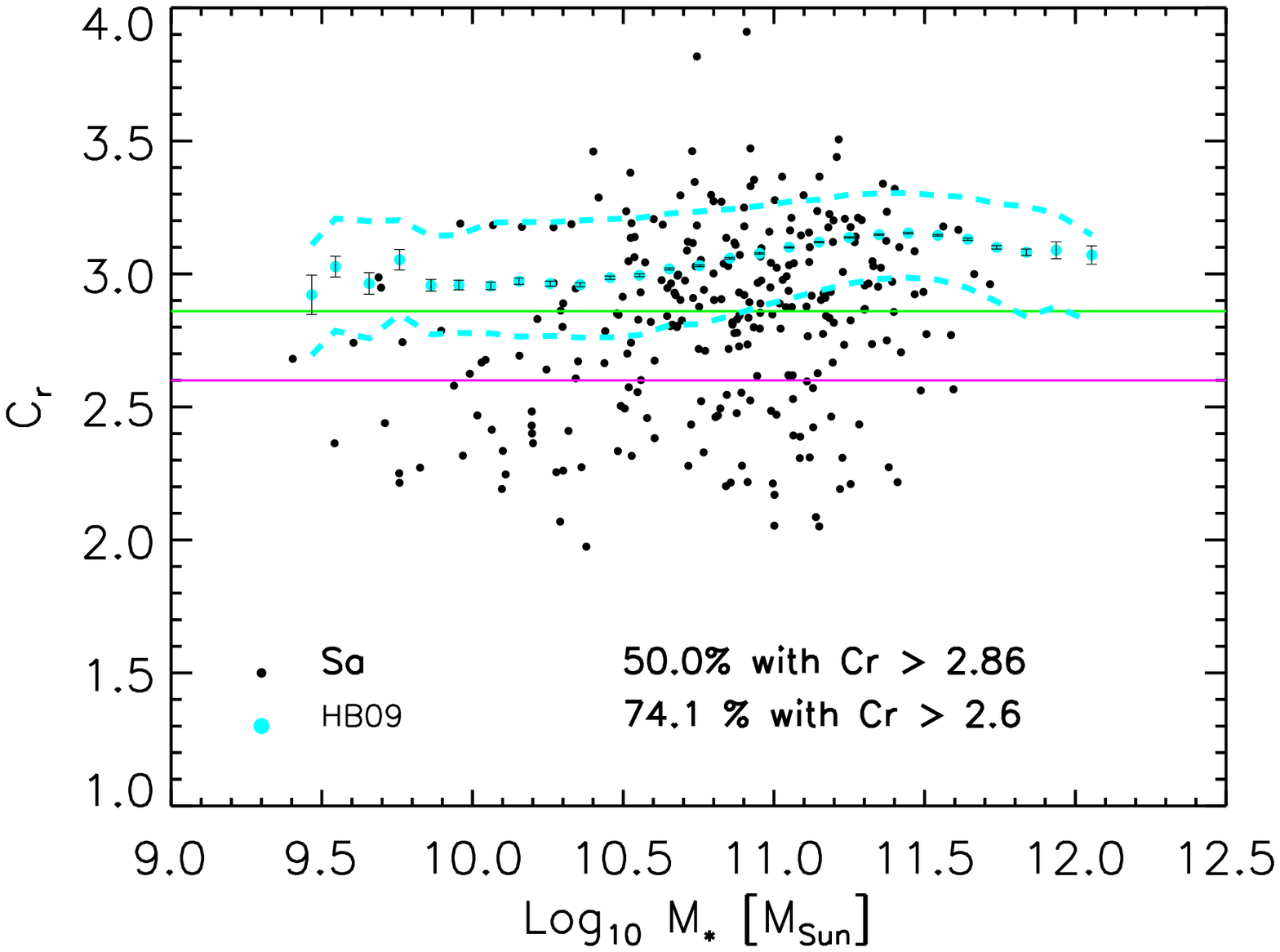}
 \includegraphics[width=0.45\hsize]{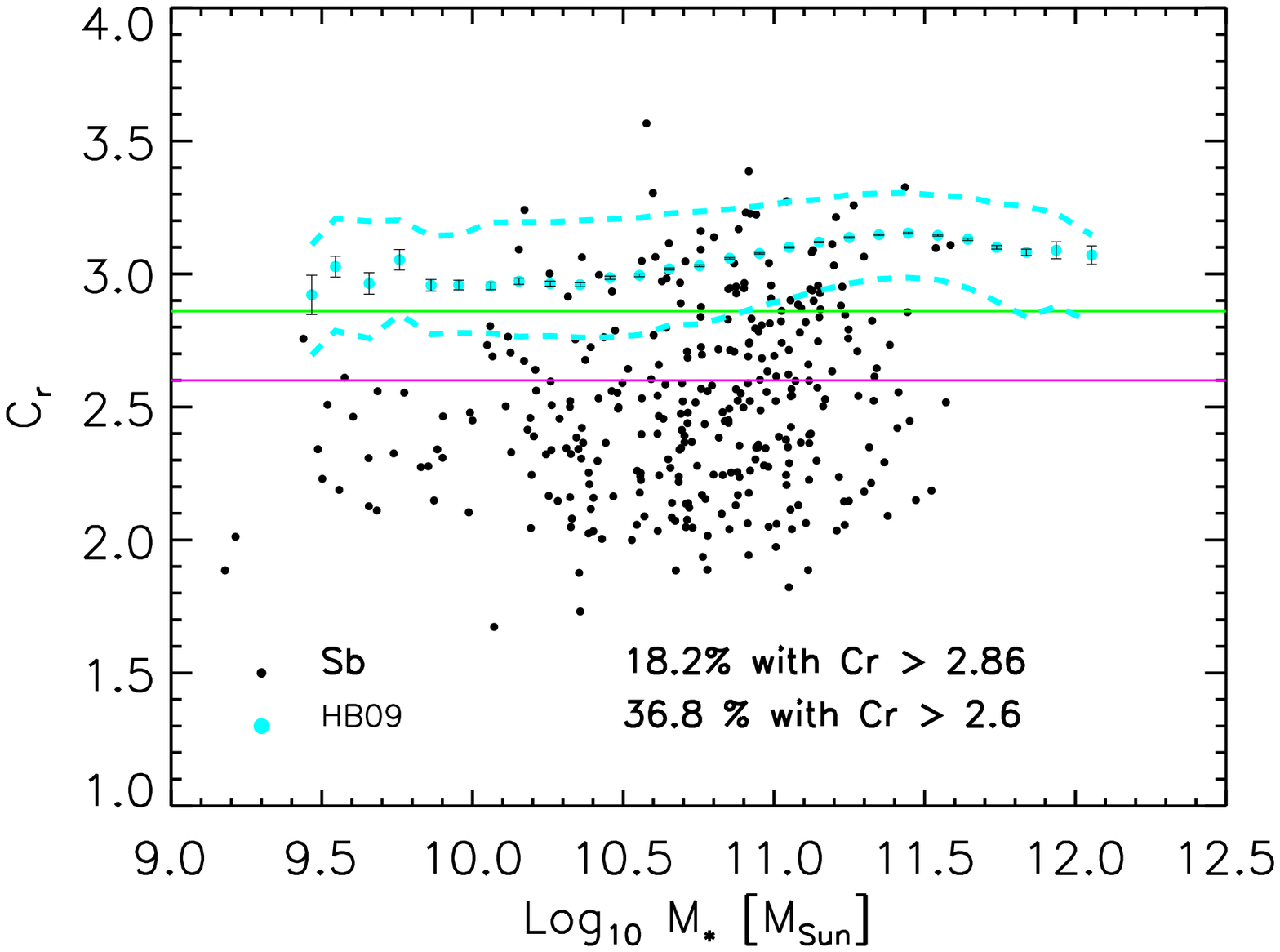}
 \includegraphics[width=0.45\hsize]{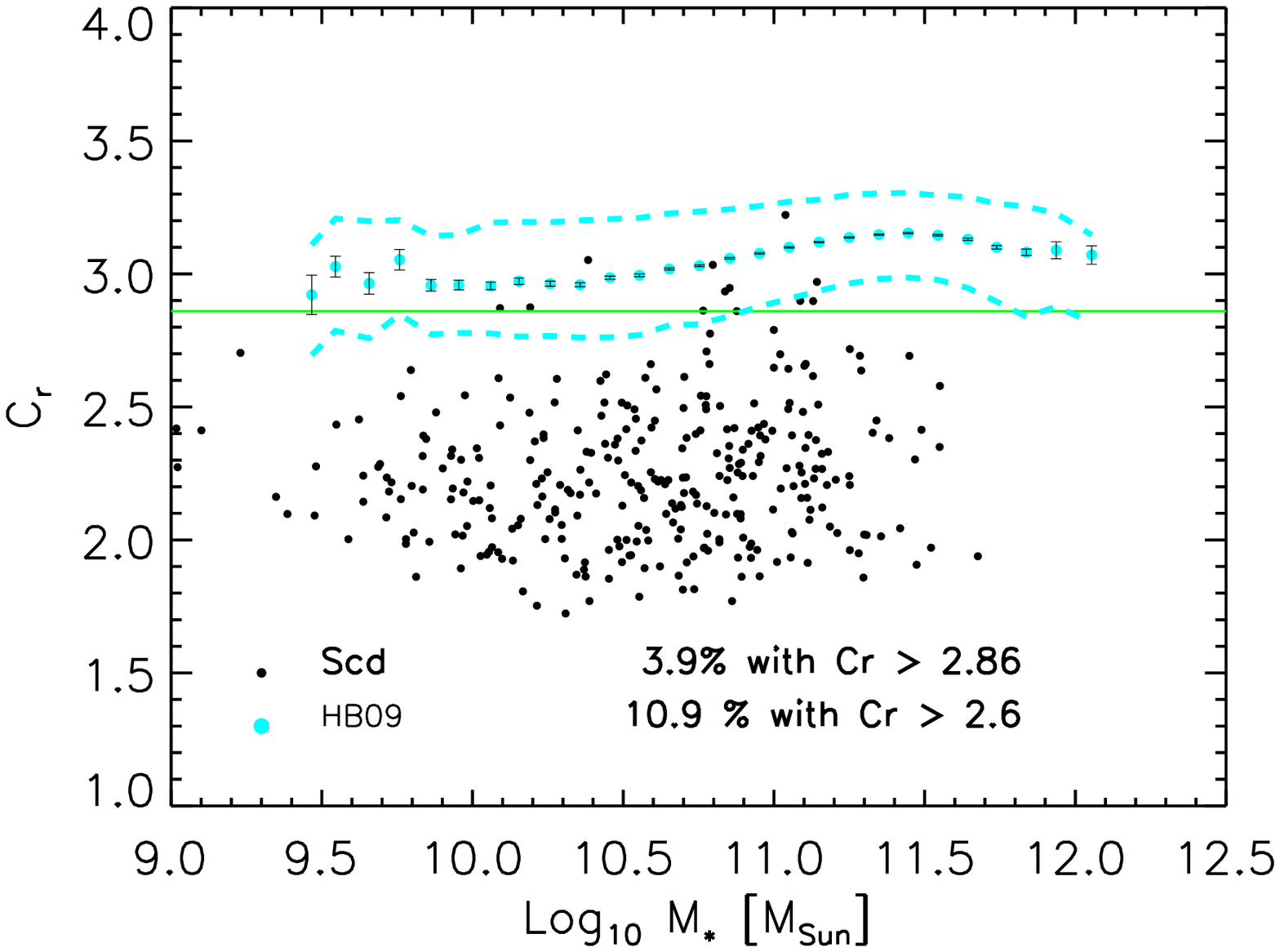}
 \includegraphics[width=0.45\hsize]{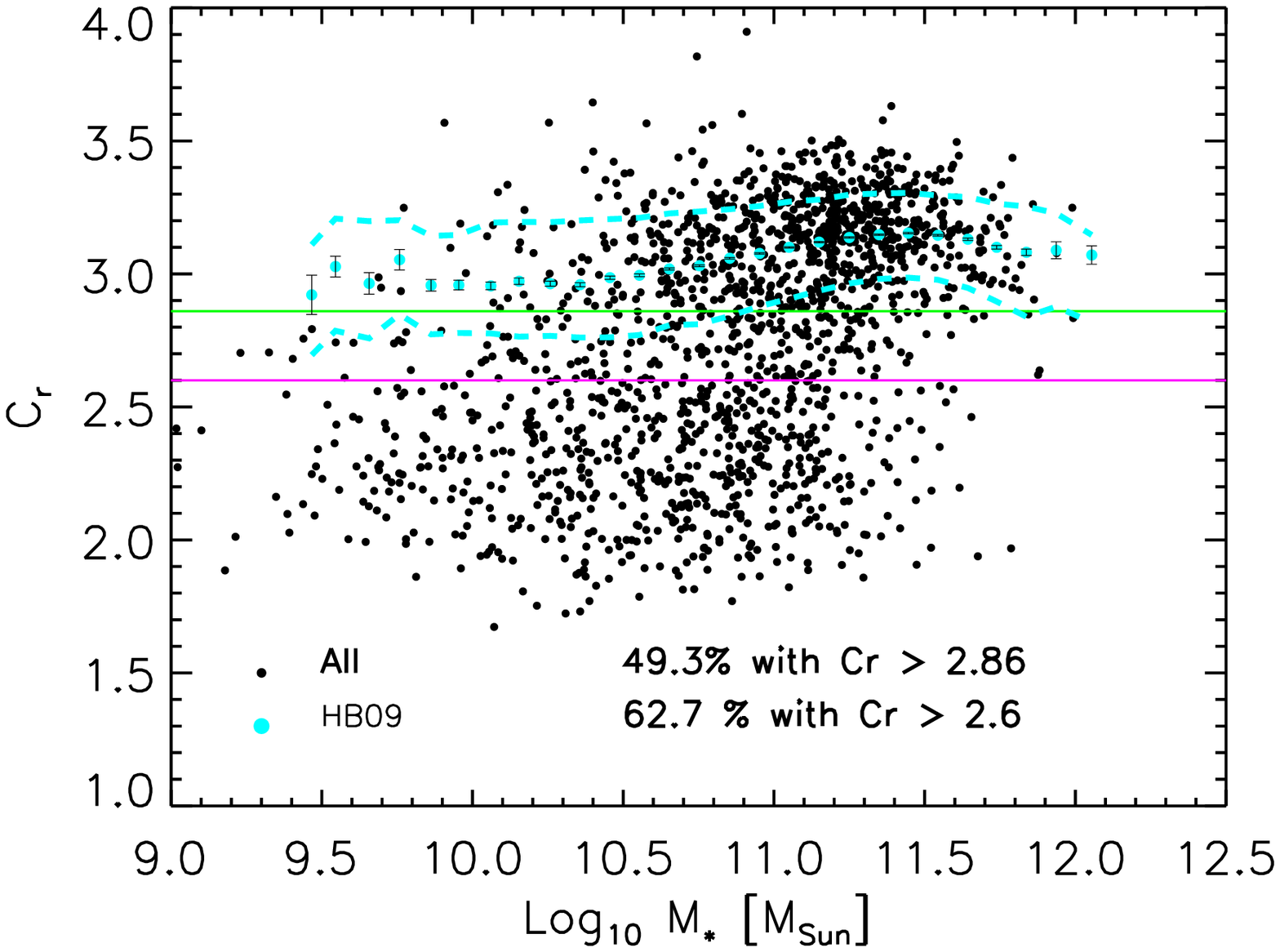}
 \caption{Same as previous figure, but as a function of stellar mass 
          rather than luminosity.}
 \label{cmorphMs}
\end{figure*}

\begin{figure*}
 \centering
 \includegraphics[width=0.45\hsize]{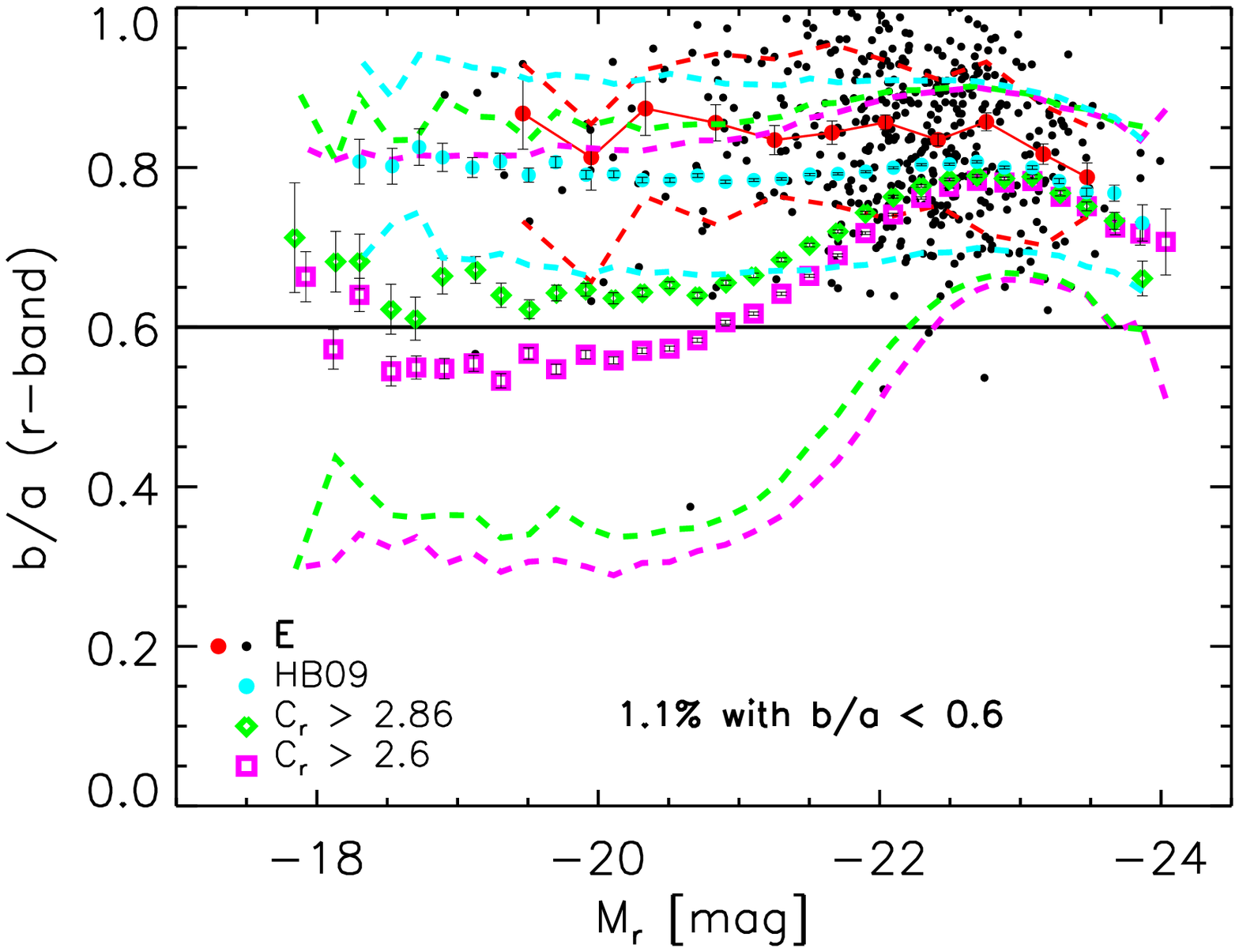}
 \includegraphics[width=0.45\hsize]{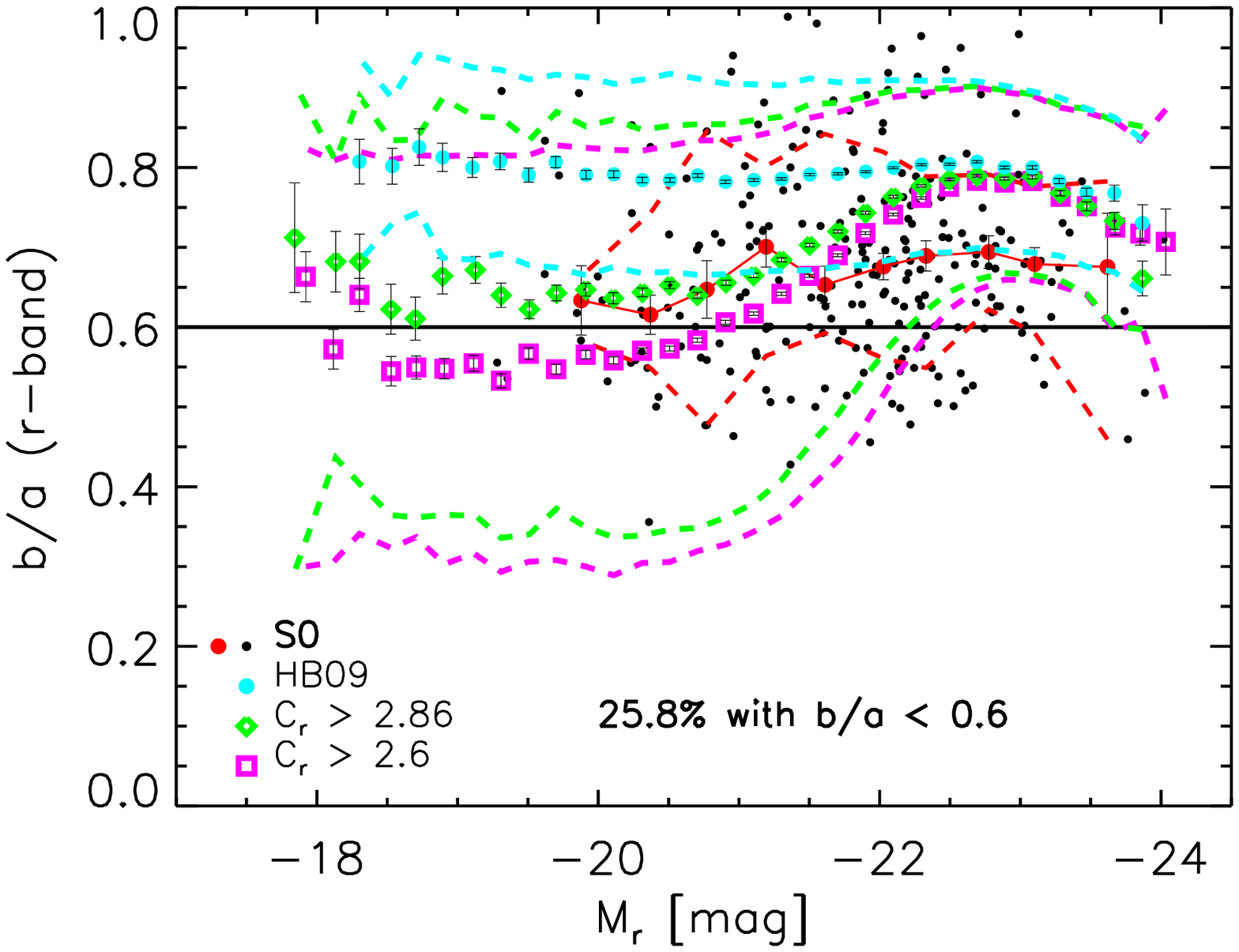}
 \includegraphics[width=0.45\hsize]{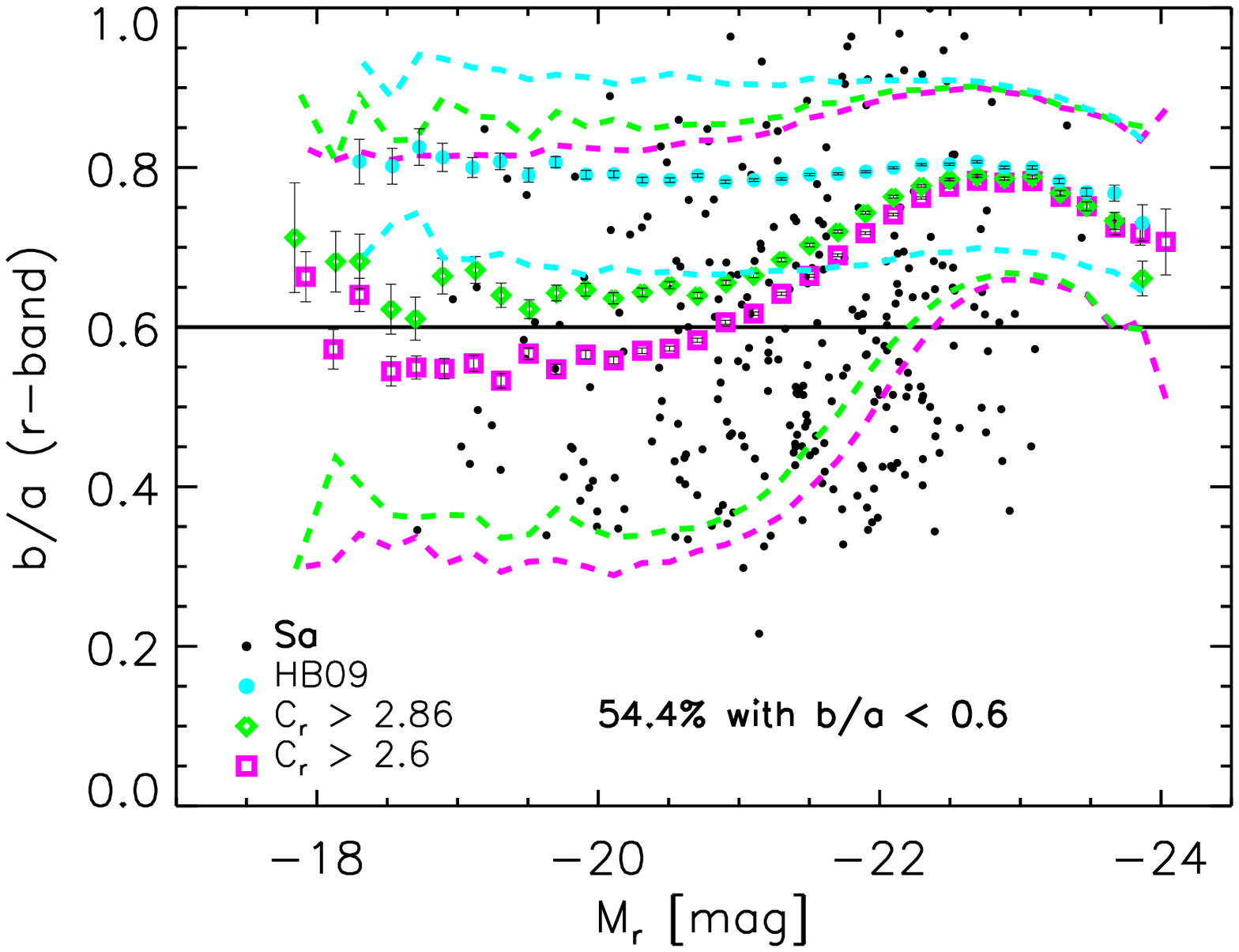}
 \includegraphics[width=0.45\hsize]{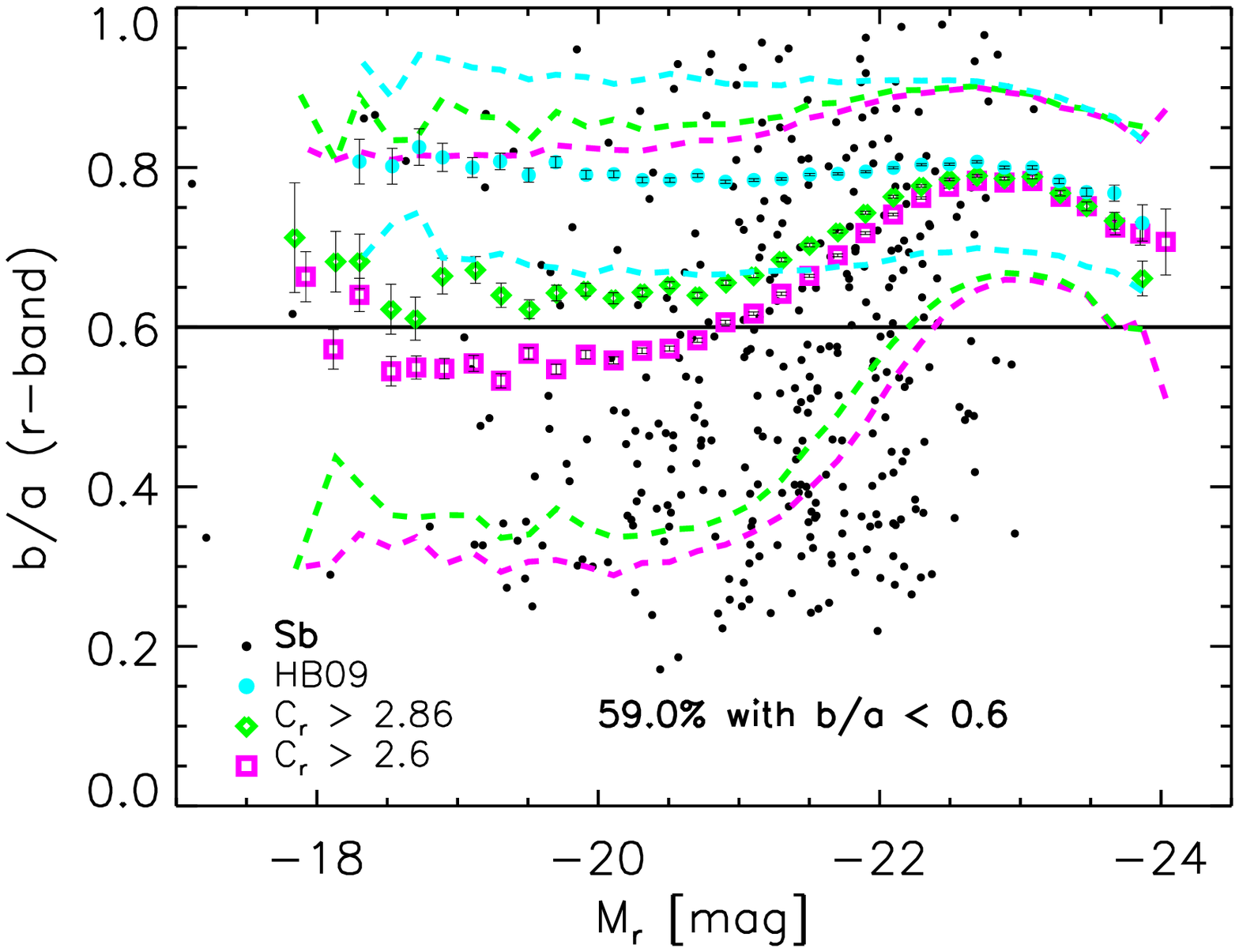}
 \includegraphics[width=0.45\hsize]{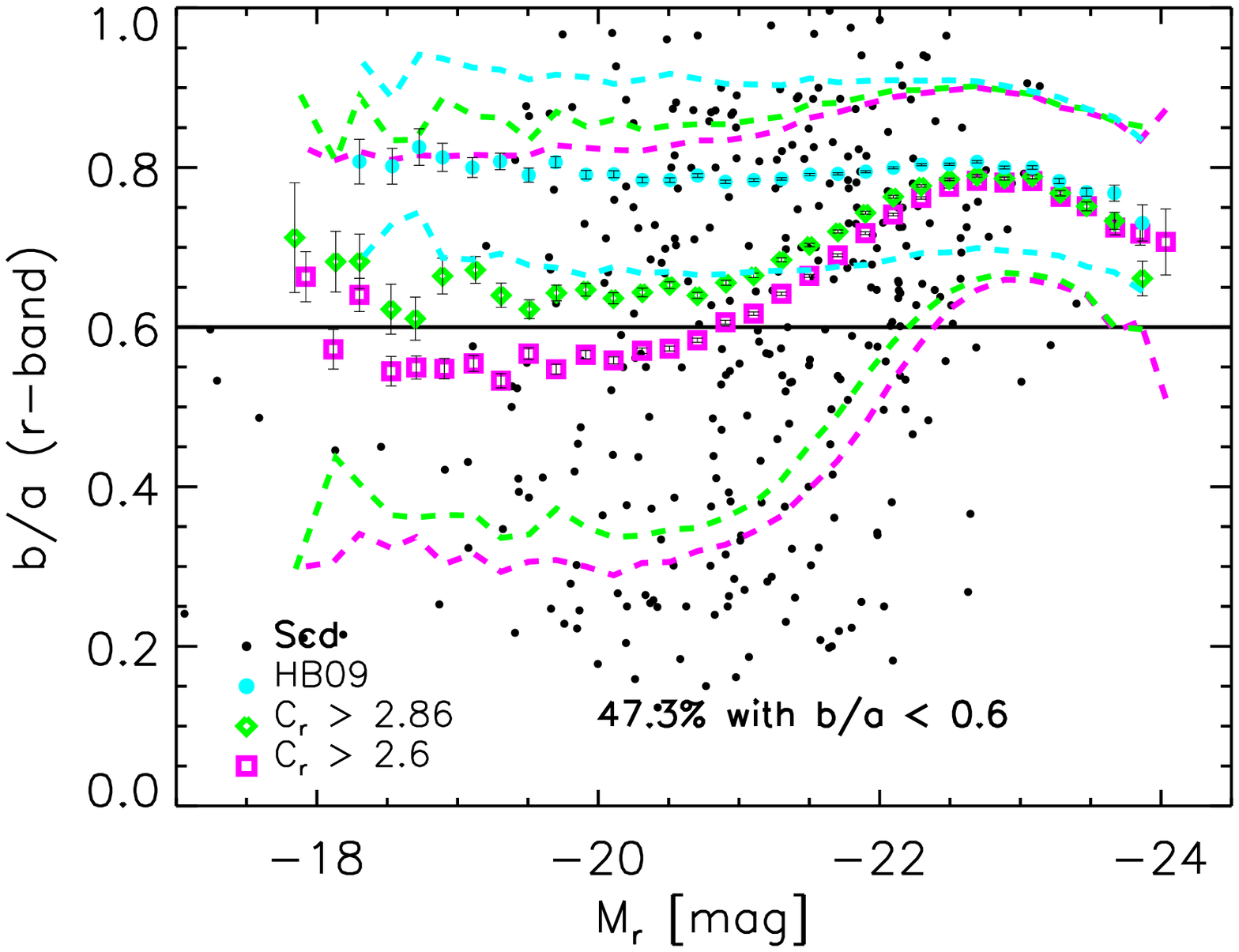}
 \includegraphics[width=0.45\hsize]{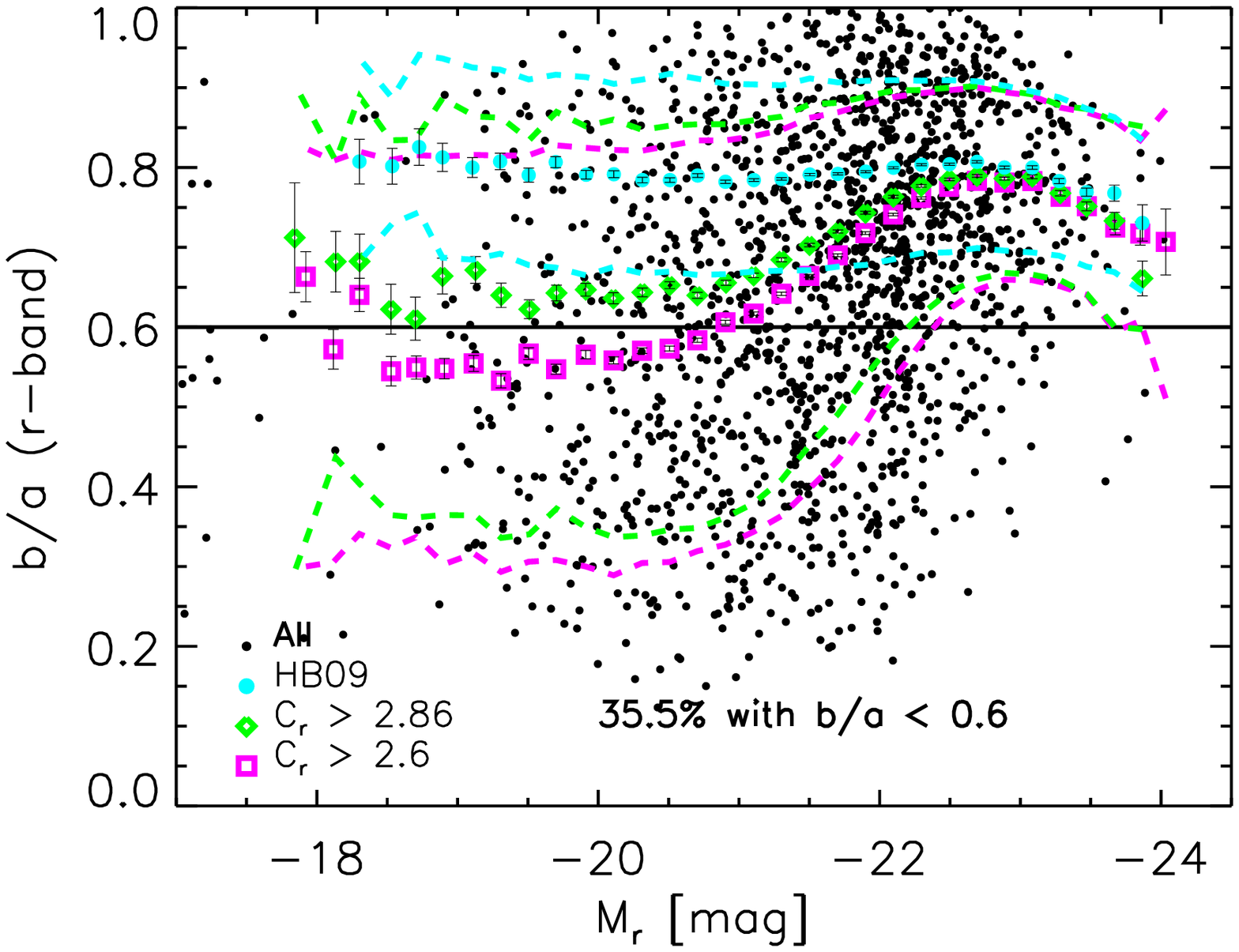}
 \caption{Similar to Figure~\ref{cmorph}, but now in the space of 
          $b/a$ vs luminosity. Magenta open squares and green diamonds, 
          each with error bars and flanked by dashed curves (same in 
          all panels), show the samples selected using cuts in $C_r$ 
          larger of $2.6$ and $2.86$, respectively.}
 \label{bamorph}
\end{figure*}

\begin{figure*}
 \centering
 \includegraphics[width=0.45\hsize]{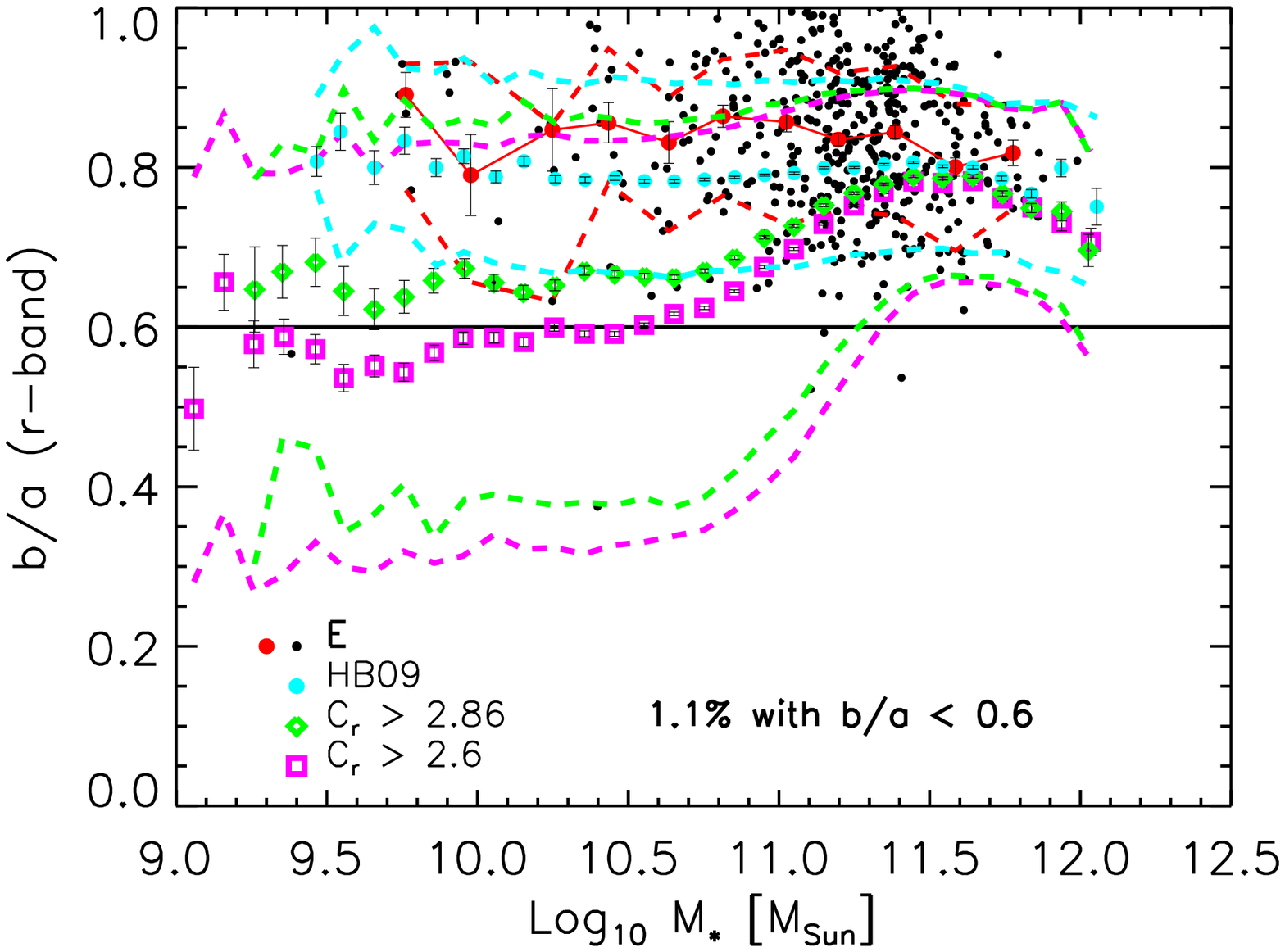}
 \includegraphics[width=0.45\hsize]{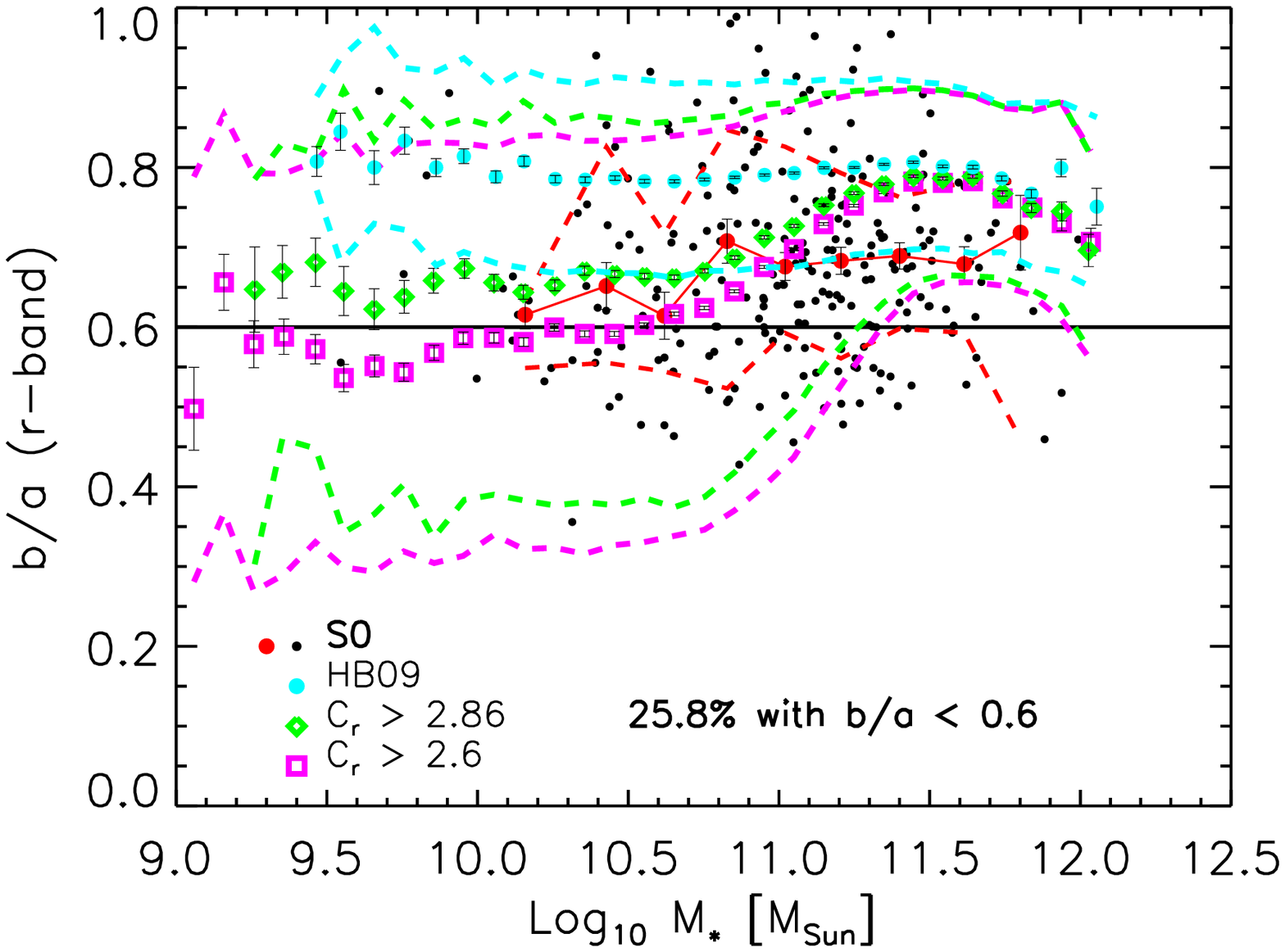}
 \includegraphics[width=0.45\hsize]{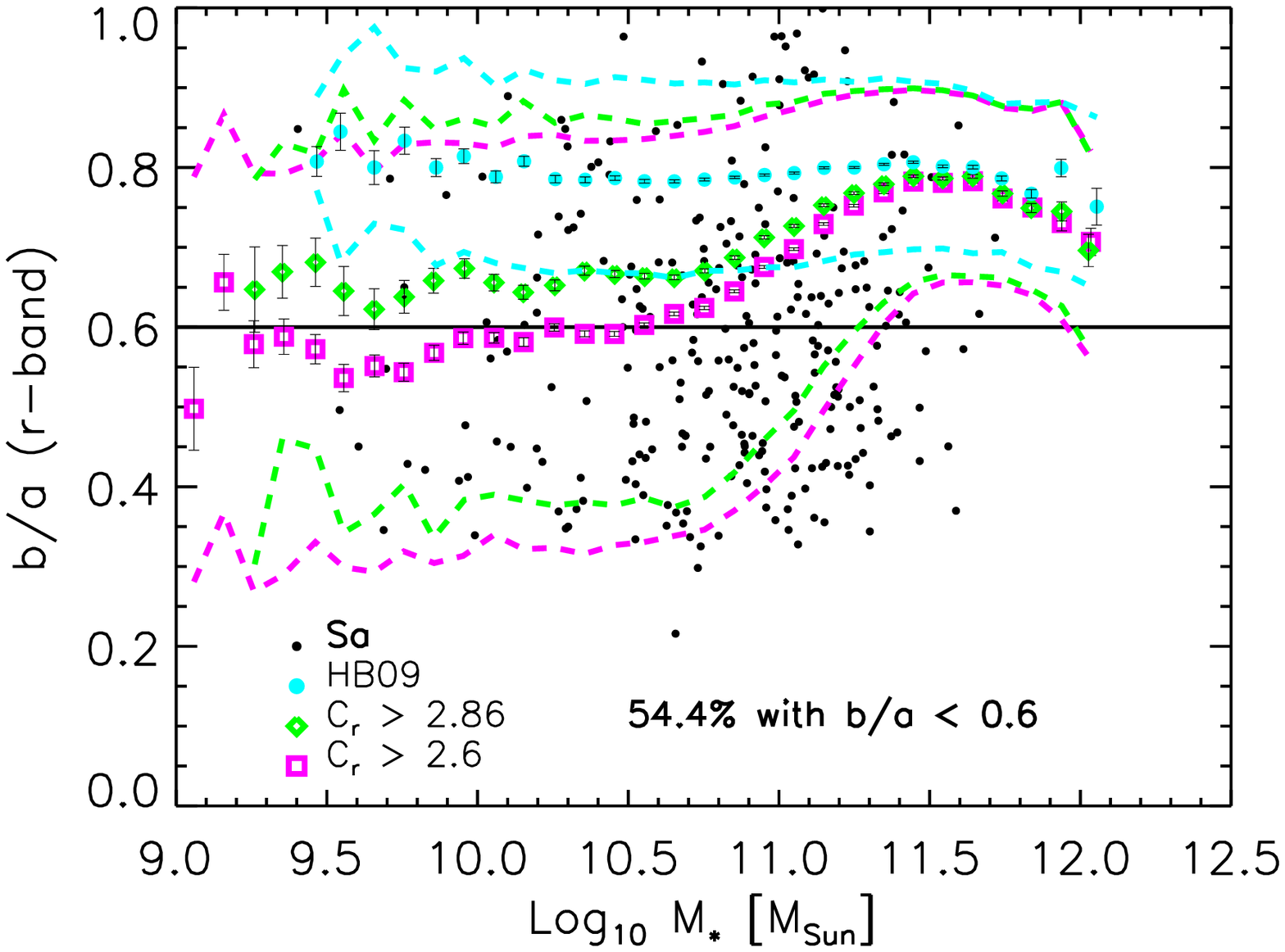}
 \includegraphics[width=0.45\hsize]{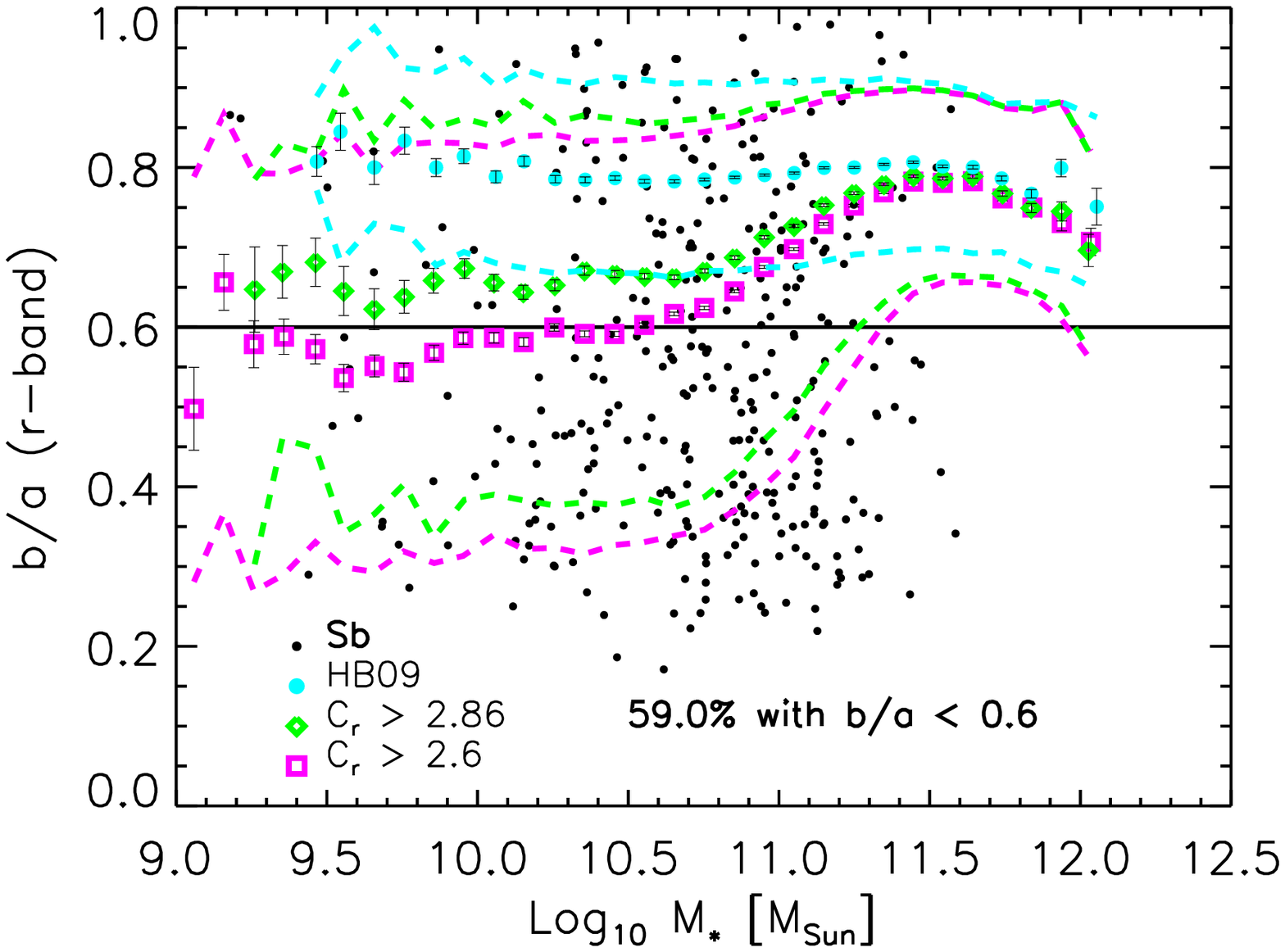}
 \includegraphics[width=0.45\hsize]{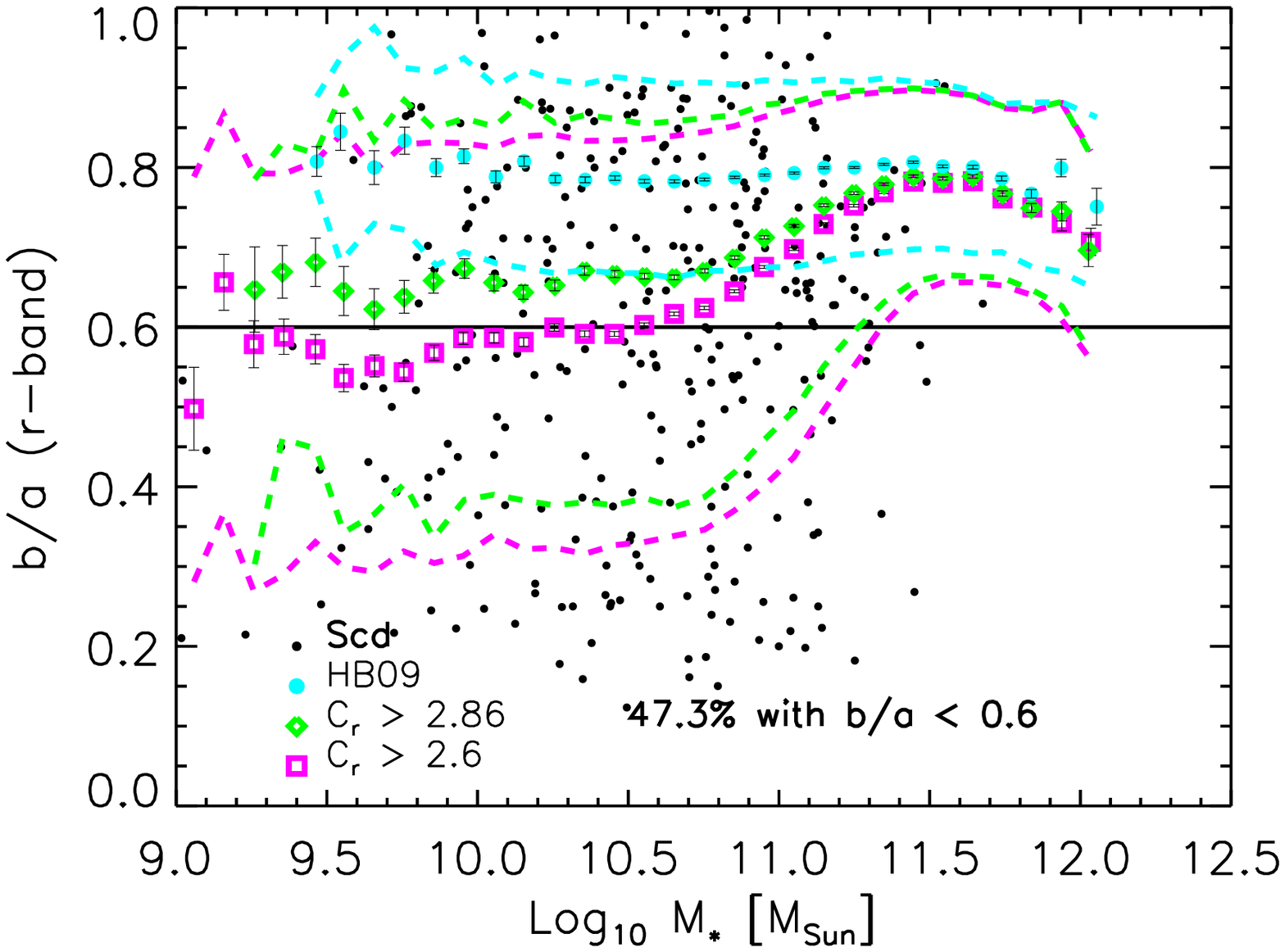}
 \includegraphics[width=0.45\hsize]{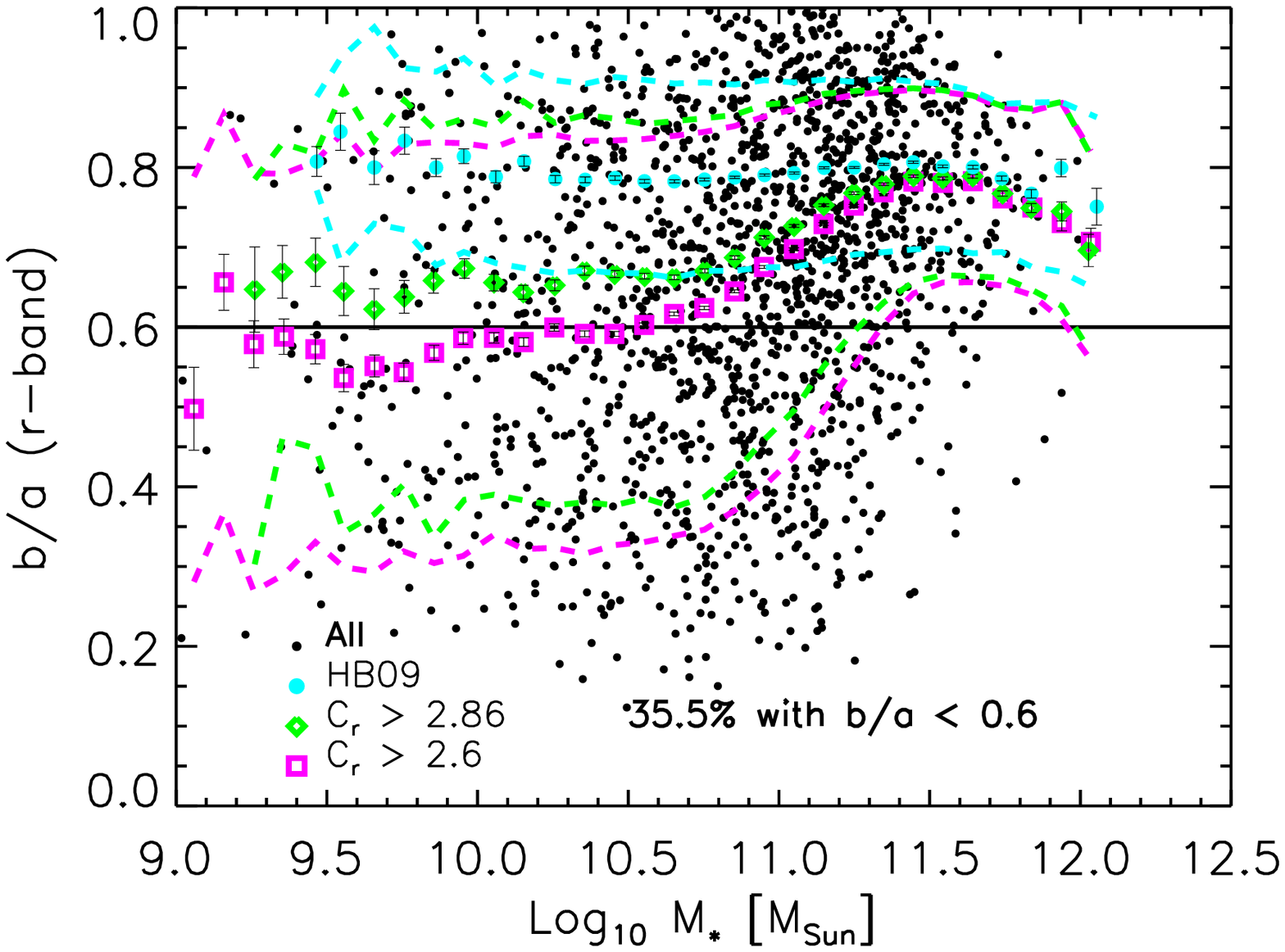}
 \caption{Same as previous figure, but as a function of stellar mass 
          rather than luminosity.}
 \label{bamorphMs}
\end{figure*}

\section{Morphology and sample selection}\label{sec:mcba}
This section compares a number of ways in which early-type samples 
have been defined in the recent literature, with the morphological 
classifications of Fukugita et al., and discusses what this implies 
for the `red' fraction.  When we show luminosities, they have been 
corrected for evolution by assuming that the magnitudes brighten with 
redshift as $1.3z$.  

\subsection{Simple measures of the light profile}\label{mcba}
Concentration index, axis ratio, and {\tt fracdev} have all been 
used as proxies for selecting red, massive, early-type galaxies.  
So it is interesting to see how these quantities correlate with 
morphological type. 

The bottom right panel of Figure~\ref{cmorph} shows the 
distribution of all objects in the Fukugita et al. sample in 
the space of concentration versus luminosity.  The two horizontal 
lines show $C_r = 2.6$ and $C_r = 2.86$, the two most popular choices 
for selecting early-type samples.  The symbols with error bars 
show the mean concentration index at each $L$ if the sample is 
selected following Hyde \& Bernardi (2009): i.e. 
{\tt fracDev} $= 1$ in $g$- and $r$-, and $r$-band $b/a > 0.6$. 
To this we add the condition log$_{10} (r_{e,g}/r_{e,r}) < 0.15$, 
which is essentially a cut on color gradient (Roche et al. 2009b).  
This removes a small fraction ($<2$\%) of late-type galaxies which 
survive the other cuts.  
Dashed curves show the scatter around the $C_r$-luminosity relation 
in the Hyde \& Bernardi sample.  

The other panels show the result of separating out the various 
morphological types.  Whereas Es and S0s occupy approximately the 
same region in this space, most Es (93\%) have $C_r\ge 2.86$, whereas 
the distribution of $C_r$ for S0s is somewhat less peaked.  (Here, 
the percentages we quote are per morphological type, in the Fukugita 
et. al. sample -- meaning a sample that is magnitude-limited to 
$m_{r,{\tt Pet}}<16$, with no $1/V_{\rm max}$ weighting applied.)  
Samples restricted to $C_r\ge 2.6$ have a substantial contribution 
from both Sas ($74\%$ of which satisfy this cut) and S0s 
(for which this fraction is $93\%$), and this remains true even if 
$C_r\ge 2.86$ ($50\%$ of Sas and $77\%$ of S0s).  
Thus, it is difficult to select a sample of Es on the basis of 
concentration index alone.  On the other hand, the larger mean 
concentration of the Hyde \& Bernardi selection cuts suggest that 
they produce a sample that is dominated by ellipticals/S0s, and less 
contaminated by Sas (see Section~\ref{puritysel} and 
Table~\ref{purity} below).  
Figure~\ref{cmorphMs} shows that replacing luminosity with 
stellar mass leads to similar conclusions.  

Before moving on, notice that although the mean concentration 
increases with luminosity and stellar mass in the Hyde \& Bernardi 
sample, this is no longer the case at the highest $L_r$ or $M_*$:  
we will have more to say about this shortly.

Figures~\ref{bamorph} and~\ref{bamorphMs} show a similar analysis of 
the axis ratio $b/a$.  
The different symbols with error bars (same in all panels) show 
samples selected to have $C_r\ge 2.6$, $C_r\ge 2.86$, and following 
Hyde \& Bernardi.  At $M_r\le -19$ or so, the mean $b/a$ in the 
first two cases increases with luminosity upto $M_r=-22.7$ or so; 
it decreases for the brightest objects.  
At low $L$, the sample with $C_r\ge 2.6$ has smaller values of $b/a$ 
on average, though the scatter around the mean is large.  
Morevoer, while there are essentially no Es with $b/a<0.6$ about 
$26\%$ of S0s have $b/a<0.6$. On the other hand, a little less than 
half the Sa's and many Scd's also have $b/a>0.6$ (because they are 
face on).  Evidently, just as $C_r$ alone is not a good way to select 
a pure sample of ellipticals, selecting on $b/a$ alone is not good either.  

The filled red circles (with error bars) in the top left panel show 
that, for Es, $b/a\approx 0.85$ independent of $M_r$, except at the 
highest luminosities where $b/a$ decreases.  This independence of 
$M_r$ differs markedly from that in either of the $C_r$ samples, but 
is reproduced by the Hyde \& Bernardi sample, for which $b/a =0.8$ 
except at $M_r\le -22.7$ where it decreases.  The difference of about 
0.05 in $b/a$ arises because the Hyde \& Bernardi sample includes 
some S0s (we quantify this in Table~\ref{purity} below), for which 
$b/a \sim 0.7$ (filled red circles in top right panel).  
This leads to an important point.  While it has long been known that 
$b/a$ tends to increase with luminosity, even in `early-type' samples, 
our Figure~\ref{bamorph} shows that this increase is driven by the 
changing morphological mix -- the change from S0s to Es -- at $M_r>-23$.  
Whether this is due to environmental or pure secular evolution effects 
is an open question.  

On the other hand, there is a plausible, environmentally driven model 
for the decrease in $b/a$ at the highest $L$.  This decrease has been 
expected for some time (see Gonz{\'a}lez-Garc{\'i}a \& van Albada 2005; 
Boylan-Kolchin et al. 2006 and references therein) -- it was first found 
by Bernardi et al. (2008).  This is thought to indicate an increasing 
incidence of radial mergers, since these would tend to result in more 
prolate objects.  
The decrease in concentrations at these high luminosities 
(Figure~\ref{cmorph}) is consistent with this picture, as is the 
fact that most of these high luminosity objects are found in clusters.  
All of the preceding discussion remains true if one replaces luminosity 
with stellar mass.  

\begin{figure*}
 \centering
 \includegraphics[width=0.45\hsize]{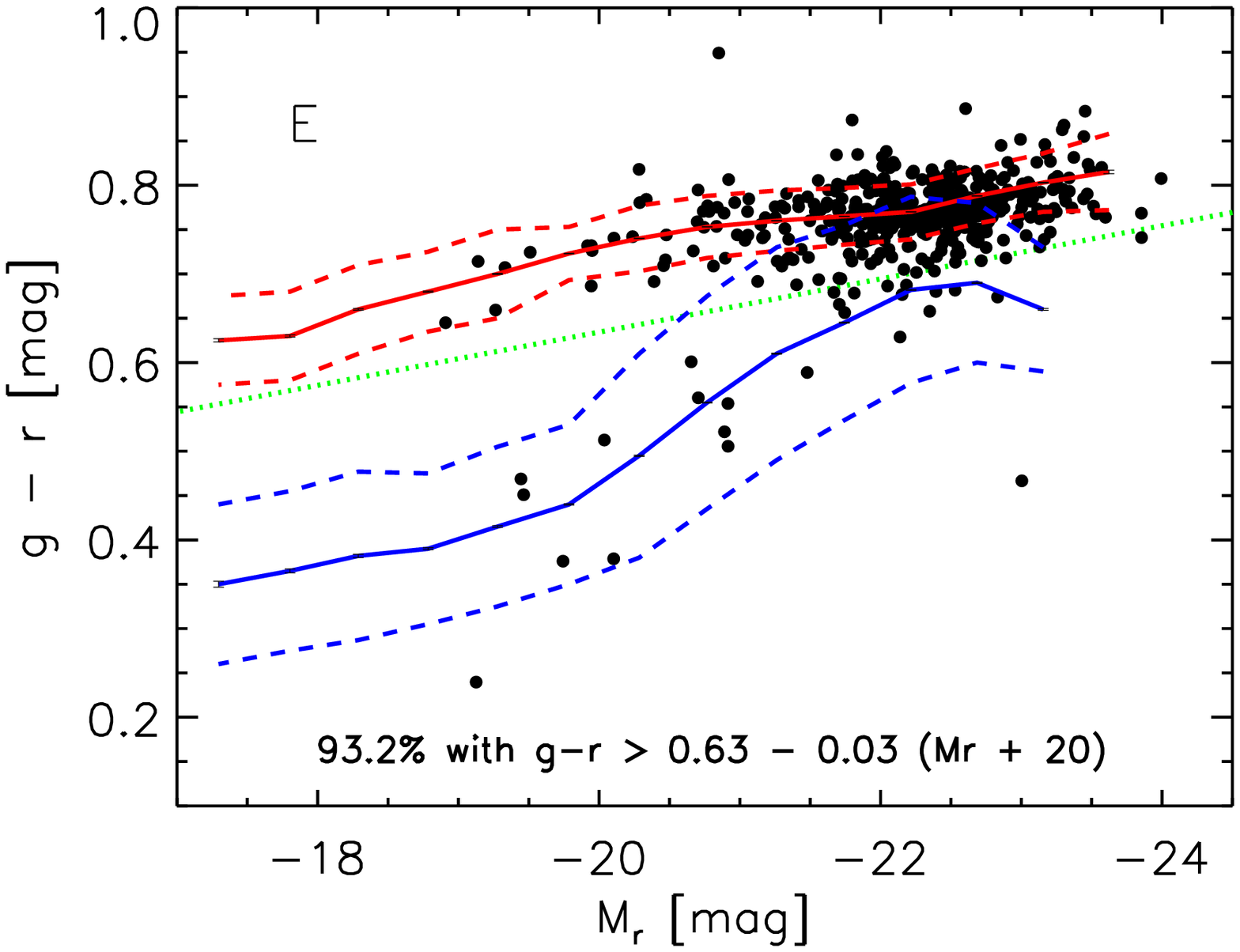}
 \includegraphics[width=0.45\hsize]{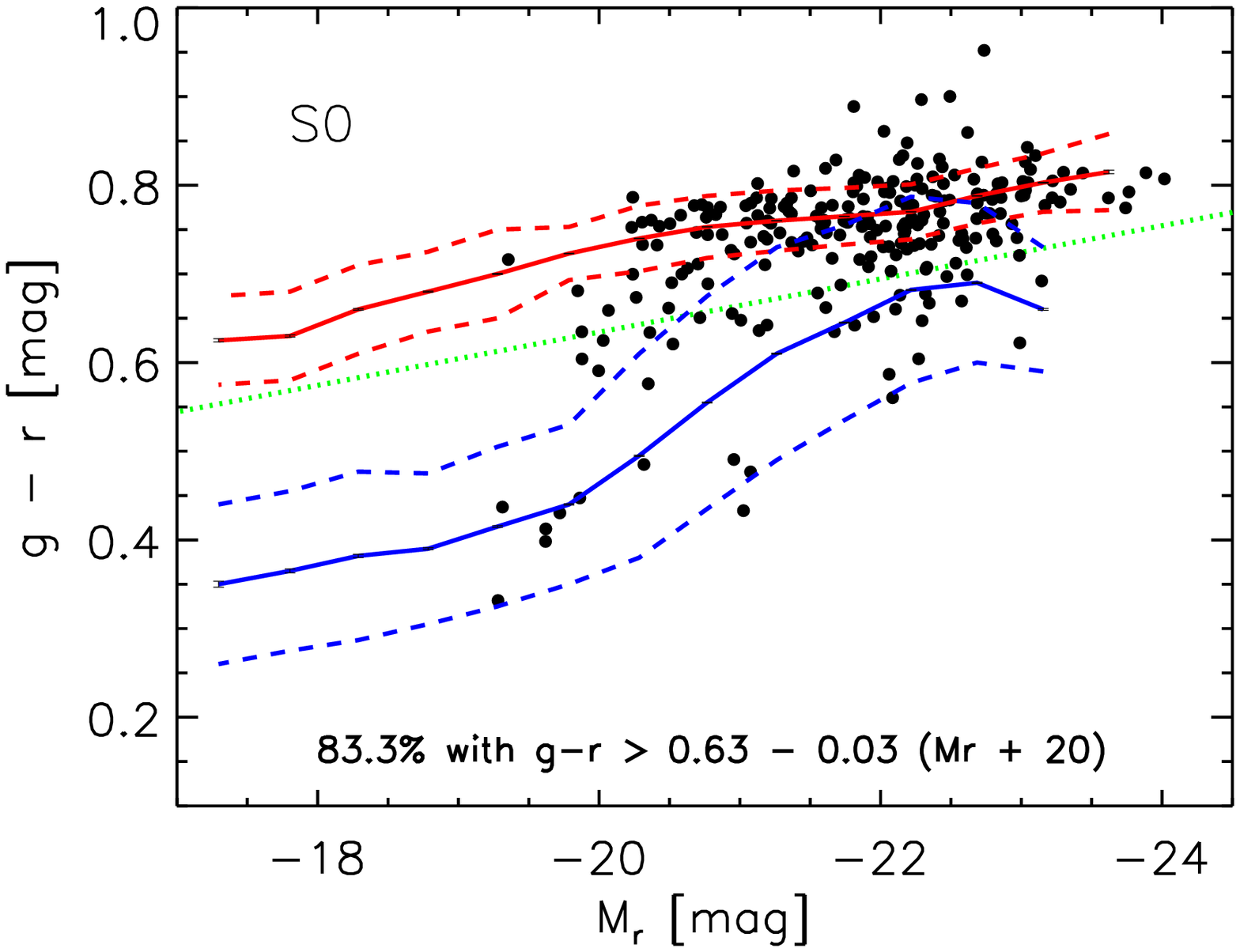} 
 \includegraphics[width=0.45\hsize]{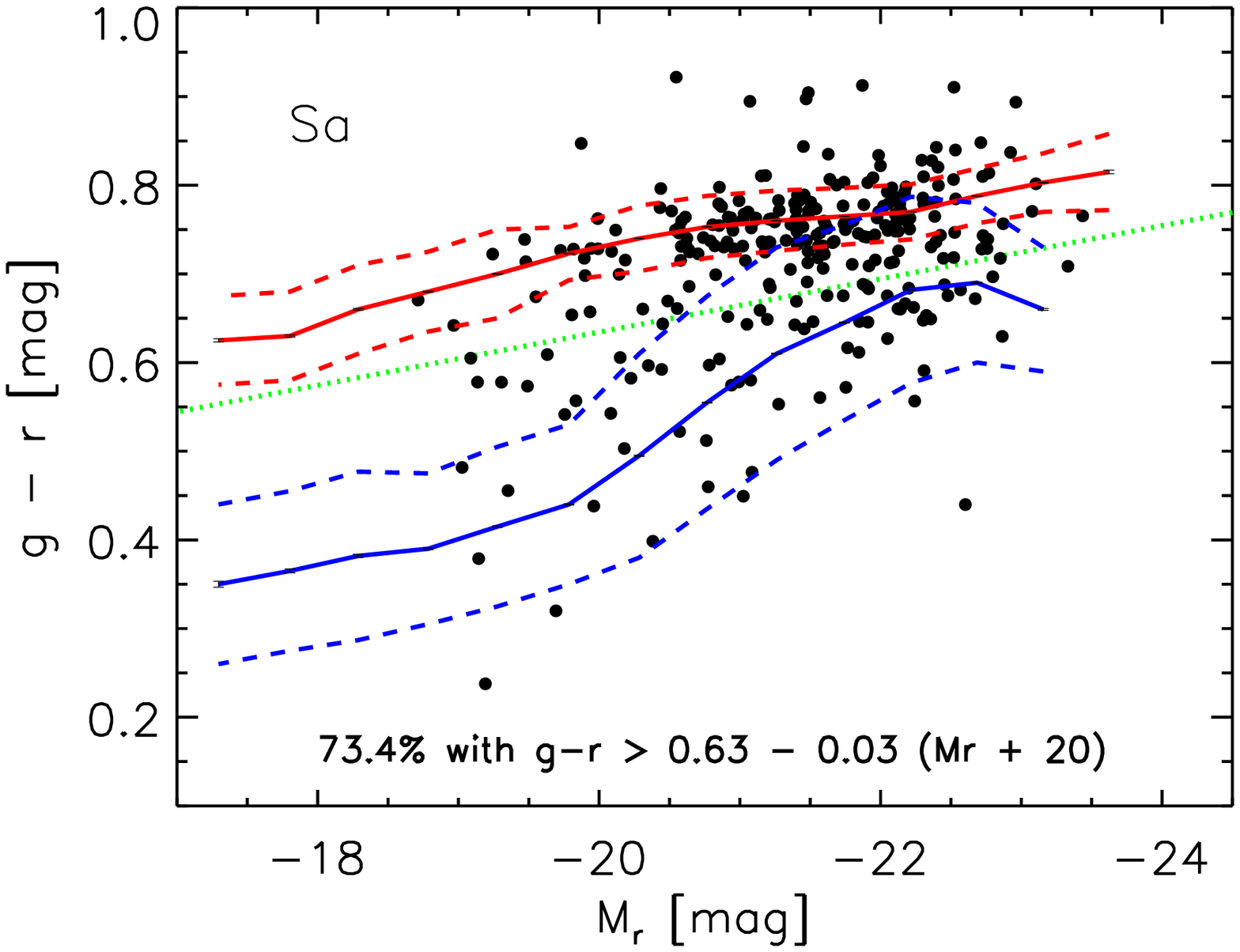}
 \includegraphics[width=0.45\hsize]{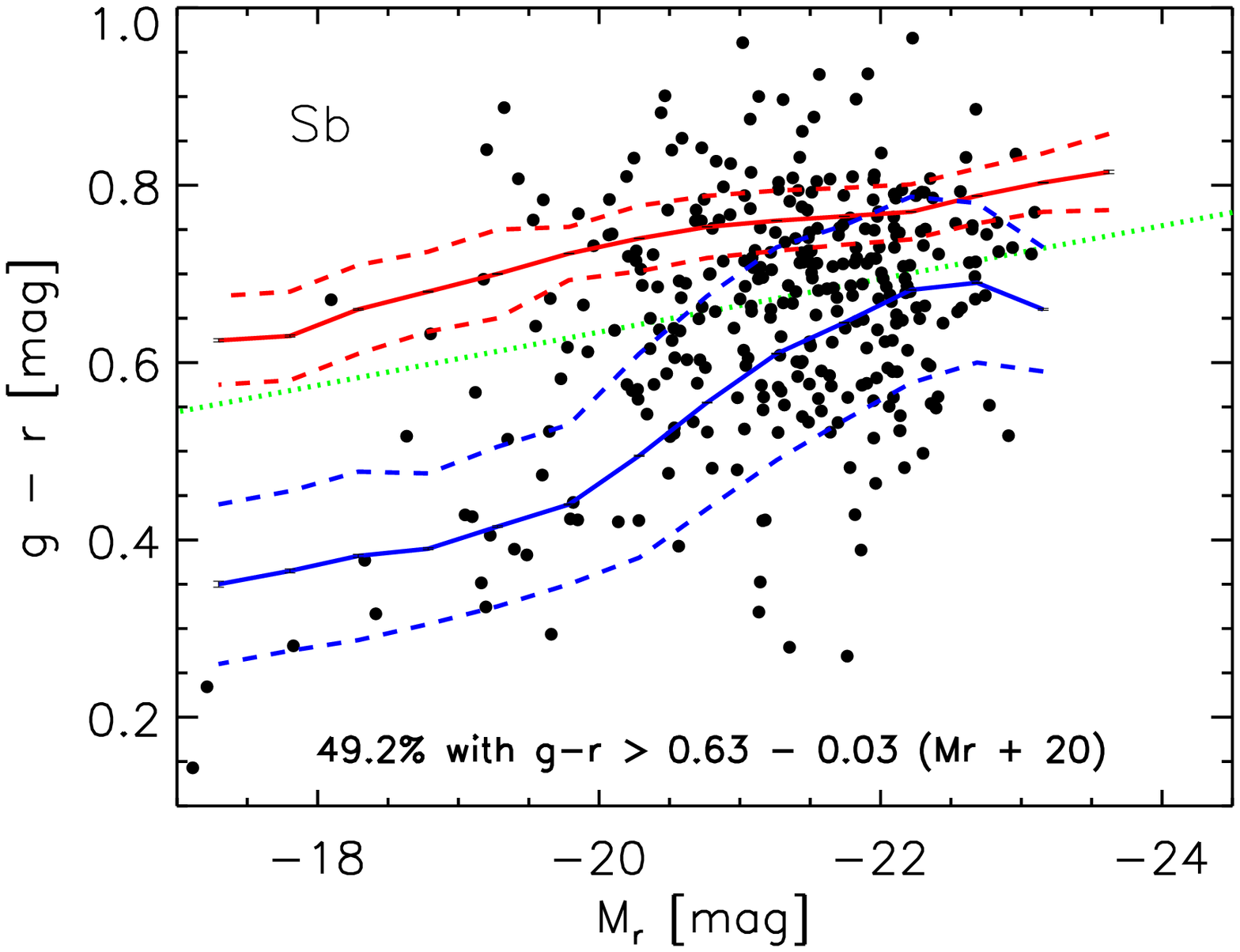}  
 \includegraphics[width=0.45\hsize]{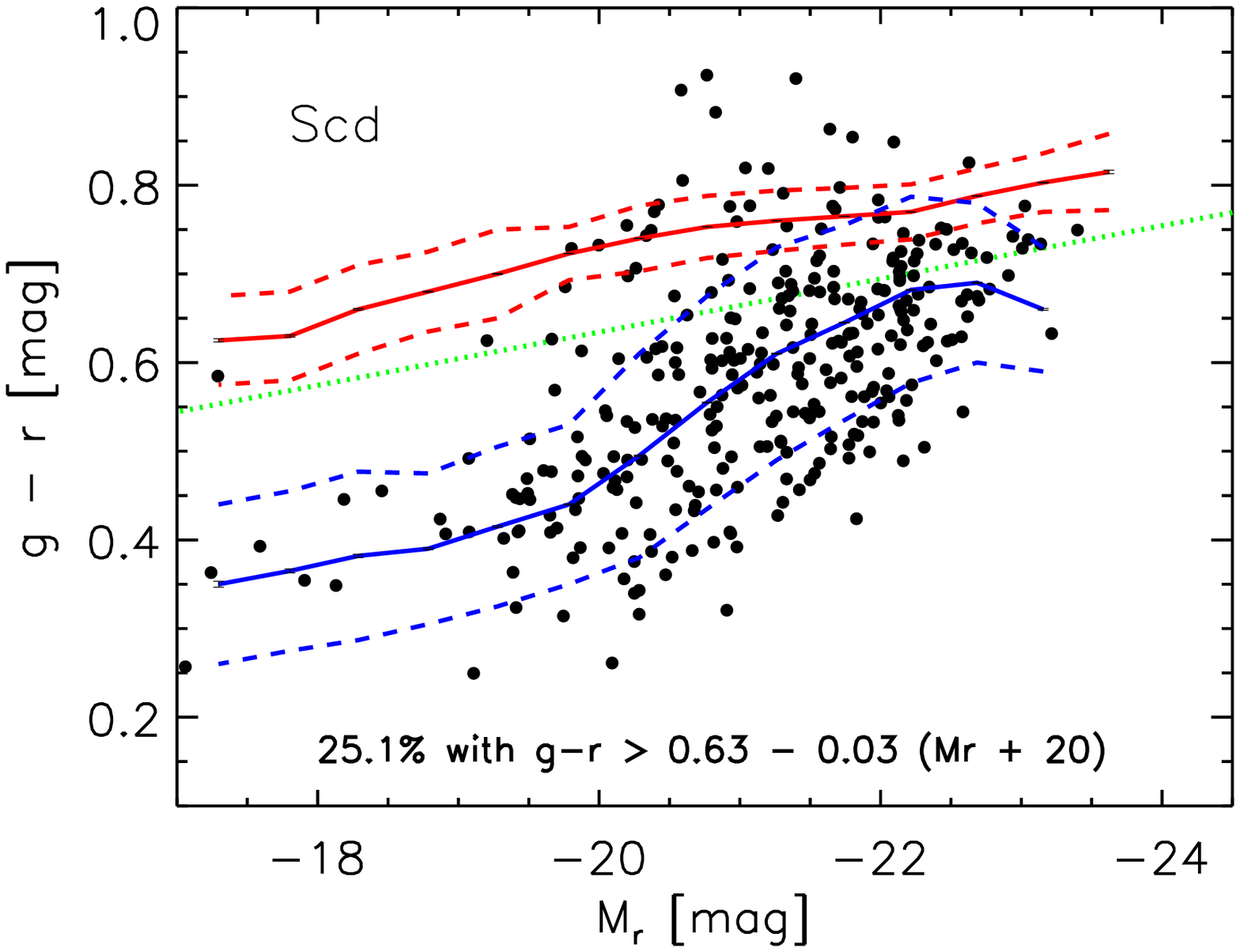}
 \includegraphics[width=0.45\hsize]{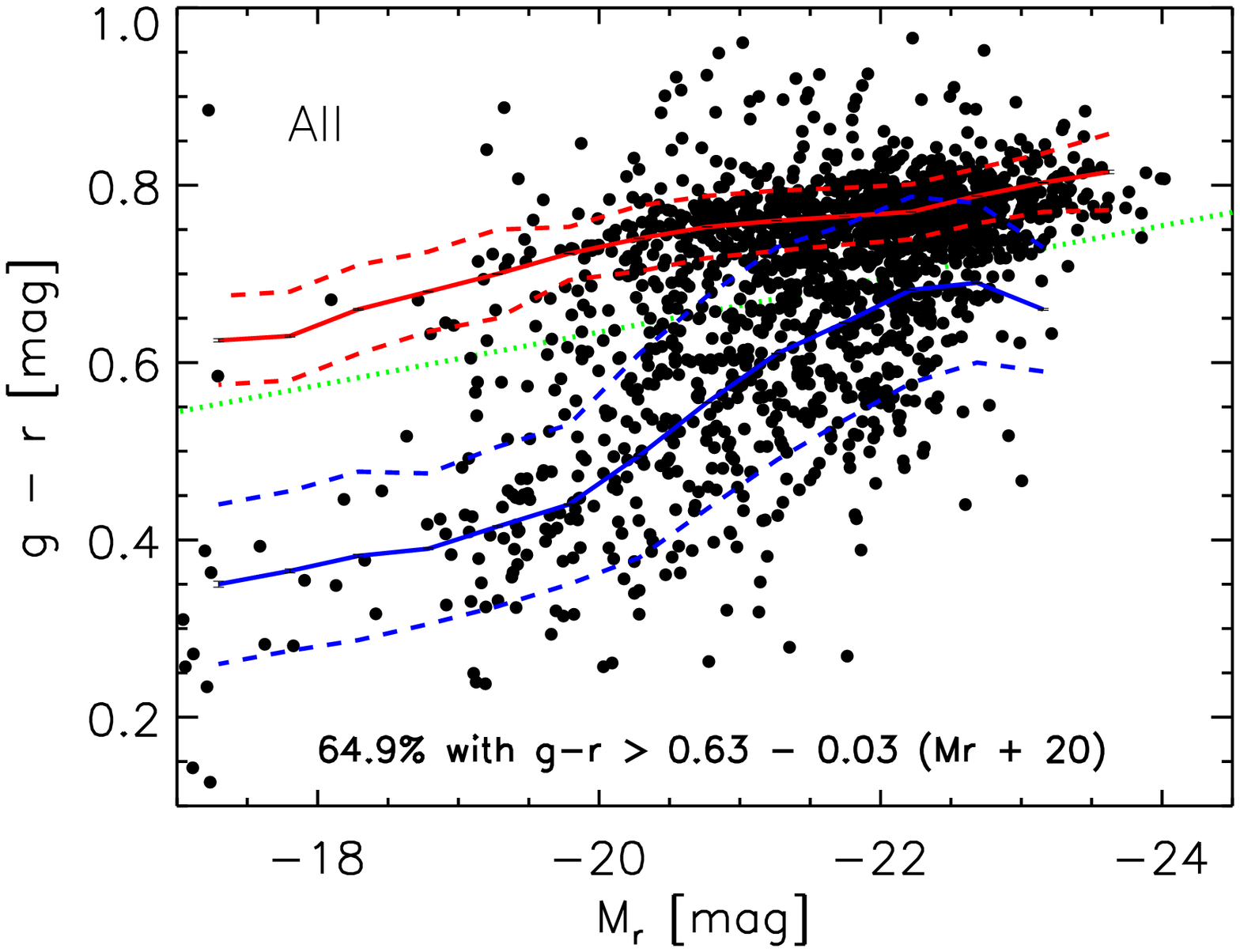}    
 \caption{Color-magnitude relation for the different morphological 
          types in the Fukugita et al. sample.
          Solid and dashed curves (same in each panel) 
          show the mean  of the `red' and `blue' sequences, 
          and their thickness, which result from performing 
          double-Gaussian fits to the color distribution at fixed 
          magnitude, of the full SDSS galaxy sample 
          (from Bernardi et al. 2010, in preparation).
          Dotted green line shows the luminosity dependent threshold 
          used to separate `red' from `blue' galaxies 
          (equation~\ref{redcut}). }
 \label{gmrType}
\end{figure*}

\begin{figure*}
 \centering
 \includegraphics[width=0.45\hsize]{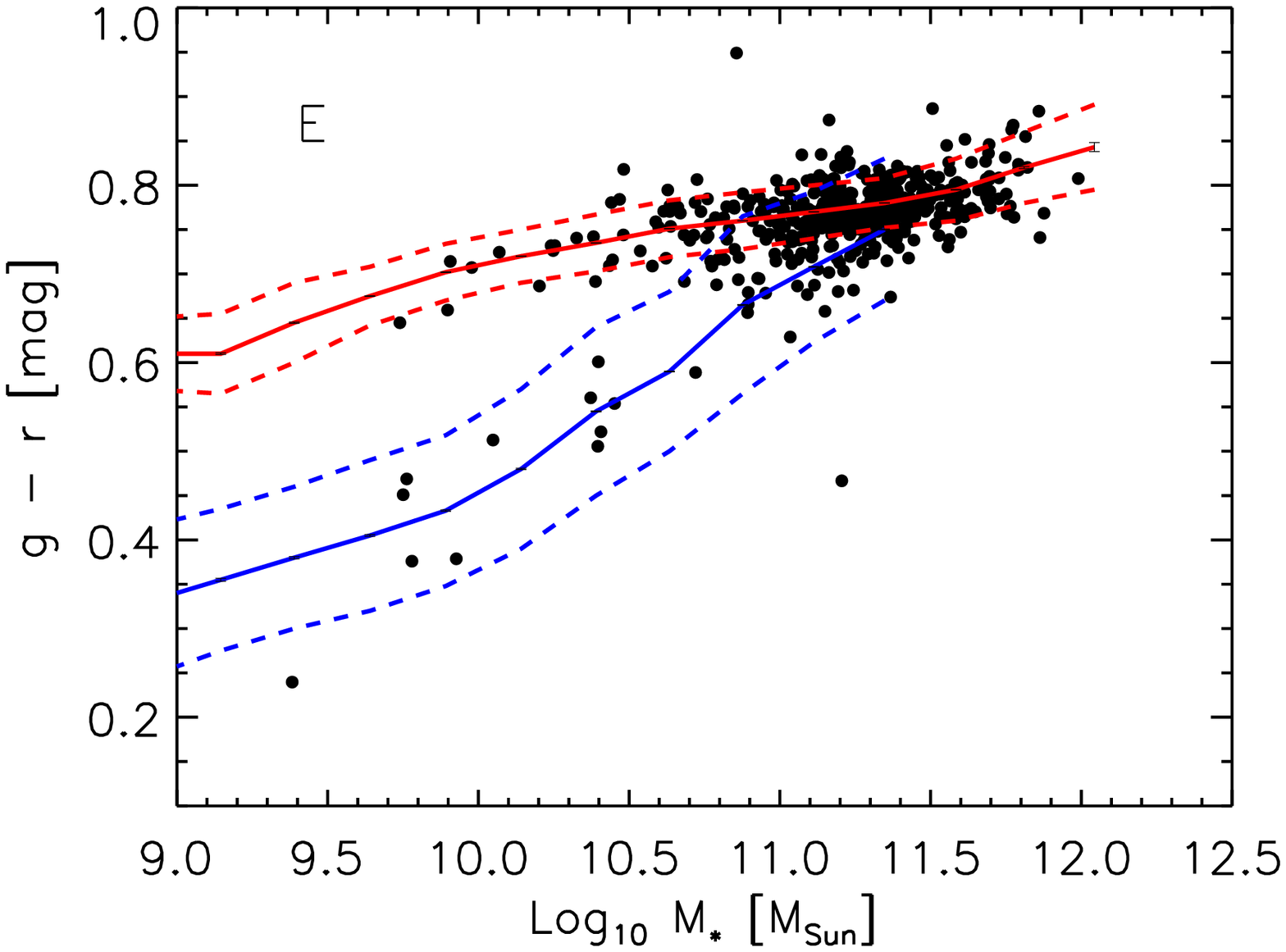}
 \includegraphics[width=0.45\hsize]{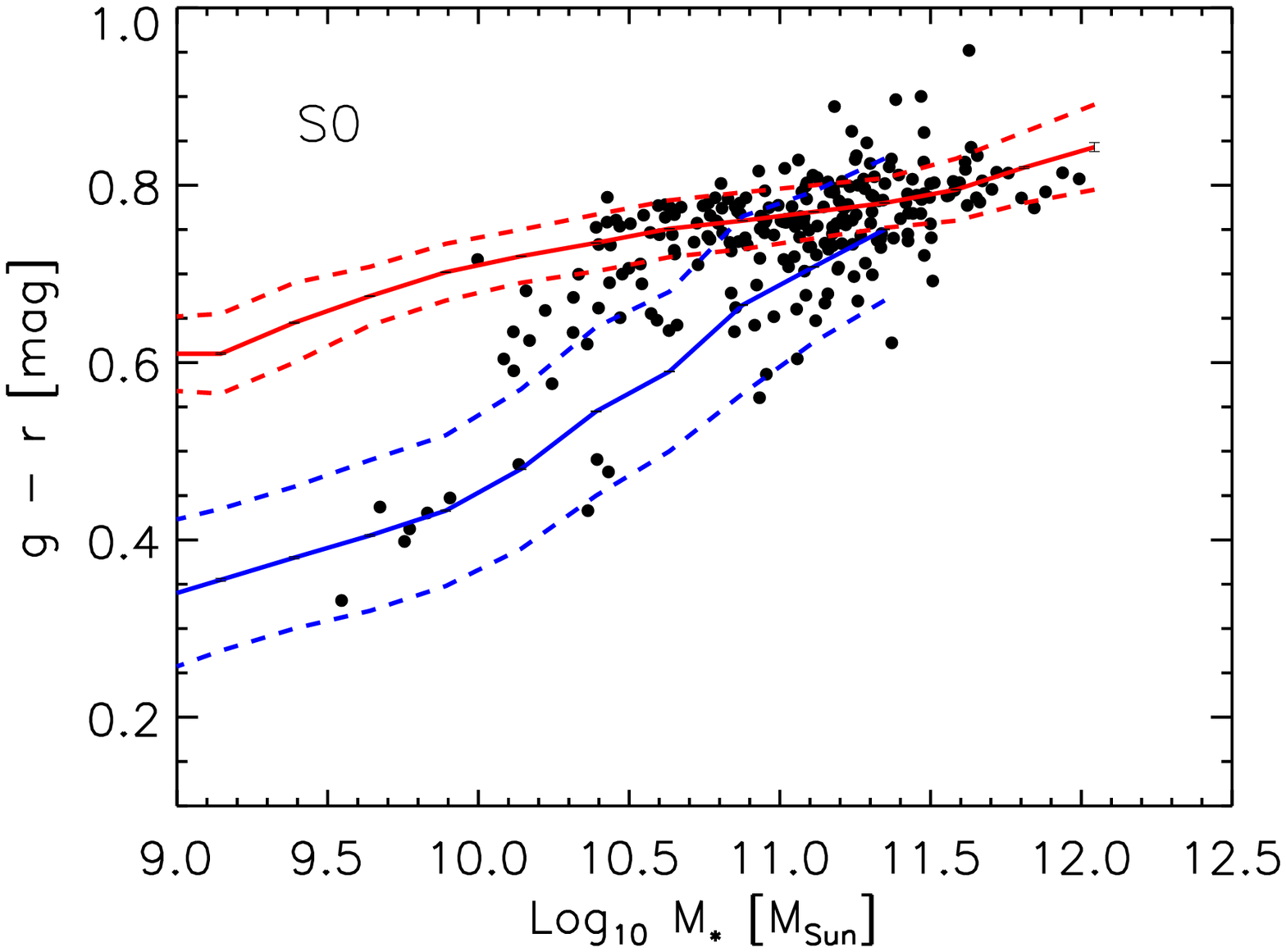} 
 \includegraphics[width=0.45\hsize]{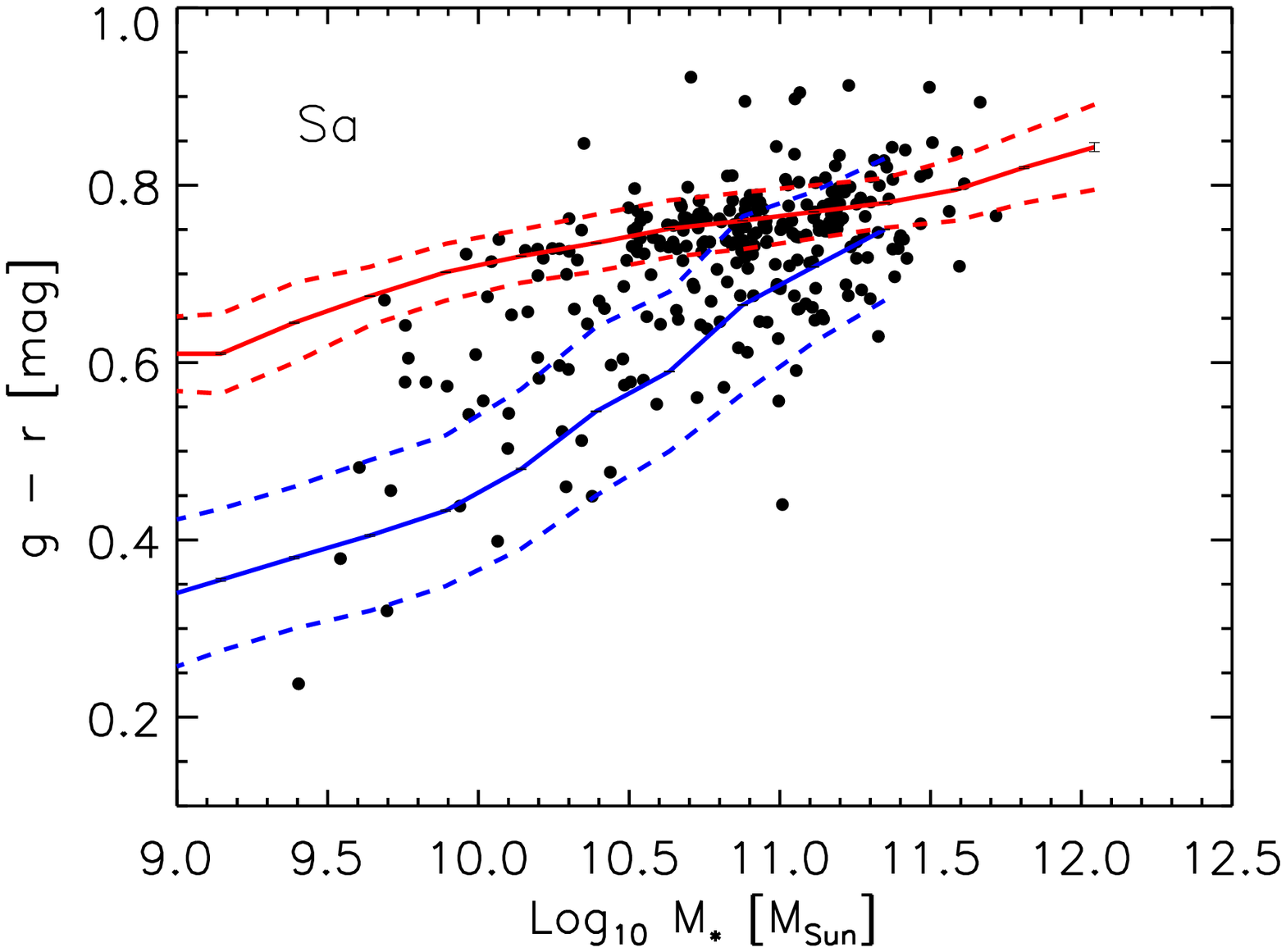}
 \includegraphics[width=0.45\hsize]{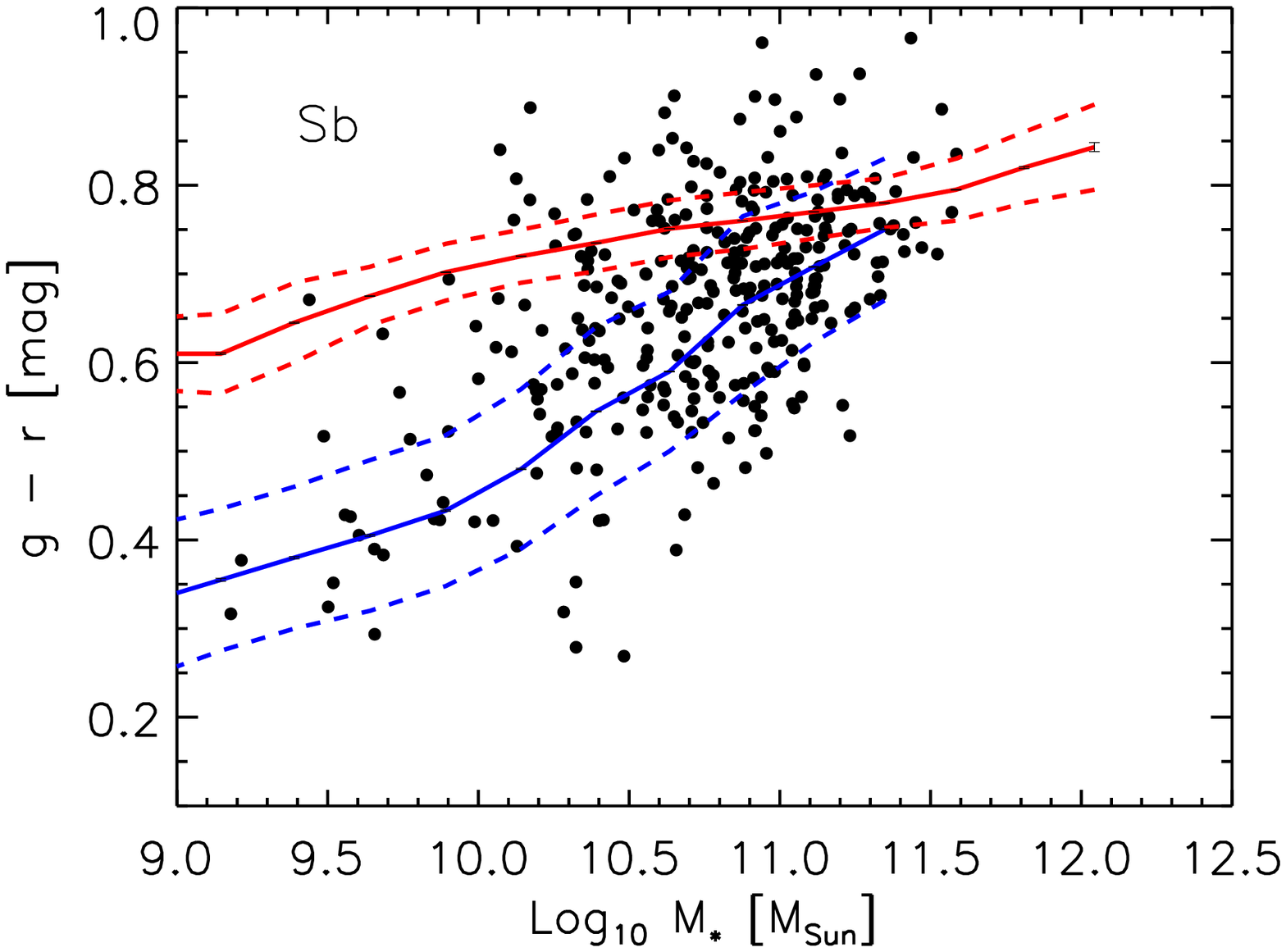}  
 \includegraphics[width=0.45\hsize]{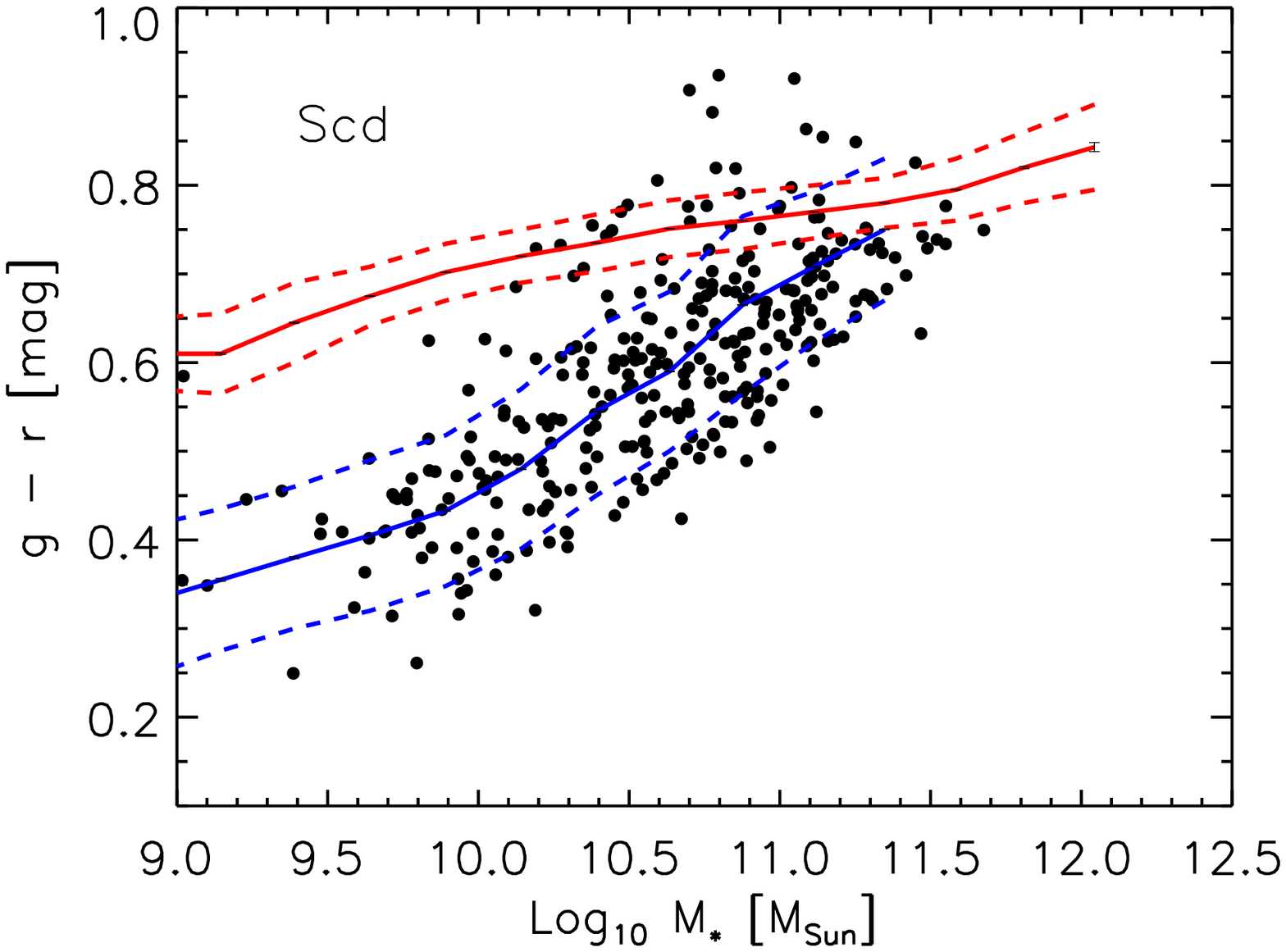}
 \includegraphics[width=0.45\hsize]{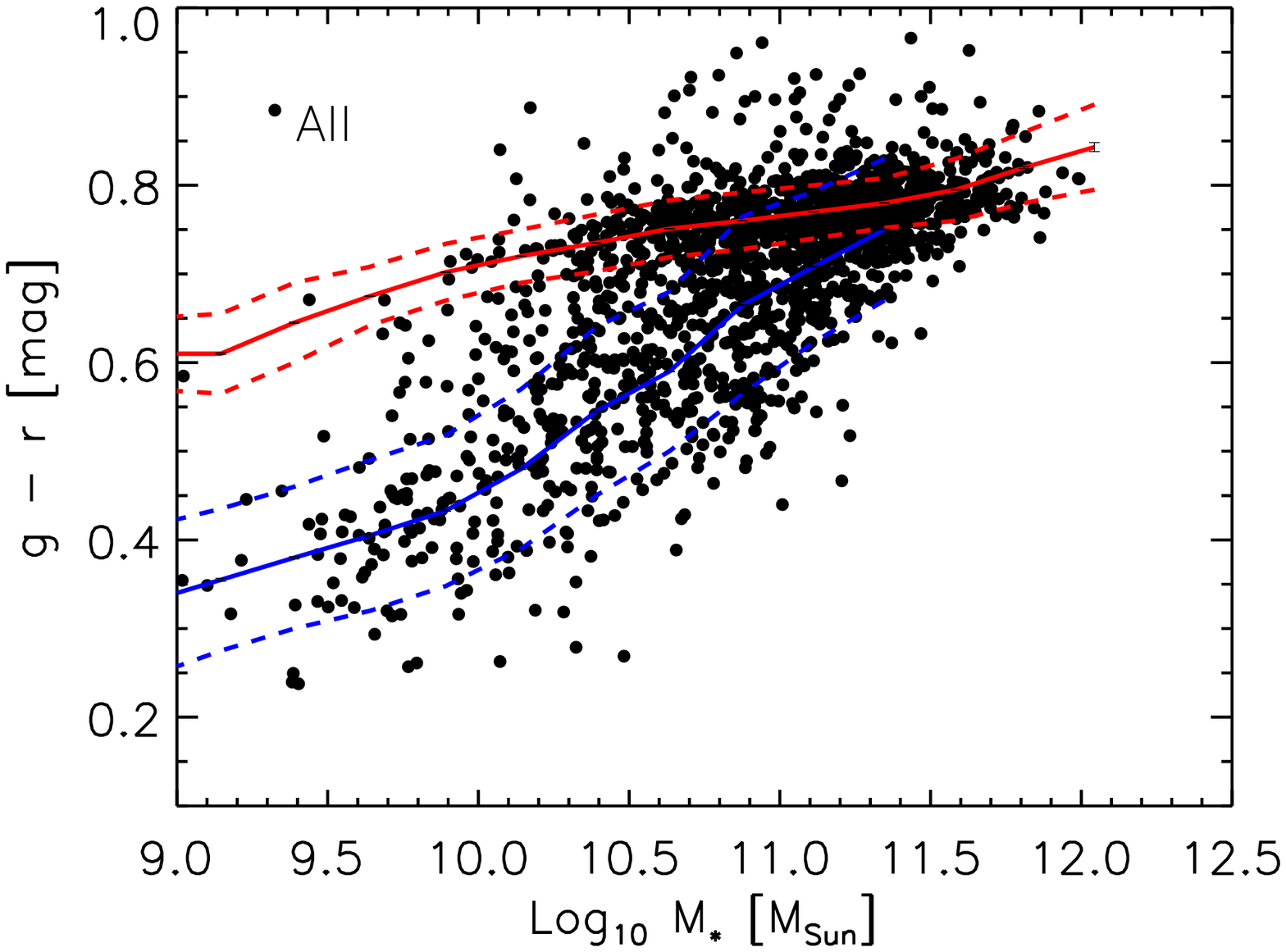}    
 \caption{Same as previous figure, but now for stellar mass in place 
          of luminosity.
          }
 \label{gmrTypeMs}
\end{figure*}

Before concluding this section, we note that we have also considered 
the quantity {\tt fracdev} which plays an important role in the 
selection cuts used by Hyde \& Bernardi (2009).  The vast majority 
of ellipticals (85\%) have {\tt fracdev}=1, with only a percent or 
so having {\tt fracdev}$\le 0.8$.  The distribution of {\tt fracdev}  
has a broader peak for S0s, but they otherwise cover the same 
range as Es: only 37\% have {\tt fracdev}$<1$.  However, 70\% of 
Sas have {\tt fracdev}$<1$, whereas for Scs's, only 10\% have 
{\tt fracdev}$\ge 0.4$. (Note that the above percentages were computed
in the magnitude limited catalog, i.e. were not weighted by 
$1/V_{\rm max}$.)  

\subsection{Colors}
In addition to simple measures based on the light profile, or more 
commonly, as an alternative to such methods, color is sometimes used 
as a way to selecting early types.  
This is typically done by noting that the color-magnitude distribution 
is bimodal (e.g. Baldry et al. 2005), and then adopting a crude 
approximation to this bimodality (e.g. Zehavi et al. 2005; 
Blanton \& Berlind 2007; Skibba \& Sheth 2009).  
Figure~\ref{gmrType} shows this bimodal distribution in the 
Fukugita et al. sample (Figure~\ref{gmrTypeMs} shows the 
corresponding color-$M_*$ relation).  The dotted green line in 
Figure~\ref{gmrType}, shows 
\begin{equation}
 g - r = 0.63 - 0.03\,(M_r + 20).
 \label{redcut}
\end{equation}  
It runs approximately parallel to the `red' sequence, and is 
similar to that obtained by subtracting $-0.17$~mags from 
equation~(4) in Skibba \& Sheth (2009); it is shallower than 
equation~(7) in Skibba \& Sheth (2009) or equation~(1) of 
Young et al. (2009) (note that we $k$-correct to $z=0$ and we use 
$h=0.7$).  
`Red' galaxies are those which lie above this line; `blue' lie below it.  

\begin{figure*}
 \centering
 \includegraphics[width=0.45\hsize]{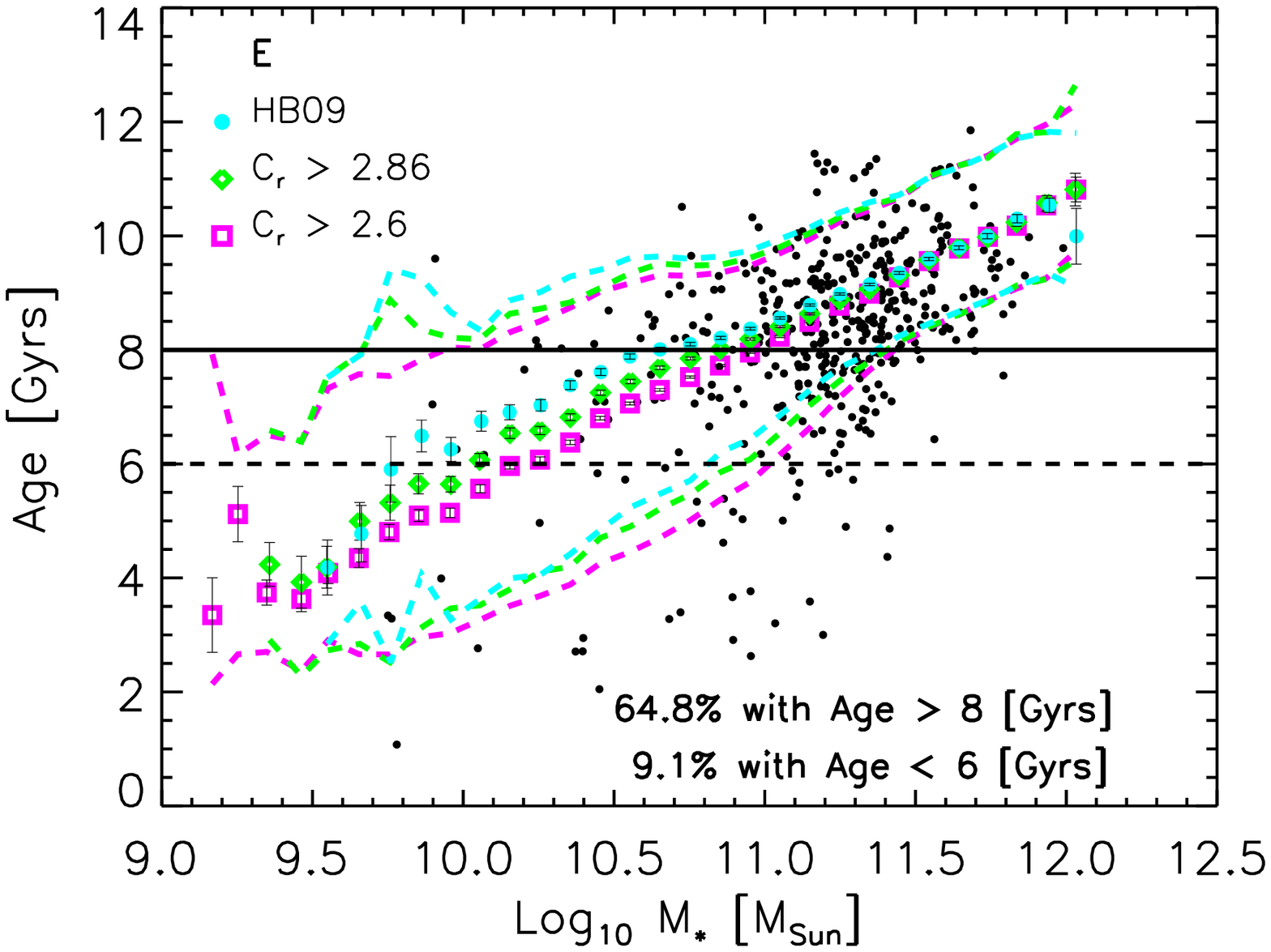}
 \includegraphics[width=0.45\hsize]{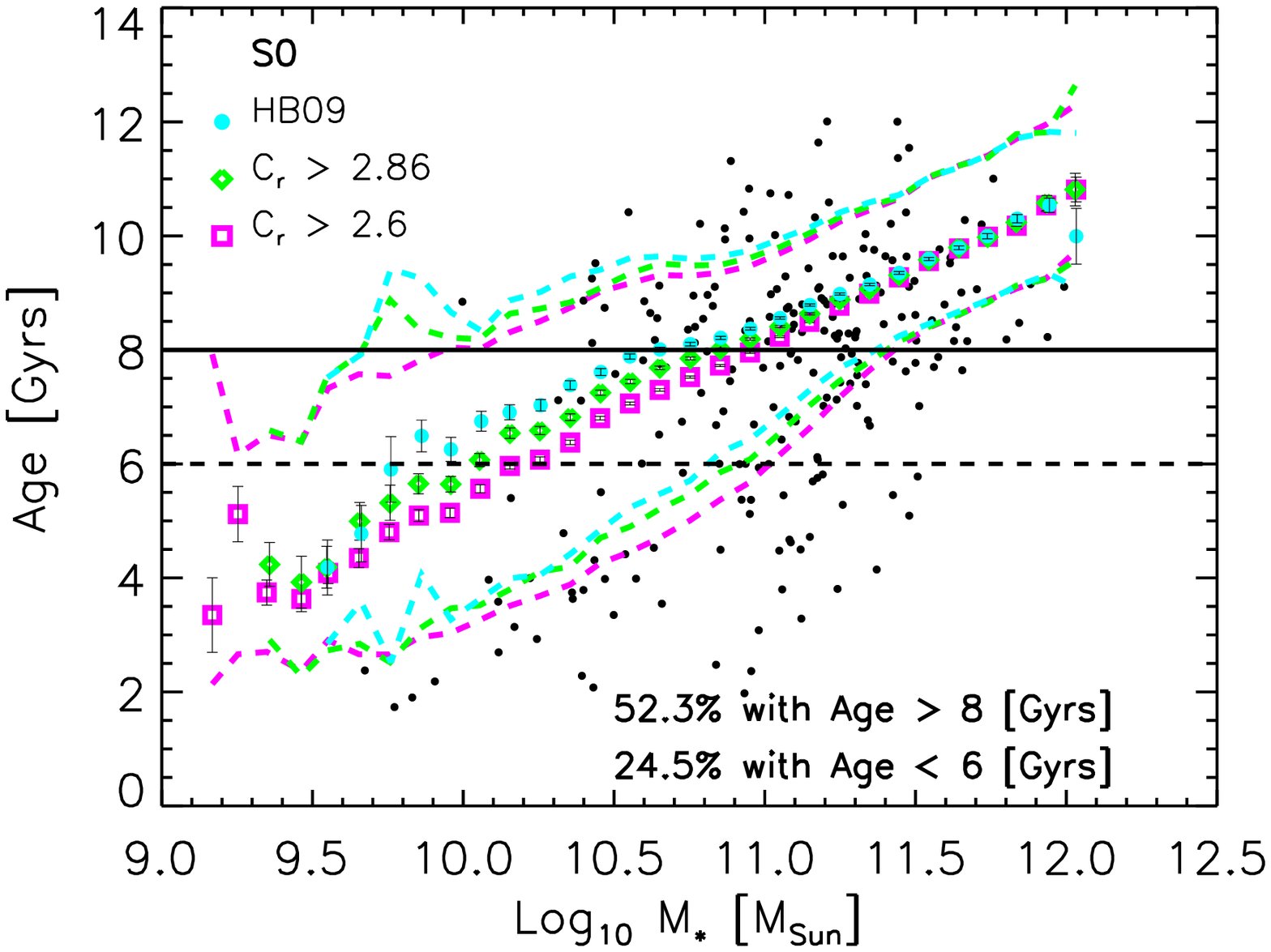}
 \includegraphics[width=0.45\hsize]{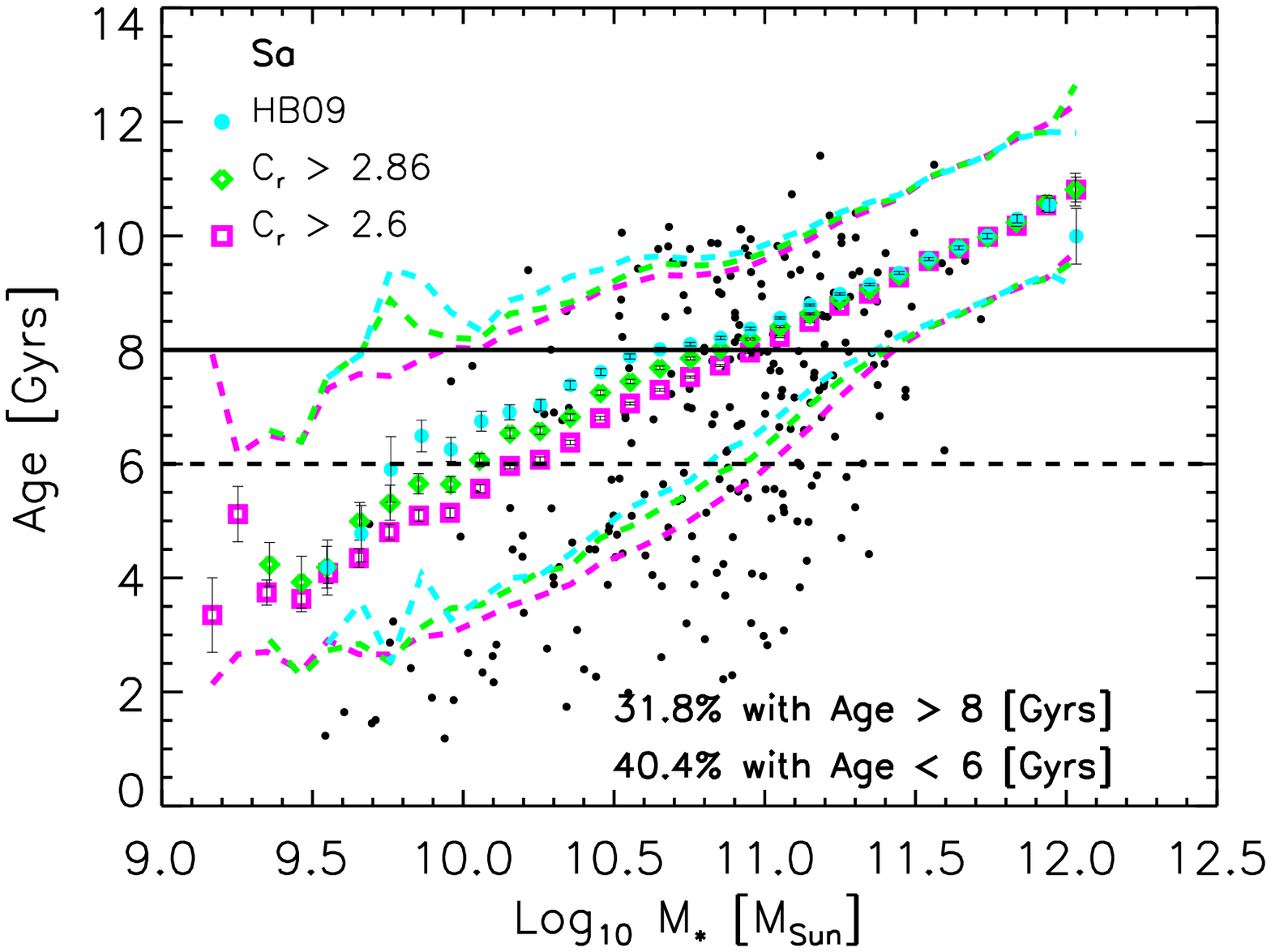}
 \includegraphics[width=0.45\hsize]{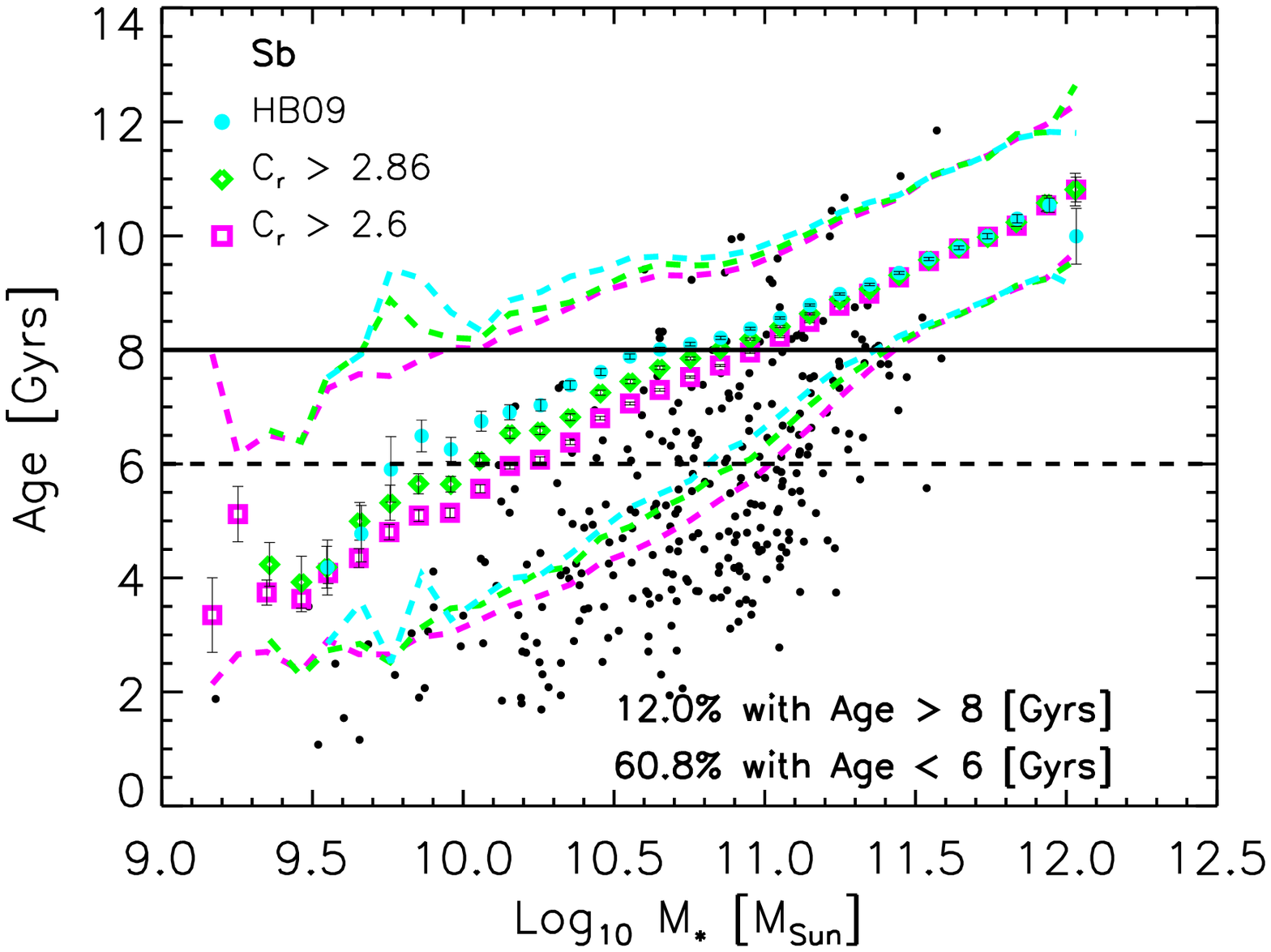}
 \includegraphics[width=0.45\hsize]{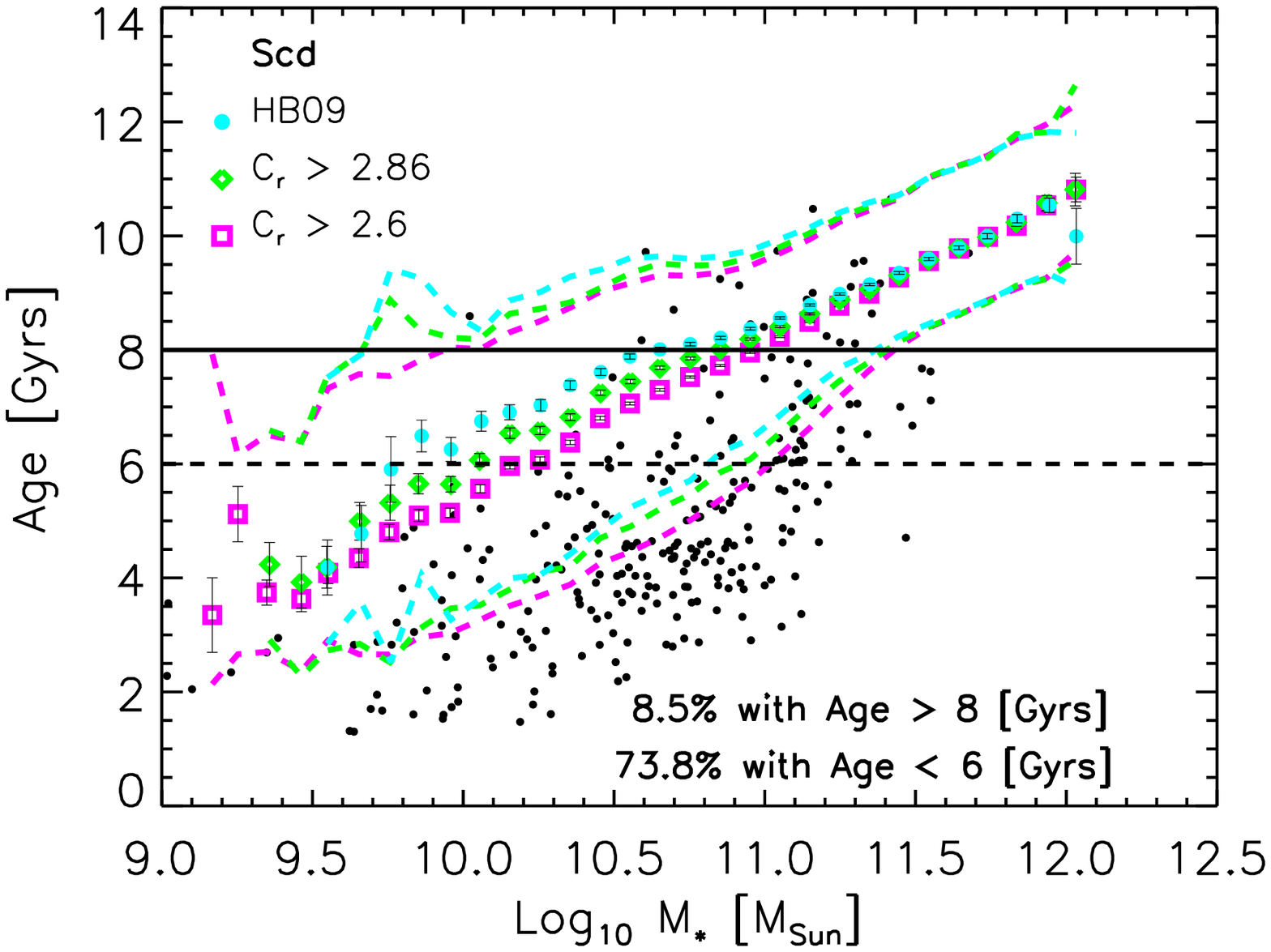}
 \includegraphics[width=0.45\hsize]{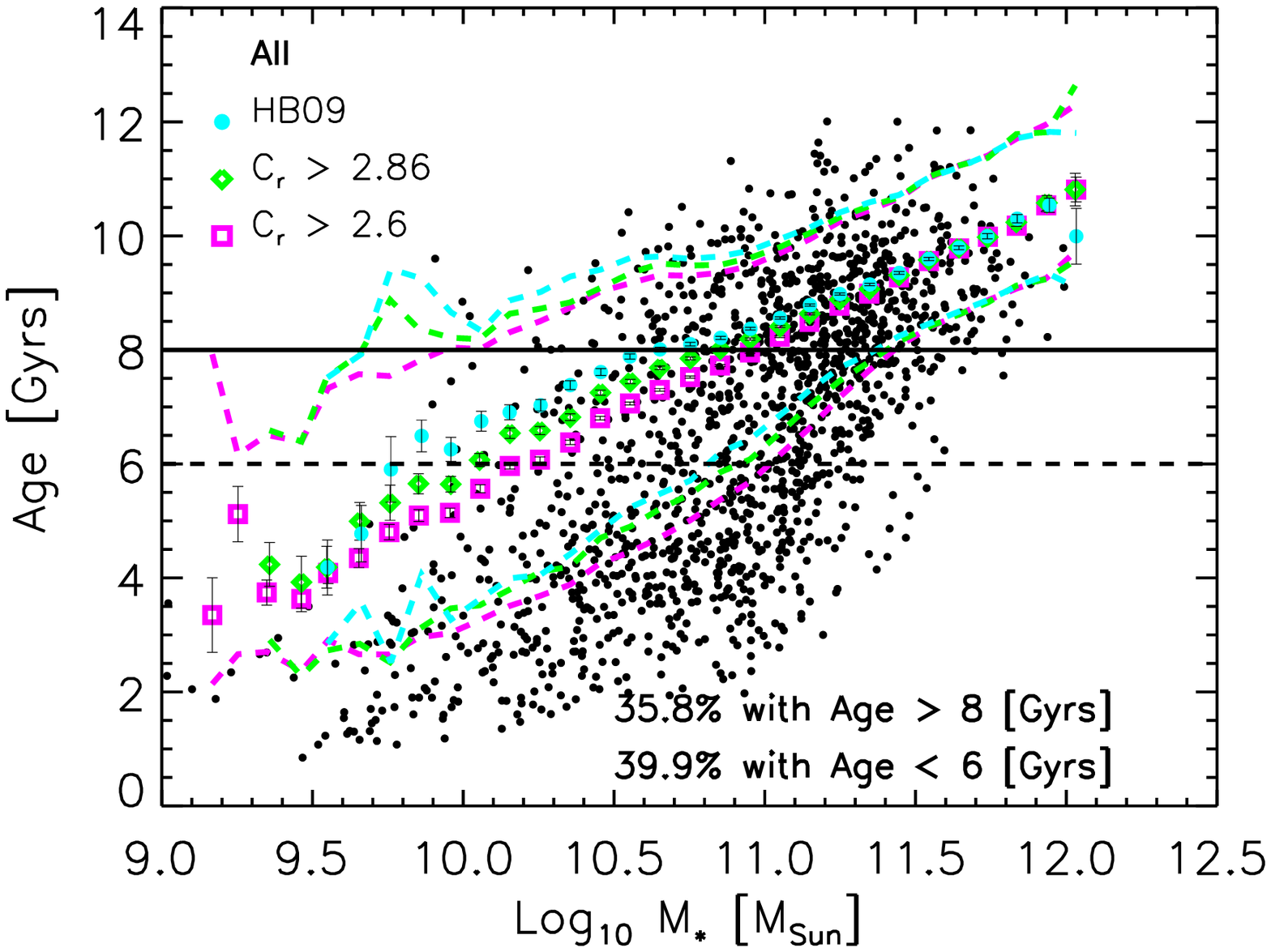}
 \caption{Joint distribution of age and stellar mass in the 
          Fukugita et al. (2007) sample.  Cyan filled circles, green diamonds 
          and magenta squares show the median age at fixed stellar mass
          for a subsample selected following Hyde \& Bernardi (2009), 
          a subsample with $C_r > 2.86$, and a subsample with 
          $C_r > 2.6$, respectively. The dashed lines show the 
          $1\sigma$ range around the median.}
 \label{agemorph}
\end{figure*}

\begin{figure*}
 \centering
 \includegraphics[width=0.7\hsize]{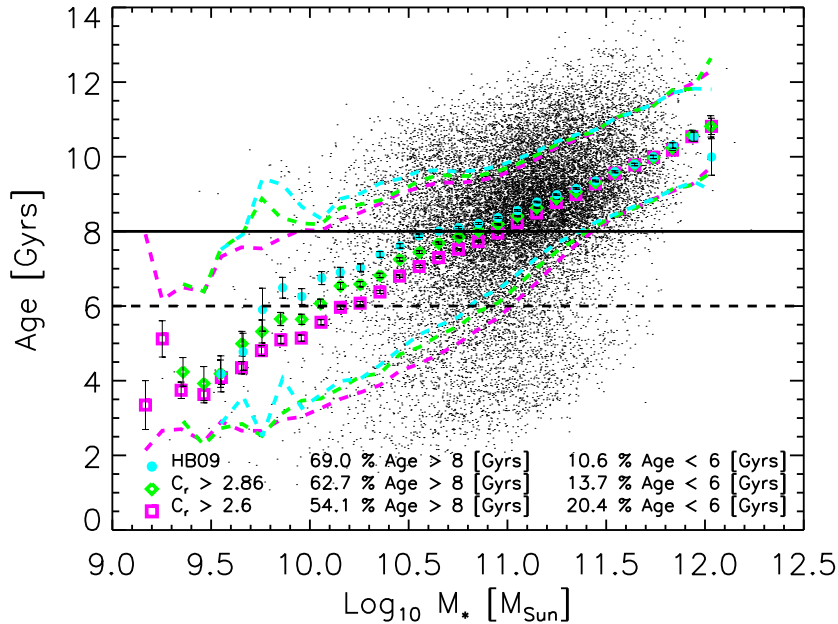}
 \caption{Joint distribution of age and stellar mass in the full sample. 
          Small dots show a random subsample of the galaxies selected following
          Hyde \& Bernardi (2009). Cyan filled circles, green diamonds 
          and magenta squares show the median age at fixed stellar mass
          for a subsample selected following Hyde \& Bernardi (2009), 
          a subsample with $C_r > 2.86$, and a subsample with 
          $C_r > 2.6$, respectively. The dashed lines show the 
          $1\sigma$ range around the median.
          The fraction of galaxies with luminosity weighted ages older 
          than 8~Gyrs and younger than 6~Gyrs, for each of the 
          selection methods, are shown. }
 \label{ageMs}
\end{figure*}

\begin{figure*}
 \centering
 \includegraphics[width=0.425\hsize]{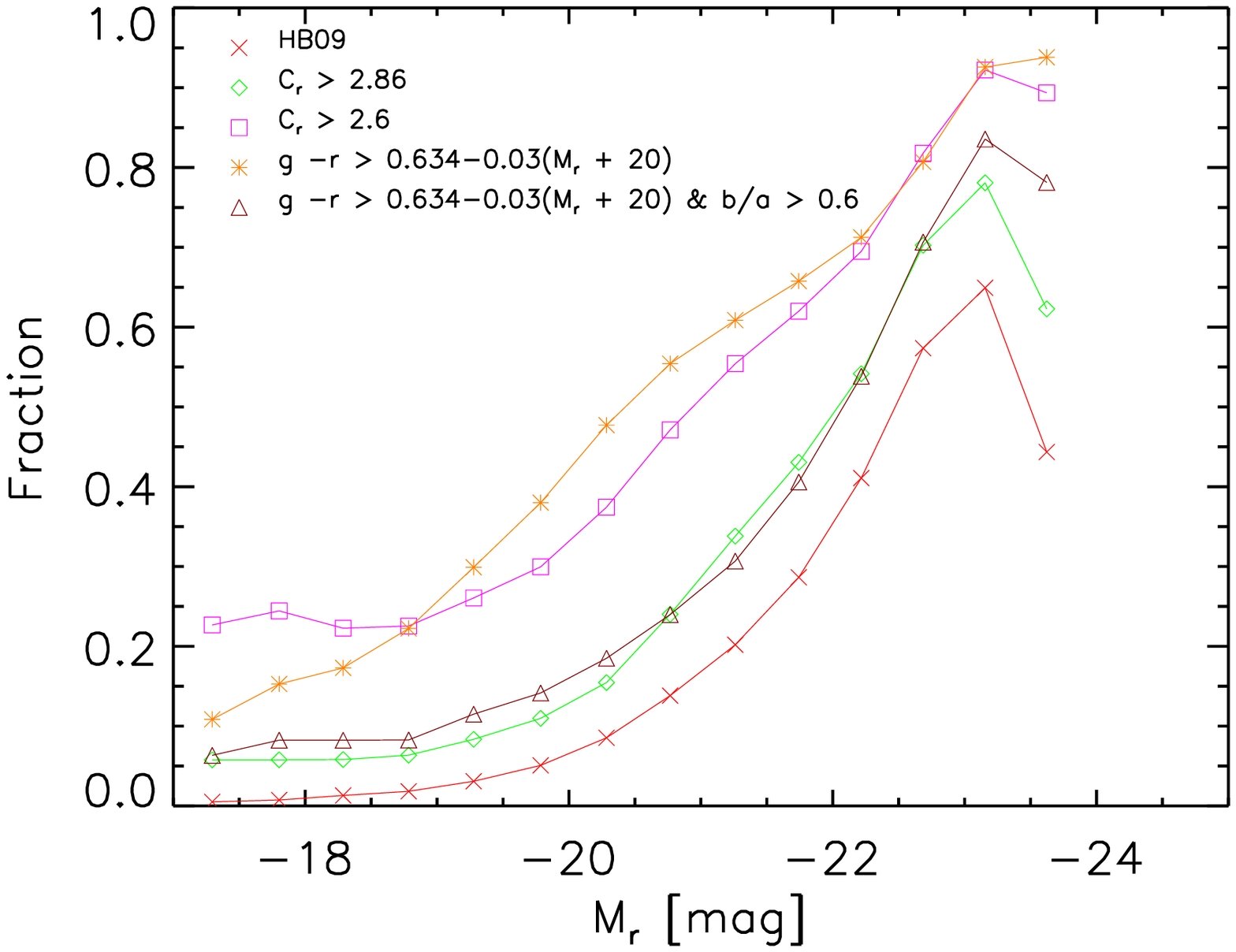}
 \includegraphics[width=0.425\hsize]{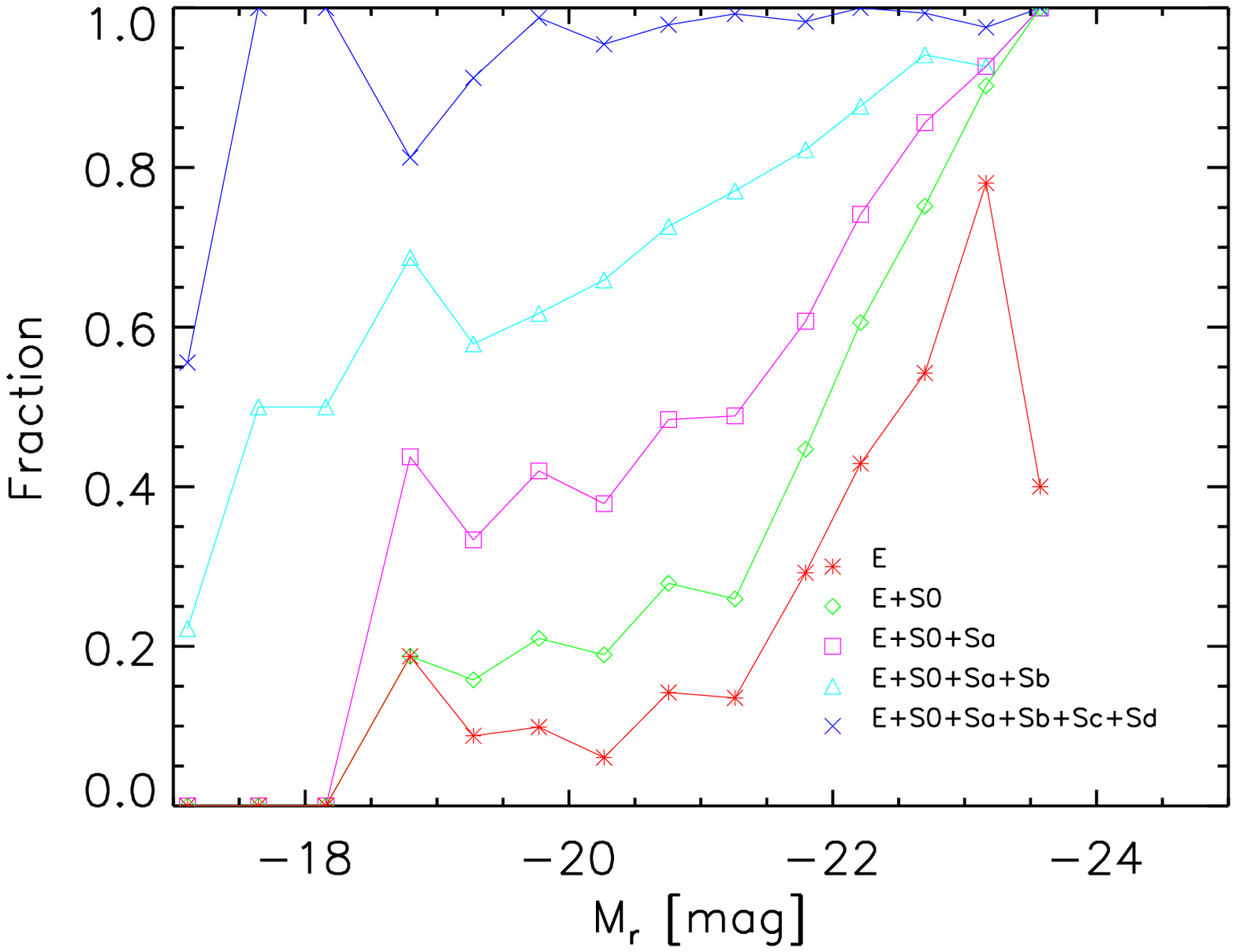}
 \includegraphics[width=0.425\hsize]{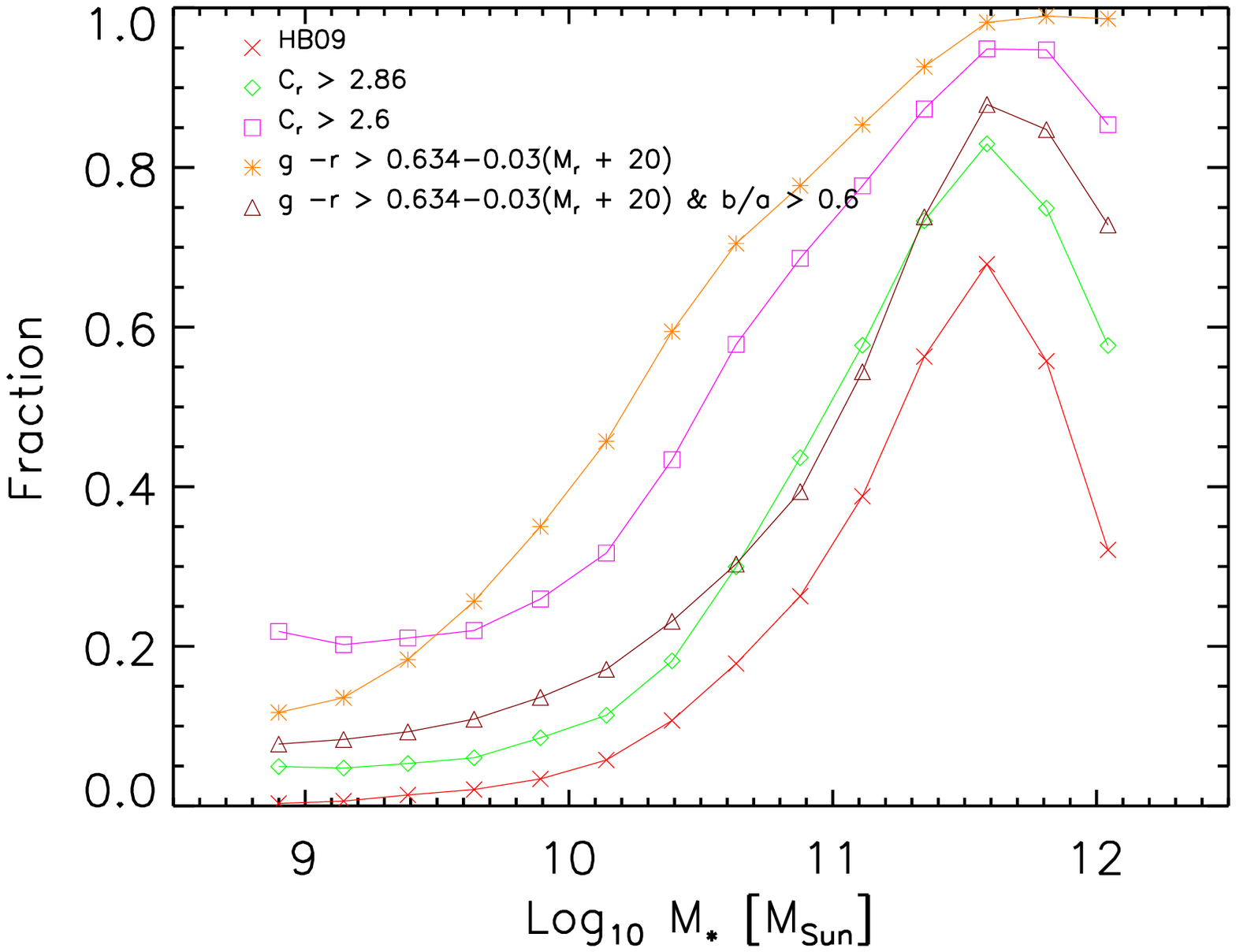}
 \includegraphics[width=0.425\hsize]{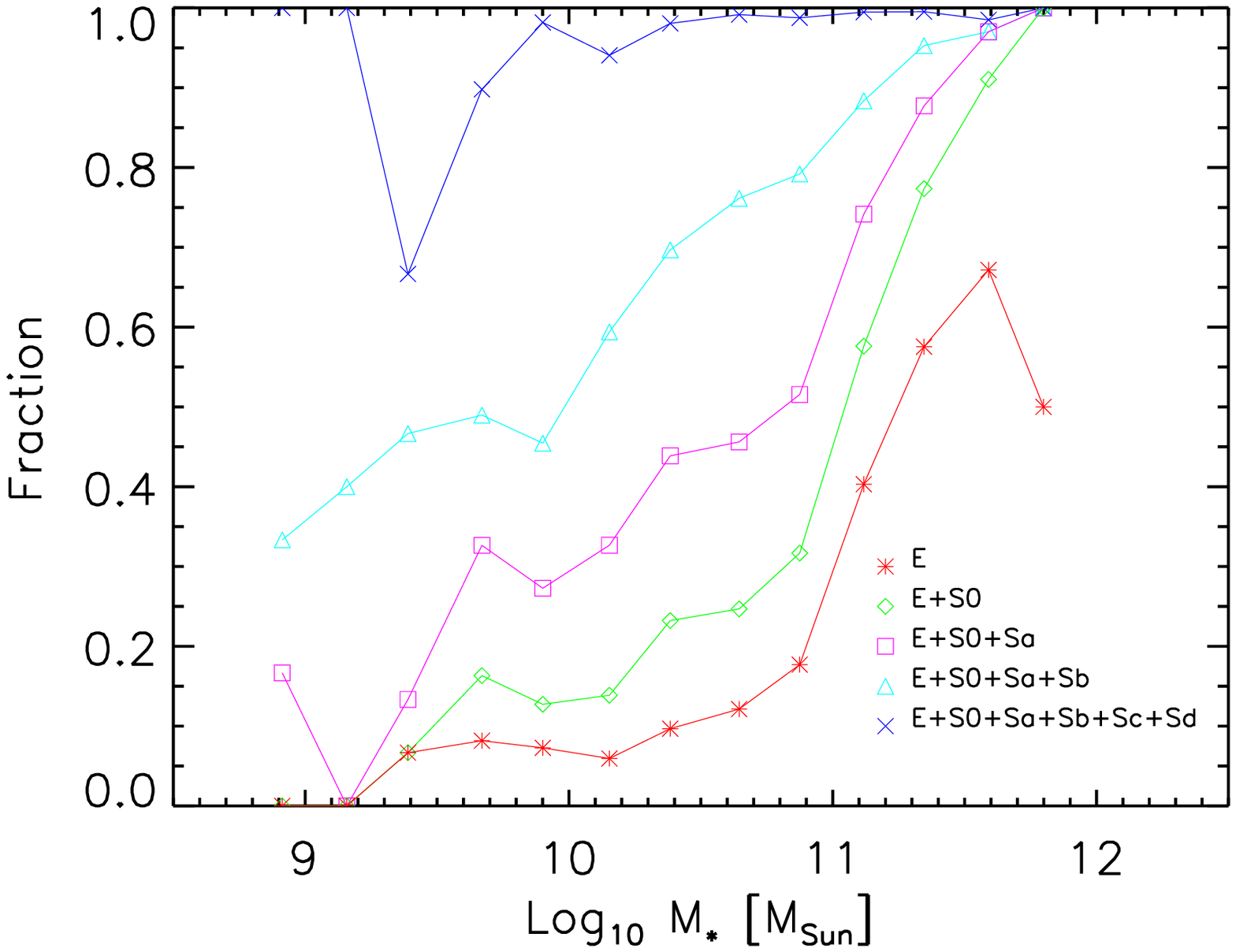}
 \includegraphics[width=0.425\hsize]{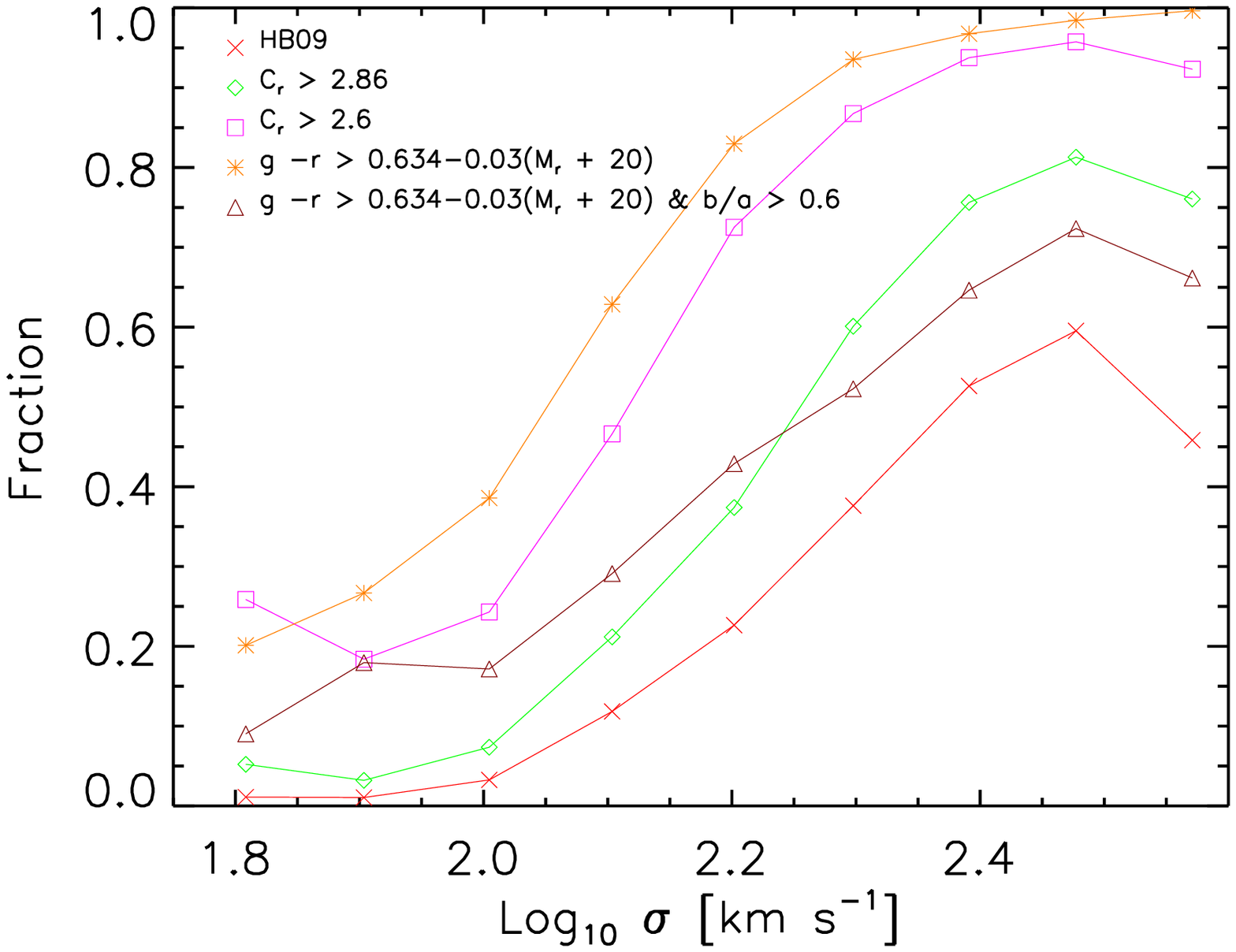}
 \includegraphics[width=0.425\hsize]{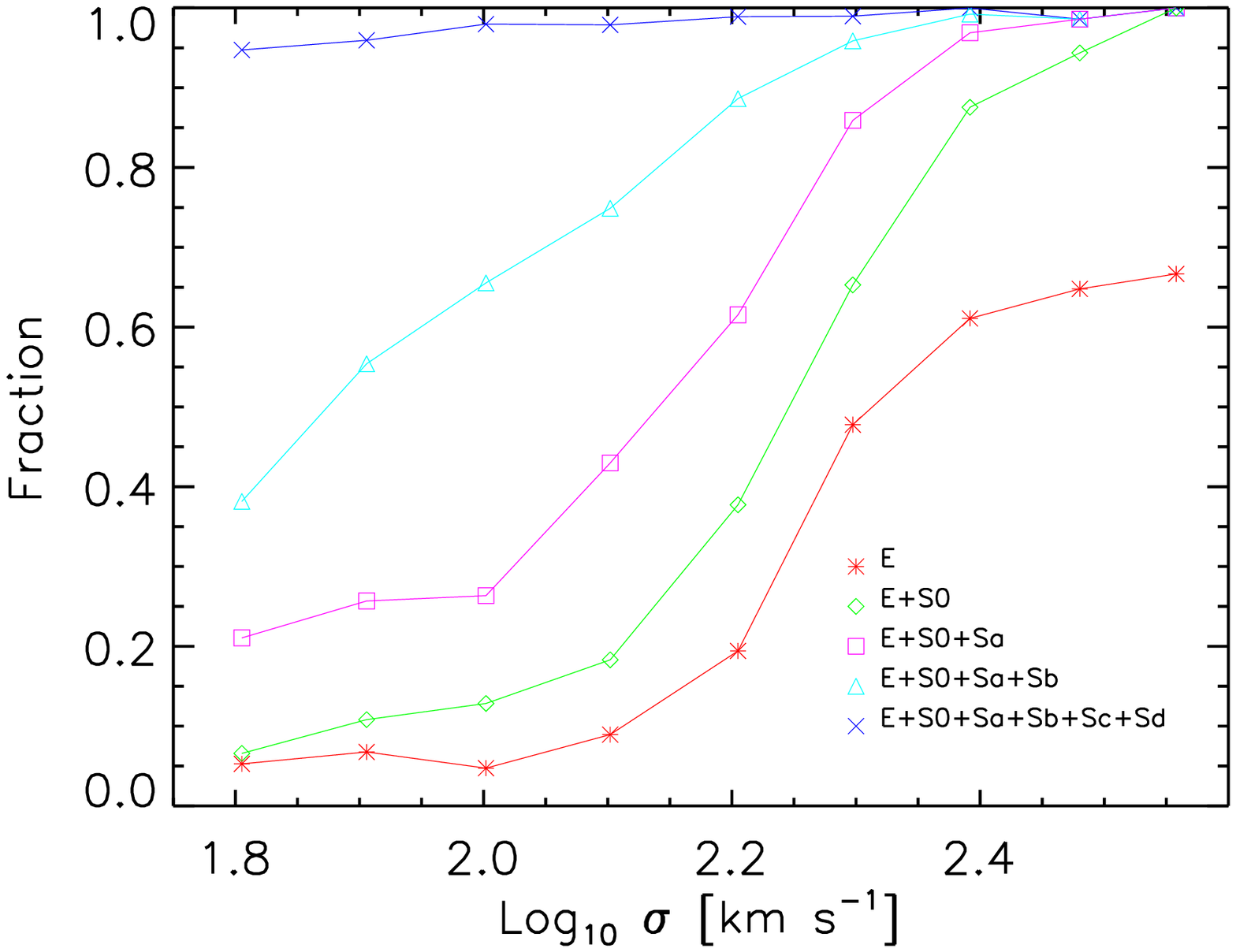}
 \caption{Left panels: The `red' fraction in the full sample 
  ($m_{r,{\tt Pet}}<17.5$), where being red means the object lies 
  redward of a luminosity dependent threshold which runs parallel to 
  the `red' sequence (orange stars); 
  or has concentration greater than 2.6 (magenta squares); 
  or greater than 2.86 (green diamonds); 
  or satisfied the Hyde \& Bernardi selection cuts (red crosses).  
  Also shown is the result of combining the color cut with one on 
  $b/a$ (brown triangles). 
  Right panels: The fraction of objects of a given $L$, 
  $M_*$, $\sigma$ and $R_e$, as later and later morphological types 
  are added to the Fukugita et al. sample ($m_{r,{\tt Pet}}<16$).  
  In all cases, each object was weighted by $V_{\rm max}^{-1}(L)$, 
  the maximum volume to which it could have been seen, given the
  apparent magnitude limit.}  
 \label{fredC}
\end{figure*}

\begin{figure*}
 \centering
 \includegraphics[width=0.425\hsize]{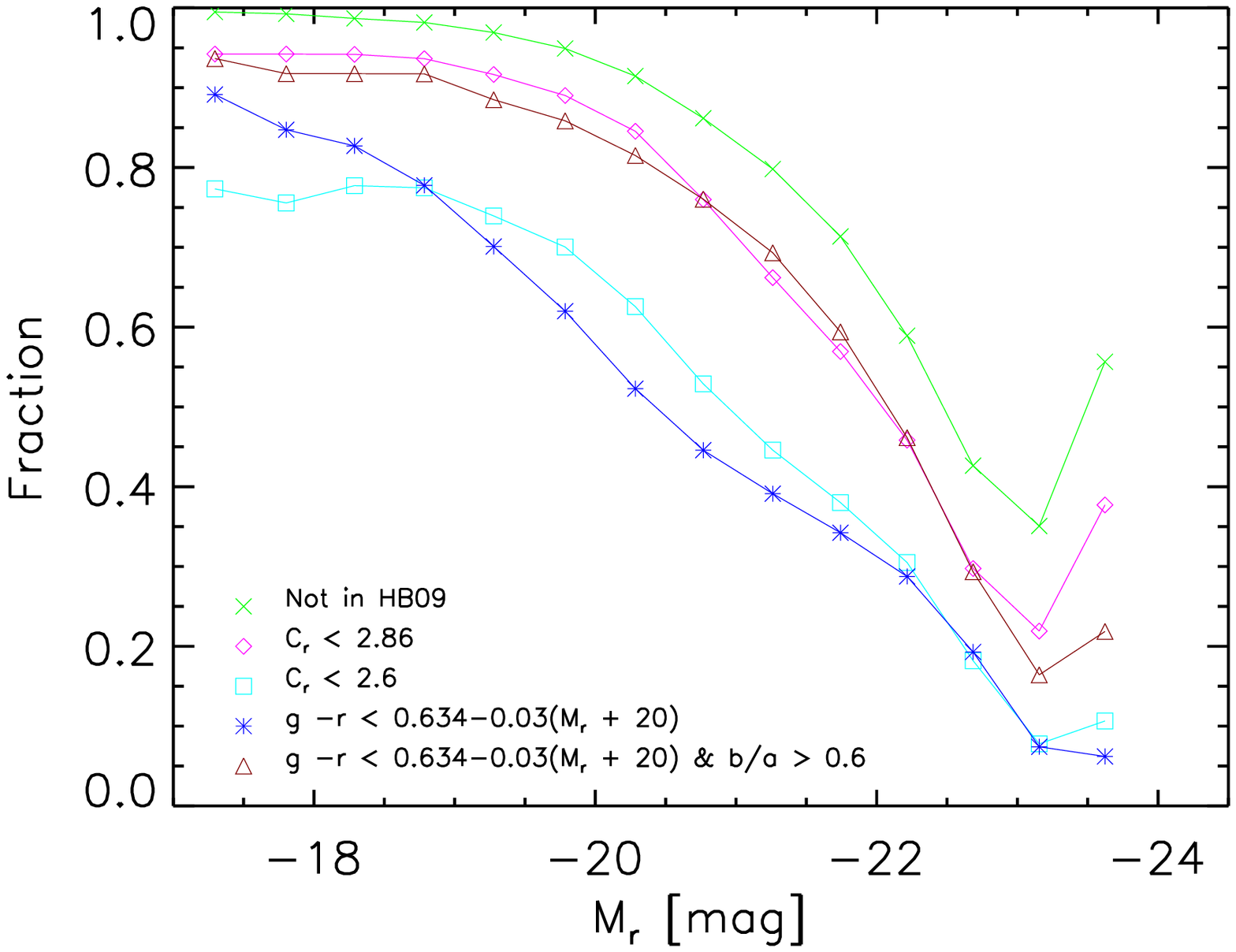}
 \includegraphics[width=0.425\hsize]{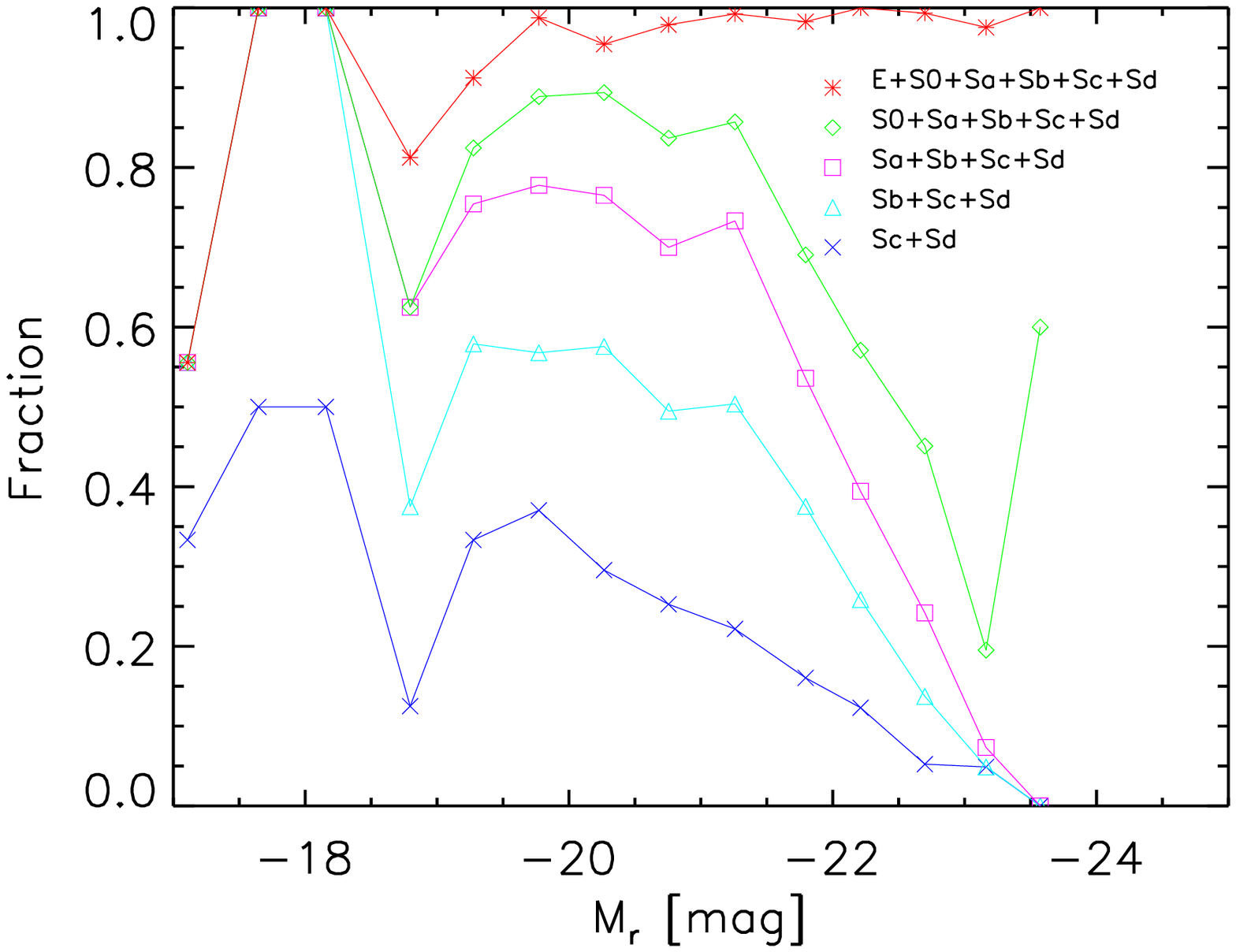}
 \includegraphics[width=0.425\hsize]{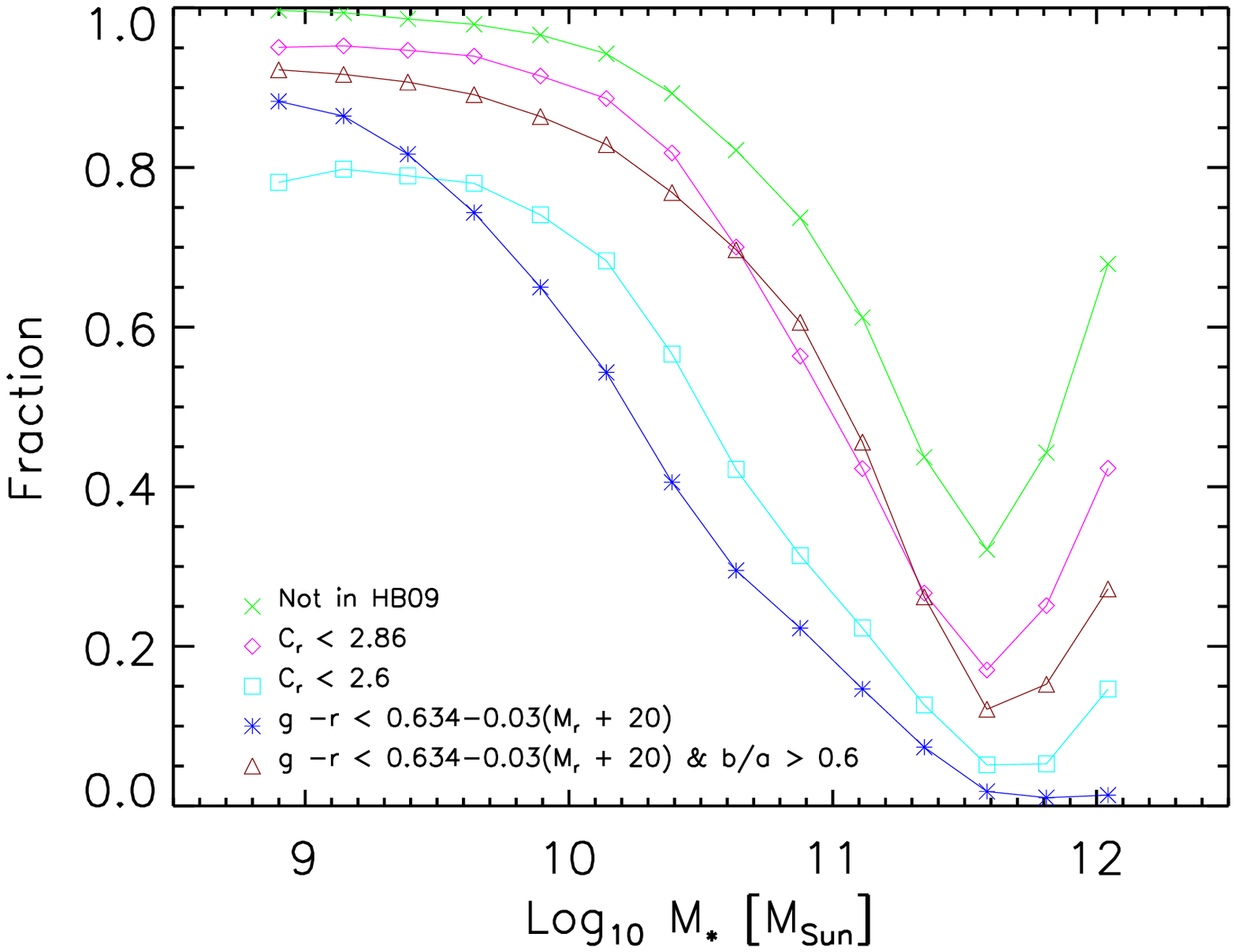}
 \includegraphics[width=0.425\hsize]{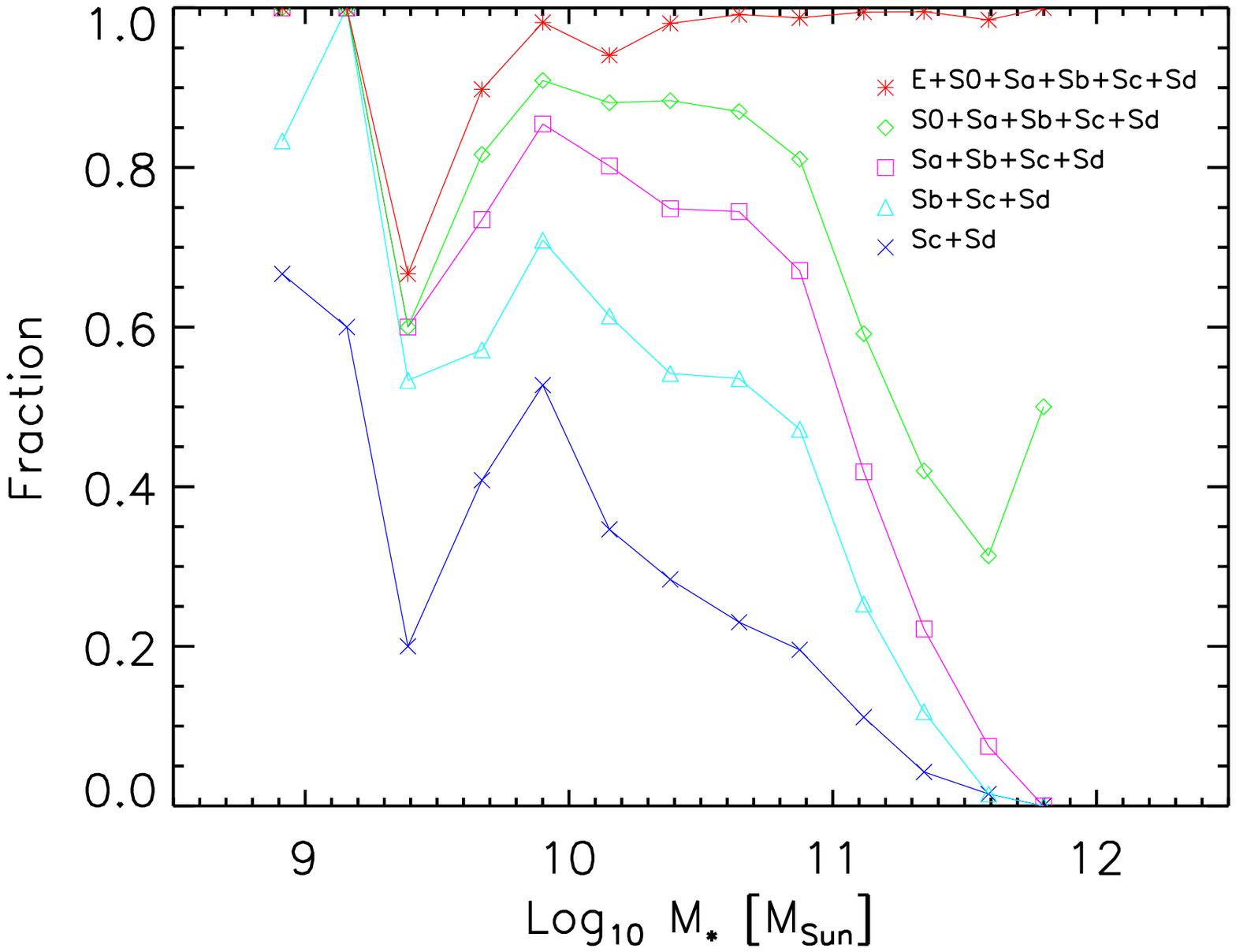}
 \includegraphics[width=0.425\hsize]{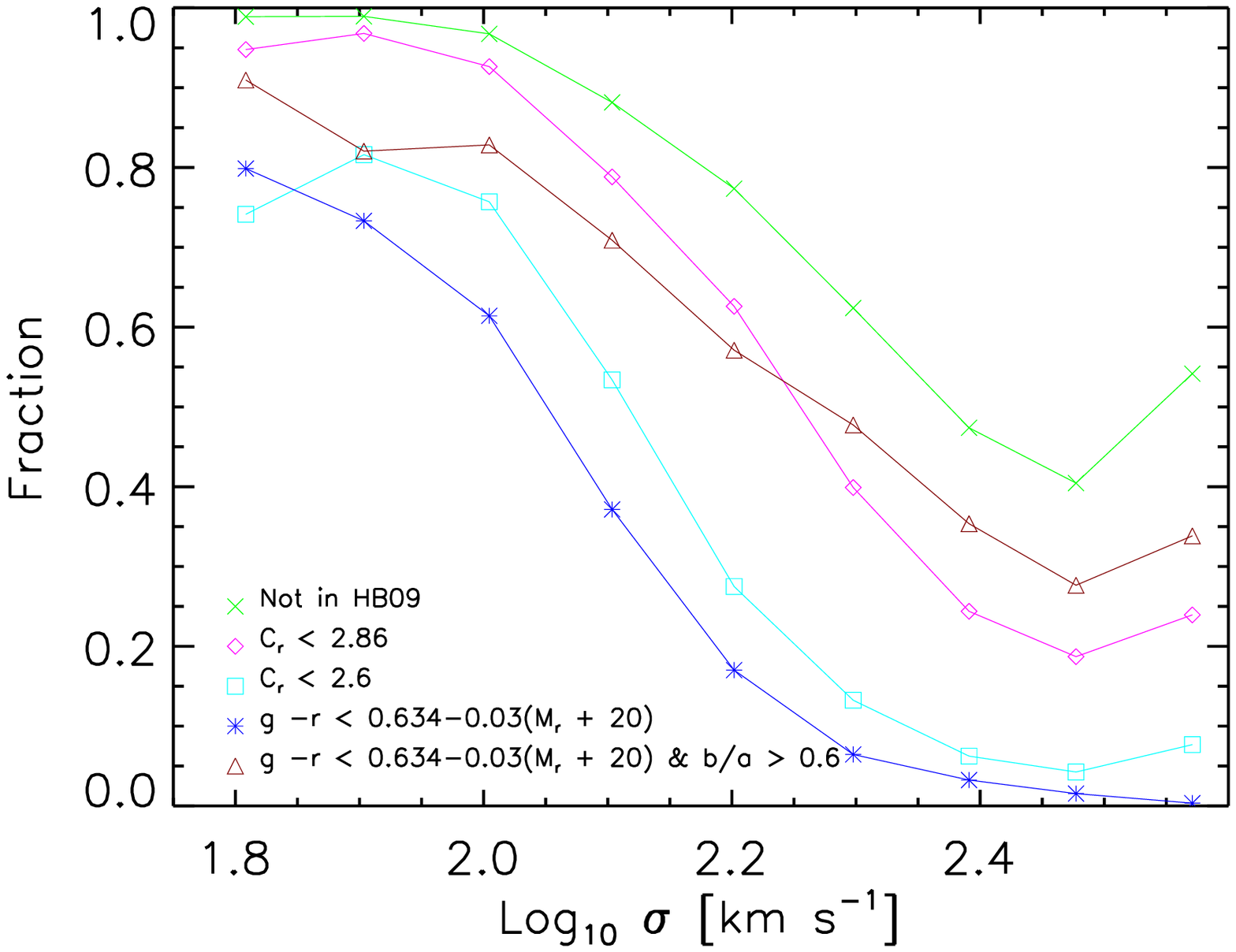}
 \includegraphics[width=0.425\hsize]{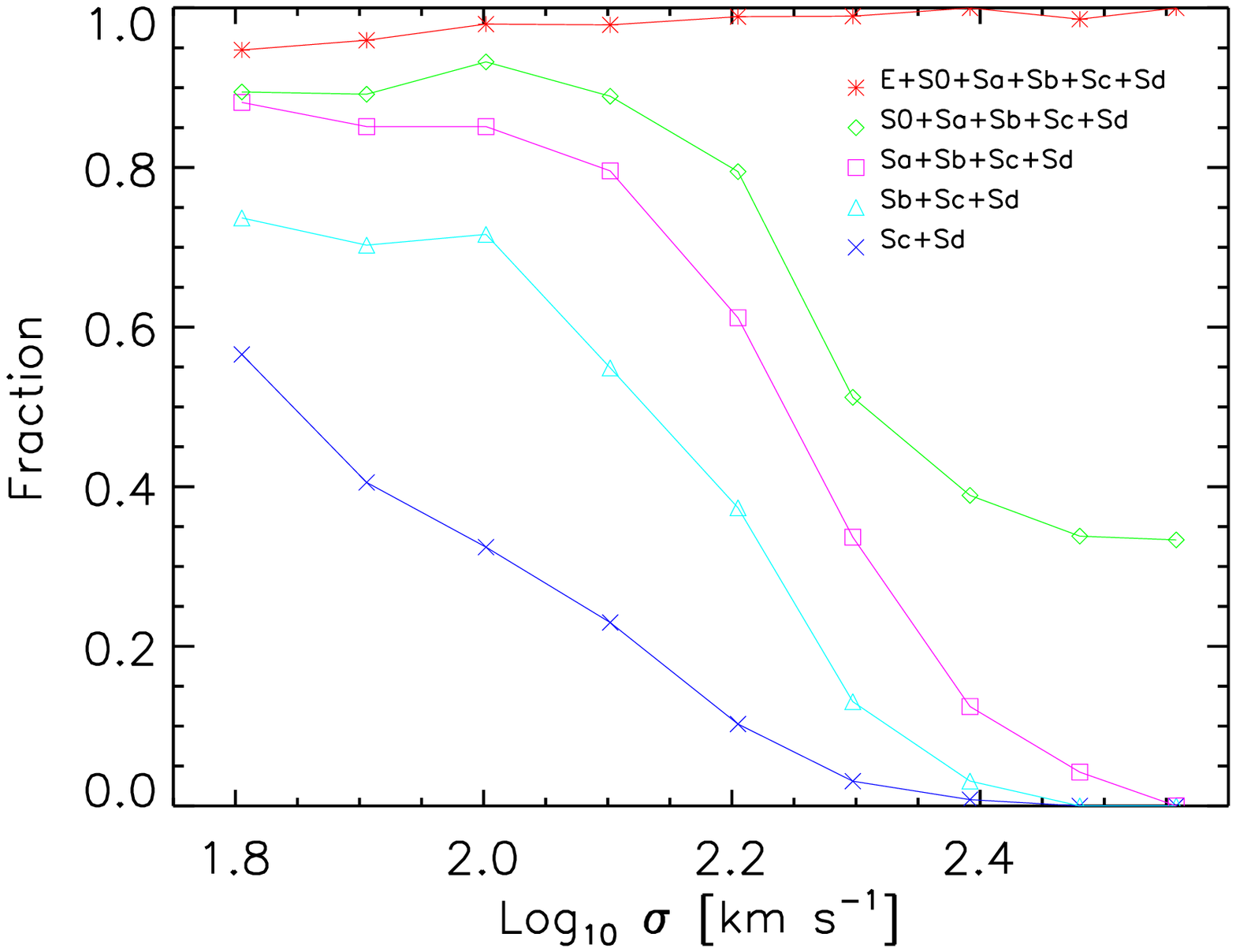}
 \caption{Left panels: Same as left hand panel of previous figure, but now 
  showing the `blue' fraction -- the objects which did not qualify 
  as being `red'.
  Right panels:  Similar to right hand panel of previous figure, but now 
  starting with later types and adding succesively more early-types.}
 \label{fblueC}
\end{figure*}

\begin{figure*}
 \centering
 \includegraphics[width=0.95\hsize]{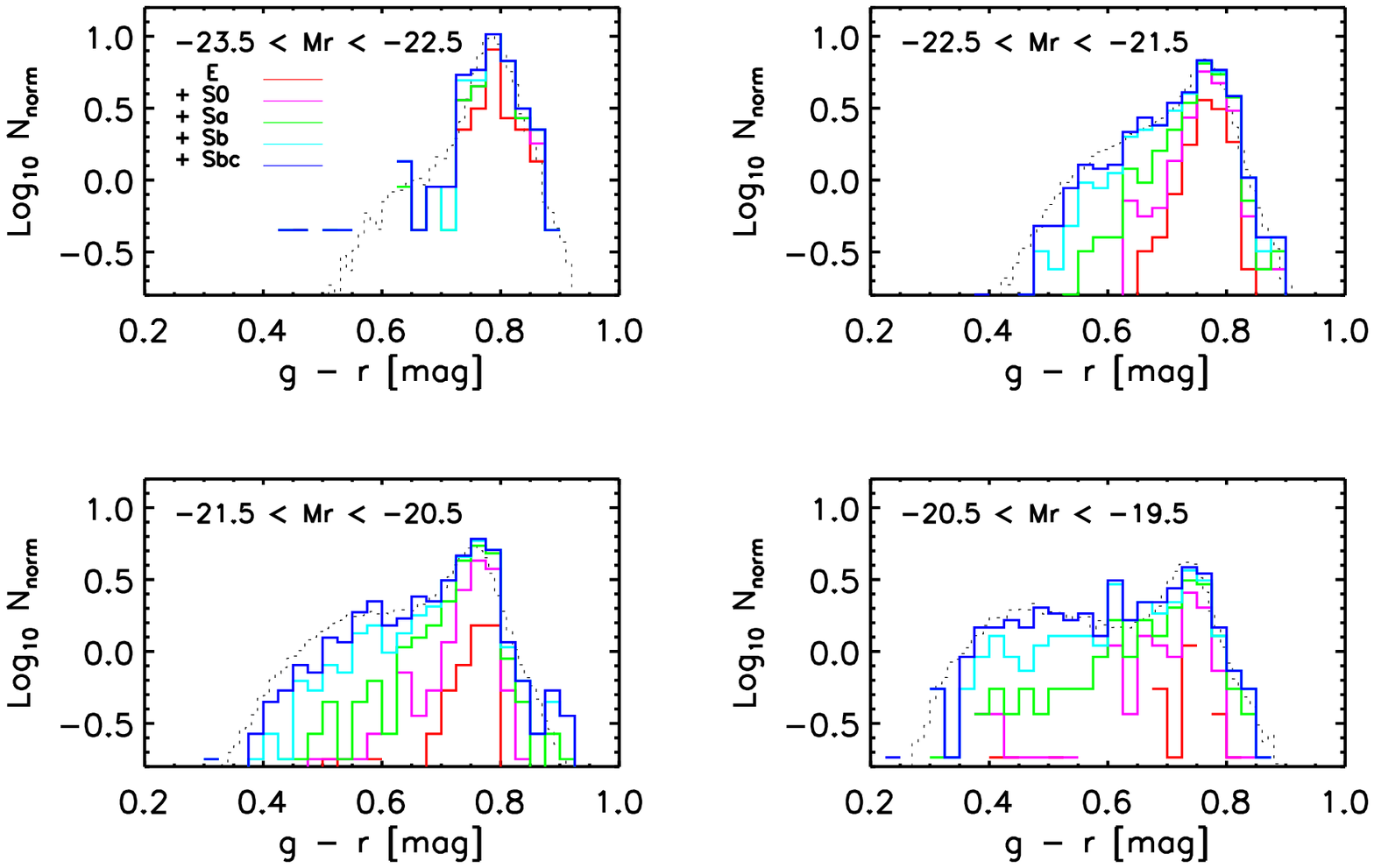}
 \caption{Depedence of the bimodality in the color distribution 
          on morphological type, for a few bins in luminosity.  
          Dashed histogram shows the distribution of the SDSS sample
          ($m_{r,{\tt Pet}}<17.5$),
          while solid histograms show the distribution in the 
          Fukugita et al. sample ($m_{r,{\tt Pet}}<16$) as later and 
          later morphological types are added.
          Note that the reddest objects at intermediate luminosity 
          are late-type galaxies.}
 \label{gmrmorph}
\end{figure*}

However, note that many late types (Sb and later) lie above this line -- 
these tend to be edge-on disks.  
In addition, some Es lie below it.  (See Huang \& Gu 2009 for a more 
detailed analysis of such objects, which show either a star forming, 
AGN or post-starburst spectrum.)  
We intend to present a more detailed study of the morphological 
dependence of the color-magnitude, stellar mass and velocity dispersion 
relations in a future paper (Bernardi et al. 2009, in preparation).  
For our purposes here, we simply note that 
Figures~\ref{gmrType} and~\ref{gmrTypeMs} illustrate that cuts in color 
are not a good way to select early-type galaxies.

\subsection{Ages}
Later in this paper, we will also study correlations between 
stellar age and galaxy mass, size and morphology.  The age estimates 
we use, from Gallazzi et al. (2005), are based on a detailed analysis 
of spectral features.  Since this is the same analysis that provided 
$M_{*{\rm Gallazzi}}$, errors in age and stellar mass are correlated 
(see Bernardi 2009 for a detailed discussion):  erroneously large 
$M_*$ will tend to have erroneously large age as well.  

Figure~\ref{agemorph} shows the age-$M_*$ correlation for the objects 
in the Fukugita et al. sample for which age estimates are available.  
This shows that, for any given morphological type, massive galaxies 
tend to be older (this correlation is not due to correlated errors).  
However, as expected, the later types tend to be substantially younger:  
Whereas two-thirds of the Es in this sample are older than 8~Gyrs, 
only half of S0s, one-quarter of Sas, and fewer than 10~\% of the 
later types (Sb, Sc, etc.) are this old.  The bottom right panel 
suggests that the age-$M_*$ distribution separates into two 
populations -- one which is younger than about 7 Gyrs and another 
which is older.  However, this is not simply correlated with 
morphological type:  the top panels shows that this bimodality is 
also present in the E and S0s samples.  

To study this further Figure~\ref{ageMs} shows the age-$M_*$ 
distribution in a random subsample of the galaxies selected 
following Hyde \& Bernardi (2009) from the full SDSS catalog.  
Note that although 90\% of the objects are older than 6~Gyrs, 
this selection clearly includes a population of younger objects.  
This population of `rejuvenated' early-type galaxies has been the 
subject of some recent interest (e.g. Huang \& Gu 2009; Thomas et al. 2010).  
Cyan filled circles show that the median age increases with stellar 
mass and dashed lines show the $1\sigma$ range around the median.  
Green diamonds and magenta squares show the median age at fixed 
stellar mass for objects with $C_r > 2.86$ and $C_r > 2.6$, 
respectively.  At large $M_*$, both these 
samples produce similar age-$M_*$ relations to the Hyde-Bernardi 
sample; at smaller $M_*$, allowing smaller $C_r$ includes younger 
galaxies.  These median relations are superimposed on the 
panels of Figure~\ref{agemorph}.

\subsection{The red and blue fractions}\label{redblue}
There is considerable interest in the `build-up' of the red sequence, 
and the possibility that some of the objects in the blue cloud were 
`transformed' into redder objects.  
We now compare estimates of the red or blue fraction that are 
based on color and concentration, with ones based on morphology.  
In particular, we show how these fractions vary as a function of 
$L$, $M_*$, and $\sigma$.

All the results which follow are based on samples which are 
Petrosian magnitude limited, so in all the statistics we present, 
each object is weighted by $V_{\rm max}^{-1}(L)$, the inverse of 
the maximum volume to which it could have been seen.  This magnitude 
limit is fainter for the full sample ($m_{\tt Pet}=17.75$) than it is 
for the Fukugita et al. subsample ($m_{\tt Pet}=16$), and note that 
$V_{\rm max}$ depends on our model for how the luminosities evolve 
(absolute magnitudes brighten as $1.3z$).  

The panels on the left of Figure~\ref{fredC} show how the mix of 
objects changes as a function of luminosity, stellar mass, and velocity 
dispersion, for the crude but popular hard cuts in concentration and 
color (as described in the the previous sections).

Figure~\ref{fredC} shows that the fraction of objects which satisfies 
the criteria used by Hyde \& Bernardi (2009) increases with increasing 
$L$, $M_*$ and $\sigma$, except at the largest values (about which, 
more later).  Requiring $C_r\ge 2.86$ instead results in approximately 5\% 
to 10\% more objects (compared to the Hyde \& Bernardi cuts) at 
each $L$ or $M_*$; although, in the case of $\sigma\sim 300$~km~s$^{-1}$, 
this cut allows about 20\% more objects.  Relaxing the cut to 
$C_r\ge 2.6$ allows an additional 15\%, with slightly more at 
intermediate $L$ and $M_*$. 

Selecting objects redder than a luminosity dependent threshold 
(equation~\ref{redcut}) which runs parallel to the `red' sequence 
allows even more objects into the sample, but combining the color cut 
with one on $b/a$ reduces the sample to one which resembles $C_r\le 2.86$ 
rather well.  The cut in $b/a$ is easy to understand, since edge-on discs 
will lie redward of the color cut even though they are not 
early-types -- the additional cut on $b/a$ is an easy (but rarely 
used!) way to remove them.

It is interesting to compare these panels with their counterparts 
on the right of Figure~\ref{fredC}, in which later and later 
morphological types are added to the Fukugita et al. subsample which 
initially only contains Es.  
This suggests that the Hyde \& Bernardi selection will be dominated 
by E, $C_r\ge 2.86$ will be dominated by E+S0s, 
and $C_r\ge 2.6$ will be dominated by E+S0+Sas.  We quantify this in the next 
subsection.  
Figure~\ref{fblueC} shows a similar comparison with the blue fraction.  
Note that the contamination of the red fraction by edge-on discs is 
a large effect:  60\% of the objects at $\log_{10}(M_*/M_\odot) = 10.5$ 
are classified as being red, when E+S0+Sas sum to only 40\%.  
Figure~\ref{gmrmorph} shows this more directly:  
the reddest objects at intermediate luminosities are late-, not 
early-type galaxies.  

Before concluding this section, we note that both concentration 
cuts greatly underpredict the red fraction of the most luminous or 
massive objects, as does the application of a $b/a$ cut to the 
straight color selection or the Hyde \& Bernardi selection. 
The most luminous or massive elliptical galaxies in the
Fukugita et al. sample show the same behavior. 
I.e., the most massive objects are less concentrated for 
their luminosities than one might have expected by extrapolation 
from lower luminosities and their light
profile is not well represented by a pure deVaucoleur law.
This is consistent with results in the previous section where,
at the highest luminosities, $b/a$ tends to decrease with 
luminosity (Figure~\ref{bamorph}).  
These trends suggest an increasing incidence of recent radial 
mergers for the most luminous and massive galaxies.

\subsection{Distribution of morphological types in differently 
            selected samples}\label{puritysel}

\begin{table}
\caption{The morphological mix in differently selected samples, 
         from $V_{\rm max}^{-1}$ weighted counts in 
         the Fukugita et al. (2007) sample restricted to $M_r<-19$; 
         numbers in brackets are from the raw counts. }
\begin{tabular}{lccc}
 \hline
 Type & HB09 & $C_r>2.86$ & $C_r>2.6$ \\
 \hline
 E   & 0.69 (0.73) & 0.38 (0.51) & 0.26 (0.43) \\
 S0  & 0.23 (0.20) & 0.22 (0.23) & 0.20 (0.22) \\
 Sa  & 0.07 (0.06) & 0.25 (0.17) & 0.30 (0.20) \\ 
 Sb  & 0.01 (0.01) & 0.12 (0.07) & 0.19 (0.12) \\
 Scd &     -       & 0.03 (0.02) & 0.05 (0.03) \\
 \hline
\end{tabular}
\label{purity}
\end{table}

Much of the previous analysis suggests that the Hyde \& Bernardi 
selection will produce a sample that is dominated by Es, 
$C_r\ge 2.86$ will include more S0s and Sas, and $C_r\ge 2.6$ 
will include Sas and later types.  
Table~\ref{purity} shows the distribution of types in subsamples 
selected from the Fukugita et al. (2007) catalog to have 
$C_r\ge 2.6$, $C_r\ge 2.86$ and following Hyde \& Bernardi (2009).  
Of the 1596 objects in the magnitude limited catalog, 
1009, 802 and 470 satisfy these cuts.  The Table shows that, 
in samples where $C_r\ge 2.6$, 54\% of the objects are Sa or later.  
This fraction falls to 40\% for $C_r\ge 2.86$ and to less than 10\% 
for the Hyde-Bernardi cuts (these numbers are obtained after 
weighting by $V_{\rm max}^{-1}$, so they do not depend on the 
selection effect associated with the apparent magnitude limit of 
the catalog).  

These numbers indicate that Es comprise at least two-thirds of the 
Hyde-Bernardi sample, but in a sample where $C_r\ge 2.86$, to reach 
this fraction one must include S0s, and if $C_r\ge 2.6$, then 
reaching this fraction requires adding Sas as well.  
Stated differently, E's comprise more than two-thirds of a 
Hyde-Bernardi sample, but about one-third of a sample with 
$C_r\ge 2.86$ and one quarter of a sample with $C_r\ge 2.6$.  
If we weight each object by its stellar mass, then 
(E+S0)s account for (72+21)\% of the total stellar mass in a 
Hyde \& Bernardi sample, (47+23)\% in a sample with $C_r\ge 2.86$, 
and (39+23)\% if $C_r\ge 2.6$. 
These differences will be important in Section~\ref{ageRe}.  

\section{Distributions for samples cut by morphology or concentration}
\label{DFs}

We now show how the luminosity, stellar mass, size and 
velocity dispersion distributions --
 $\phi(L)$, $\phi(M_*)$, $\phi(R_e)$ and $\phi(\sigma)$ 
 --  depend on how the sample was defined.  We use the same 
popular cuts in concentration as in the previous Sections, 
$C_r\ge 2.86$ and $C_r\ge 2.6$, which we suggested might be similar 
to selecting early-type samples which, in addition to Es, include
S0s + Sas, and S0s + Sas + Sbs, respectively.  
We then make similar measurements in the Fukugita et al. subsample, 
to see if this correspondence is indeed good.  

In this Section, we use {\tt cmodel} rather than Petrosian quantities, 
for the reasons stated earlier.  
The only place where we continue to use a Petrosian-based quantity is 
when we define subsamples based on concentration, since $C_r$ is 
the ratio of two Petrosian-based sizes, {\rm or for comparison 
with results from previous work}.

The results which follow are based on samples which are 
Petrosian magnitude limited, so in all the statistics we present, 
each object is weighted by $V_{\rm max}^{-1}(L)$, the inverse of 
the maximum volume to which it could have been seen.  In addition 
to depending on the magnitude limit ($m_{\tt Pet}\le 17.5$ for the 
full sample, and $m_{\tt Pet}\le 16$ for the Fukugita subsample), 
the weight $V_{\rm max}^{-1}(L)$ also depends on our model for how 
the luminosities have evolved.  A common test of the accuracy of the 
evolution model is to see how $\langle V/V_{\rm max}\rangle$, the 
ratio of the volume to which an object was seen to that which it 
could have been seen, averaged over all objects, differs from 0.5.  
In the full sample, it is $0.506$ for our assumption that the 
absolute magnitudes evolve as $1.3z$; had we used $1.62z$ 
(Blanton et al. 2003), it would have been $0.509$.  On the other
hand, if we had ignored evolution entirely, it would have been 0.527.

In addition, SDSS fiber collisions mean that spectra were not taken 
for about 7\% of the objects which satisfy $m_{\tt Pet}\le 17.5$.  
We account for this by dividing our $V_{\rm max}^{-1}$ weighted 
counts by a factor of 0.93.  This ignores the fact that fiber 
collisions matter more in crowded fields (such as clusters); so 
in principle, this correction factor has some scatter, which may 
depend on morphological type.  
We show below that, when we ignore this scatter, then our 
analysis of the full sample produces results that are in good 
agreement with those of Blanton et al. (2003), who account for 
the fact that this factor varies spatially.  This suggests that 
the spatial dependence is small, so, in what follows, we ignore 
the fact that it (almost certainly) depends on morphological type.  

\subsection{Parametric form for the intrinsic distribution}
We will summarize the shapes of the distributions we find by 
using the functional form 
\begin{equation}
 \phi(X)\,{\rm d}X = \phi_*\,\left(\frac{X}{X_*}\right)^\alpha\, 
                     \frac{{\rm e}^{-(X/X_*)^\beta}}{\Gamma(\alpha/\beta)}\,
                     \beta\,\frac{{\rm d}X}{X}.
 \label{phiX}
\end{equation}
This is the form used by Sheth et al. (2003) to fit the distribution 
of velocity dispersions; it is a generalization of the Schechter 
function commonly fit to the luminosity function (which has $\beta=1$, 
a slightly different definition of $\alpha$).  
We have found that the increased flexibility which $\beta\ne 1$ allows 
is necessary for most of the distributions which follow.  This is 
not unexpected:  at fixed luminosity, most of the observables we 
study below scatter around a mean value which scales as a power-law 
in luminosity (e.g. Bernardi et al. 2003; Hyde \& Bernardi 2009).  
Because this mean does not scale linearly with $L$, and because the 
scatter around the mean can be significant, then if $\phi(L)$ is 
well-fit by a Schechter function, it makes little physical or 
statistical sense to fit the other observables with a Schechter 
function as well.  

\subsection{Effect of measurement errors}
In practice, we will also be interested in the effect of measurement 
errors on the shape of the distribution.  If the errors are Lognormal 
(Gaussian in $\ln X$) with a small dispersion, then the observed 
distribution is related to the intrinsic one by 
\begin{eqnarray}
 O\psi(O) &=& \int d\ln X\, X\phi(X)\,p(\ln O|\ln X) \nonumber\\
  &\approx& O\phi(O)\left(1 + \frac{\sigma_{\rm err}^2}{2}\, C\right)
  \label{psiO}\\
 C &=& \Bigl(\alpha - \beta\, (O/O*)^\beta\Bigr)^2  - \beta^2 (O/O*)^\beta
 \label{errorC}
\end{eqnarray}
The peak of $X\phi(X)$ occurs at 
$X_{\rm max} = X_*\,(\alpha/\beta)^{1/\beta}$, where 
$C_{\rm max} = -\alpha\beta$.  Since $\alpha$ and $\beta$ are usually 
positive, errors typically act to decrease the height of the peak.   
Since the net effect of errors is to broaden the distribution, 
hence extending the tails, errors also tend to decrease $\beta$.  
The expression above shows that, in the $O/O_*\gg 1$ tail, errors 
matter more when $\beta$ is large.  

Fitting to equation~\eqref{psiO} rather than to equation~\eqref{phiX} 
is a crude but effective way to estimate the intrinsic shape (i.e., 
to remove the effect of measurement errors on the fitted 
parameters), provided the measurement errors are small.  
The rms errors on 
$(\ln L_r, \ln M_*, \ln R_e, \ln \sigma)$ are indeed 
small:   
$\sigma_{\rm err} = (0.05,0.25,0.15,0.15)$, 
and so it is the results of these fits which we report in what 
follows.  However, to illustrate which distributions are most 
affected by measurement error, we also show results from fitting to 
equation~(\ref{phiX}); in most cases, the differences between the 
returned parameters are small, except when $\beta>1$.  

In practice, the fitting was done by minimizing 
\begin{equation}
 \chi^2 \equiv
 \sum_i \left[y_i - \log_{10} \Bigl(\ln(10)\, O\psi(O)\Bigr)\right]^2,
\end{equation}
where $y_i$ was $\log_{10}$ of the $V_{\rm max}^{-1}$ weighted count 
in the $i$th logarithmically spaced bin (and recall that, 
because of fiber collisions, the weight is actually 
$V_{\rm max}^{-1}/0.93$).

\begin{figure*}
 \centering
 \includegraphics[width=0.49\hsize]{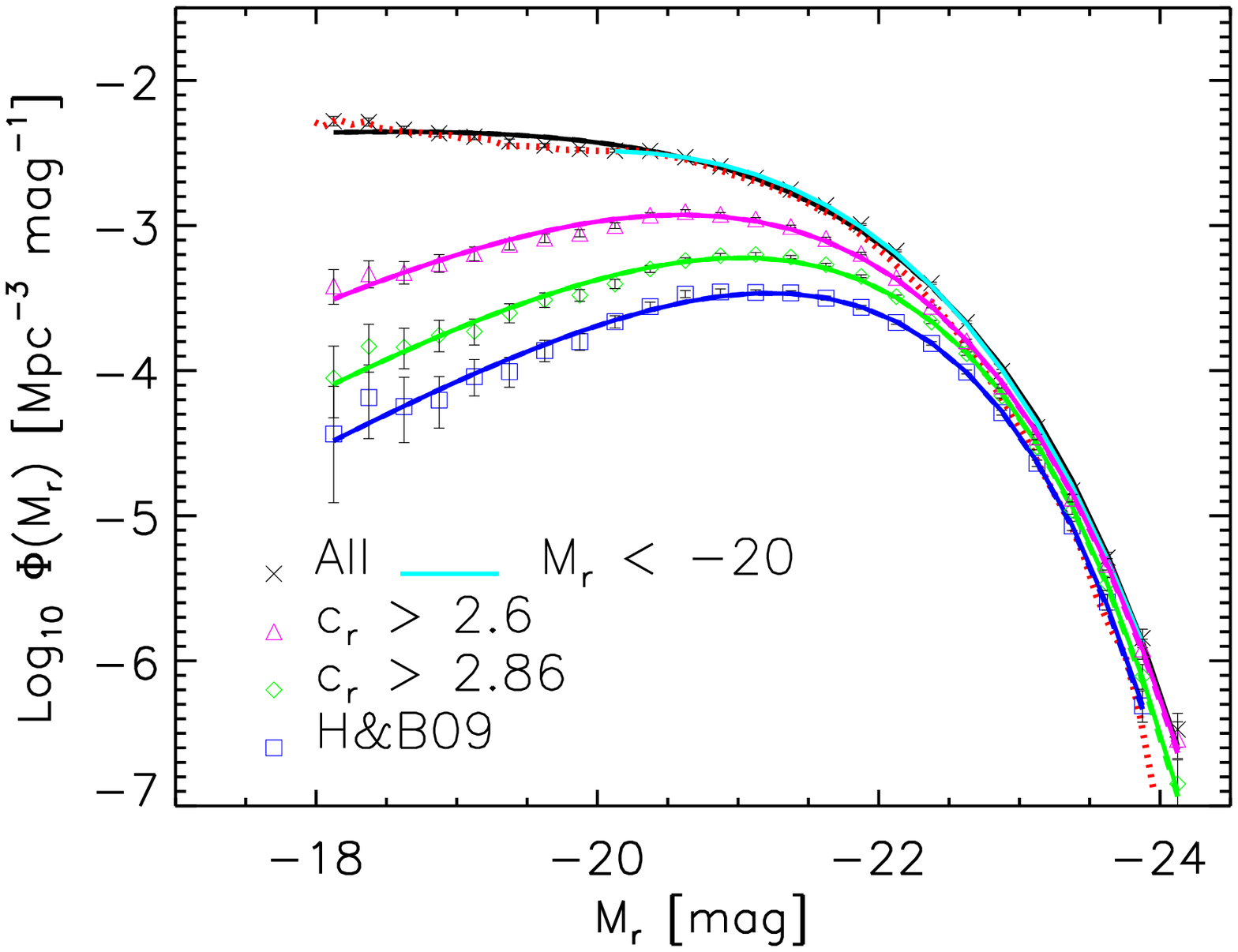}
 \includegraphics[width=0.49\hsize]{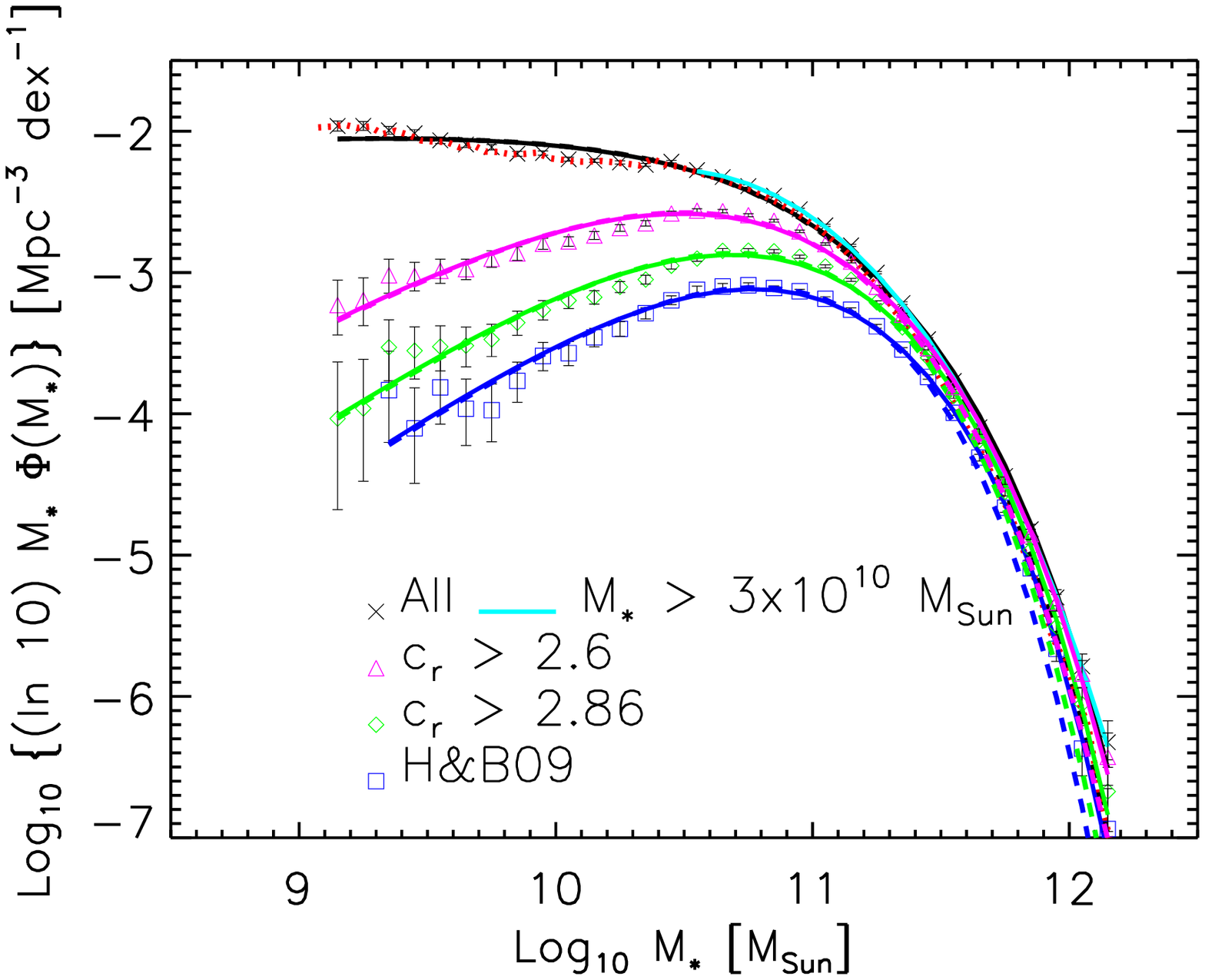} 
 \includegraphics[width=0.49\hsize]{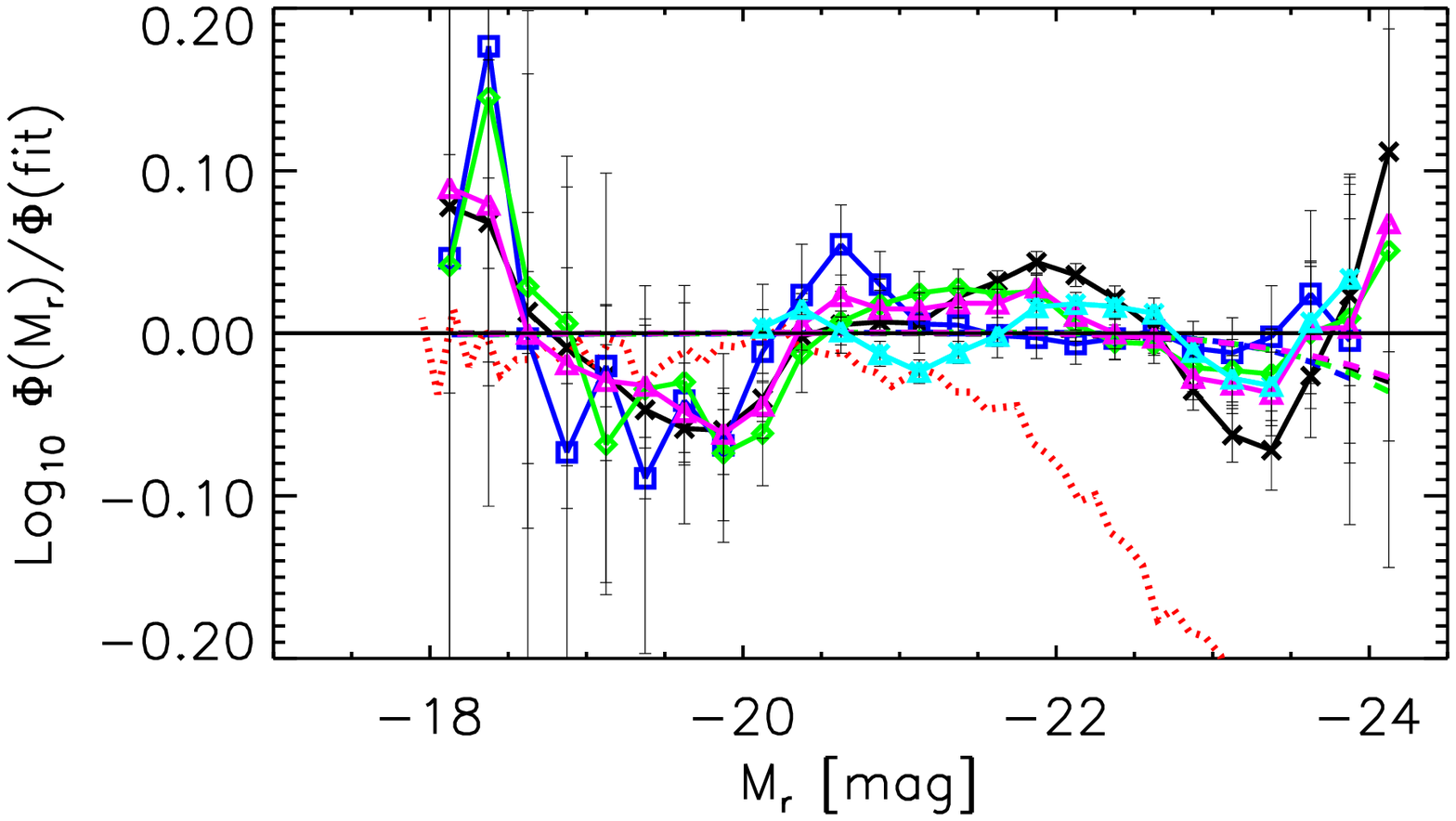}
 \includegraphics[width=0.49\hsize]{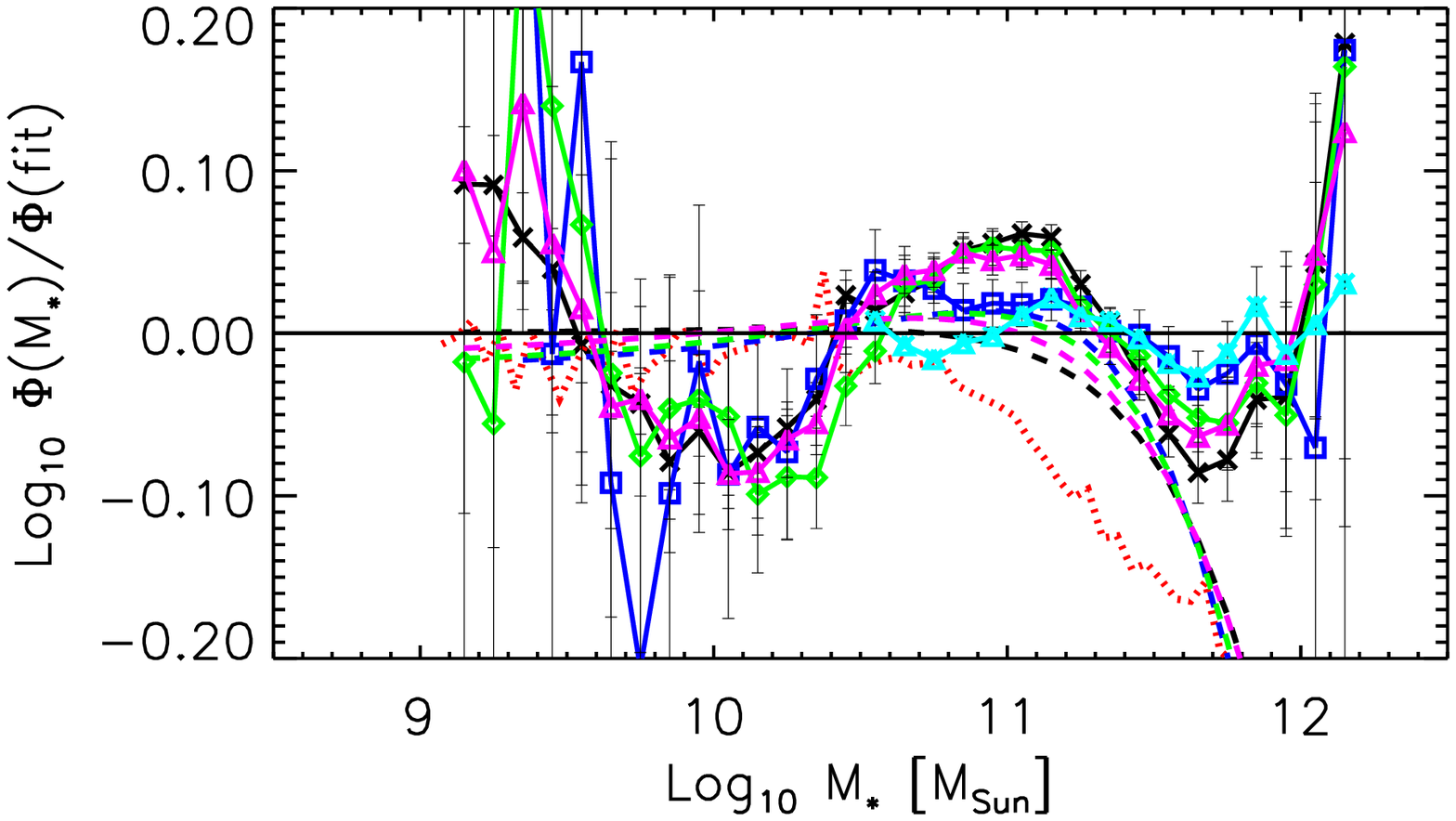} 
 \caption{The effect of different selection cuts on the 
  {\tt cmodel} luminosity and stellar mass functions:
  black crosses, magenta triangles, green diamonds and blue squares 
  show the full sample, a subsample with $C_r > 2.6$, a subsample 
  with $C_r > 2.86$, and a subsample selected following 
  Hyde \& Bernardi (2009).  
  Top panels:  Smooth solid curves show the fits to the observed distributions 
  (equation~\ref{phiX}) while dashed curves (almost indistinguishable 
  from solid curves) show the intrinsic distributions (equation~\ref{psiO}).  
  Cyan solid line shows our fit to the full sample for $M_r < -20$ and 
  $M_* > 10^{10} M_\odot$. 
  The parameters of these fits are reported in Tables~\ref{tabL}--\ref{tabS}.
  Red dotted line shows the measurement associated with Petrosian 
  quantities (i.e. from Figures~\ref{LF2} and~\ref{MsFpet}). 
  Bottom panels:  Same as top panel, but now each set of data points and dashed 
  curve are shown after dividing by their associated solid curve.}
 \label{differentCIa}
\end{figure*}

\begin{figure*}
 \centering
 \includegraphics[width=0.49\hsize]{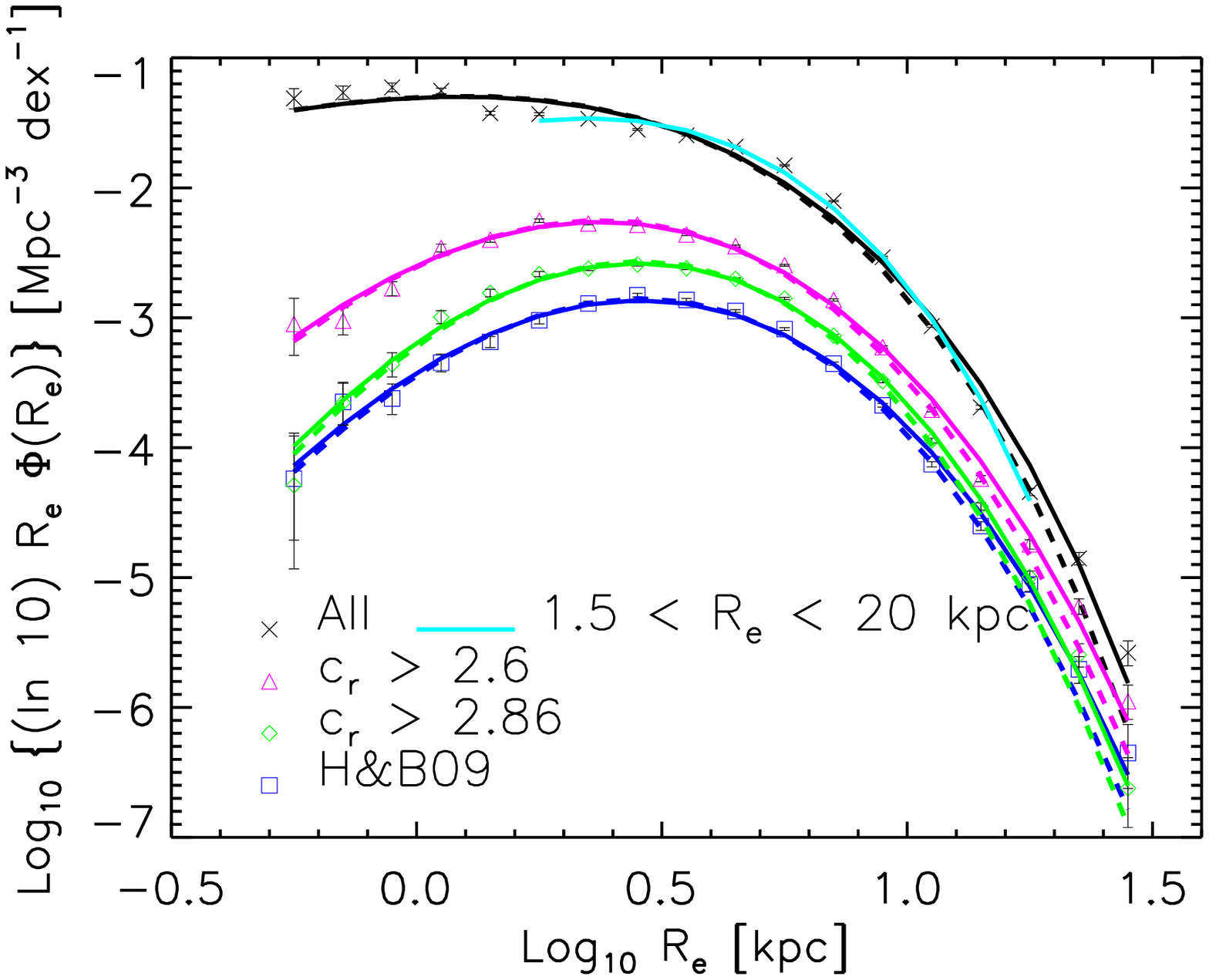}
 \includegraphics[width=0.49\hsize]{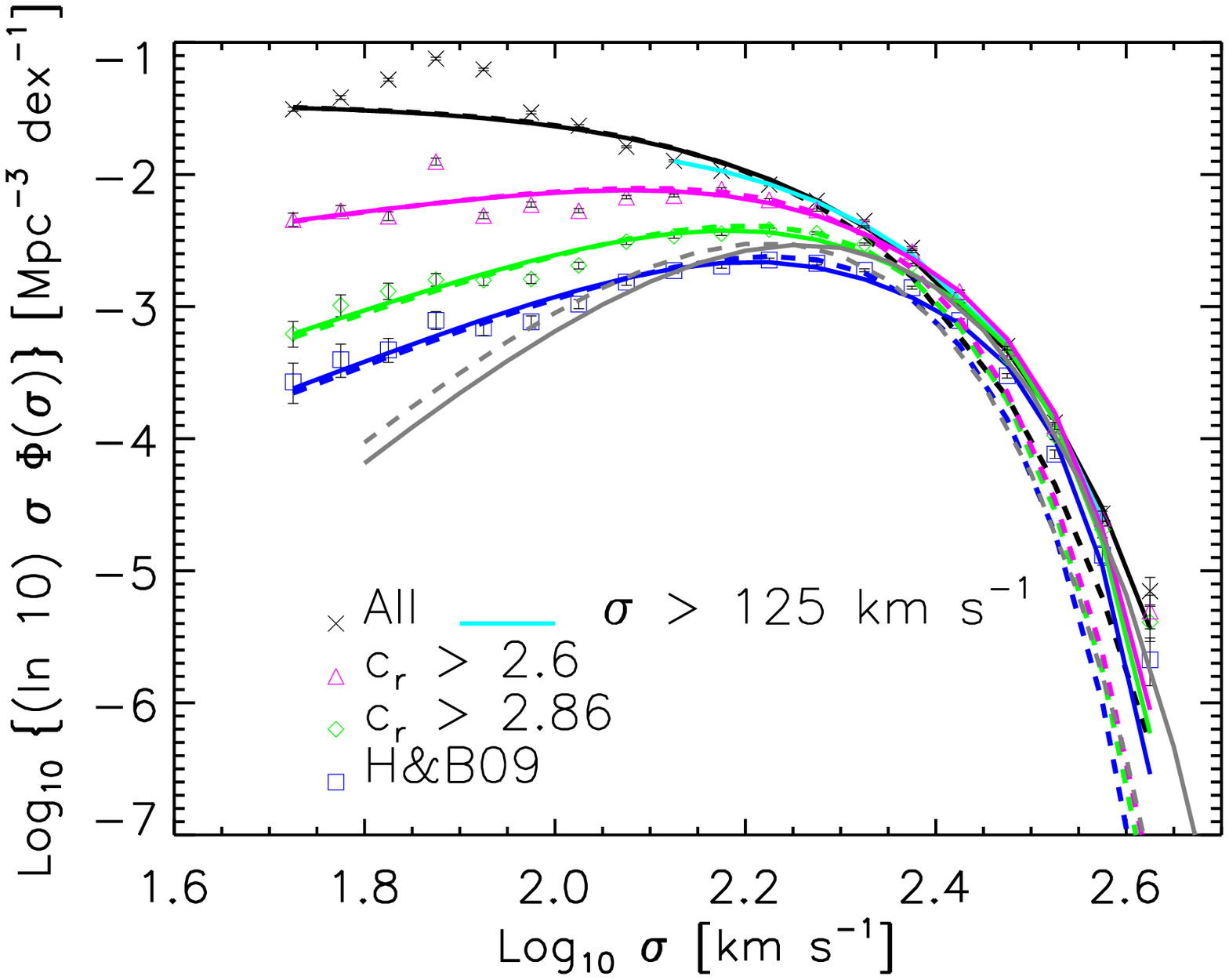}  
 \includegraphics[width=0.49\hsize]{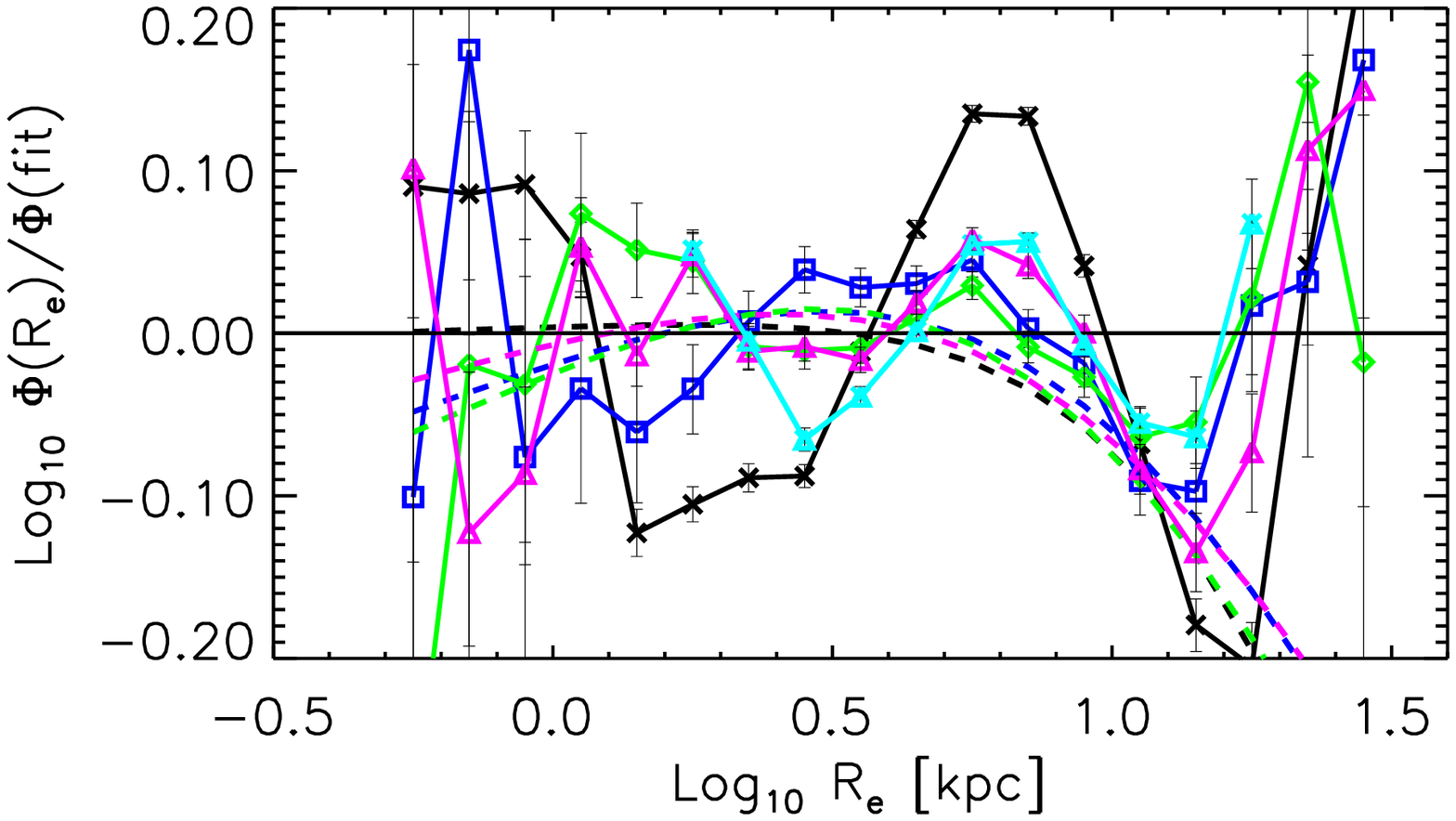}
 \includegraphics[width=0.49\hsize]{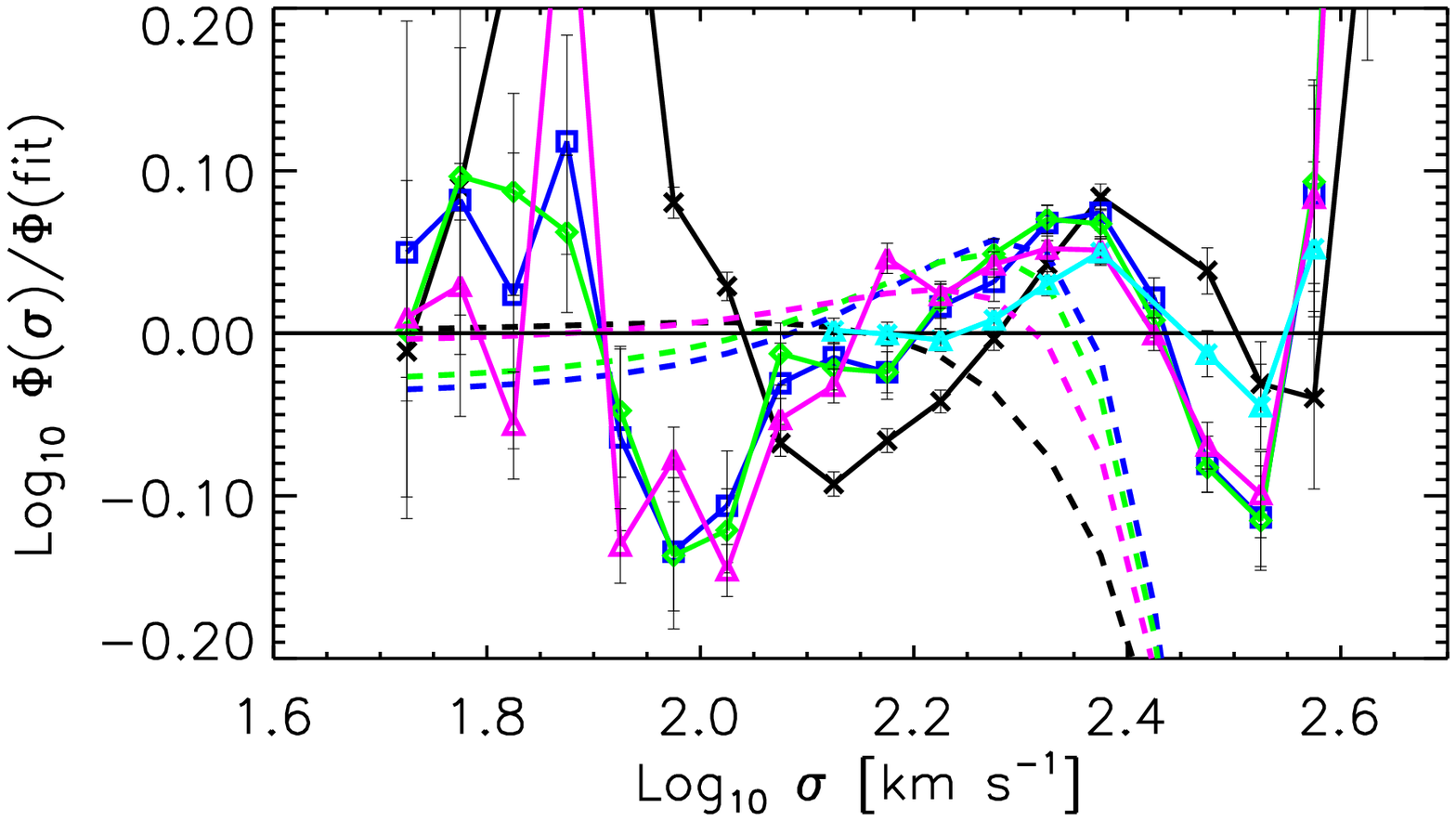}  
\caption{Same as previous figure, but now for the size and 
 velocity dispersion. Cyan solid line shows our
  fit to the full sample for $1.5 < R_e/{\rm kpc} < 20$ and 
  $\sigma > 125$ km s$^{-1}$.
  The top right panel also shows the 
  fits obtained by Sheth et al. (2003) to the observed 
 (solid grey line) and intrinsic (dashed grey) $\phi(\sigma)$ 
 distribution.}
 \label{differentCIb}
\end{figure*}

\begin{figure*}
 \centering
 \includegraphics[width=0.42\hsize]{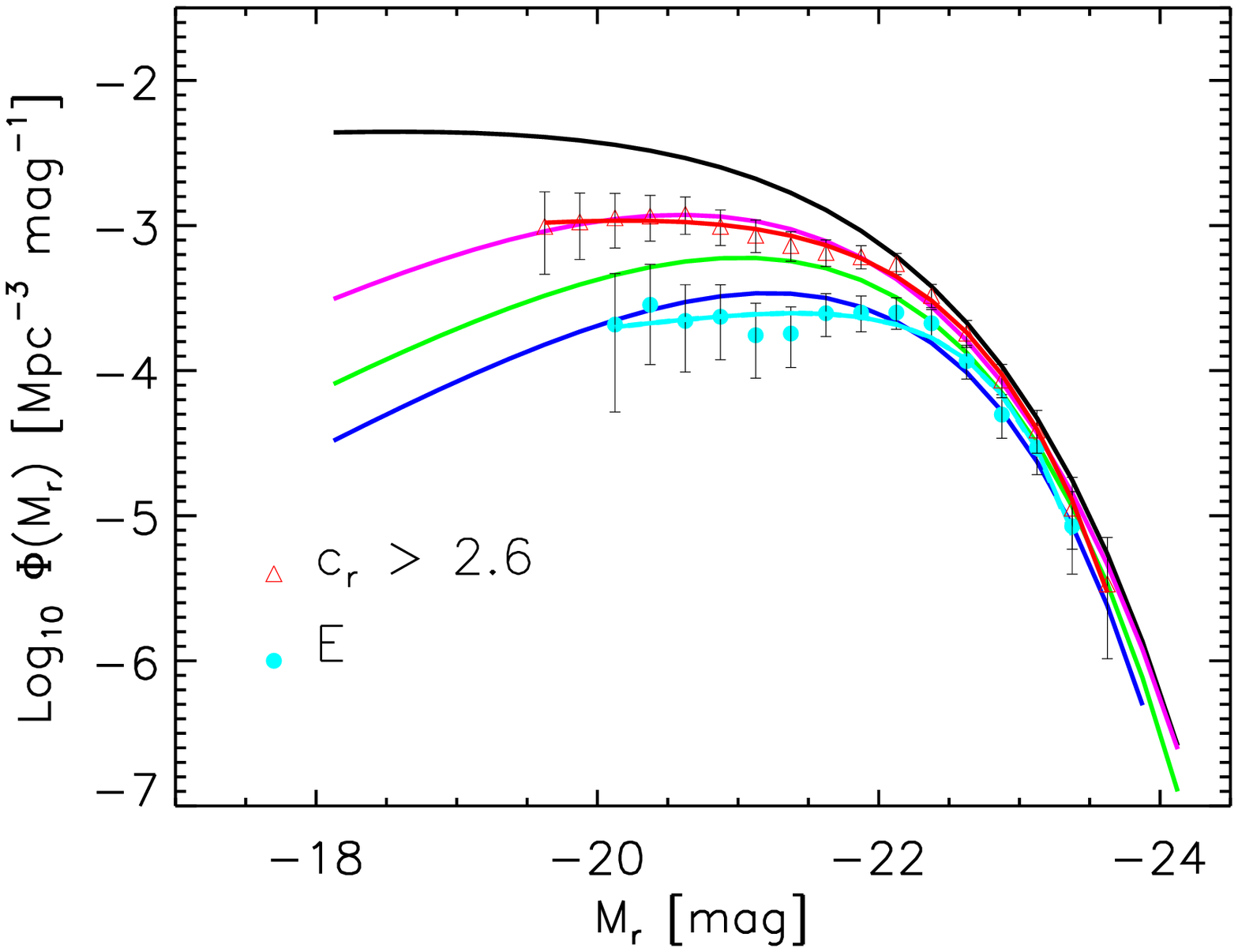}
 \includegraphics[width=0.42\hsize]{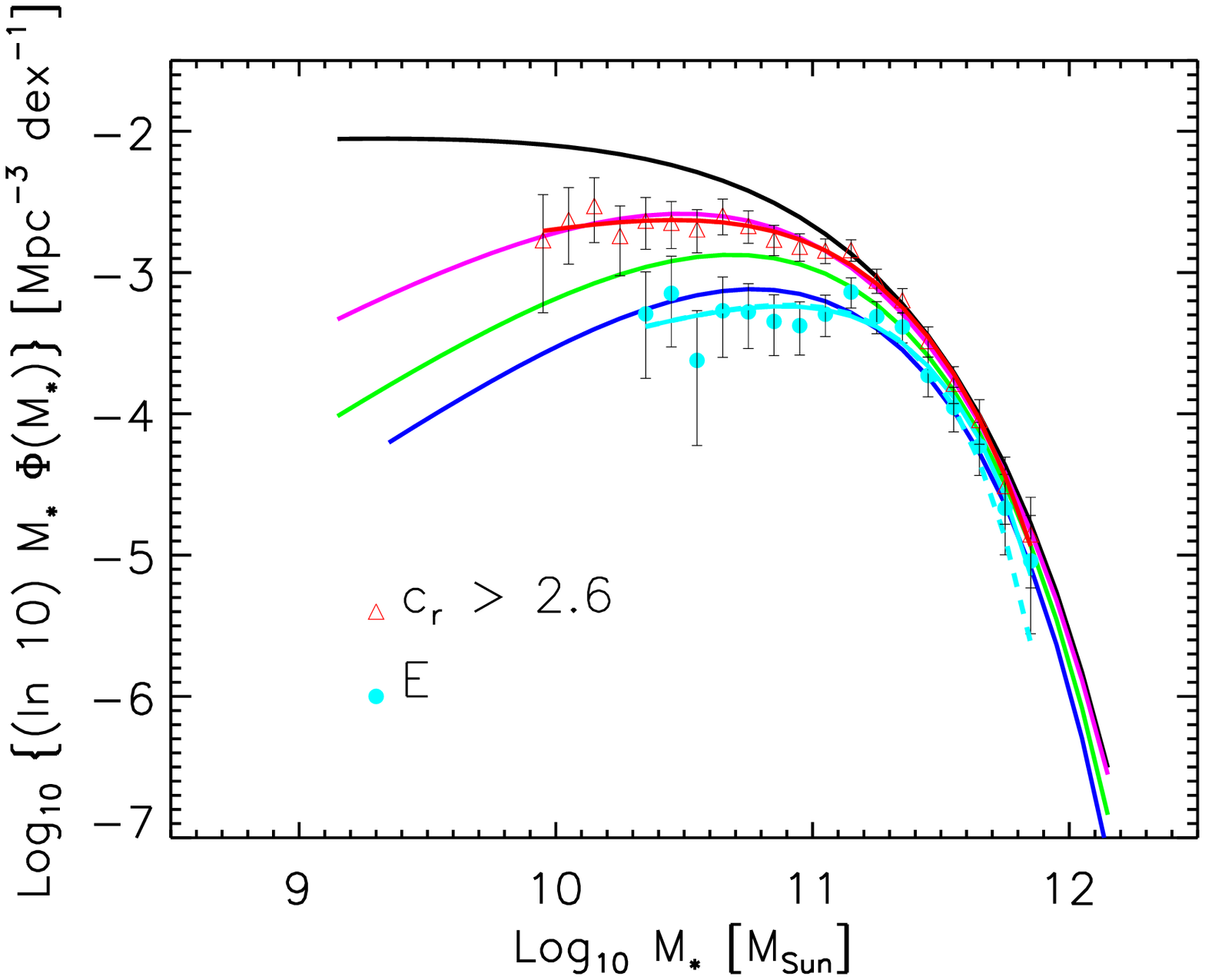} 
 \includegraphics[width=0.42\hsize]{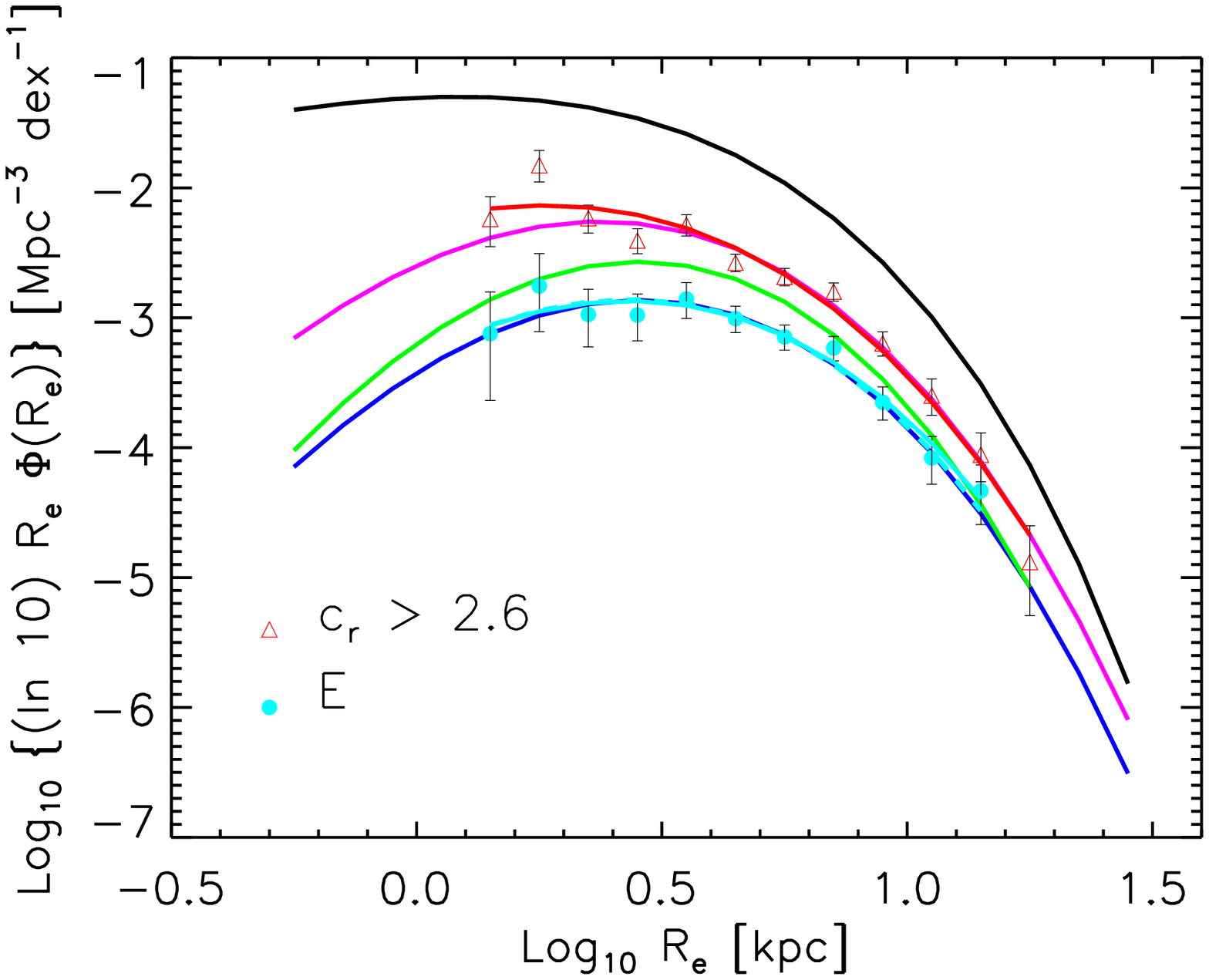}
 \includegraphics[width=0.42\hsize]{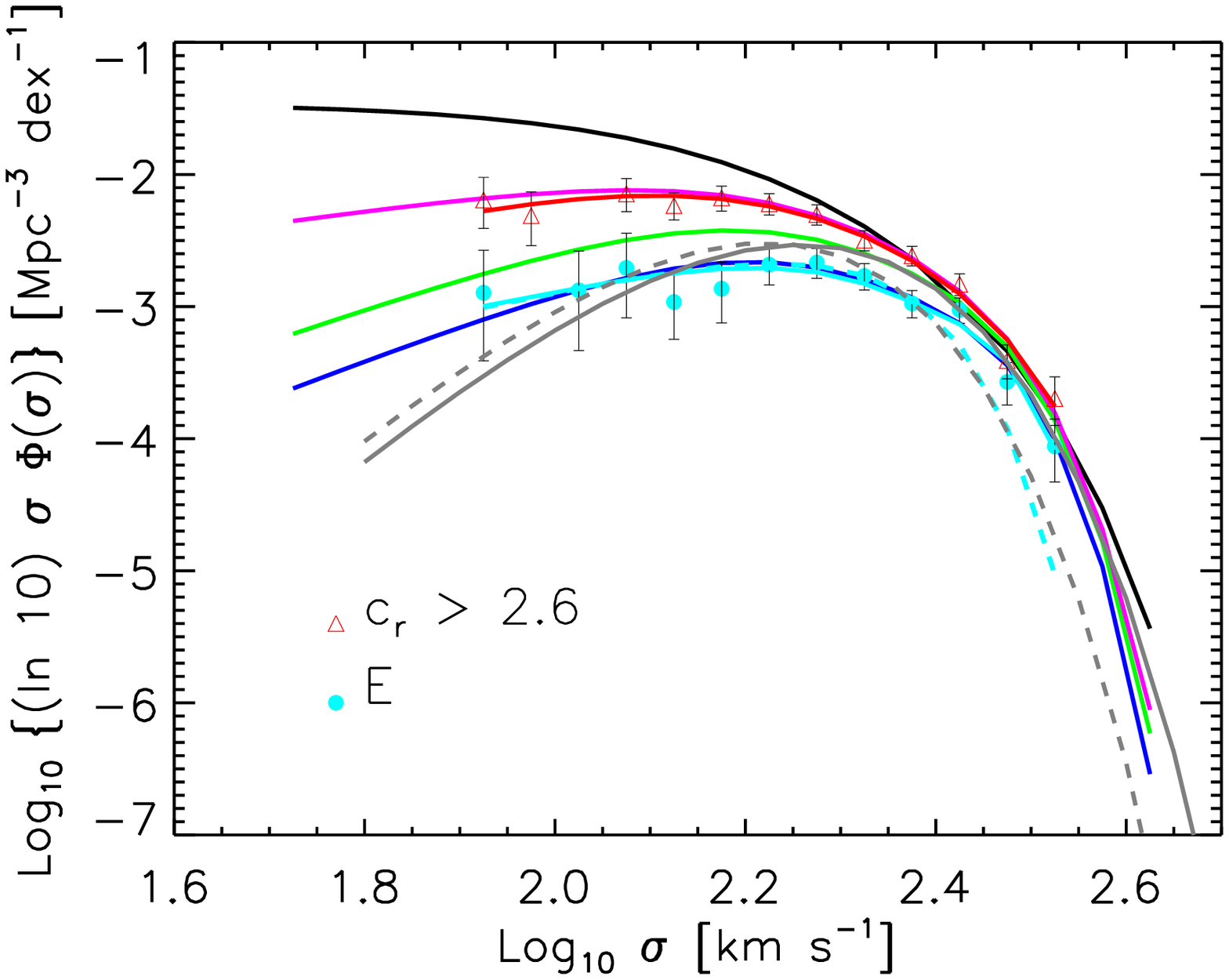}  
 \includegraphics[width=0.42\hsize]{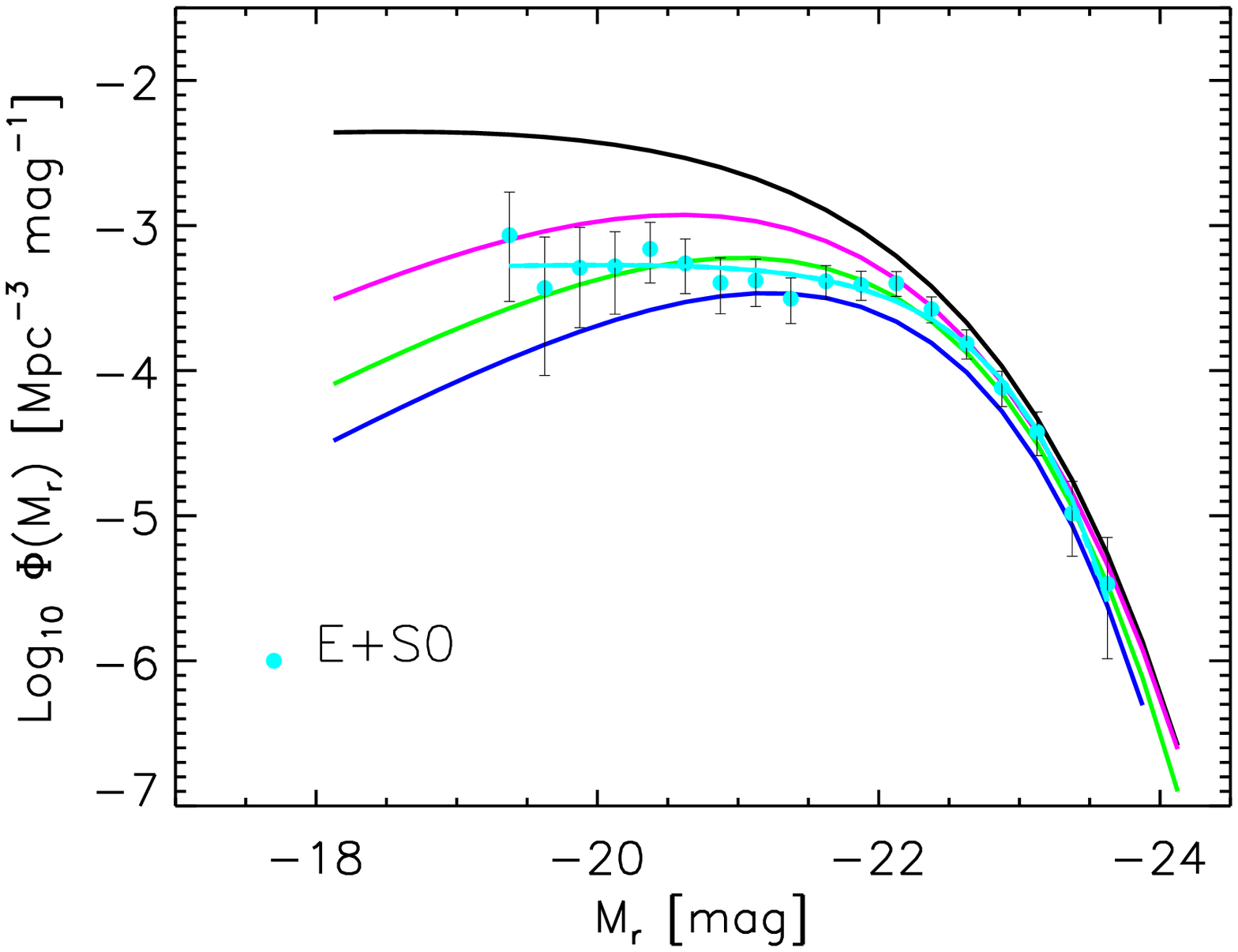}
 \includegraphics[width=0.42\hsize]{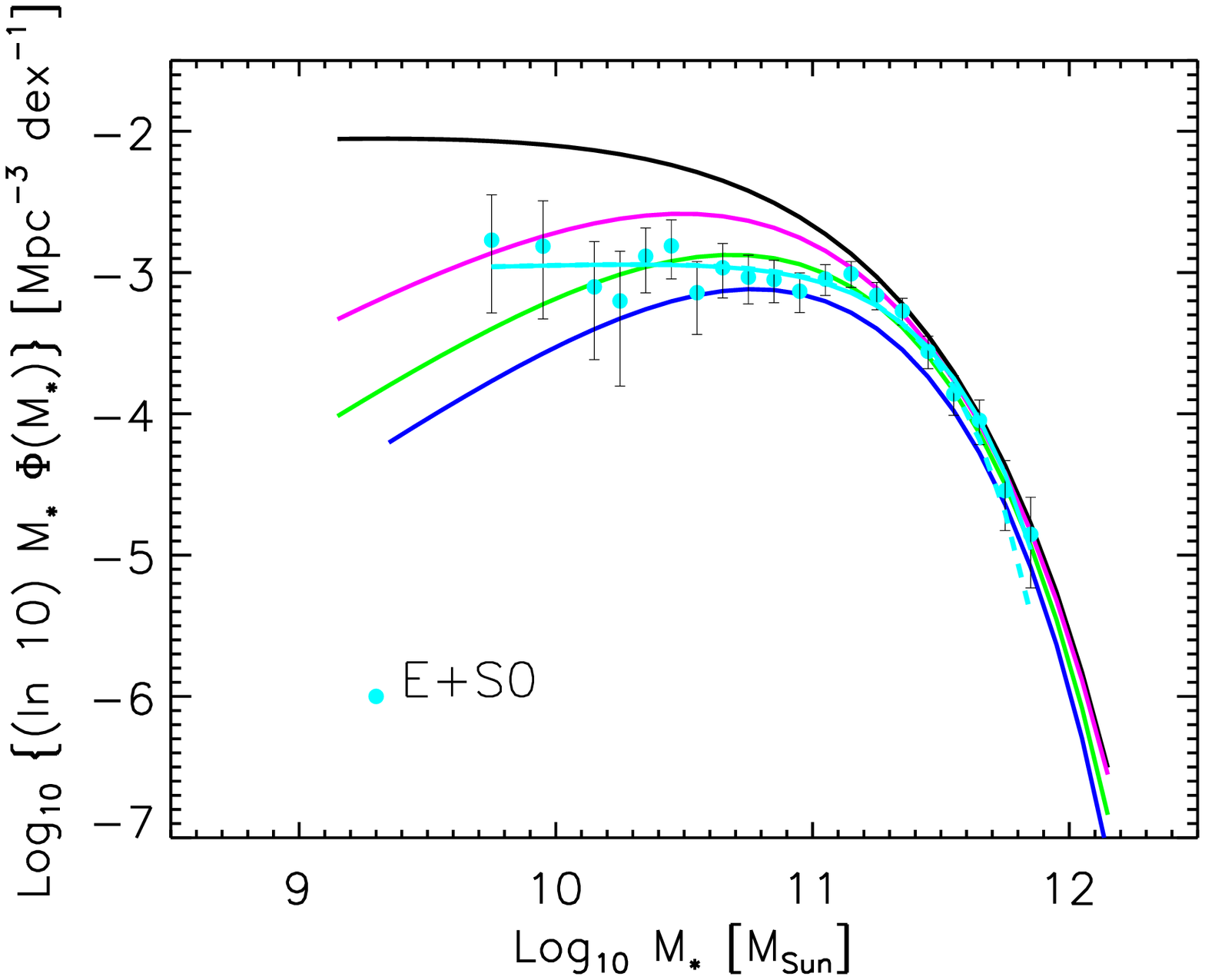} 
 \includegraphics[width=0.42\hsize]{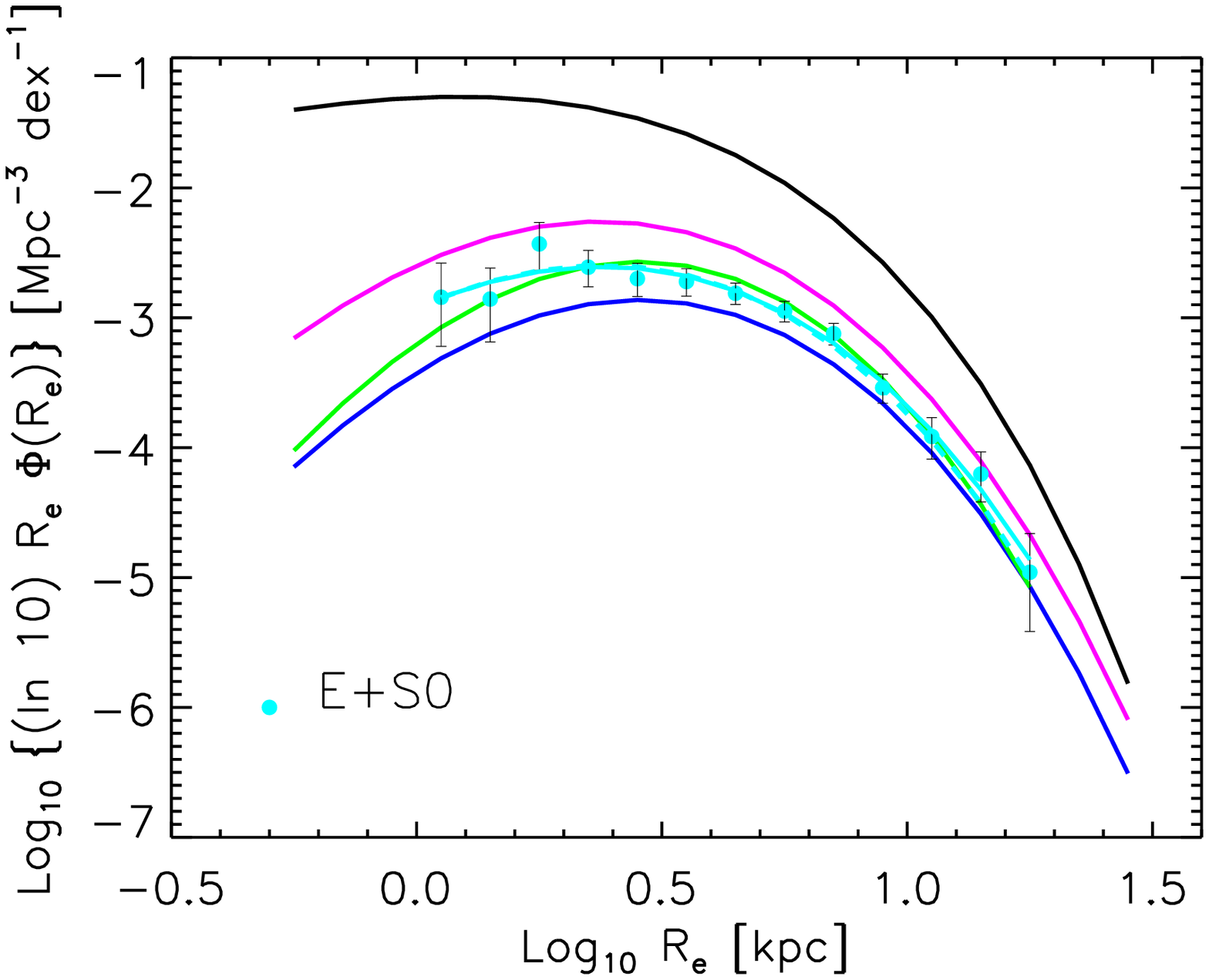}
 \includegraphics[width=0.42\hsize]{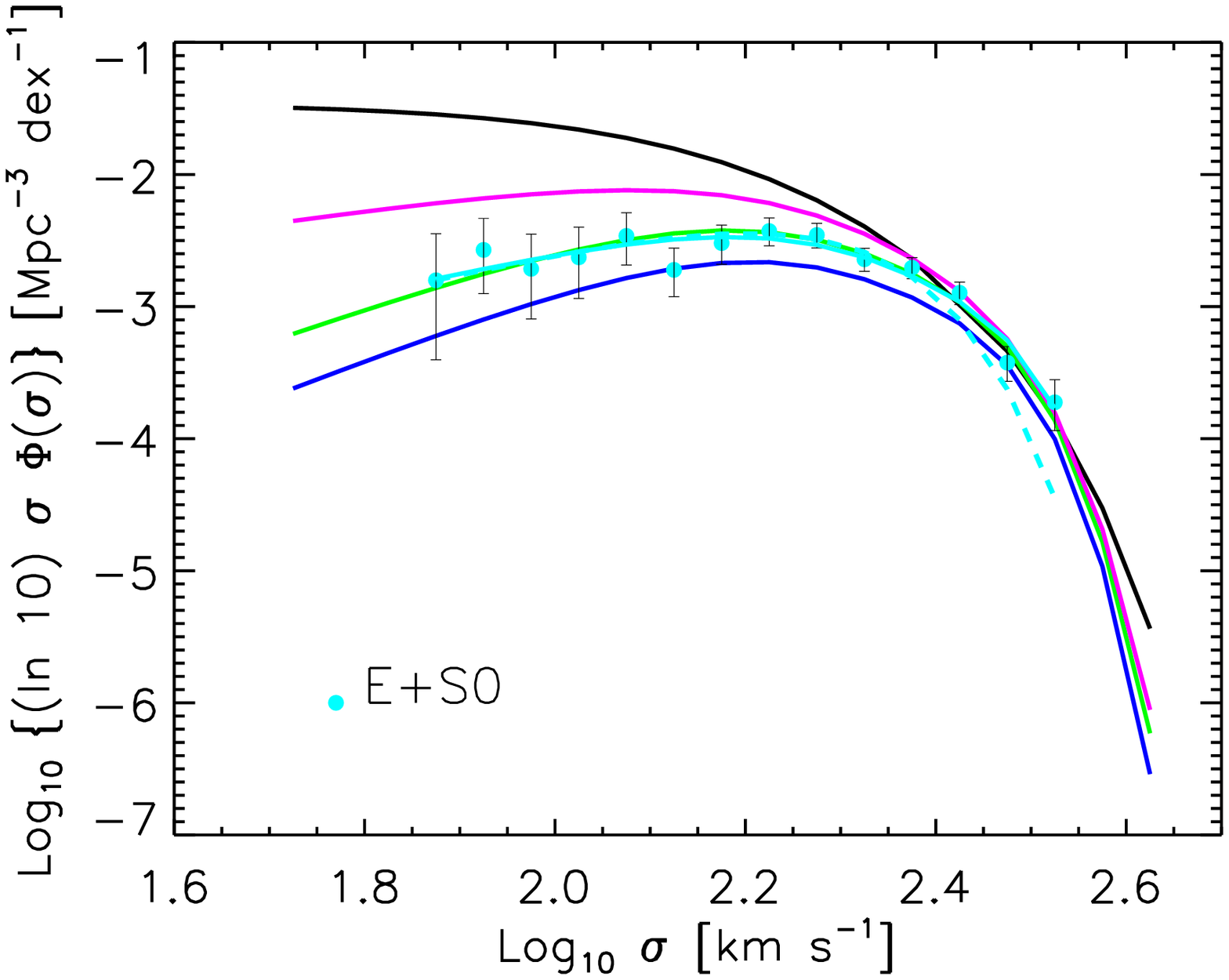}  
 \caption{Distributions for objects of different morphological types
    in the Fukugita et al. (2007) sample (cyan filled circles with error bars).
    A subsample with $C_r\ge 2.6$ is also shown in the `E' panels (red triangles).
    Smooth solid curves are same as in Figure\ref{differentCIa} and \ref{differentCIb}:
    from top to bottom in each panel, they show 
    the full SDSS sample (black), a subsample with $C_r > 2.6$ (magenta), 
    a subsample with $C_r > 2.86$ (green), and a subsample selected 
    following Hyde \& Bernardi (2009) (blue).}
 \label{E}
\end{figure*}

\begin{figure*}
 \centering
 \includegraphics[width=0.42\hsize]{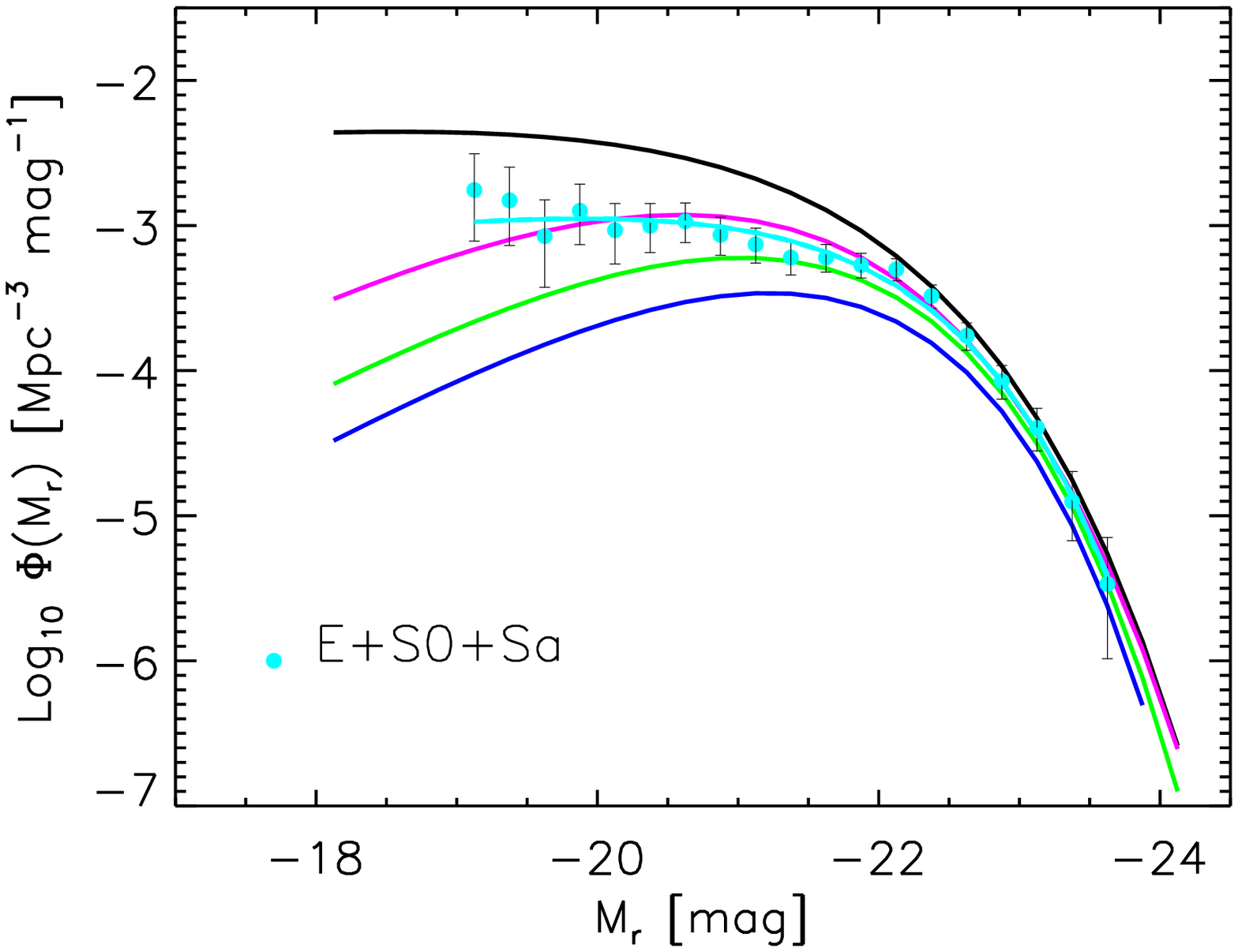}
 \includegraphics[width=0.42\hsize]{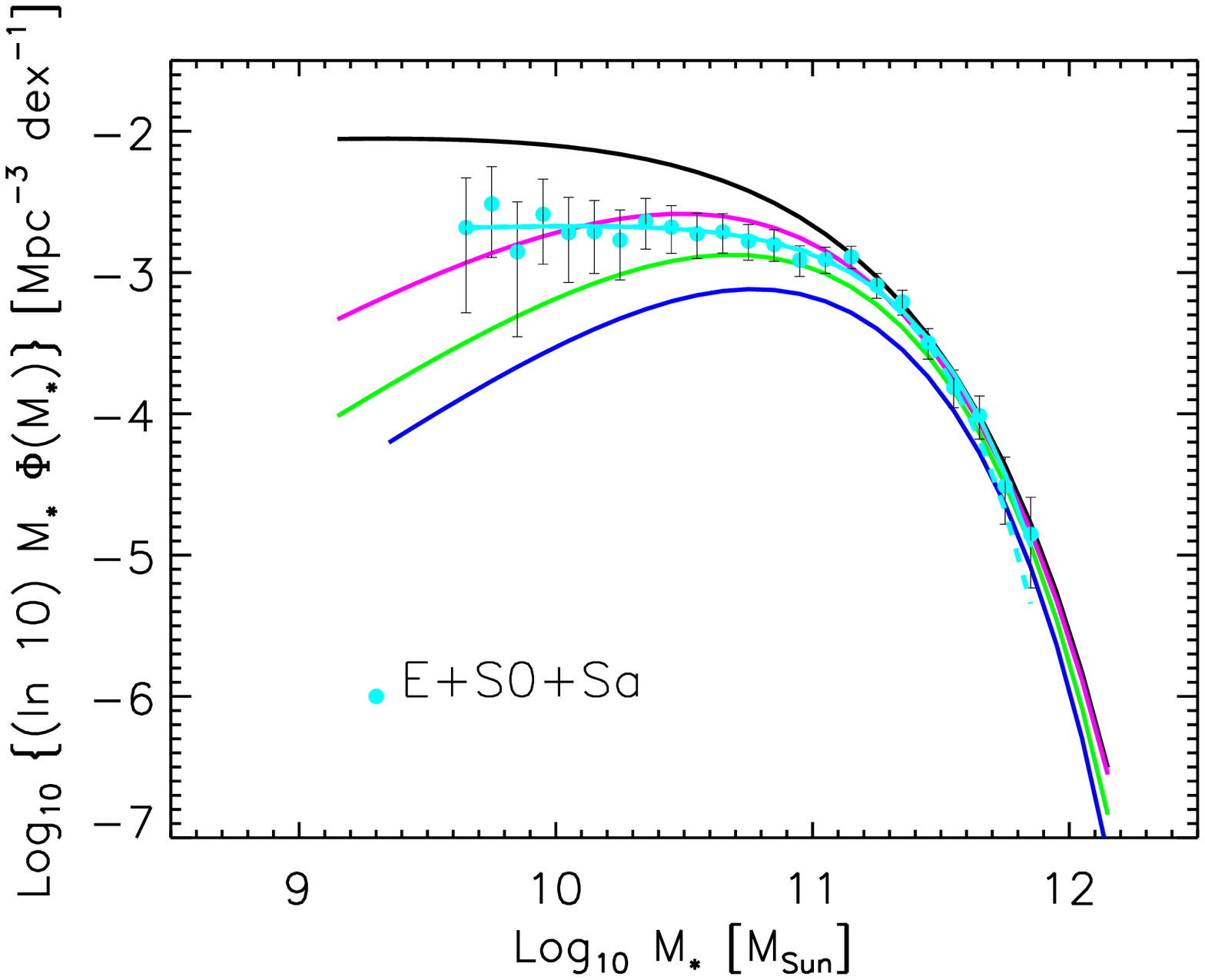} 
 \includegraphics[width=0.42\hsize]{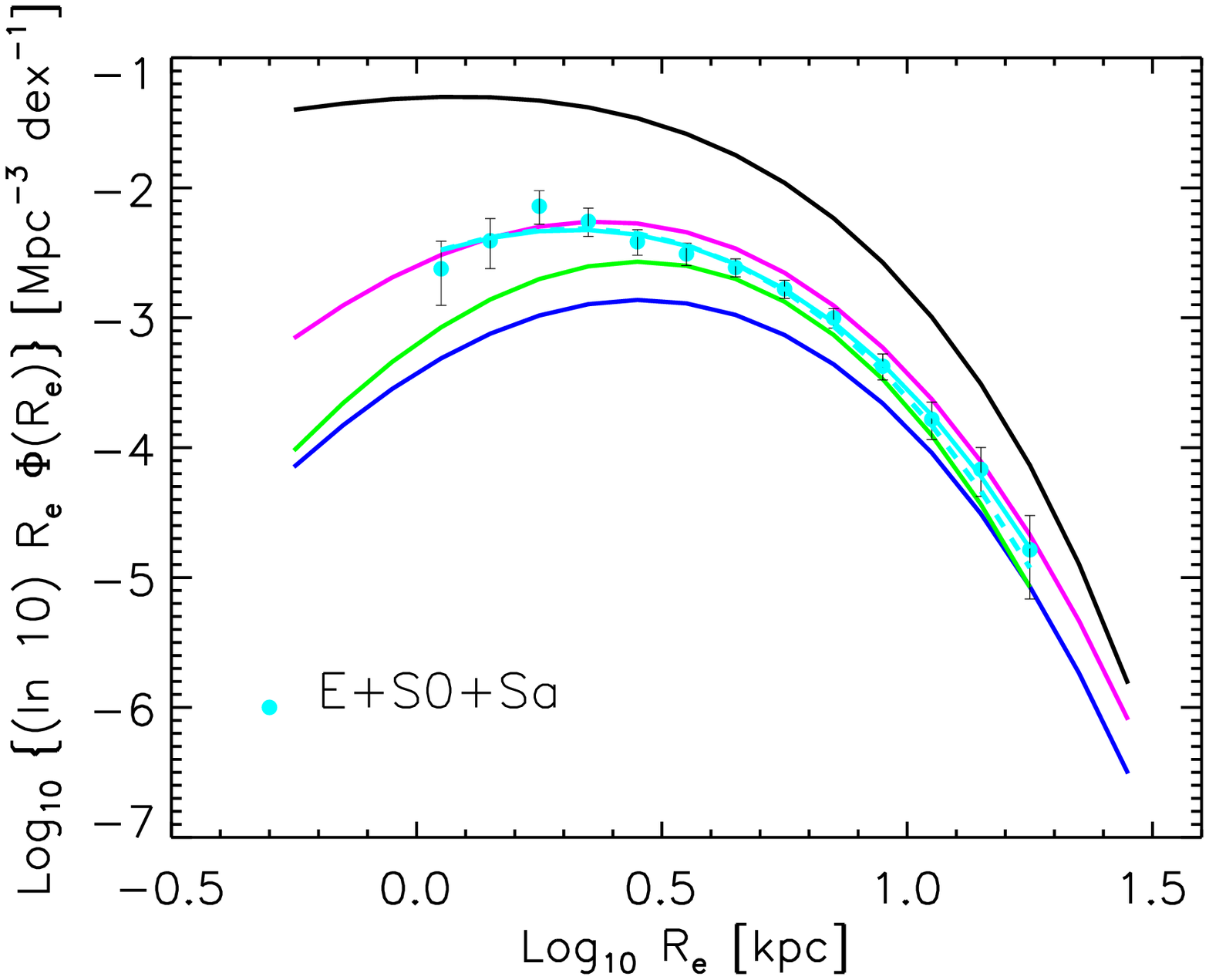}
 \includegraphics[width=0.42\hsize]{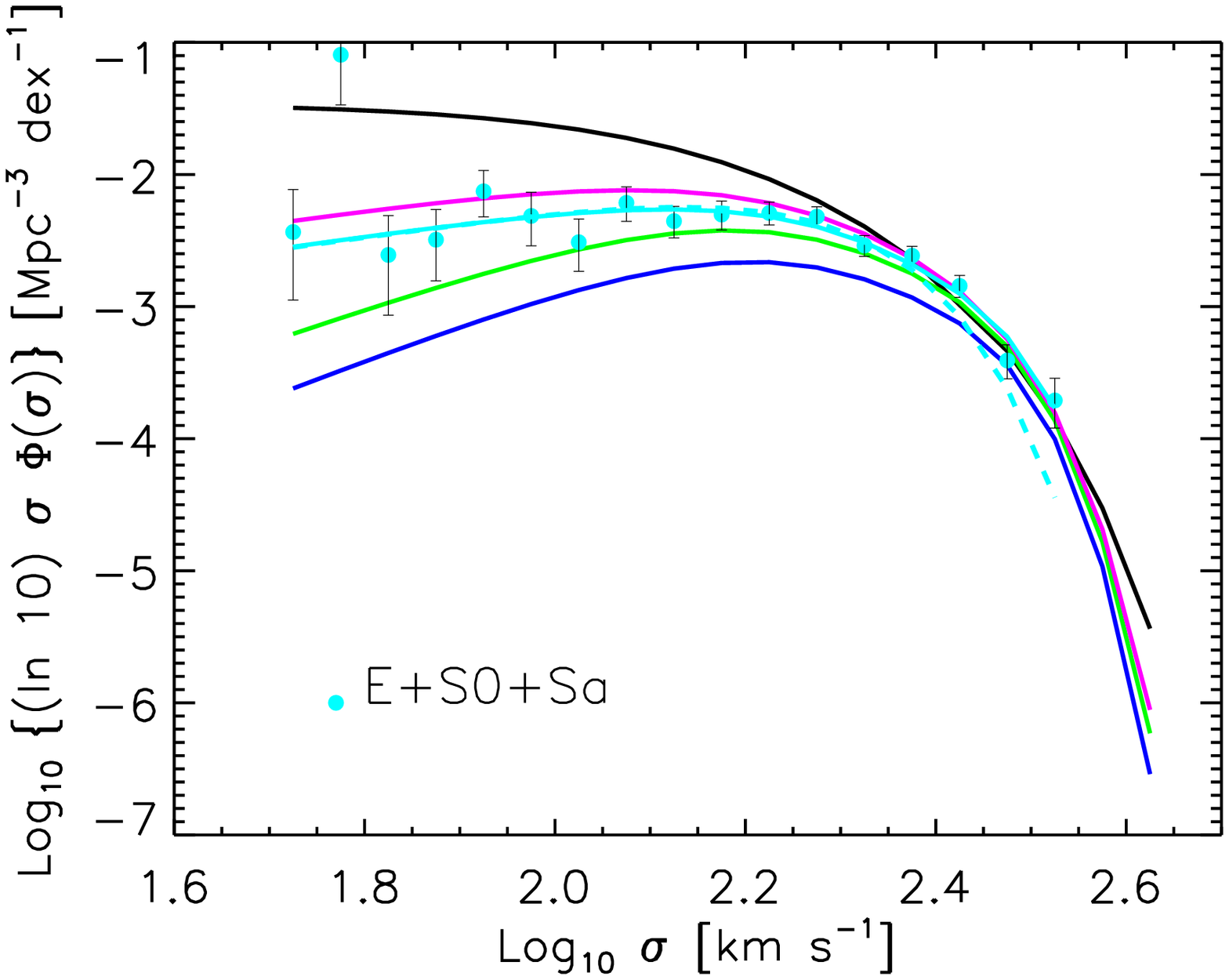}  
 \includegraphics[width=0.42\hsize]{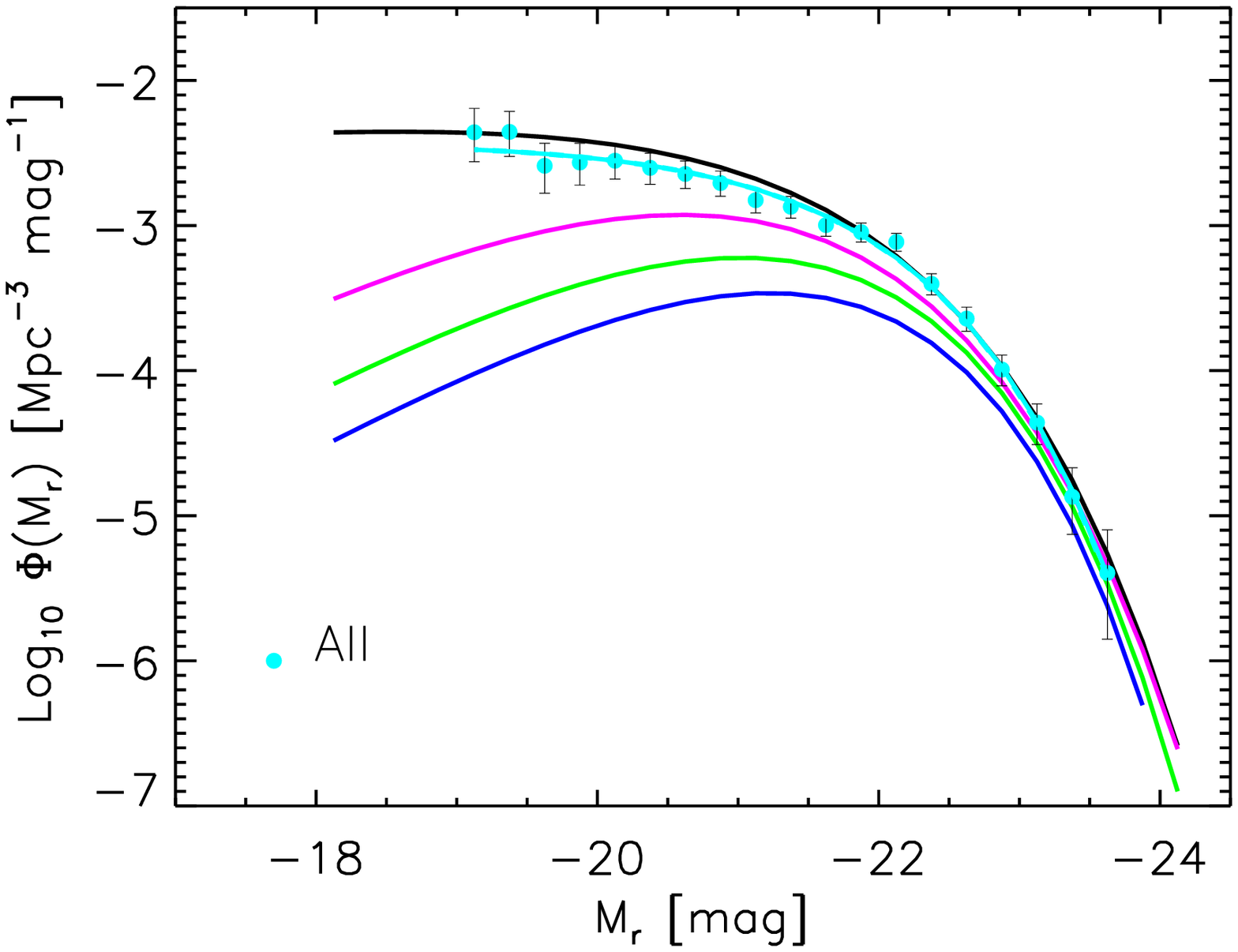}
 \includegraphics[width=0.42\hsize]{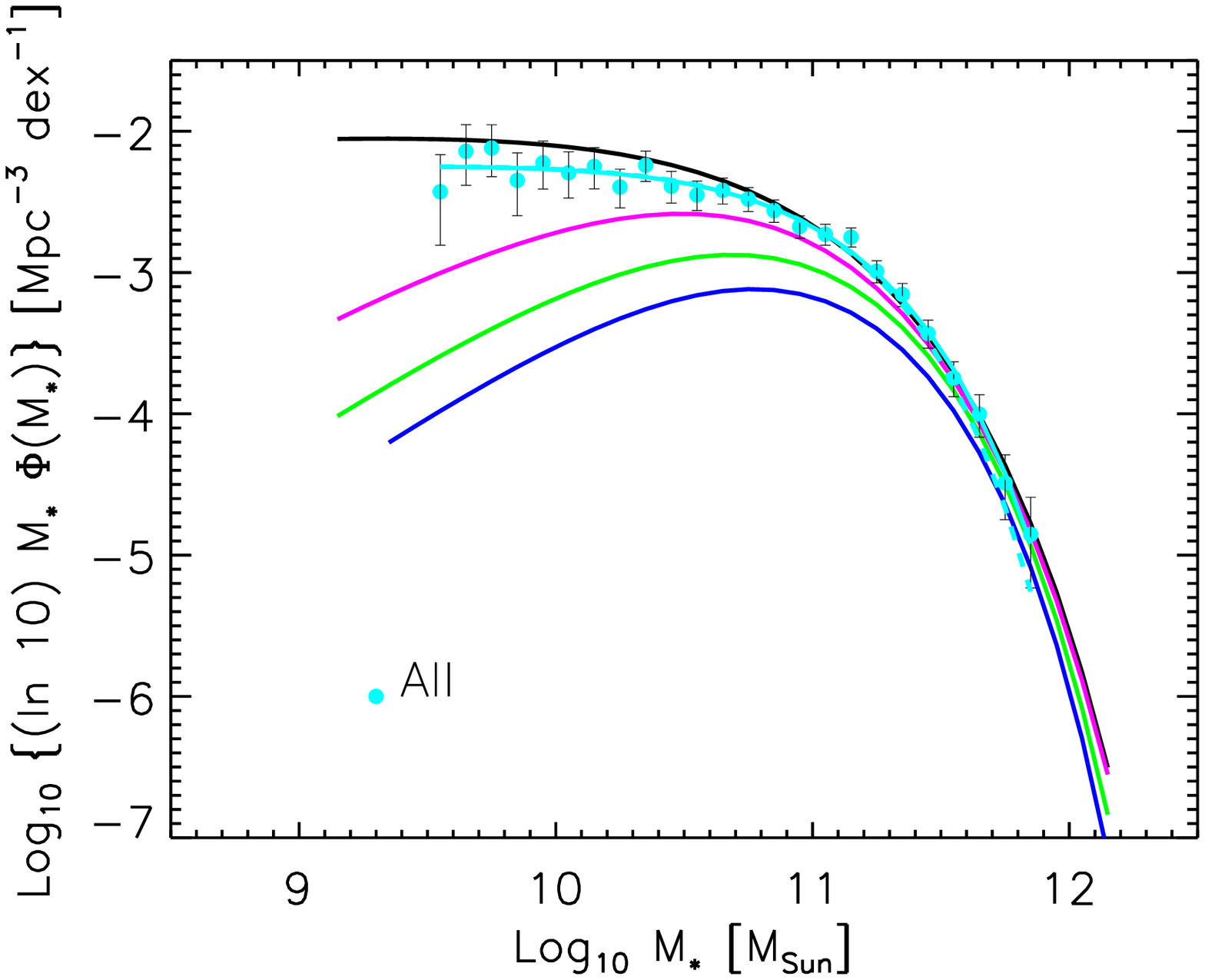} 
 \includegraphics[width=0.42\hsize]{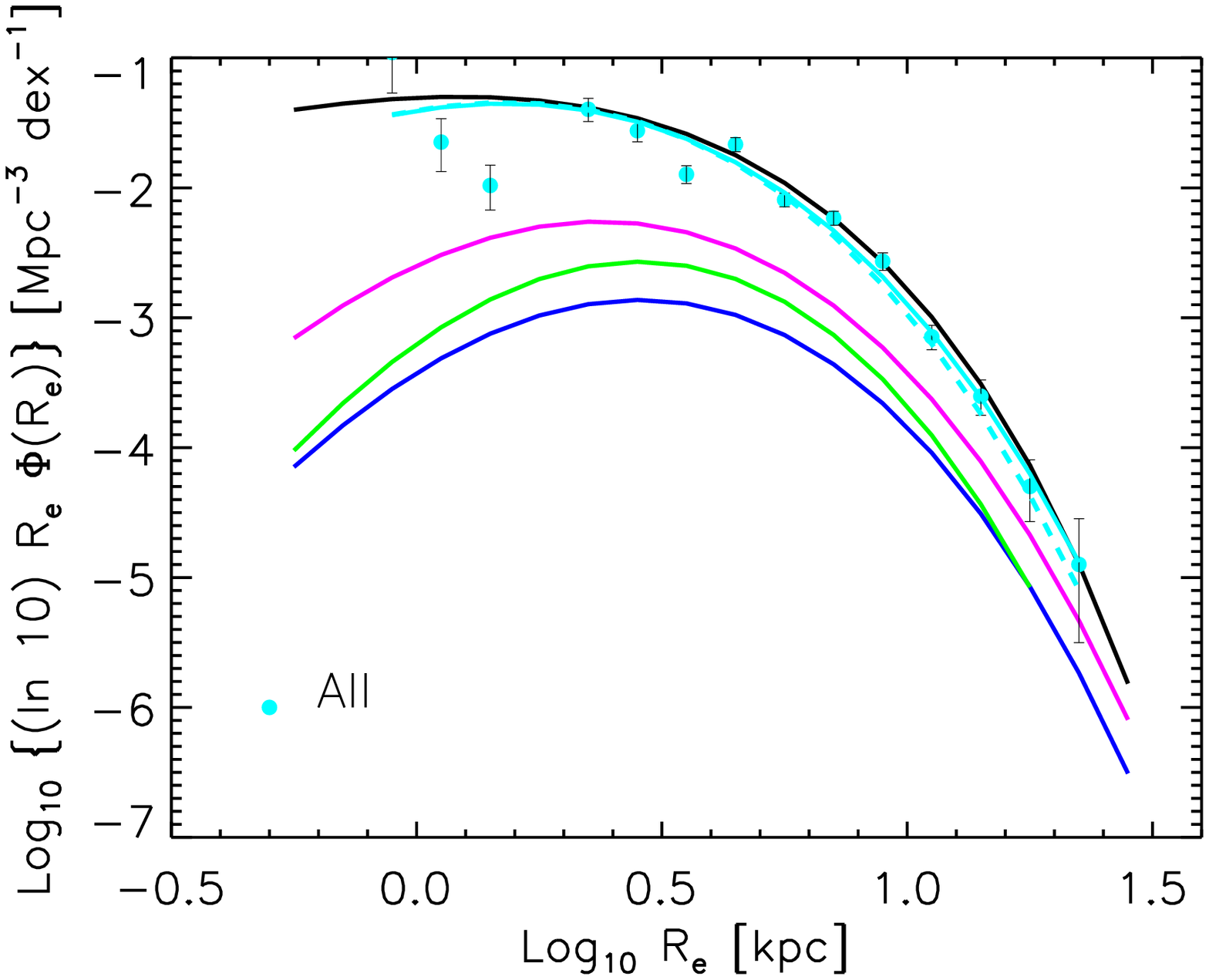}
 \includegraphics[width=0.42\hsize]{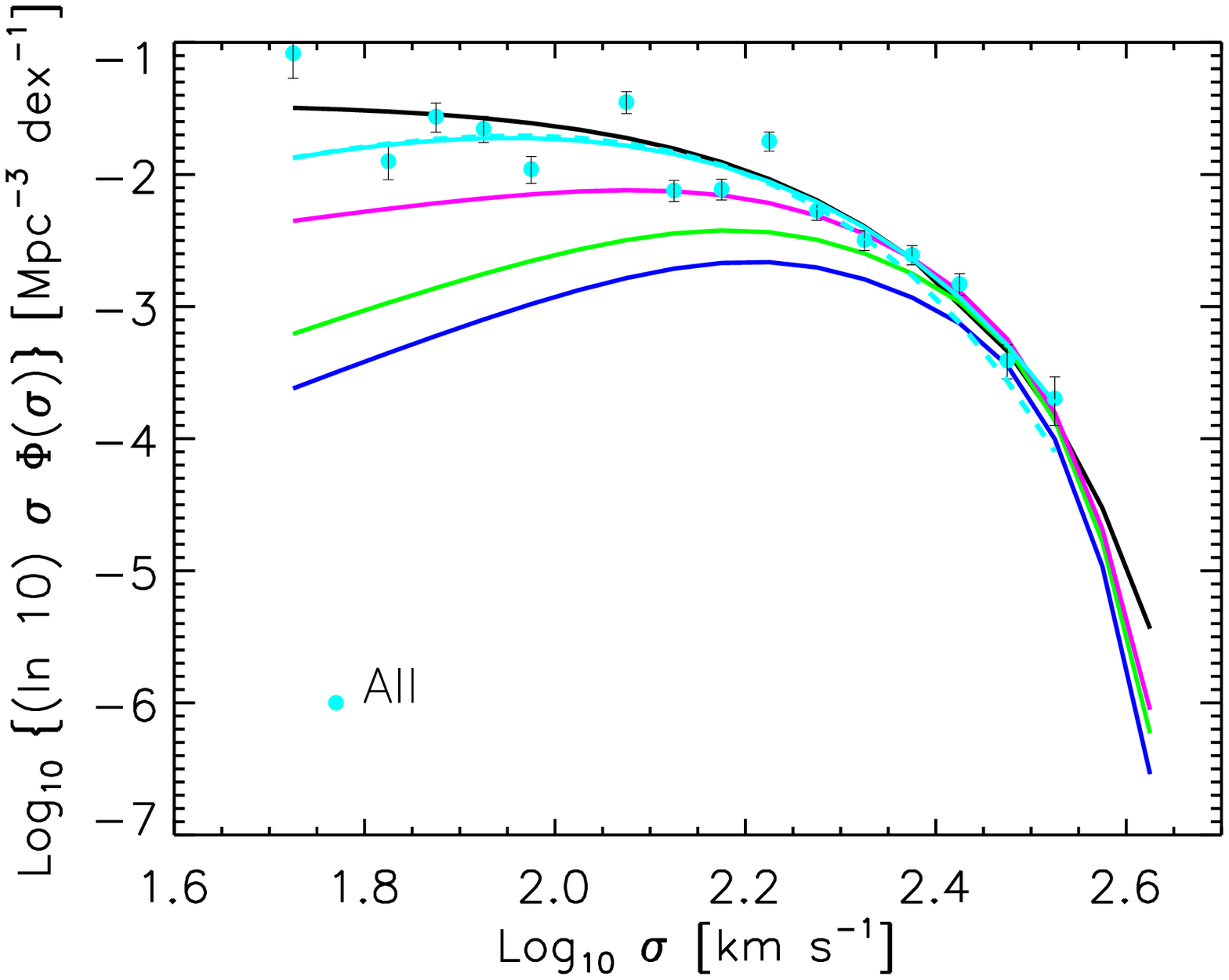}  
 \caption{Continued.}
 \label{ES0Sa}
\end{figure*}

\subsection{Covariances between fitted parameters}
When fitting to equation~\eqref{phiX}, a reasonable understanding 
of the covariance between the fitted parameters can be got by asking 
that all parameter combinations give the same mean density 
(Sheth et al. 2003):
\begin{equation}
 \rho_X = \phi_*\,X_* \,\frac{\Gamma[(1+\alpha)/\beta]}{\Gamma(\alpha/\beta)}.
 \label{rhoX}
\end{equation}
In practice, $\phi_*$ is determined essentially independently of 
the other parameters, so it is the other three parameters which 
are covariant.  Hence, it is convenient to define 
\begin{equation}
 \langle X\rangle = \rho_X/\phi_*.
 \label{meanX}
\end{equation}
If $\beta$ is fixed to unity, then this becomes 
 $\langle X\rangle = X_*\,\alpha$.  
But if $\beta$ is not fixed, then a further constraint equation can be 
got by asking that all fits return the same peak position or height.  
For distributions with broad peaks, it may be better to instead 
require that the second central moment, 
\begin{equation}
 \sigma_X^2 = \langle X^2\rangle - \langle X\rangle^2
 \quad {\rm where} \quad \ \langle X^2\rangle = X_*^2 \,
   \frac{\Gamma[(2+\alpha)/\beta]}{\Gamma(\alpha/\beta)} ,
 \label{meanXsq}
\end{equation}
be well reproduced.  Thus, the covariance between $\alpha$ and 
$\beta$ is given by requiring that $\sigma_X/\langle X\rangle$ 
equal the measured value for this ratio.  
The changes to these correlations are sufficiently small if we 
instead fit to equation~\eqref{psiO}, so we have not  
presented the algebra here.  

\subsection{Distributions for samples cut by concentration}
Figure~\ref{differentCIa} shows $\phi(L)$ and $\phi(M_*)$ 
in the full sample (top curve in top panel), 
when one removes objects with $C_r<2.6$ (second from top), 
objects with $C_r<2.86$ (third from top), and when one selects 
early-types on the basis of a number of other criteria (bottom, 
following Hyde \& Bernardi 2009).  For comparison, the dotted 
curves show the measurement in the full sample when Petrosian 
quantities are used (i.e. from Figures~\ref{LF2} and~\ref{MsFpet}).  

The solid curves show the result of fitting to equation~\eqref{phiX}, 
and the dashed curves (almost indistinguishable from the solid ones, 
except at the most massive end) result from fitting to 
equation~\eqref{psiO} instead, so as to remove the effects of 
measurement error on our estimate of the shape of the intrinsic 
distribution.  To reduce the dynamic range, bottom panels show 
each set of curves divided by the associated solid curve (i.e., 
by the fit to the observed sample).  The dotted lines in these 
bottom panels show that Petrosian based counts lie well below 
those based on {\tt cmodel} quantities at $M_r\le -20$ or  
$\log_{10}M_*/M_\odot\ge 10.6$.  The dashed lines in the bottom 
right panel show that the intrinsic distribution has been 
noticably broadened by errors above $\log_{10}M_*/M_\odot\ge 11$.  

Figure~\ref{differentCIb} shows a similar analysis of 
$\phi(R_e)$ and $\phi(\sigma)$.  For $\phi(\sigma)$ we also 
compare our results with those of Sheth et al. (2003).  The 
Sheth et al. analysis was based on a sample selected by 
Bernardi et al. (2003), which was more like $C_r>2.86$ 
at large masses, but because of cuts on emission lines and 
S/N in the spectra, had few low mass objects.  Indeed, at 
large $\sigma$, our measured $\phi(\sigma)$ is similar to 
theirs.

Tables~\ref{tabL}--\ref{tabS} report the parameters of the 
fits (to equations~\ref{phiX} and~\ref{psiO}) shown in 
Figures~\ref{differentCIa} and~\ref{differentCIb}. 
It is worth noting that, for a given sample, the fits return 
essentially the same value of $\phi_*$ in all the tables, even 
though this was not explicitly required.  
And note that $\phi(M_*)$ and $\phi(\sigma)$ are the 
distributions that are most sensitive to errors; the former because 
the errors are large, and the latter because $\beta$ is large.  


\subsection{Comparison with morphological selection}\label{morphs}
Figures~\ref{E} and \ref{ES0Sa} show 
 $\phi(L)$, $\phi(M_*)$, $\phi(R_e)$ and $\phi(\sigma)$ 
for the Fukugita et al. sample as one adds more and more morphological 
types.  The smooth curves (same in each panel) show the fits to the 
samples shown in Figures~\ref{differentCIa} and~\ref{differentCIb}.  
In the top panels of Figure~\ref{E} only, we also show the result 
of removing objects with $C_r<2.6$ before making the measurements.  
This allows a direct comparison with one of the curves from the 
previous Figure.  Notice that this gives results which are very 
similar to those from the larger (fainter, deeper) full sample; 
the different magnitude limits do not matter very much.  

The results of fitting equation~(\ref{psiO}) to the Fukugita 
subsamples are provided in Tables~\ref{tabL}--\ref{tabS}, which 
also show the associated luminosity and stellar mass densities, 
and the mean sizes.  
Ellipticals account for about 19\% of the $r$-band {\tt cmodel} 
luminosity density and 25\% of the stellar mass density; 
including S0s increases these numbers to 33\% and 41\%, and adding 
Sas brings the contributions to 50\% and 60\% respectively.  

Note that the number, luminosity and stellar mass densities of Es -- 
about $10^{-3}$Mpc$^{-3}$, $0.2\times 10^8L_\odot$Mpc$^{-3}$ and
$0.6\times 10^8M_\odot$Mpc$^{-3}$ respectively -- are very similar to 
those of the early-type sample selected following 
Hyde \& Bernardi (2009), as is the mean half-light radius of 3.2~kpc.  
Some of this match is fortuitous -- we showed before that E's 
account for about 70\% of this sample, not 100\%.  However, this 
lack of purity is balanced by the fact that Hyde \& Bernardi 
select about 75\% of the Es, not 100\%:  the purity and 
completeness effects approximately cancel.  Similarly, although 
requiring $C_r\ge 2.86$ produces counts which are similar to those 
of E+S0s, about 40\% of the sample is made of Sas and later types, 
but the purity again approximately cancels the incompleteness.  
Finally, the counts when $C_r\ge 2.6$ are similar to those in the 
E+S0+Sa sample, although 25\% of the objects are of later type.


\section{The stellar mass function in the full sample}\label{phiMass}
Our stellar mass function has considerably more objects at large 
$M_*$ than reported in previous work.  Before we quantify this, 
we show the results of a variety of tests we performed to check 
that the discrepancy with the literature is real.  This is important, 
since the high mass end has been the subject of much recent attention 
(e.g., in the context of the build-up of the red sequence).

\subsection{Consistency checks}
\begin{figure}
 \centering
 \includegraphics[width=0.99\hsize]{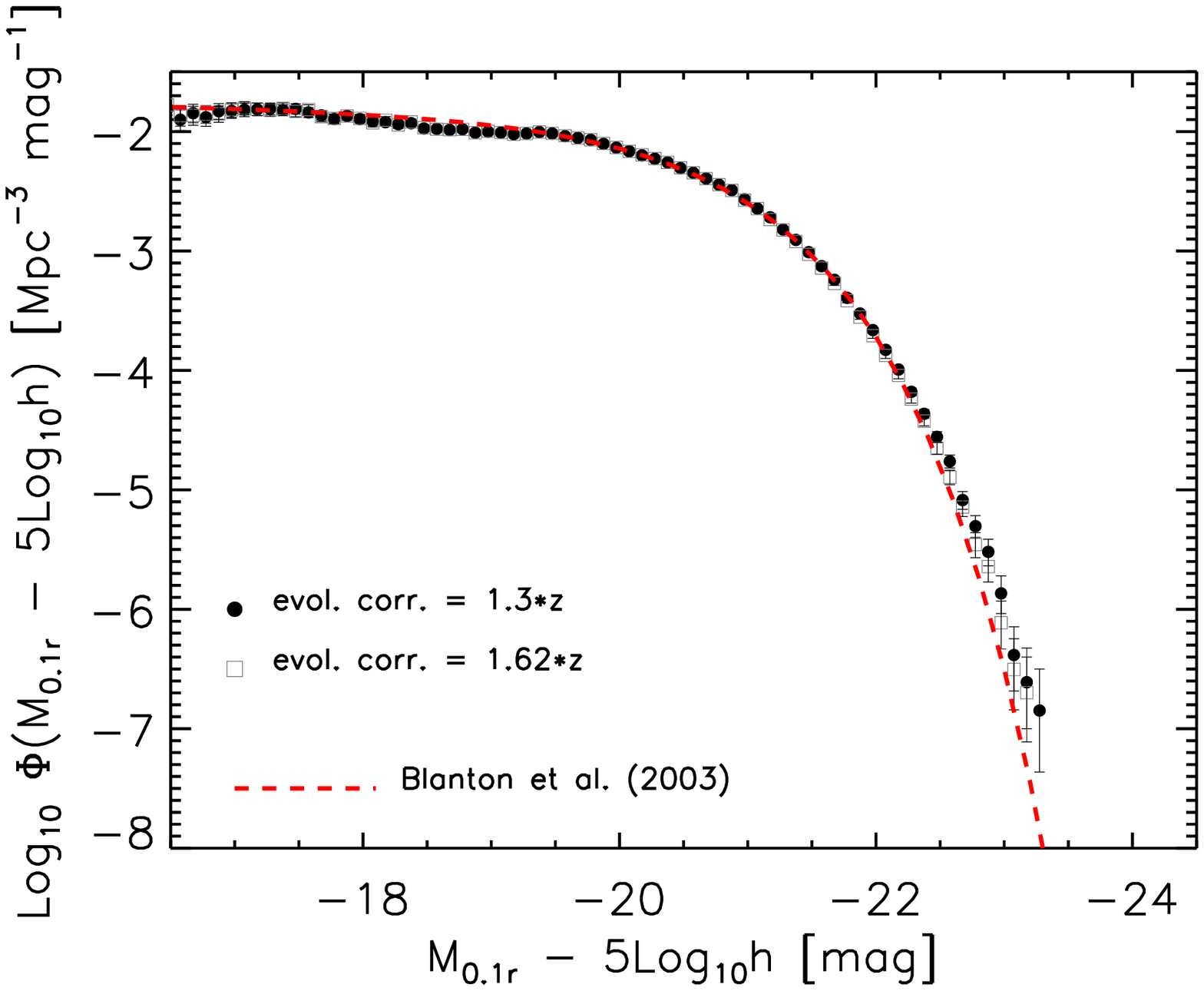}
 \includegraphics[width=0.99\hsize]{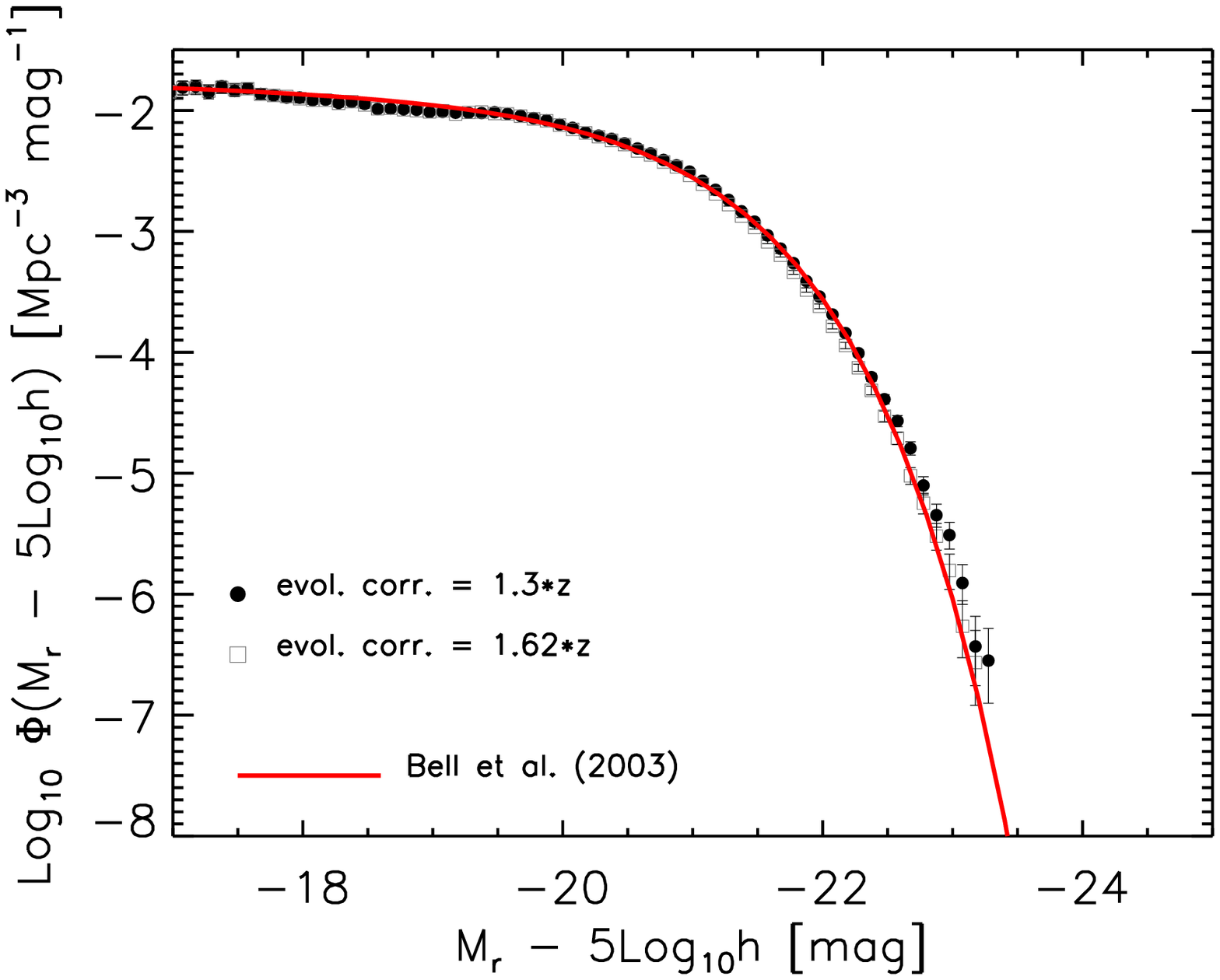}
 \caption{Luminosity function in the $^{0.1}r-$ and $r-$bands (top 
          and bottom panels), determined from $r-$band Petrosian 
          apparent magnitudes in the range $14.5-17.5$, for the two 
          choices of the pure luminosity evolution parameter advocated 
          by Blanton et al. (2003) and Bell et al. (2003):  $1.6z$ and
          $1.3z$, respectively.  
          Dashed line in top panel shows the fit reported by 
          Blanton et al. (2003); 
          solid line in bottom panel shows that reported by 
          Bell et al. (2003).}
 \label{LFs}
\end{figure}

\begin{figure}
 \centering
 \includegraphics[width=0.99\hsize]{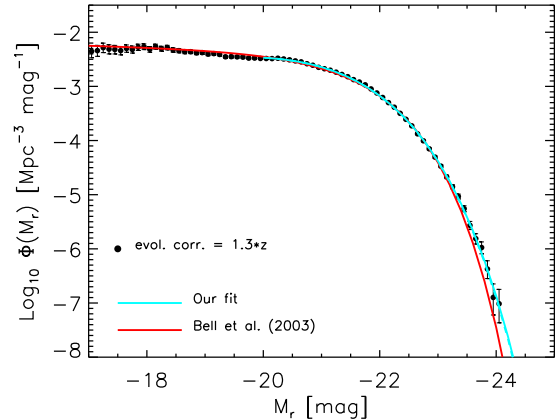}
 \caption{Luminosity function in the $r-$band determined from 
          Petrosian apparent magnitudes in the range
          $14.5\le m_{r,{\tt Pet}}\le 17.5$. 
          Red line shows the fit reported by Bell et al. (2003). 
          Cyan line shows our fit to equation~\eqref{phiX} 
          for $\log_{10} M_*/M_\odot > 10.5$, with parameters 
          reported in Table~\ref{tabLpetro}. }
 \label{LF2}
\end{figure}

We first checked that we were able to reproduce previous results for 
the luminosity function.  These have typically used Petrosian rather 
than {\tt cmodel} magnitudes, and different $H_0$ conventions.  
The top panel in Figure~\ref{LFs} shows the result of estimating 
the luminosity function using Petrosian magnitudes (using the 
$V_{\rm max}$ method and code as in the previous sections) and  
the curve show the Schechter function fits reported by 
Blanton et al. (2003).  This agreement shows that our algorithms 
correctly transform between different $H_0$ conventions, and 
between different definitions of $k$-corrections.  It also shows 
that varying the evolution correction between the value reported 
by Blanton et al. ($1.6z$) and that from Bell et al. ($1.3z$), 
which we use when estimating stellar masses, makes little difference.  

The bottom panel shows the Schechter function fit reported by 
Bell et al. (2003); there is good agreement. However, note that 
here we have not shifted the Petrosian magnitudes brightwards by 
0.1~mags for galaxies with $C_r > 2.6$ (Bell et al. did this to 
account crudely for the fact that Petrosian magnitudes underestimate 
the luminosity of early-types).  At faint luminosities, the measurements 
oscillate around the fits, suggesting that fits to the sum of two 
Schechter functions will provide better agreement, but we do not 
pursue this further here.  
At the bright end, we find slightly more objects than either of 
the Bell et al. (top) or Blanton et al. (bottom) fits, but a glance 
at Fig.~5 in Blanton et al. shows that the fit they report slightly 
underestimates the counts in the high luminosity tail.  

\begin{figure*}
 \centering
 \includegraphics[width=0.85\hsize]{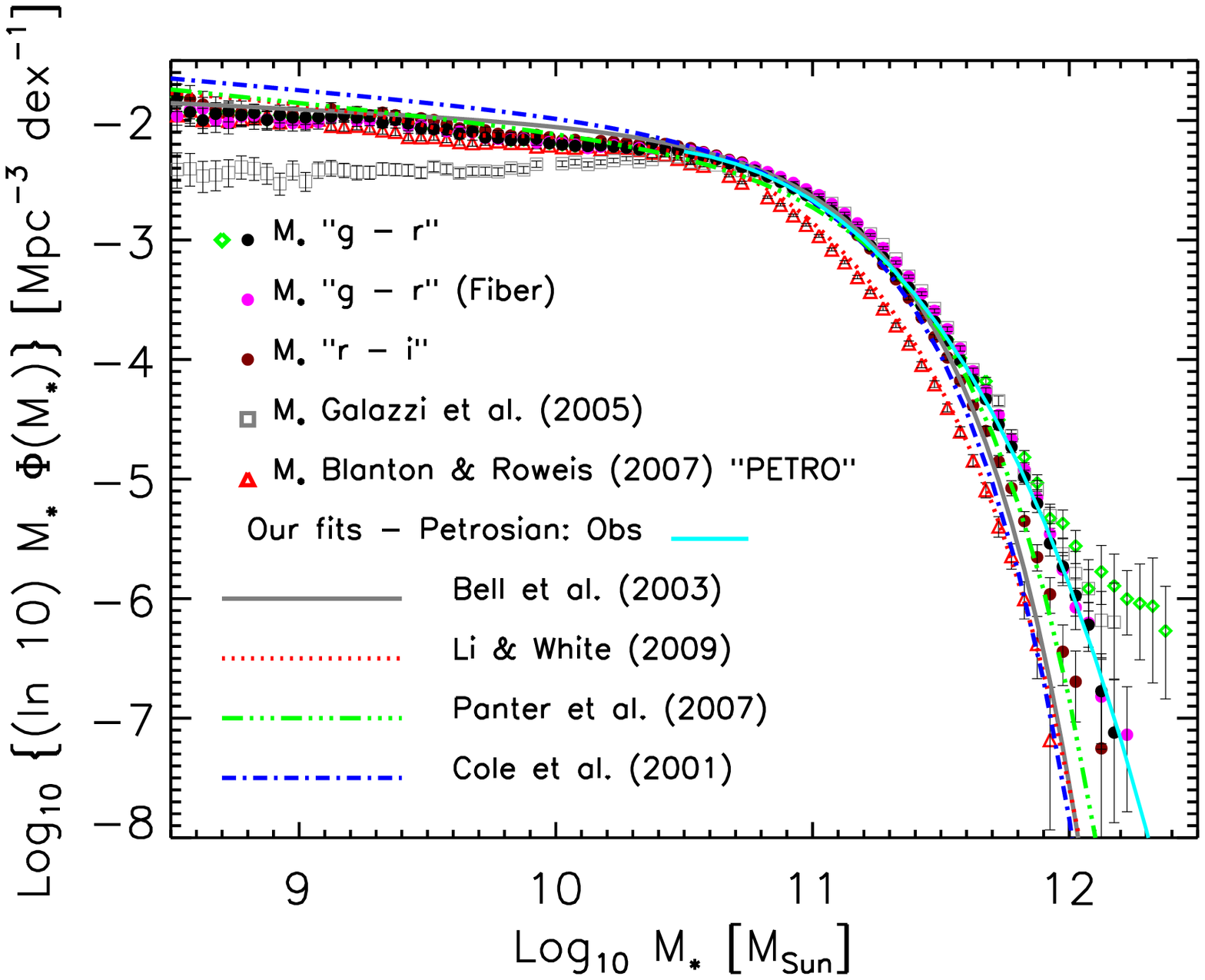}\\
 \vspace{-1cm}
 \includegraphics[width=0.85\hsize]{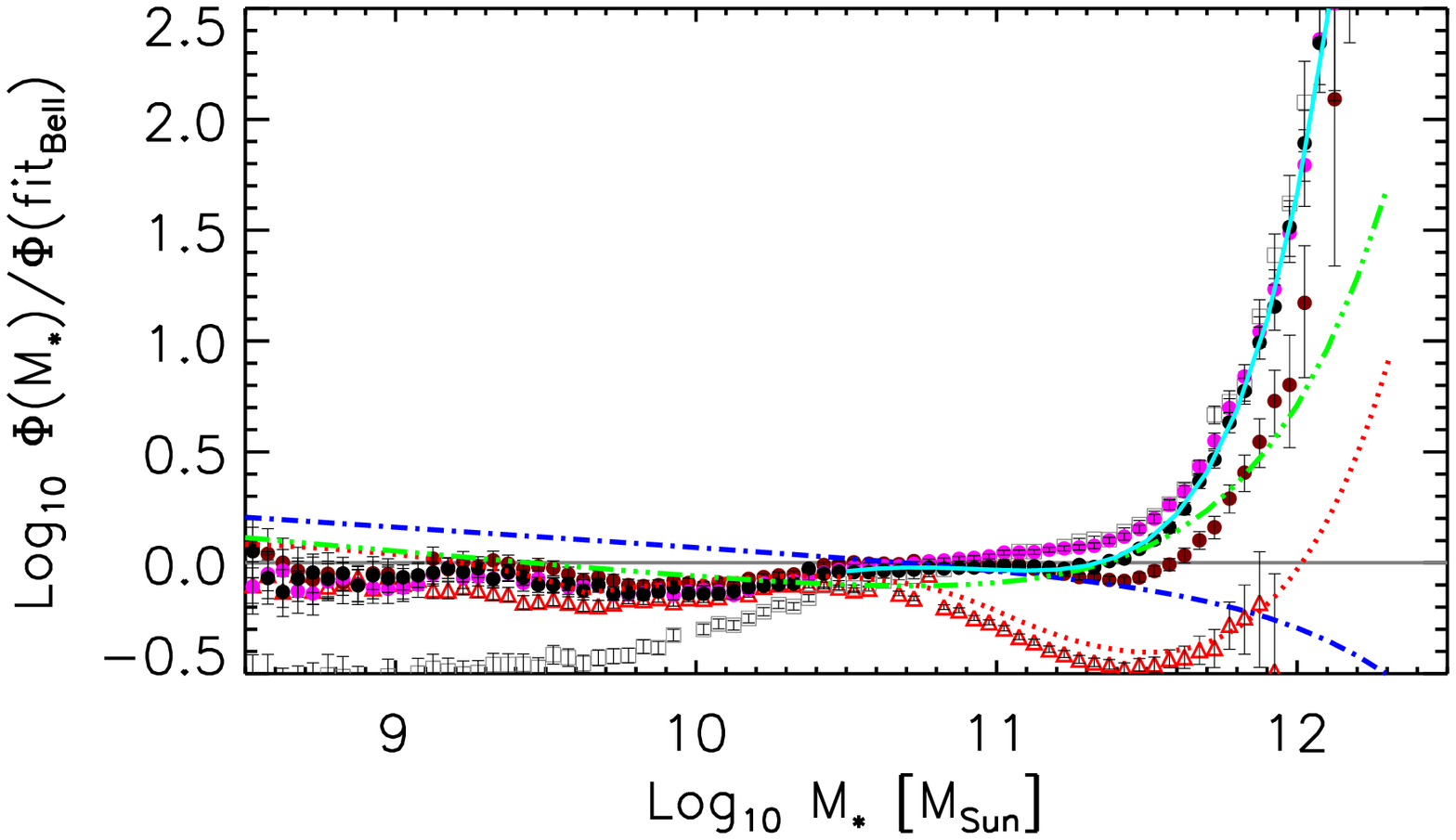}  
 \caption{Stellar mass functions estimated from Petrosian magnitudes.  
    Top:  Green open diamonds show the distribution of $M_*$ estimated 
    from $g-r$ colors (equation~\ref{gmrBell}) 
    for the full SDSS sample; 
    filled black circles show the same after removing outliers 
    at $L\ge 10^{11}L_\odot$ or $M_*\ge 10^{11.5}M_\odot$ 
    (see Figure~\ref{outliers}), and filled grey circles use 
    $r-i$ (equation~\ref{rmiBell}) instead.
    Grey open squares use $M_*$ estimates from Gallazzi et al. (2005).  
    Open red triangles show $M_{*{\rm Petro}}$ 
    from Blanton \& Roweis (2007).
    Solid cyan line shows our fit to equation~\eqref{phiX} 
    with parameters reported in Table~\ref{tabLpetro} 
    for $\log_{10} M_*/M_\odot > 10.5$.
    Solid grey curve shows the fit reported by Bell et al. (2003), 
    dashed-dot blue curve that of Cole et al. (2001), 
    dashed-dot-dot green curve that of Panter et al. (2007),
    and dotted red curve the fit from Li \& White (2009), 
    all transformed to $H_0=70$~km~s$^{-1}$Mpc$^{-1}$ and 
    Chabrier (2003) IMF.
    Bottom:  Same as top panel, but now all quantities have been 
    normalized by the Bell et al. (2003) fit (and the results 
    corresponding to the green open diamonds are not shown). }
 \label{MsFpet}
\end{figure*}

\begin{figure}
 \centering
 \includegraphics[width=0.92\hsize]{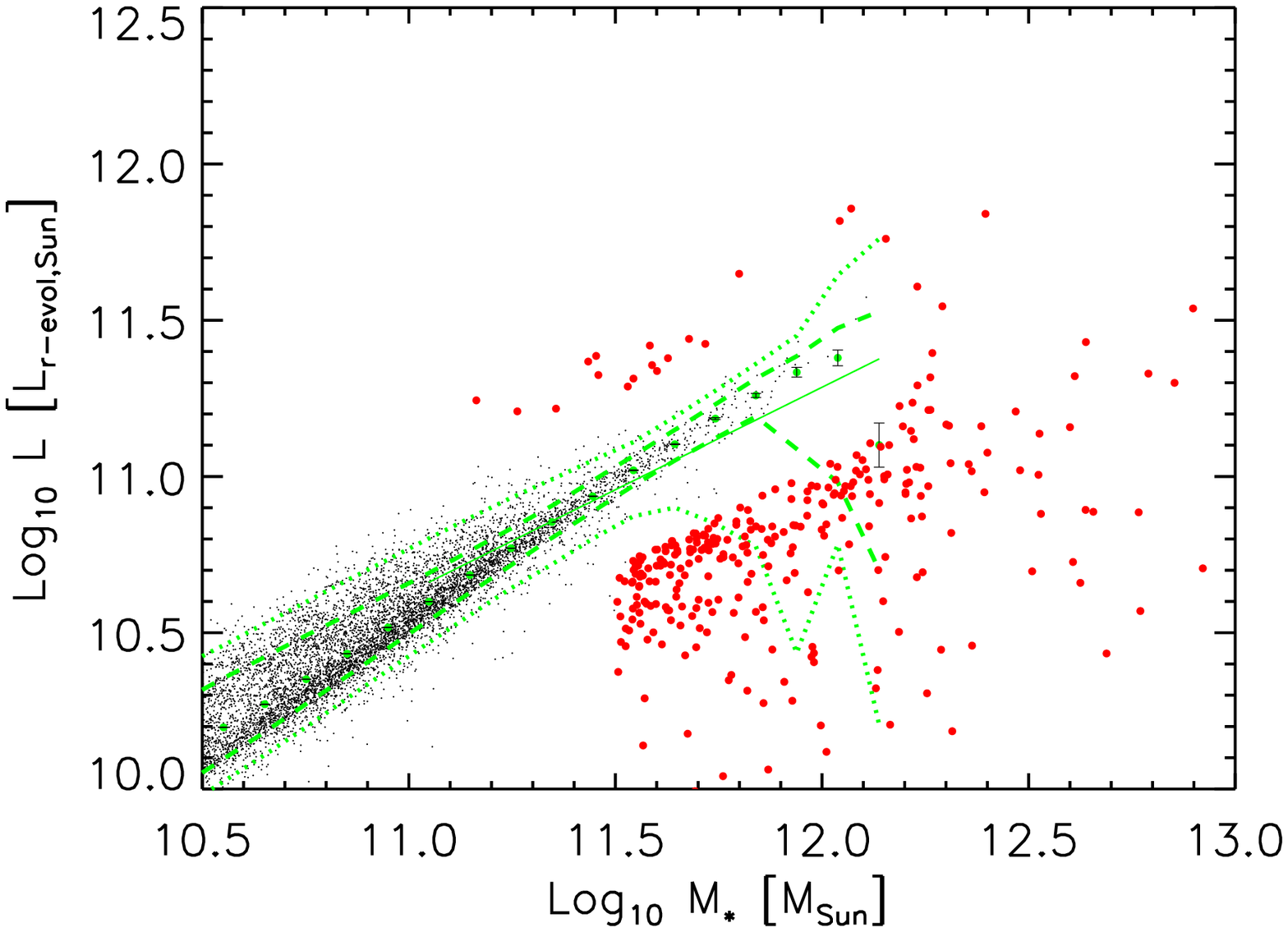}
\caption{Petrosian $L_r$ vs $M_*$ ($M_*$ computed using equation~\ref{gmrBell}). 
         Small dots show a representative subsample of the galaxies. 
         Solid green line shows the linear fit measured for 
         $M_*\ge 10^{11}M_\odot$.
         Dashed and dotted green lines show the $1$ and $2-\sigma$ 
         range around the median, respectively.  
         Red filled circles show outliers that were removed when 
         determining $\phi(L)$ and $\phi(M_*)$. 
         }
 \label{outliers}
\end{figure}

In contrast to the good agreement for the luminosity function, 
our estimates of the stellar mass function (Figure~\ref{MsFpet}) 
show a significant excess relative to previous work at 
$M_*\ge 10^{11.5}M_\odot$.  (Note that, to compare with previous 
work, we convert from $M_*/L_r$ to $M_*$ by using the Petrosian 
magnitude. However, we do not shift the Petrosian magnitudes 
brightwards by 0.1~mags for galaxies with $C_r>2.6$ -- a shift 
that was made by Bell et al. (2003).  If we had done so the excess 
would be even larger).
To ensure that the discrepancy with previous fits is not caused 
by outliers, we removed galaxies with $L\ge 10^{11}L_\odot$ or 
$M_*\ge 10^{11.5}M_\odot$ which differ by more than 0.3~dex from 
the linear fit (solid line in Figure~\ref{outliers}), in luminosity 
or $M_*$.  Except for the $g-r$ based $M_*$, where we show results 
before and after removal of these outliers, all the other 
measurements shown in Figure~\ref{MsFpet} are from the sample in 
which these outliers have been removed.  Note that while removing 
outliers makes the plot slightly cleaner (see Figure~\ref{MsFpet}), 
the discrepancy at high $M_*$ remains.

The discrepancy is most severe if we use stellar masses from 
Gallazzi et al. (2005).  (Note that they do not provide stellar 
mass estimates for fainter, typically lower mass objects.)  
The discrepancy is slightly smaller if we use equation~(\ref{gmrBell})
to translate $g-r$ color into $M_*/L_r$ (for which both $g-r$ and 
$L_r$ are evolution corrected; if we had used restframe quantities 
without correcting for evolution the discrepancy would be even worse). 
Using $r-i$ instead (equation~\ref{rmiBell}) yields results which 
are more similar to the original Bell et al. (2003) fit.  
And finally, $\phi(M_*)$ based on  $M_{*{\rm Petro}}$ of 
Blanton \& Roweis (2007) has the lowest abundances of all. 

The dotted line shows that our measurement of the distribution of 
$M_{*{\rm Petro}}$ is well-fit (except for a small offset) by the 
formula reported by Li \& White (2009), which was based on their 
own estimate of $\phi(M_*)$ from Blanton \& Roweis $M_{*{\rm Petro}}$.
The fact that we find good agreement with their fit suggests that 
our algorithm for estimating $\phi(M_*)$ from a given list of $M_*$ 
values is accurate. 

The dashed-dot-dot green line shows the fit reported by 
Panter et al. (2007) who computed stellar masses for a sample of 
$3\times 10^5$ SDSS galaxies based on the analysis of the spectral 
energy distribution of the SDSS spectra.  While this fit lies slightly 
below our data at high $M_*$, the discrepancy is smaller than it is 
for most of the other fits we show.

We argued previously that aperture effects may have inflated the 
Gallazzi et al. masses slightly.  As a check, we recomputed masses 
from the $g-r$ color using equation~(\ref{gmrBell}), but now, using 
the {\tt Fiber} color output by the SDSS pipeline.  In contrast to 
the {\tt model} color, which measures the light on a scale which 
is proportional to the half-light radius, this measures the light 
in an aperture which has the same size as the SDSS fiber.  
The spectro-photometry of the survey is sufficiently accurate that 
this is a meaningful comparison (e.g. Roche et al. 2009a).  Note that 
this gives abundances which are larger than those based on the 
{\tt model} colors.  Moreover, they are almost indistinguishable 
from those of $M_{*{\rm Gallazzi}}$.

The cyan solid curve shows the result of fitting equation~(\ref{phiX}) 
to our measurements of $\phi(M_*)$ based on  $g-r$ color 
(we do not show fits to equation~\ref{psiO}, because none of the other 
fits in the literature account for errors in the stellar mass estimate) 
when $\log_{10} M_*/M_\odot > 10.5$.  The best-fit parameters are 
reported in Table~\ref{tabLpetro}. The Table also reports the fit 
to the full sample (i.e. $\log_{10} M_*/M_\odot > 8.5$). 
(These differ from those reported in the previous section, because 
here they are based on Petrosian rather than {\tt cmodel} magnitudes.)
The Figures show the results for $M_r < -20$ and 
$\log_{10} M_*/M_\odot > 10.5$ because this is the regime of most 
interest here.

Our estimates depart from the Bell et al. fit at densities of order 
$10^{-4}$~Mpc$^{-3}$.  However, their analysis was based on only 
412~deg$^2$ of sky, for which the expected number of objects on the 
mass scale where we begin to see a discrepancy 
($M_*\sim 10^{11.5}M_\odot$) is of order tens.  This, we suspect, 
is the origin of the discrepancy between our results and theirs -- 
we are extrapolating their fit beyond its regime of validity.  

Bell et al. (2003) did not account for errors, so the most 
straightforward way to quantify the increase we find is to compare 
our measured counts with their fit.  If we do this for our 
Petrosian-based counts, then the stellar mass density in objects 
more massive than 
 $(1,2,3)\times 10^{11}\,M_\odot$ 
is $\sim$~$(3,24,86)$ percent larger than one would infer 
from the Bell et al. fit.  
However, Bell et al. attempted to account for the fact that 
Petrosian magnitudes underestimate the total light by shifting 
galaxies with $C_r > 2.6$ brightwards by 0.1~mags.  Since we have 
not performed such a shift, the appropriate comparison with their 
fit is really to use our measured {\tt cmodel} based counts, so the 
difference between our counts and Bell et al. are actually larger.  
We do this in the next section.

If we compare our estimate of the stellar mass density 
in objects more massive than 
 $(1,2,3)\times 10^{11}\,M_\odot$ with those
from the Li \& White (2009) fit, then our values
are $\sim$~$(109,202,352)$ percent larger.

\subsection{Towards greater accuracy at large $M_*$}

\begin{figure*}
 \centering
 \includegraphics[width=0.85\hsize]{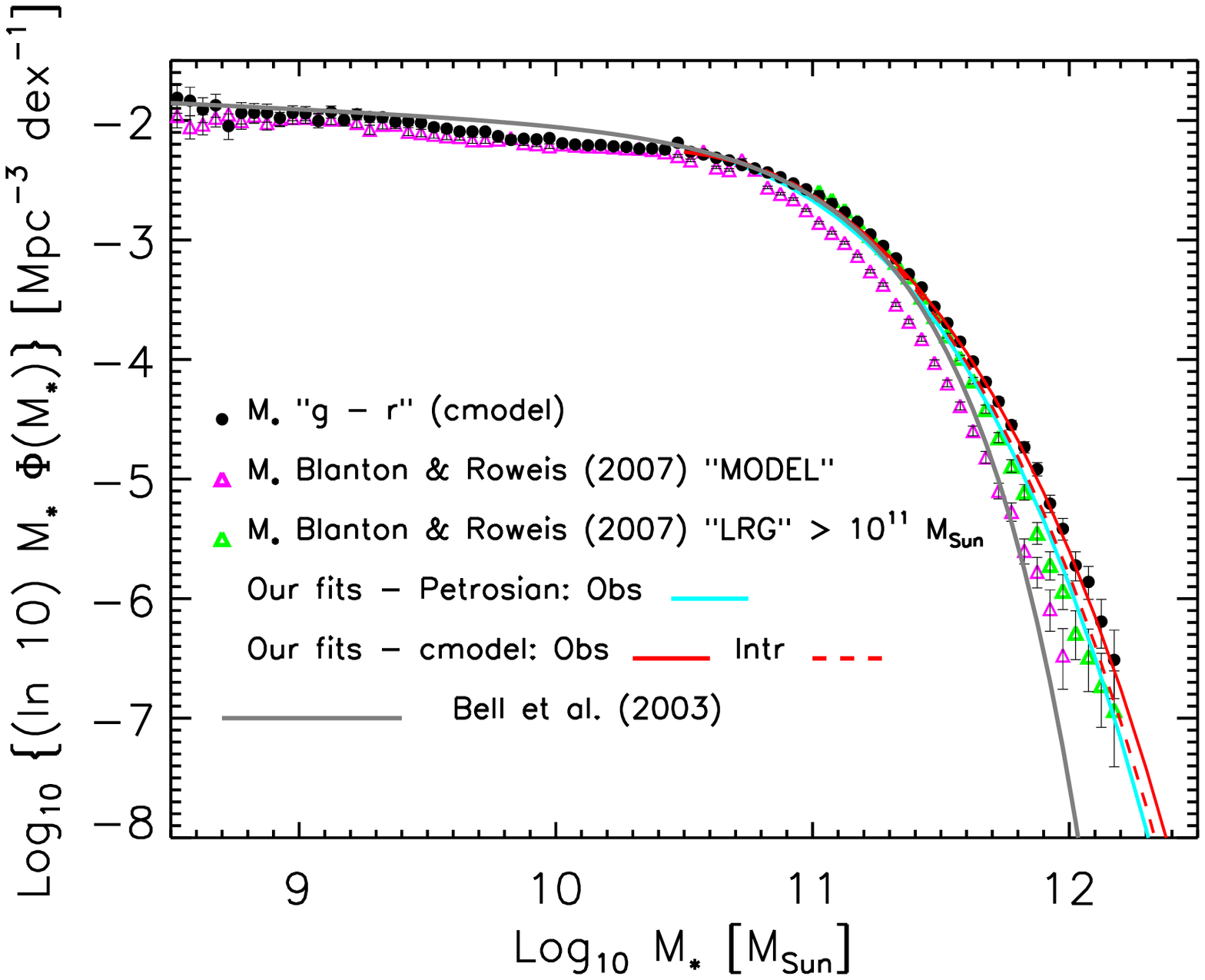}\\
 \vspace{-1cm}
 \includegraphics[width=0.85\hsize]{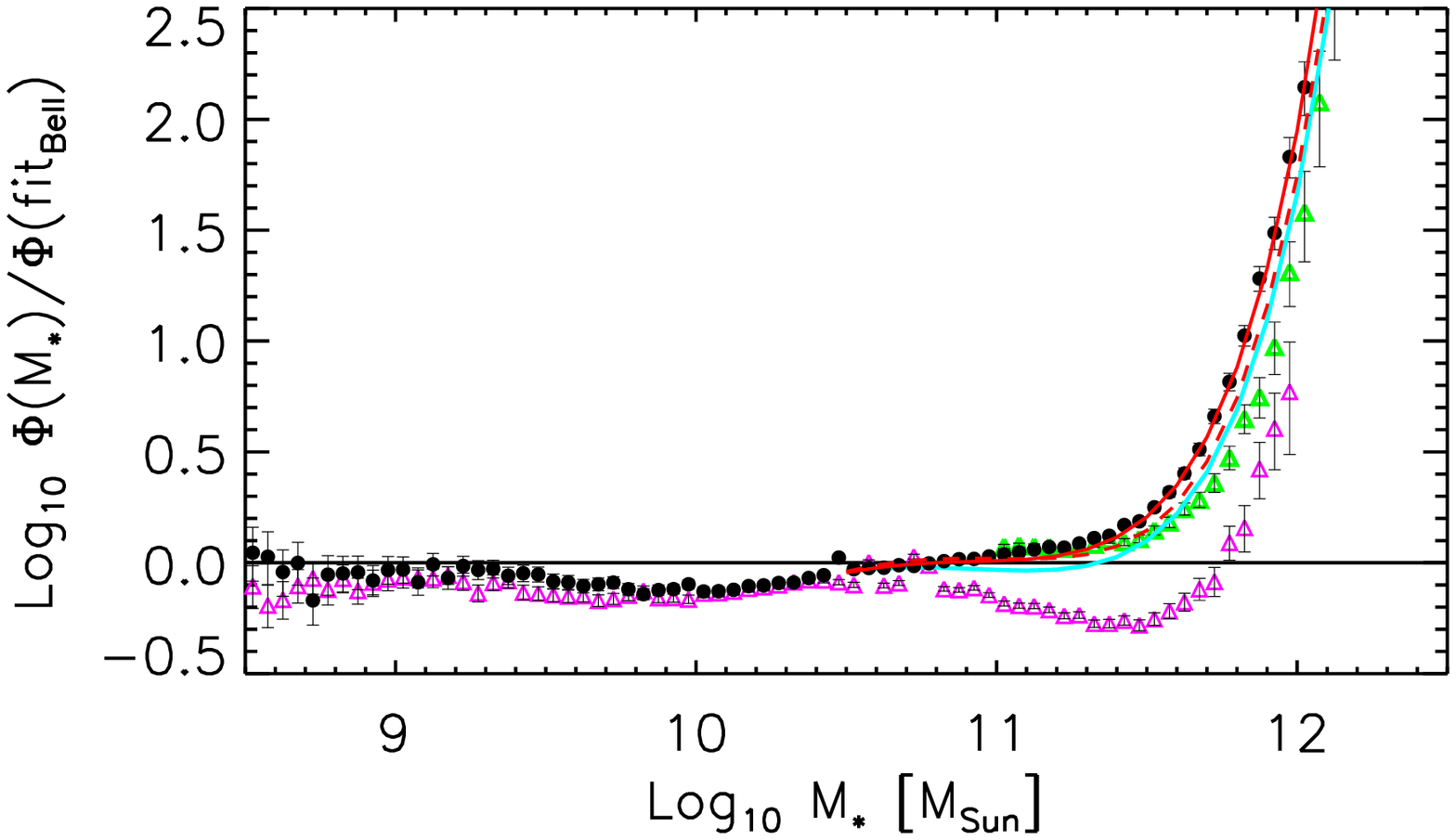}  
 \caption{Stellar mass functions estimated from {\tt cmodel} 
          magnitudes.
    Top: Filled black circles show $M_*$ estimated from $g-r$ 
    colors (equation~\ref{gmrBell}).
    Open magenta and green triangles show  
    $M_{*{\rm Model}}$ and $M_{*{\rm LRG}}$ from Blanton \& Roweis (2007).
    Solid and dashed red lines show our fits to equations~\eqref{phiX} 
    and~\eqref{psiO} with parameters reported in Table~\ref{tabLpetro} 
    for $\log_{10} M_*/M_\odot > 10.5$ estimated from {\tt cmodel} magnitudes.
    For comparison solid cyan line shows our fit to equation~\eqref{phiX} 
    with parameters reported in Table~\ref{tabLpetro} 
    for $\log_{10} M_*/M_\odot > 10.5$ estimated from Petrosian magnitudes
    (as in Figure~\ref{MsFpet}).  
    Solid grey curve shows the fit reported by Bell et al. (2003) 
    (transformed to $H_0=70$~km~s$^{-1}$Mpc$^{-1}$).
    Bottom:  Same as top panel, but now all quantities have been 
    normalized by the Bell et al. (2003) fit. }
 \label{MsFmod}
\end{figure*}

It is well known that the {\tt cmodel} luminosities are more reliable 
at the large masses where the discrepancy in $\phi(M_*)$ is largest.  
Therefore, Figure~\ref{MsFmod} compares various estimates of 
$\phi(M_*)$ based on {\tt cmodel} magnitudes.  In this case, the 
estimates based on $M_{*{\rm Model}}$ of Blanton \& Roweis (2007) 
produce the lowest abundances, (but note they are larger than those 
based on $M_{*{\rm Petro}}$ in Figure~\ref{MsFpet}), whereas those based on $M_{*{\rm LRG}}$ 
are substantially larger, as one might expect (c.f. Figure~\ref{BG-BR}).
The $M_{*{\rm LRG}}$ abundances are also in good agreement (slightly smaller)
with those based on $g-r$ color (equation~\ref{gmrBell}, with {\tt cmodel} 
magnitudes), except at smaller masses. 
(Although it is not apparent because of how we have chosen to plot 
our measurements, above $10^{11}M_\odot$, the abundances based on 
$M_{*{\rm Gallazzi}}$ and $M_{*{\rm LRG}}$ are in good agreement.)  

We argued previously that we believe the $g-r$ masses are more 
reliable than those based on $r-i$.  Therefore, we only show fits 
to the distribution of $g-r$ derived masses:
solid and dashed curves show fits to equation~(\ref{phiX}) 
and~(\ref{psiO}) respectively (the latter account for broadening 
of the distribution due to errors in the determination of $M_*$).  
Both result in larger abundances than the observed abundances 
based on Petrosian quantities -- here represented by the fit shown 
in the previous figure -- although the {\it intrinsic} distribution we 
determine for the {\tt cmodel} based masses is similar to the 
{\it observed} distribution of {\tt Petrosian}-based masses.


If we sum up the observed counts to estimate the stellar mass density 
($M_*$ from equation~\ref{gmrBell}) in objects more massive than 
 $(1,2,3)\times 10^{11}\,M_\odot$, then the result is 
 $\sim$~$(30,68,170)$ percent larger than that one infers from 
the Bell et al. (2003) fit. Using our fit to the observed distribution
(values between round brackets in Table~\ref{tabMs}, 
for $\log_{10} M_*/M_\odot > 10.5$) gives similar results:
stellar mass densities $\sim$~$(21,60,160)$ percent larger than those from 
the Bell et al. (2003) fit. 
In practice, however, the Bell et al. 
fit tends to be used as though it were the intrinsic quantity, 
rather than the one that has been broadened by measurement error.  
Our fit to the intrinsic distribution, based on {\tt cmodel} magnitudes 
(from Table~\ref{tabMs}, for $\log_{10} M_*/M_\odot > 10.5$) gives 
stellar mass densities in objects more massive than 
 $(1,2,3)\times 10^{11}\,M_\odot$ of
$(8.84, 3.00, 1.03)\times 10^7\,M_\odot$Mpc$^{-3}$.
This corresponds to an extra $\sim$~$(15,40,109)$ percent more 
than from the Bell et al. fit.  

\begin{table*}
\caption[]{Top two rows: Parameters of $\phi(L_r)$ (fit to $M_r < -17.5$) 
           and $\phi(M_*)$ (fit to $\log_{10} M_*/M_\odot > 8.5$) 
           derived from fitting equations~(\ref{phiX}) (in brackets) 
           and~(\ref{psiO}) to the observed counts based on 
           Petrosian magnitudes.
           Bottom two rows: Parameters of $\phi(L_r)$ (fit to $M_r < -20$) 
           and $\phi(M_*)$ (fit to $\log_{10} M_*/M_\odot > 10.5$). These
           second set of fits better reproduce the high luminosity and
           mass end.\\}
\begin{tabular}{lccccc}
 \hline 
  Sample & $\phi_*/10^{-2}{\rm Mpc}^{-3}$ & $X_*$ & $\alpha$ & $\beta$ & 
  $\rho_X$\\ 
 \hline
  All  $L/10^9L_\odot$ & ($8.427$) $ 8.749\pm  4.228$ & ($  15.77$) $  16.04\pm    2.18$ & ($   0.08$) $   0.08\pm    0.05$ & ($   0.827$) $   0.833\pm    0.036$ &  $   0.128  $ \\
  All  $M_*/10^9M_\odot$  & ($5.620$) $ 5.886\pm  1.839$ & ($  23.20$) $  25.86\pm    5.85$ & ($   0.14$) $   0.13\pm    0.05$ & ($   0.616$) $   0.654\pm    0.034$ &  $   0.289  $ \\
 \hline &&&&\\
  All  $L/10^9L_\odot$ & ($1.693$) $ 1.707\pm  0.432$ & ($   7.32$) $   7.56\pm    2.05$ & ($   0.55$) $   0.54\pm    0.16$ & ($   0.698$) $   0.705\pm    0.040$ &  $   0.117  $ \\
  All  $M_*/10^9M_\odot$ & ($0.888$) $ 0.857\pm  0.097$ & ($   0.94$) $   0.95\pm    0.35$ & ($   1.39$) $   1.50\pm    0.16$ & ($   0.410$) $   0.421\pm    0.016$ &  $   0.247  $ \\
 \hline &&&&\\
\end{tabular}
\label{tabLpetro} 
\end{table*}

\begin{figure}
 \centering
 \includegraphics[width=0.99\hsize]{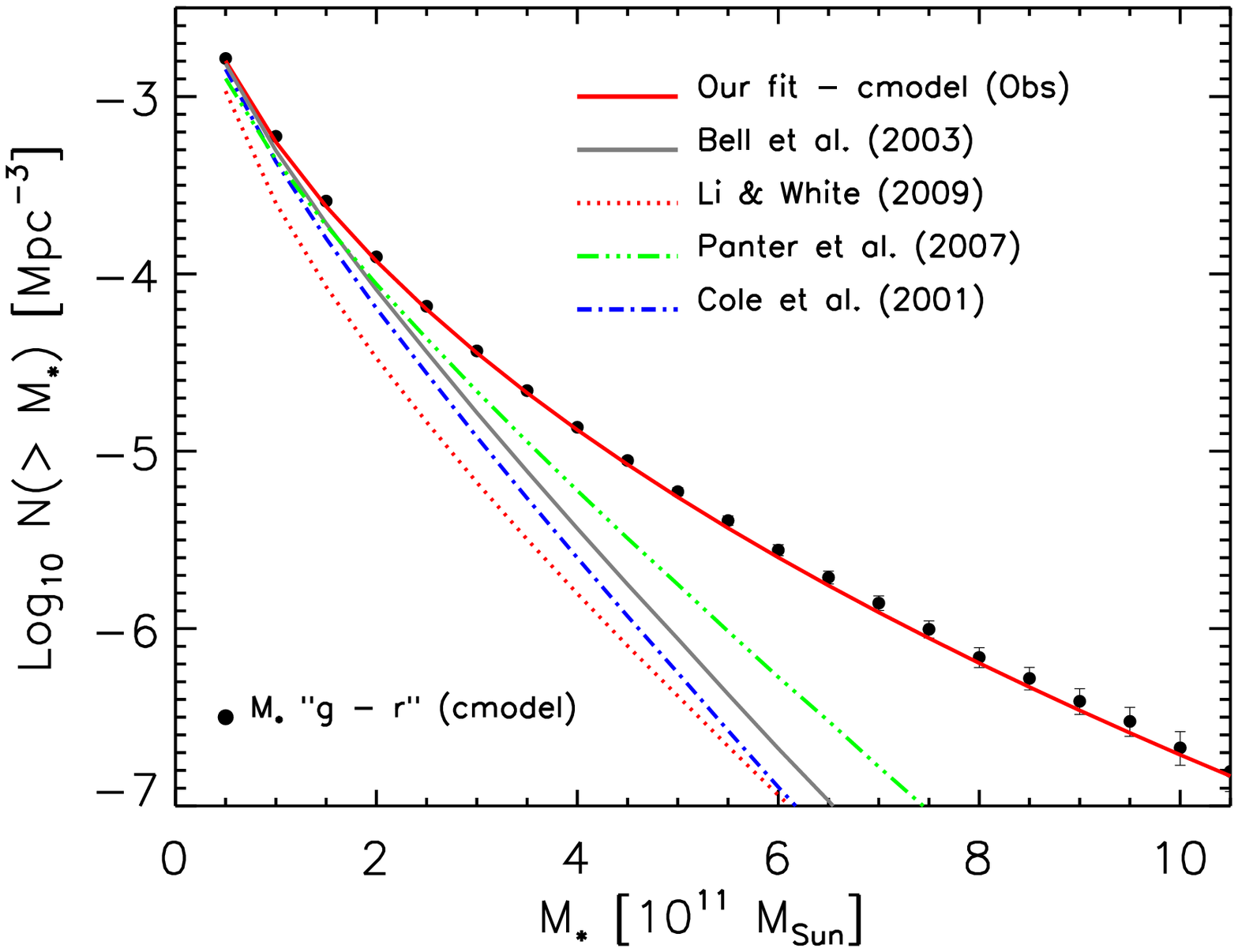}
 \includegraphics[width=0.99\hsize]{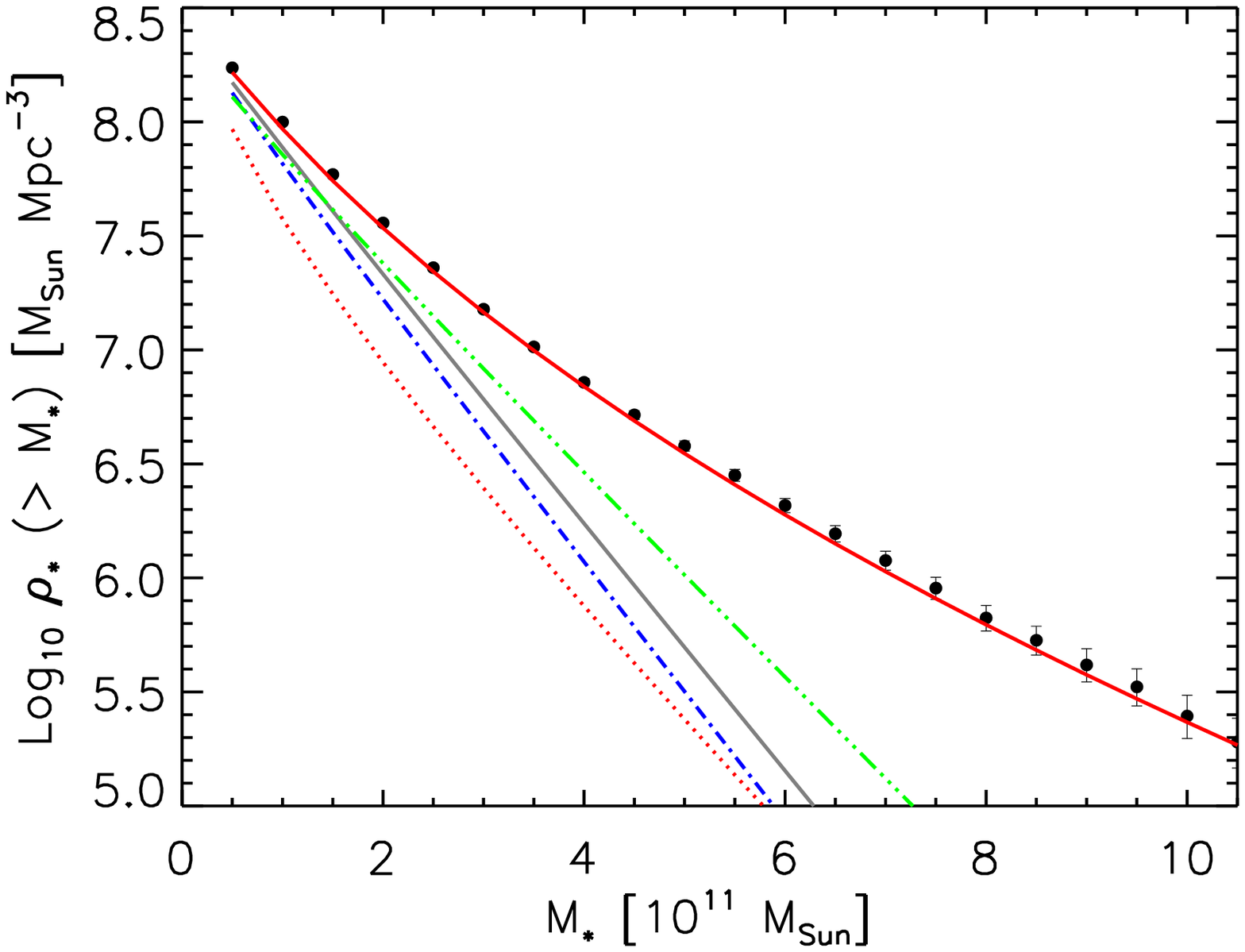}  
 \caption{Comparison of various estimates of the cumulative number density of 
          objects more massive than $M_*$ (top), and the stellar mass 
          density in these objects (bottom). All measurements were 
          transformed to $H_0=70$~km~s$^{-1}$Mpc$^{-1}$ and 
          Chabrier (2003) IMF.}
 \label{MsFcumulative}
\end{figure}

Another way to express this difference is in terms of the mass 
scale at which the integrated comoving number density of objects 
is $10^{-5}$~Mpc$^{-3}$.  For our fits to the intrinsic {\tt cmodel} 
based counts, this scale is $3.98 \times 10^{11}M_\odot$ 
($4.27 \times 10^{11}M_\odot$ for the observed fit), whereas for 
Bell et al. (2003) it is $3.31 \times 10^{11}M_\odot$.  
For $10^{-5.5}$~Mpc$^{-3}$, these scales are 
$5.25 \times 10^{11}M_\odot$ ($5.75 \times 10^{11}M_\odot$ for
the observed fit) and $4.07 \times 10^{11}M_\odot$, respectively.  
Figure~\ref{MsFcumulative} illustrates these differences graphically.  
Note that the stellar masses used by Panter et al. 
and Li \& White were obtained using Petrosian magnitudes. To account
for the difference between Petrosian and model luminosities one
could add $\sim 0.05$~dex to their values of $M_*$ (strictly 
speaking, to those objects with $C_r>2.86$) -- we have not applied 
such a shift.

\subsection{On major dry mergers at high masses}
\begin{figure}
 \centering
 \includegraphics[width=0.99\hsize]{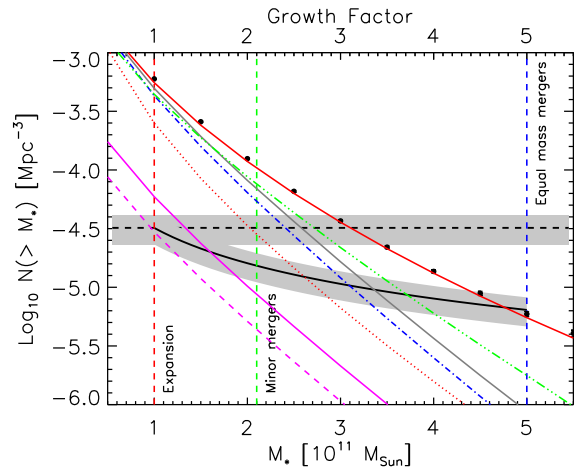}
 \caption{Cumulative number density for comparison with Fig.~3 in 
          Bezanson et al. (2009).  Most of the curves are as labeled 
          in top panel of Figure~\ref{MsFcumulative}.
          The solid magenta line shows the cumulative number density 
          from Marchesini et al. (2009) for galaxies at $2<z<3$. 
          The magenta dashed lines shows the same result but assuming that 
          the quiescent fraction of galaxies at $z\sim 2.5$ is 0.5.
          Dashed black line shows the number density of 
          galaxies with $M_* > 10^{11} M_\odot$ from the above fit as
          in Bezanson et al. (2009).
          Solid black line shows the result of having equal mass mergers.}
 \label{Bezanson}
\end{figure}

The increase in $z\sim 0$ counts at high masses matters greatly in 
studies which seek to constrain the growth histories of massive 
galaxies by comparing with counts at $z\sim 2$.  
Figure~\ref{Bezanson} shows a closer-up view of the top panel of 
Figure~\ref{MsFcumulative}.  The plot is in the same format as 
Figure~3 in Bezanson et al. (2009).  
We have added a (solid magenta) curve showing the cumulative counts 
at $2< z\le 3$ from Marchesini et al. (2009), shifted to our 
Chabrier IMF by subtracting 0.05~dex from their $M_*$ values.  The 
dashed curve below it shows the same counts shifted downwards by 
a factor of two, to reflect the fact that only perhaps half of the 
galaxies at $z\sim 2.5$ are quiescent.

\begin{table*}
\caption[]{Top two rows: Parameters of $\phi(M_*)$ 
           fit to $\log_{10} M_*/M_\odot > 8.5$ (top) and $> 10.5$ (bottom)
           derived from fitting equations~(\ref{phiX}) (in brackets) 
           and~(\ref{psiO}) to the observed counts based on {\tt cmodel} 
           magnitudes and a Salpeter IMF for elliptical galaxies, 
           an offset of -0.05 from the Salpeter IMF for S0s and 
           an offset of -0.25 for the remaining galaxies.
           Bottom two rows: Similar to top two rows but for
           $\phi(M_{\rm dyn})$, where $M_{\rm dyn} = 5R_e\sigma^2/G$.\\}
\begin{tabular}{lccccc}
 \hline  
   Sample & $\phi_*/10^{-2}$\,Mpc$^{-3}$ & $M_*/10^9\,M_\odot$ & $\alpha$ & $\beta$ & $\rho_*/10^9\,M_\odot$~Mpc$^{-3}$\\ 
 \hline
$\Delta$ IMF & ($ 35.196$) $130.824\pm  76.761$ & ($  69.09$) $  75.52\pm    7.01$ & ($   0.01$) $   0.01\pm    0.02$ & ($   0.657$) $   0.700\pm    0.024$ &  $   0.382  $ \\
$\Delta$ IMF & ($  1.797$) $  1.958\pm   0.907$ & ($  38.99$) $  49.18\pm   15.89$ & ($   0.27$) $   0.23\pm    0.12$ & ($   0.595$) $   0.647\pm    0.046$ &  $   0.365  $ \\
 \hline 
$M_{\rm dyn}$ & ($  6.066$) $  6.194\pm   2.699$ & ($  17.85$) $  21.10\pm    9.71$ & ($   0.19$) $   0.18\pm    0.10$ & ($   0.485$) $   0.512\pm    0.041$ &  $   0.617  $ \\
$M_{\rm dyn}$ & ($  2.135$) $  1.757\pm   0.474$ & ($   1.28$) $   0.45\pm    0.28$ & ($   0.82$) $   1.12\pm    0.22$ & ($   0.361$) $   0.337\pm    0.018$ &  $   0.581  $ \\
 \hline &&&&\\
\end{tabular}
\label{tabMsIMF} 
\end{table*}

Bezanson et al. (2009) argue that models in which the high redshift 
objects change their sizes but not their masses by the present time 
(e.g. Fan et al. 2008) lie well below the $z=0$ counts.  Because 
this results in an order of magnitude fewer counts than observed 
at $z=0$, such models, while viable, do not represent the primary 
growth mechanism of massive galaxies.  Other models invoke minor 
(dry) mergers (e.g., Bernardi 2009).  
If every one of the objects with $M\ge 10^{11}M_\odot$ at $z\sim 2$ 
merged with other objects of much smaller mass, then the abundance 
of these objects would not change, but their masses would:  the 
expected evolution of the population with 
 $M\ge 10^{11}M_\odot$ at $z\sim 2$ 
is shown by the horizontal shaded region.  
The fractional mass increase by a minor merger is expected to lead 
to a size increase that is larger by a factor of two 
(e.g. Bernardi 2009).  The observed 
size change suggests that the masses have not increased by more than 
a factor of about two: this is the vertical dashed line labeled 
`minor mergers'.  The horizontal shaded region intersects this 
vertical line at abundances which are about a factor of five 
smaller than our $z=0$ counts, so this model is also viable.  
On the other hand, if every one of the objects with 
$M\ge 10^{11}M_\odot$ at $z\sim 2$ merged with another of the same 
mass -- a major dry merger -- then this would shift the counts downwards 
and to the right, as shown by the shaded curved region.  In this case, 
the fractional mass and size changes are equal, so the observed 
size increase requires mass growth by a factor of five.  This is 
the vertical line labeled `equal mass mergers'.  The intersection 
of the curved shaded region with this dashed line lies above previous 
estimates of the $z=0$ abundances; this lead Bezanson et al. (2009) 
to conclude that major mergers could not be the dominant evolution 
mode at the massive end.  While we believe this an overly simplistic 
model, here we are simply pointing out that our higher abundances 
suggest that their conclusion should be revisited.  

\subsection{Morphological dependence of the IMF}

\begin{figure*}
 \centering
 \includegraphics[width=0.8\hsize]{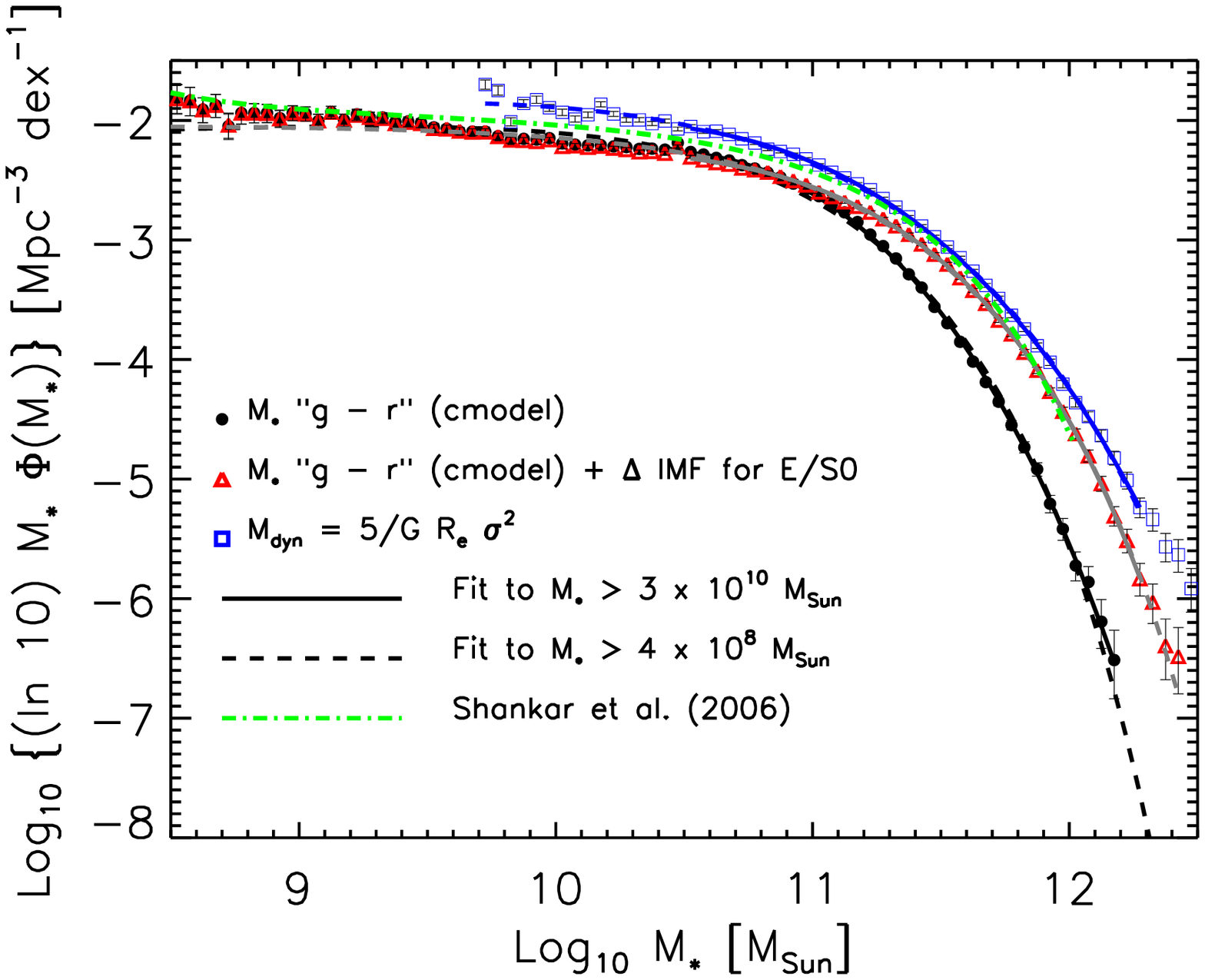}\\
 \vspace{-0.33cm}
 \includegraphics[width=0.8\hsize]{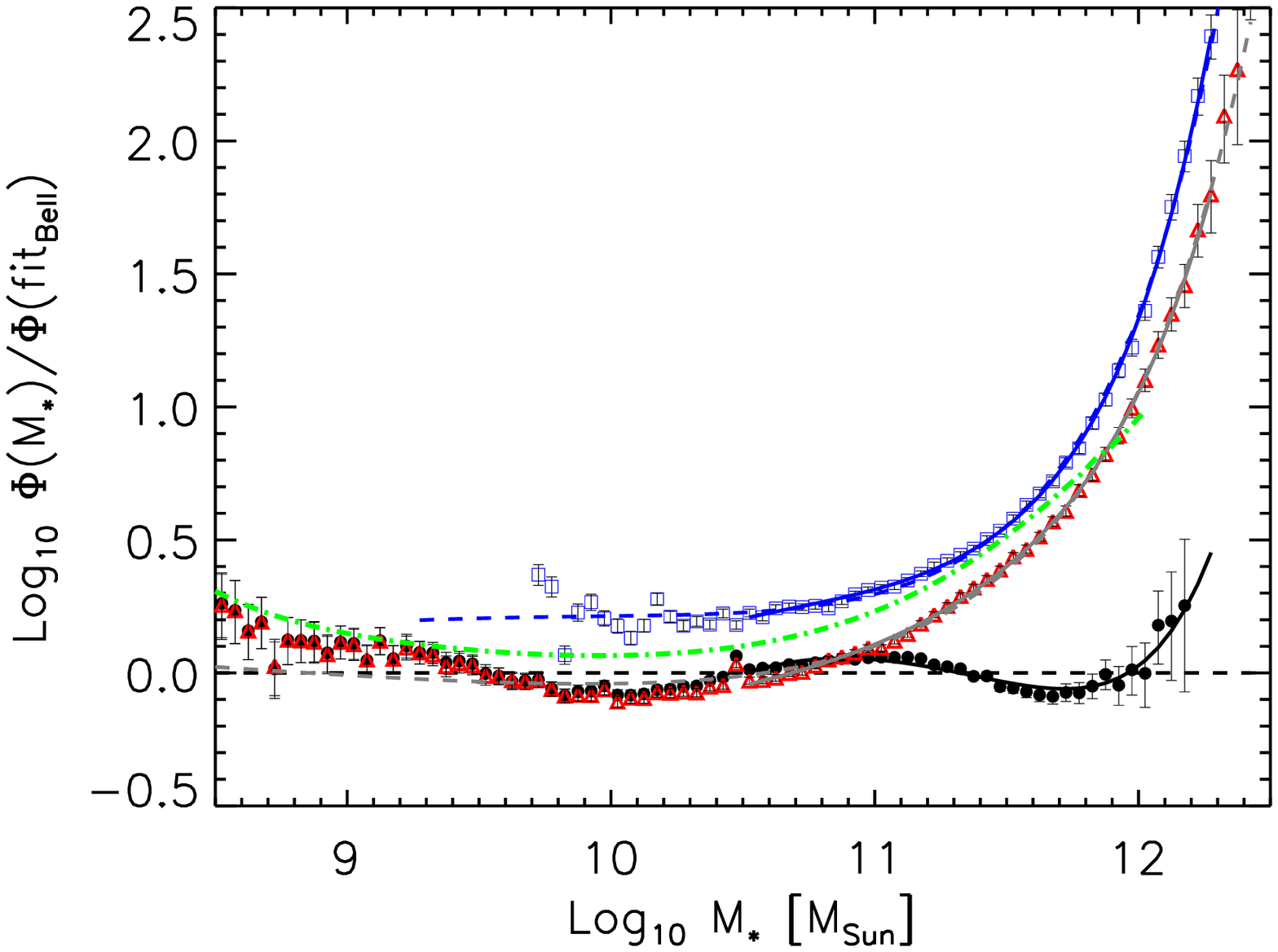}  
 \caption{Top: Solid black circles show the stellar mass function $\phi(M_*)$
         estimated from {\tt cmodel} magnitudes and $g-r$ colors 
         (equation~\ref{gmrBell}).
         Open red triangles show $\phi(M_*)$ if we use a Salpeter IMF 
         for elliptical galaxies, the Salpeter IMF offset
         by $-0.05$~dex in $M_*/L$ for a given color for S0s, 
         and the Salpeter IMF with an offset of $-0.25$ for the remaining 
         galaxies (see text for details). 
         Open blue squares show $\phi(M_{\rm dyn})$ where 
         $M_{\rm dyn} = 5R_e\sigma^2/G$.
         Solid and dashed lines show our {\it observed} fits 
         (equation~\ref{phiX}) computed using $\log_{10} M_*/M_\odot > 10.5$
         and $\log_{10} M_*/M_\odot > 8.6$, respectively.
         These fits are reported in Table~\ref{tabMsIMF}.
         Dot-dashed green line shows the dynamical mass function of 
         Shankar et al. (2006) based on dynamical mass-to-light ratios.
         Bottom:  Same as top panel, but now all quantities have been 
         normalized by the {\it observed} fit to the black circles
         (from Table~\ref{tabMs} computed for $\log_{10} M_*/M_\odot > 8.6$). }
 \label{MsFIMF}
\end{figure*}

Recently, Calura et al. (2009) have argued that a number of observations 
are better reproduced if one assumes a Salpeter (1955) IMF for ellipticals
and a Scalo (1986) IMF for spirals.  (A Salpeter IMF for ellipticals 
is also prefered by Treu et al. 2010, which appeared while our paper 
was being refereed.)  Whereas the $M_*/L$-color relation for a 
Scalo IMF is similar to that for the Chabrier IMF which we have been 
using, the relation for the Salpeter IMF is offset by 0.25~dex  
(see Table~\ref{tabIMF}).  
Since we have found a method to separate Es, S0s and Spirals we can 
incorporate such a dependence easily.  

For ellipticals, i.e., objects selected following Hyde \& Bernardi (2009), 
we compute $M_*$ by adding 0.25 to the right hand side of 
equation~(\ref{gmrBell}).  For S0s, i.e., objects with $C_r > 2.86$ that 
were not identified as ellipticals, we compute $M_*$ by adding 0.2 to 
equation~(\ref{gmrBell}), since S0s are closer to ellipticals than 
to spirals.  For all other objects, we use equation~(\ref{gmrBell})
as before.
The open red triangles in Figure~\ref{MsFIMF} show the stellar mass 
function which results.  Smooth curves show the {\it observed} 
fits to equation~(\ref{phiX});
the best-fit parameters are reported in Table~\ref{tabMsIMF}.  

Summing up the observed counts to estimate the stellar mass density
in the range log$_{10} M_*/M_\odot > 8.6$ yields 
$3.92\times 10^8M_\odot$~Mpc$^{-3}$.
Integrating the {\it intrinsic} fit (see Table~\ref{tabMsIMF}) 
over the entire range of masses gives a similar result: 
$3.82\times 10^8M_\odot$~Mpc$^{-3}$.
The intrinsic fit to log$_{10} M_*/M_\odot > 10.5$  
gives $(2.92,1.81,0.97,0.50)\times 10^8M_\odot$~Mpc$^{-3}$ 
for objects with log$_{10} M_*/M_\odot$ above $(10.5,11.0,11.3,11.5)$.  
This means that $\sim (75, 46, 25, 13\%)$ of the mass is in systems 
with $M_* > (0.3,1,2,3) \times 10^{11} M_{\odot}$.
These estimated values of the stellar mass density
(i.e. from the intrinsic fit to log$_{10} M_*/M_\odot > 10.5$), 
for objects with log$_{10} M_*/M_\odot$ above $(11.0,11.3,11.5)$, 
are $\sim$~$(105,224,388)$ percent 
larger than those infered from stellar masses computed using 
the {\tt cmodel} magnitudes but using the Chabrier IMF for all 
types (solid black circles in Figure~\ref{MsFIMF}).


Finally, we compare $\phi(M_*)$ with $\phi(M_{\rm dyn})$ (open 
blue squares in Figure~\ref{MsFIMF}), where $M_{\rm dyn} = 5R_e\sigma^2/G$ 
is the dynamical mass.  Our fits to $\phi(M_{\rm dyn})$ are reported in 
Table~\ref{tabMsIMF}.  At low $\sigma$ and $R$ the velocity dispersions 
and sizes are noisy.  This makes the mass estimate noisy below 
$\sim 4\times 10^9 M_\odot$, so we only show results above this mass. 
Using different IMFs for galaxies of different morphological type 
reduces the difference between the estimated value of the stellar 
and dynamical mass especially at larger masses.

We also find this estimate of the stellar mass function to be in 
reasonably good agreement with the one computed by Shankar et al. (2006) 
based on dynamical mass-to-light ratios calibrated following 
Salucci \& Persic (1999), Cirasuolo et al. (2005) and references 
therein, lending further support to the possibility of an 
Hubble-type dependent IMF.

\subsection{The match with the integrated star formation rate}
\label{sec|SFR}
It has been argued that a direct integration of the cosmological
star formation rate (SFR) overpredicts the local stellar mass density 
(see, e.g., Wilkins et al. 2008, and references therein).  This has led 
several authors to invoke some corrections, such as a time-variable IMF.
We now readdress this interesting issue by comparing our value for 
$\rho_*$ with that from integrating the SFR.  

The stellar mass density at redshift $z$ is given by 
\begin{equation}
 \rho_*(z) = \int_{6}^{z} dz' \frac{dt'}{dz'} 
             \dot{\rho}_*(t') (1-f_r[t(z)-t'])\, .
\label{eq|rhoz}
\end{equation}
where $\dot{\rho}_*$ is the cosmological SFR in units of
$M_{\odot}\, {\rm Mpc^{-3}\, yr^{-1}}$, and 
$f_r(t)$ is the fraction of stellar mass that has been returned to the 
interstellar medium.  For our IMF,
\begin{equation}
 f_r(t) = 0.05\,
         \ln \left(1 + \frac{t}{3\times 10^5~{\rm yr}}\right), 
\label{eq|floss}
\end{equation}
where $t$ is in years (Conroy \& Wechsler 2009).  Note that this 
results in smaller remaining mass fractions ($\sim 50\%$ at $z\sim 0$ 
instead of the usual 63-70\%), than assumed in most previous work.  
This will be important below.  

\begin{figure}
\includegraphics[width=0.99\hsize]{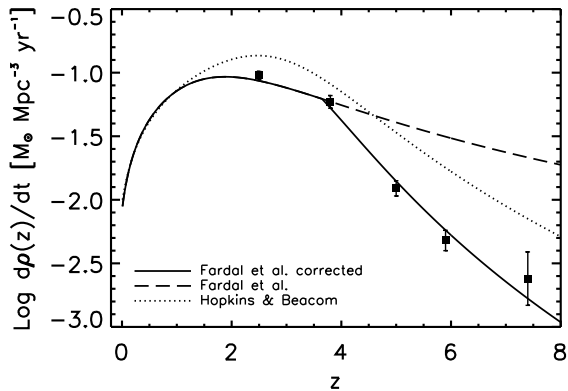}
{\caption{Cosmological SFR as given by 
 Fardal et al. (2007; long-dashed line), and by 
 Hopkins \& Beacom (2006; dotted line), compared to the 
 dust-corrected UV data by Bouwens et al. (2009; solid squares). 
 The solid line is our ``corrected'' Fardal et al. SFR fit tuned
 to match the data at $z>4$.}
\label{fig|SFRz}}
\end{figure}

\begin{figure}
\includegraphics[width=0.99\hsize]{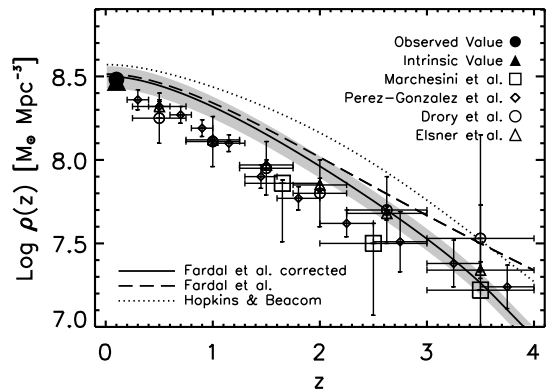}
{\caption{Comparison of the expected stellar mass density based on 
 the SFR, with the measured values at a range of redshifts.  
 Filled triangle shows our determination of the {\it intrinsic} local
 stellar mass density; filled circle shows the observed value.  
 Recent determinations at higher redshifts by other groups (as labelled) 
 are also shown.  
 Dotted, dashed and solid curves show the result of inserting the SFRs 
 shown in the previous figure in equation~(\ref{eq|rhoz}).}
\label{fig|rhoz}}
\end{figure}

We specify the SFR as follows.  
Bouwens et al. (2009) have recently calibrated the SFR over the 
range $2<z<6$ using deep optical and infrared data from ACS/NICMOS in 
the GOODS fields, \emph{UBVi} droput Lyman Break Galaxies, and ULIRGs 
data from Caputi et al. (2008).  
The solid squares in Figure~\ref{fig|SFRz} show their 
measurements, decreased by $-0.25$~dex to correct from their assumed 
Salpeter to a Chabrier IMF.  
These values of the SFR are much lower than simple extrapolations of 
the SFRs by Hopkins \& Beacom (2006) and Fardal et al. (2007), 
shown by dotted and dashed lines, respectively.  
We also note that the Fardal et al. fit matches well the updated 
SFR recent Bouwens et al. estimates in the range $2.5<z<4$.
In detail, the curves show the parameterization of Cole et al. (2001),
\begin{equation}
 \dot{\rho}_*(z)=\frac{(a+bz)\gamma}{1+(z/c)^d}\, .
 \label{eq|SFR}
\end{equation}
Fardal et al. (2007) set $a=0.0103, b=0.088, c=2.4, d=2.8, \gamma=1$, 
and we then multiply the total by 0.708 ($-0.1$~dex) to correct from 
the assumed diet-Salpeter to our Chabrier IMF.  
Hopkins \& Beacom (2006) set $a=0.014, b=0.11, c=1.4, d=2.2, \gamma=0.7$ 
(all parameters defined for $h=0.7$).  We convert from their IMF 
(from Baldry \& Glazebrook 2003) to the Chabrier IMF by multiplying by 
$1.135$ (i.e., $0.055$~dex, see Table~\ref{tabIMF}).
Based on detailed spectral modelling, 
Bouwens et al. (2009) concluded that the discrepancy is due to 
dust extinction for star forming galaxies in this redshift range being 
smaller then previously assumed.  
(But we note that GRB-based estimates from, e.g., Kistler et al. 2009, 
suggest this is not a closed issue.)
To improve the match with Bouwens et al., we use the Fardal et al. 
values at $z<3.65$, but set 
 $a=0.0134,b=0.0908,c=3.1,d=6.5,\gamma=0.7$ 
at $z>3.65$.  This is shown by the solid curve.  

The dotted, dashed and solid curves in Figure~\ref{fig|rhoz} show the 
result of inserting these three models for the SFR (Hopkins \& Beacom, 
Fardal et al., and Fardal et al. corrected) into equation~(\ref{eq|rhoz}).  
The gray band bracketing the solid curve shows the typical $\sim 15\%$ 
$1\sigma$ uncertainty (estimated by Fardal et al. 2007) associated with 
the SFR fit.  
The Figure also shows a compilation of estimates of the stellar mass 
density over a range of redshifts.  
Our own estimate of the local $\rho_*$ value is shown by the filled 
triangle (filled circle shows the observed rather than intrinsic value); 
it is in good agreement with the one from Bell et al. (2003) (corrected to 
a Chabrier IMF), and is slightly larger than those from 
Panter et al. (2007), and Li \& White (2009).  
Comparison of the measured $\rho_*(z)$ values with our new estimate of 
the integrated SFR shows that the measurements lie only slightly below 
the integrated SFR at all epochs, with the discrepancy smallest at $z>2$ 
and at $z\sim 0$. Also note that K or NIR selected high-$z$ galaxies
might be missing a significant population of highly obscured, dust 
enshrouded, forming galaxies in the range $0.5<z<2$.

The improvement with respect to previous works is due to the combined 
effects of a larger recycling factor and smaller high-$z$ SFR, both of 
which act to reduce the value of the integral 
(also see discussion in Shankar et al. 2006).
Despite the good agreement with the Fardal et al. estimate, we note 
that other SFR fits (e.g., Hopkins \& Beacom) yield substantially 
higher values for the local stellar mass density. Clearly, systematic
differences such as this one must be resolved before this issue is 
completely settled.


\begin{figure*}
 \centering
 \includegraphics[width=0.49\hsize]{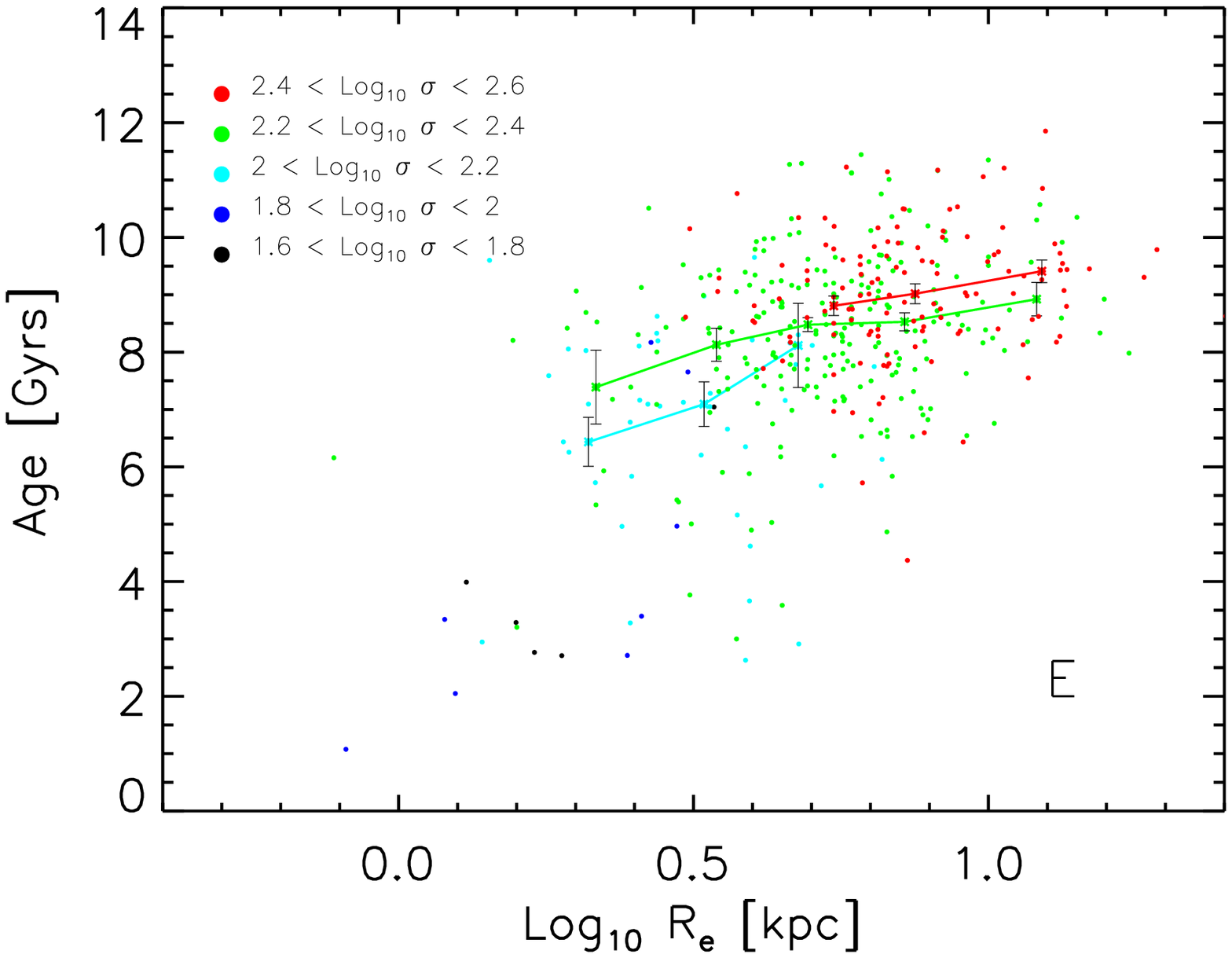}
 \includegraphics[width=0.49\hsize]{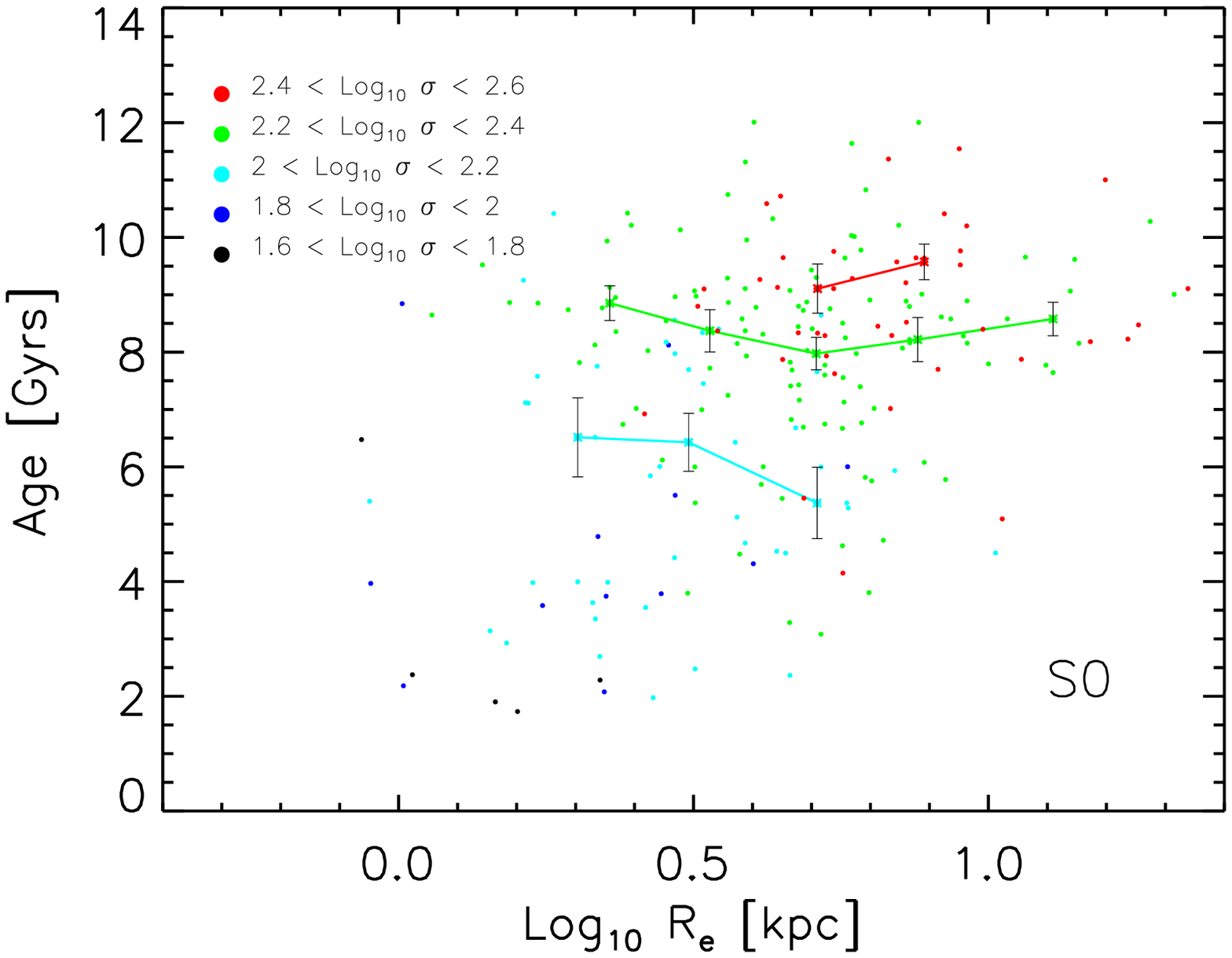} 
 \includegraphics[width=0.49\hsize]{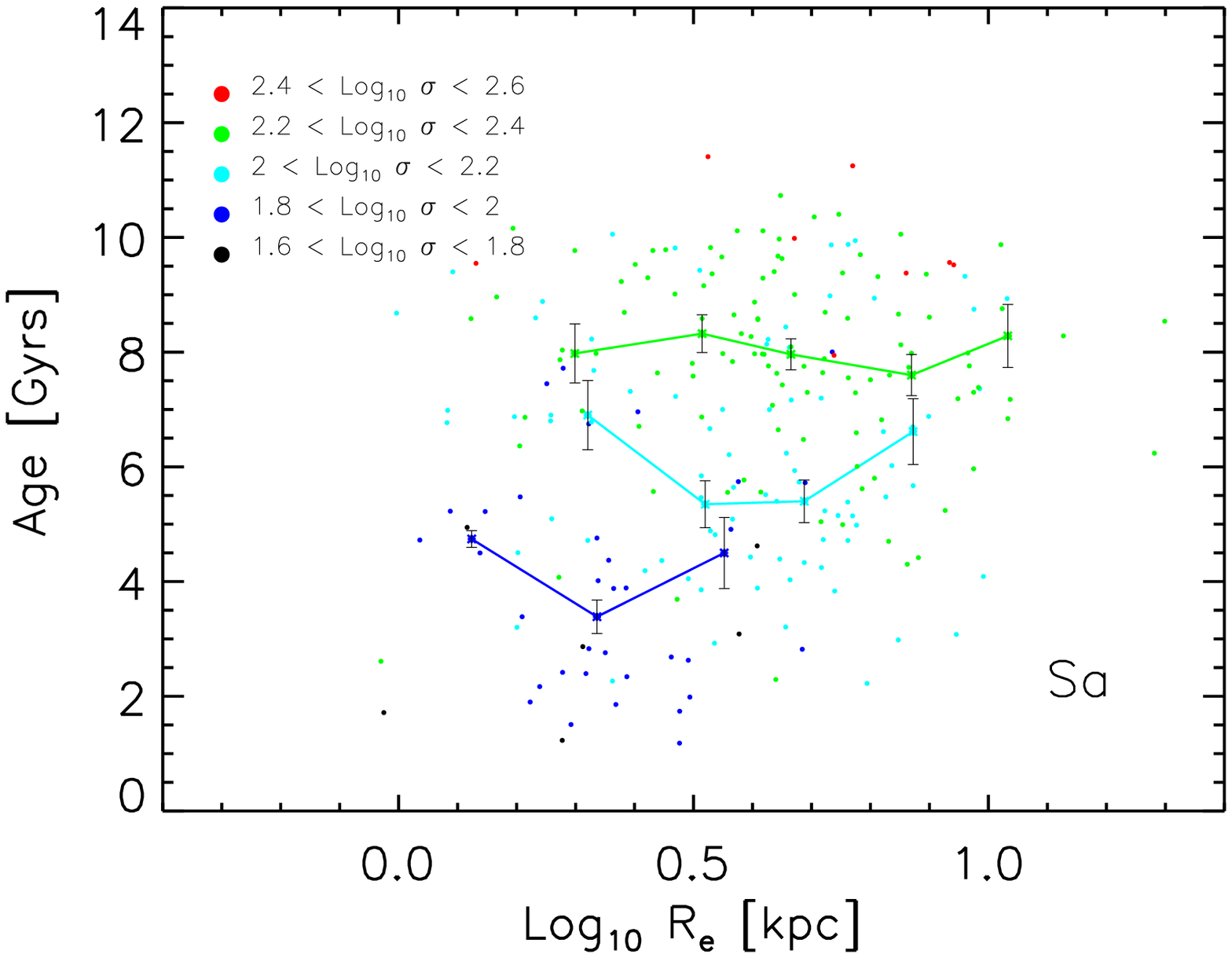}
 \includegraphics[width=0.49\hsize]{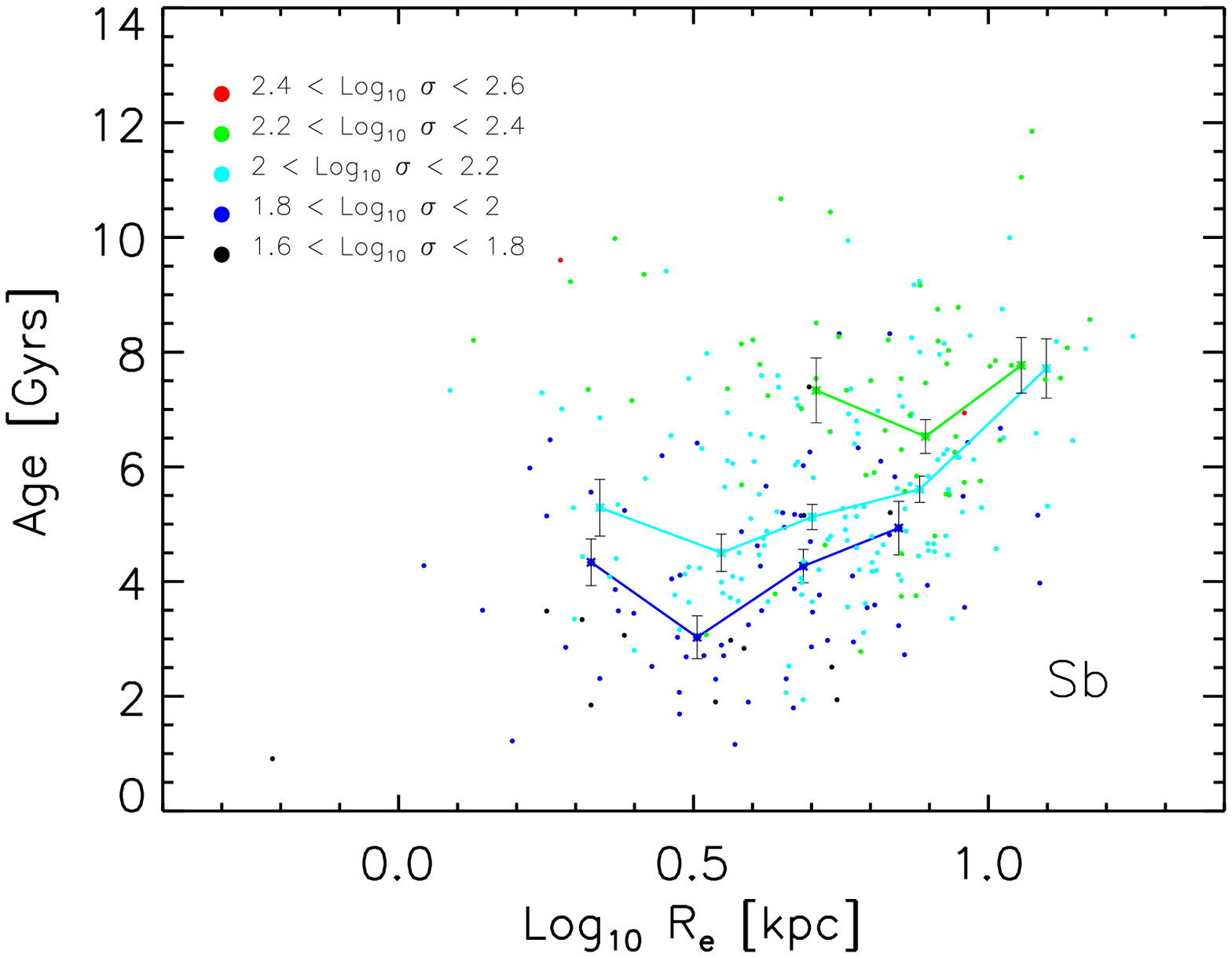}  
 \includegraphics[width=0.49\hsize]{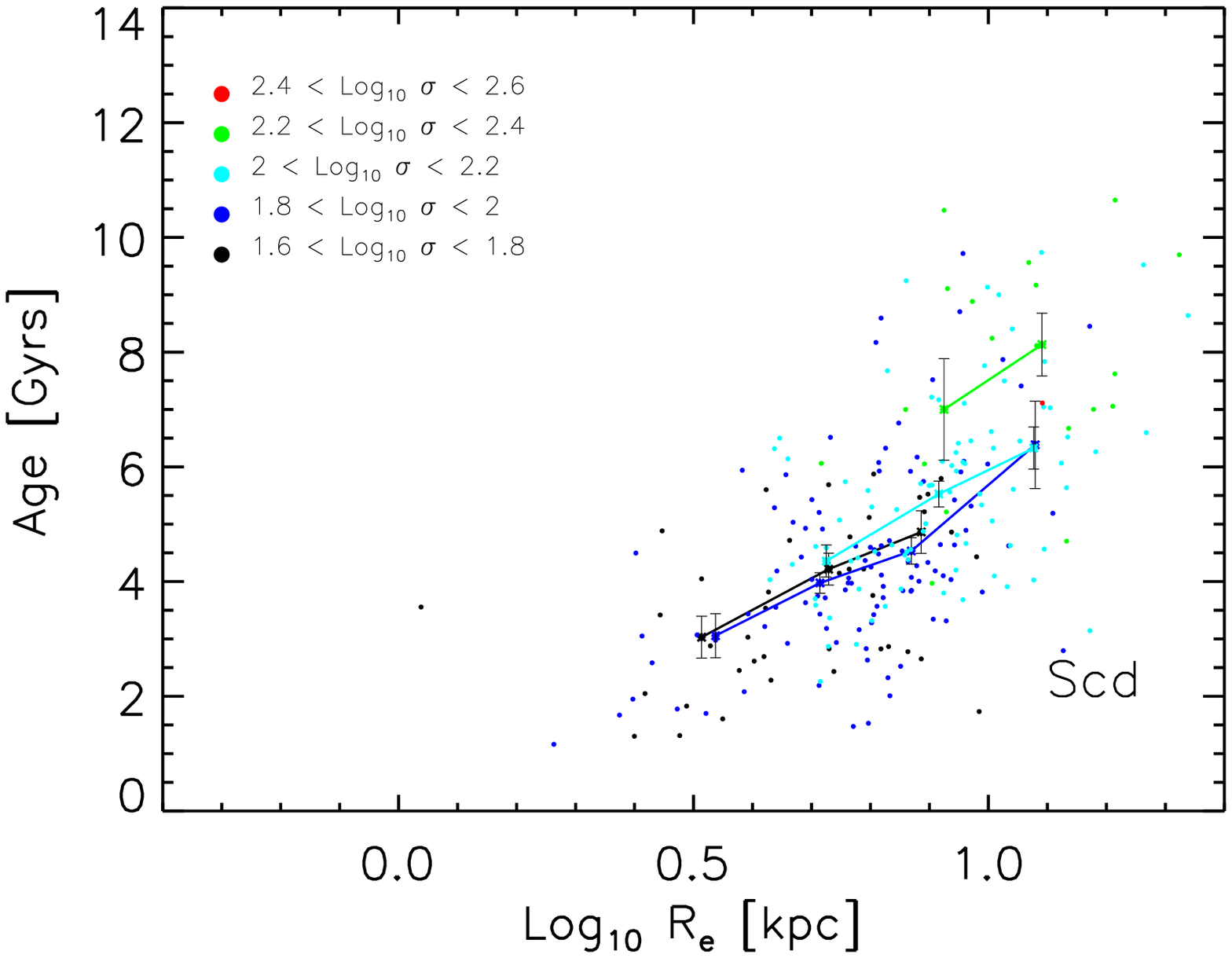}
 \includegraphics[width=0.49\hsize]{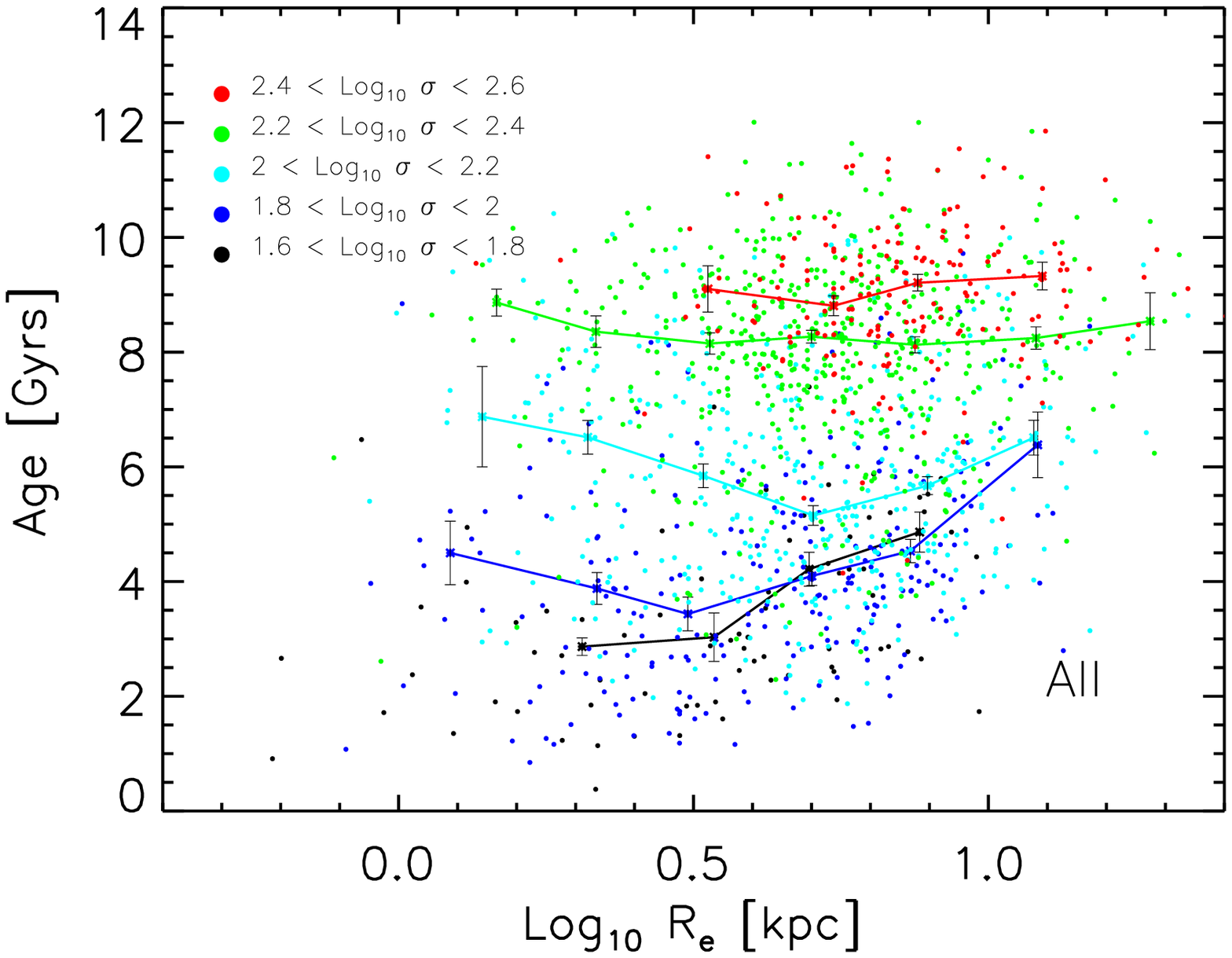}    
 \caption{Correlation between age and size at fixed velocity 
  dispersion (as labeled) for types E, S0, Sa, Sb, Sc-Sd, and All in the  
  Fukugita et al. (2007) sample. }
 \label{ageRsig}
\end{figure*}

\begin{figure*}
 \centering
 \includegraphics[width=0.49\hsize]{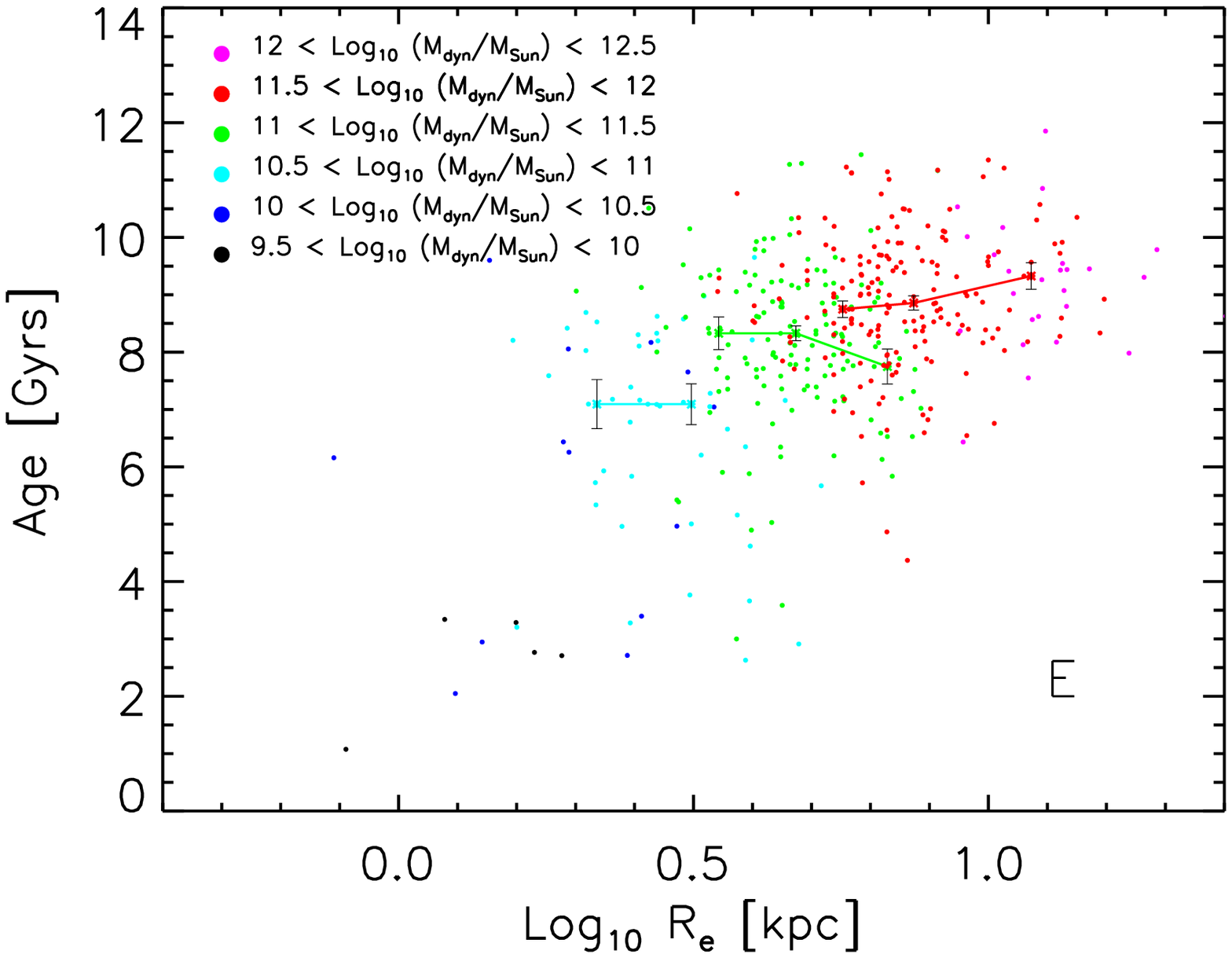}
 \includegraphics[width=0.49\hsize]{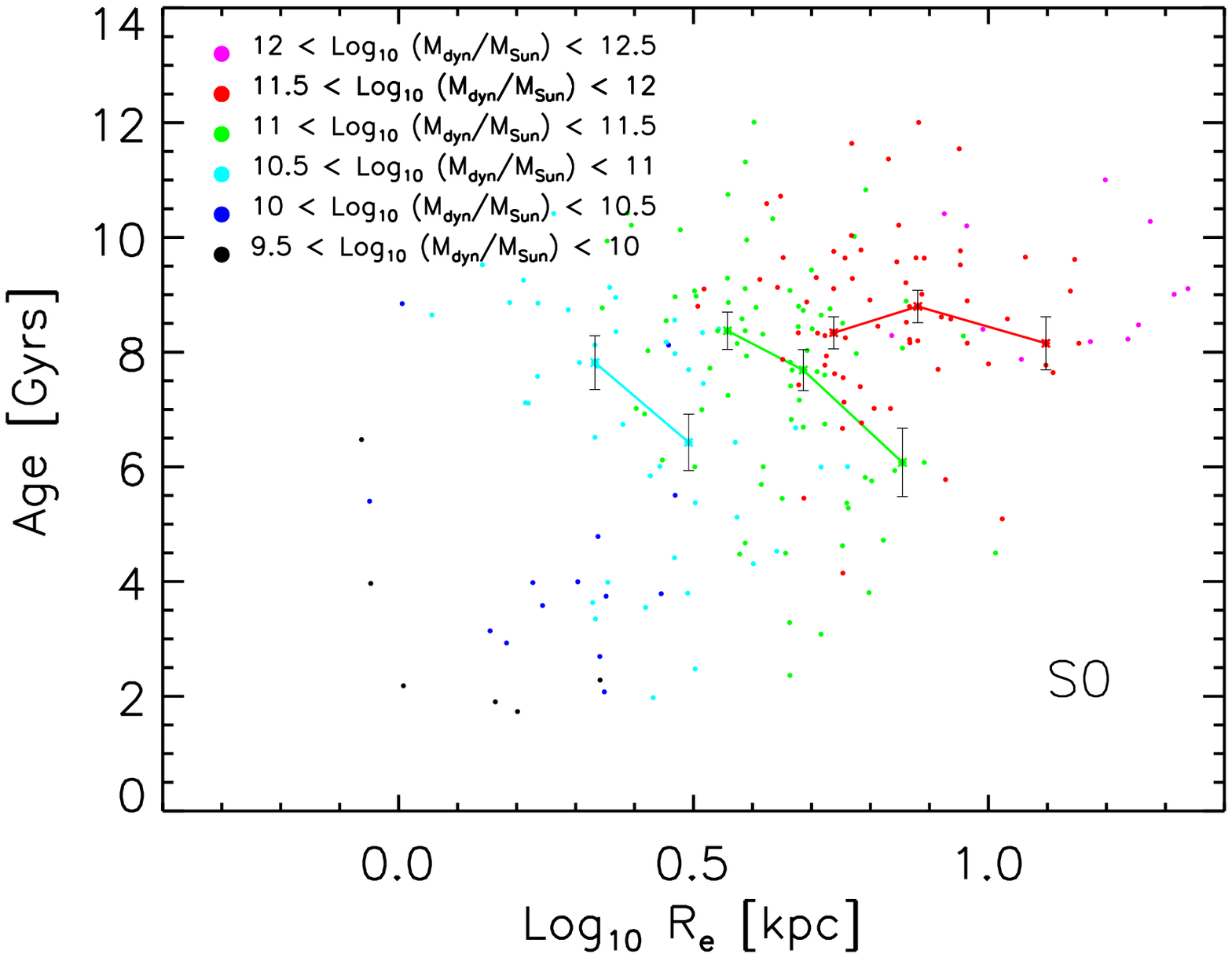} 
 \includegraphics[width=0.49\hsize]{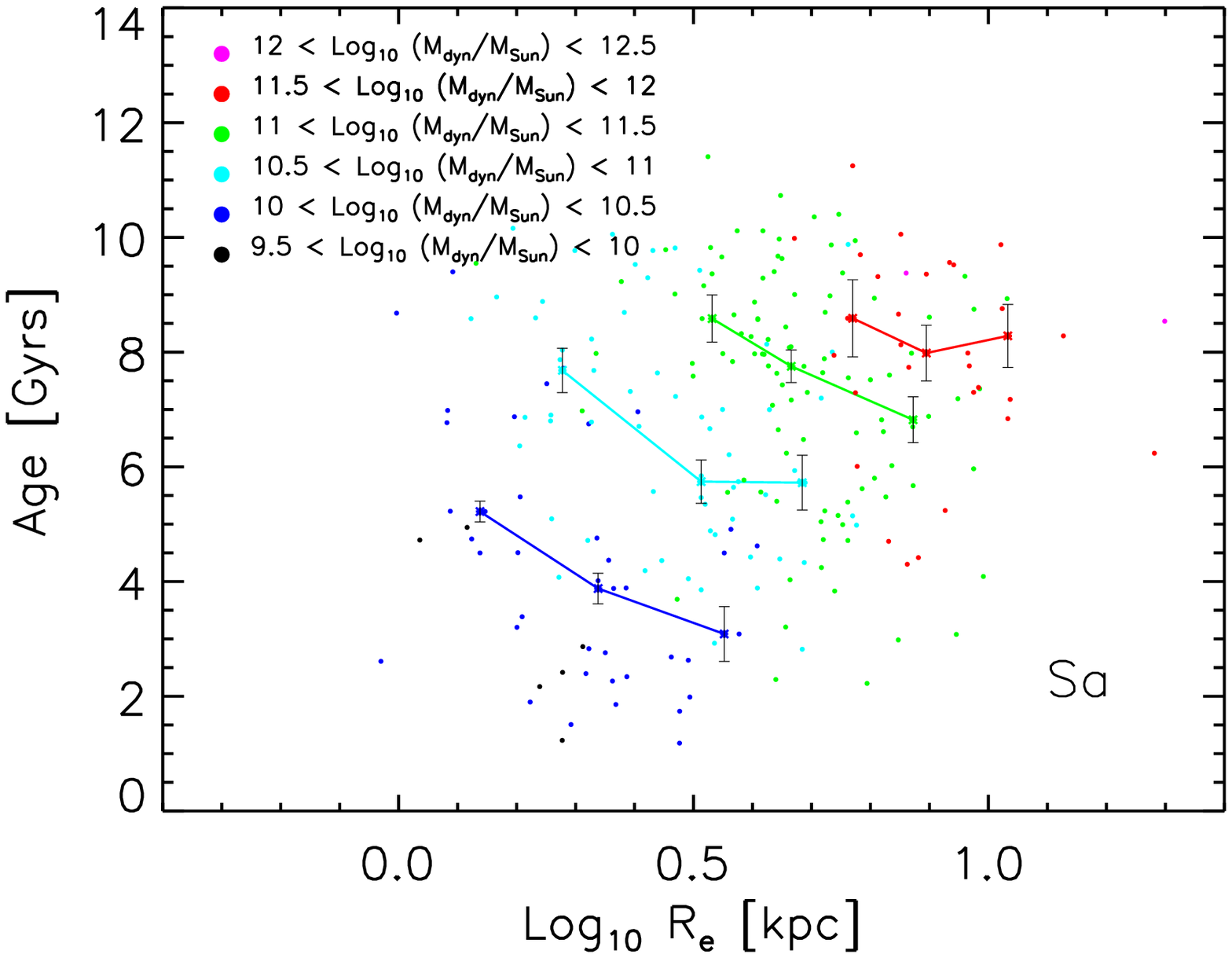}
 \includegraphics[width=0.49\hsize]{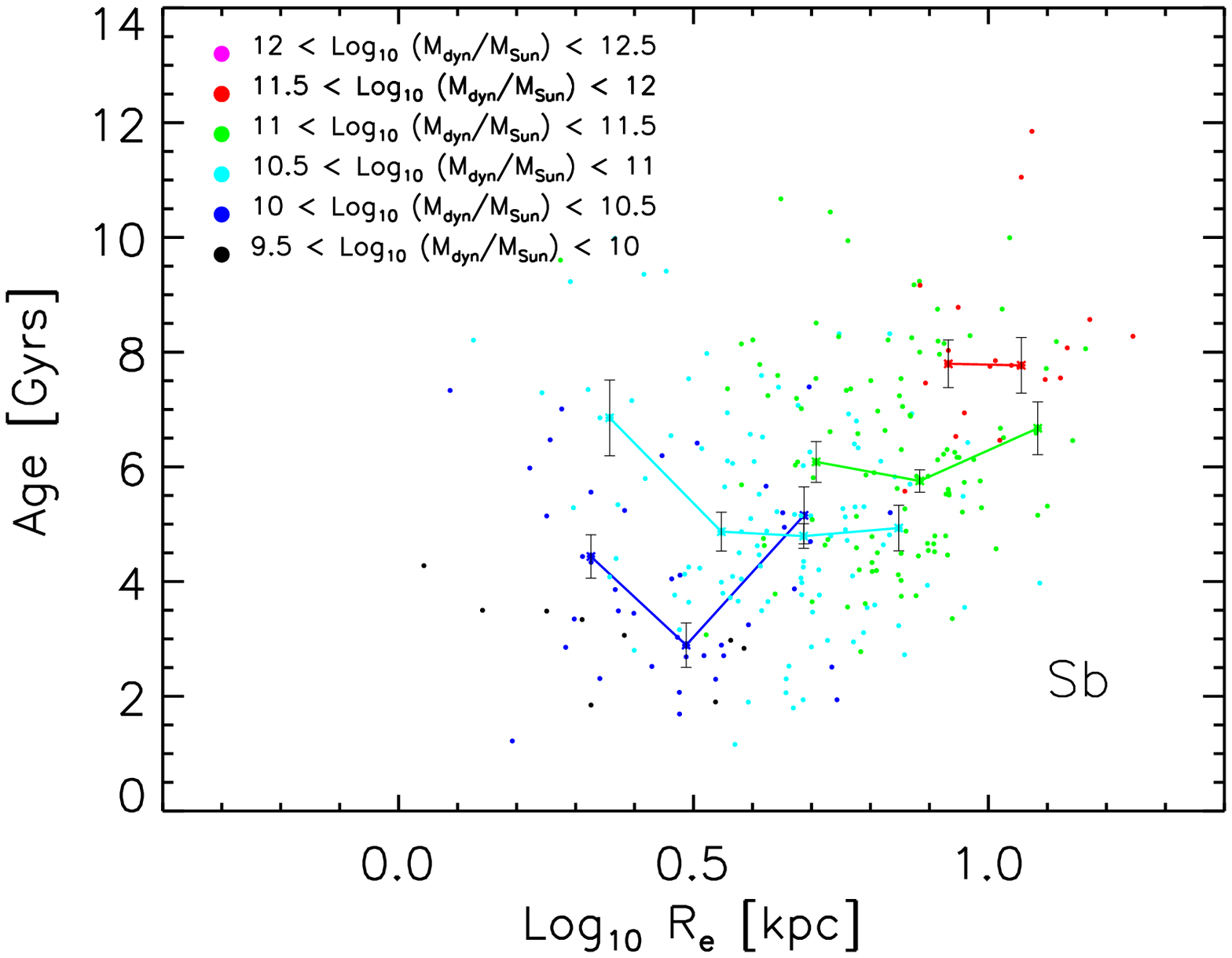}  
 \includegraphics[width=0.49\hsize]{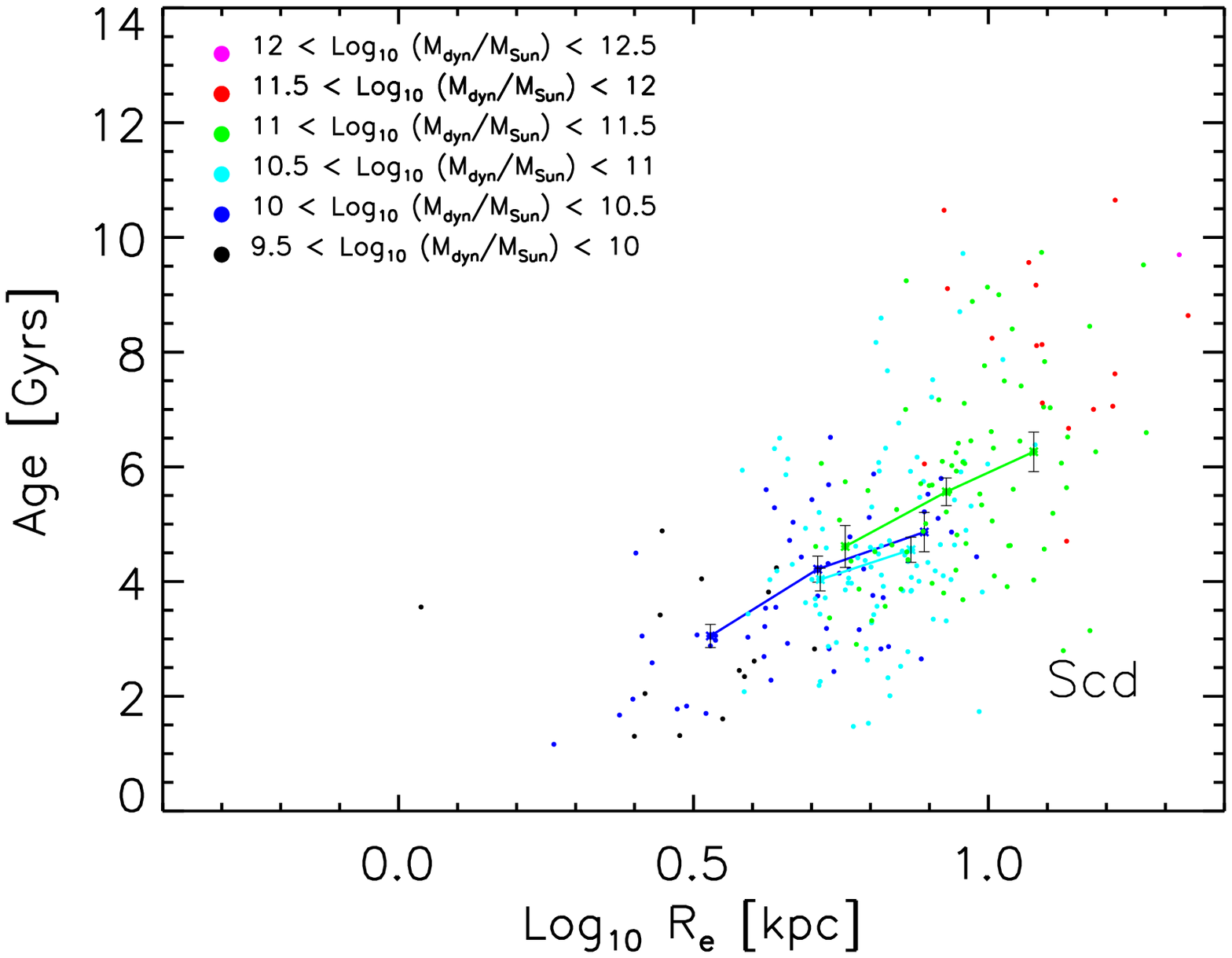}
 \includegraphics[width=0.49\hsize]{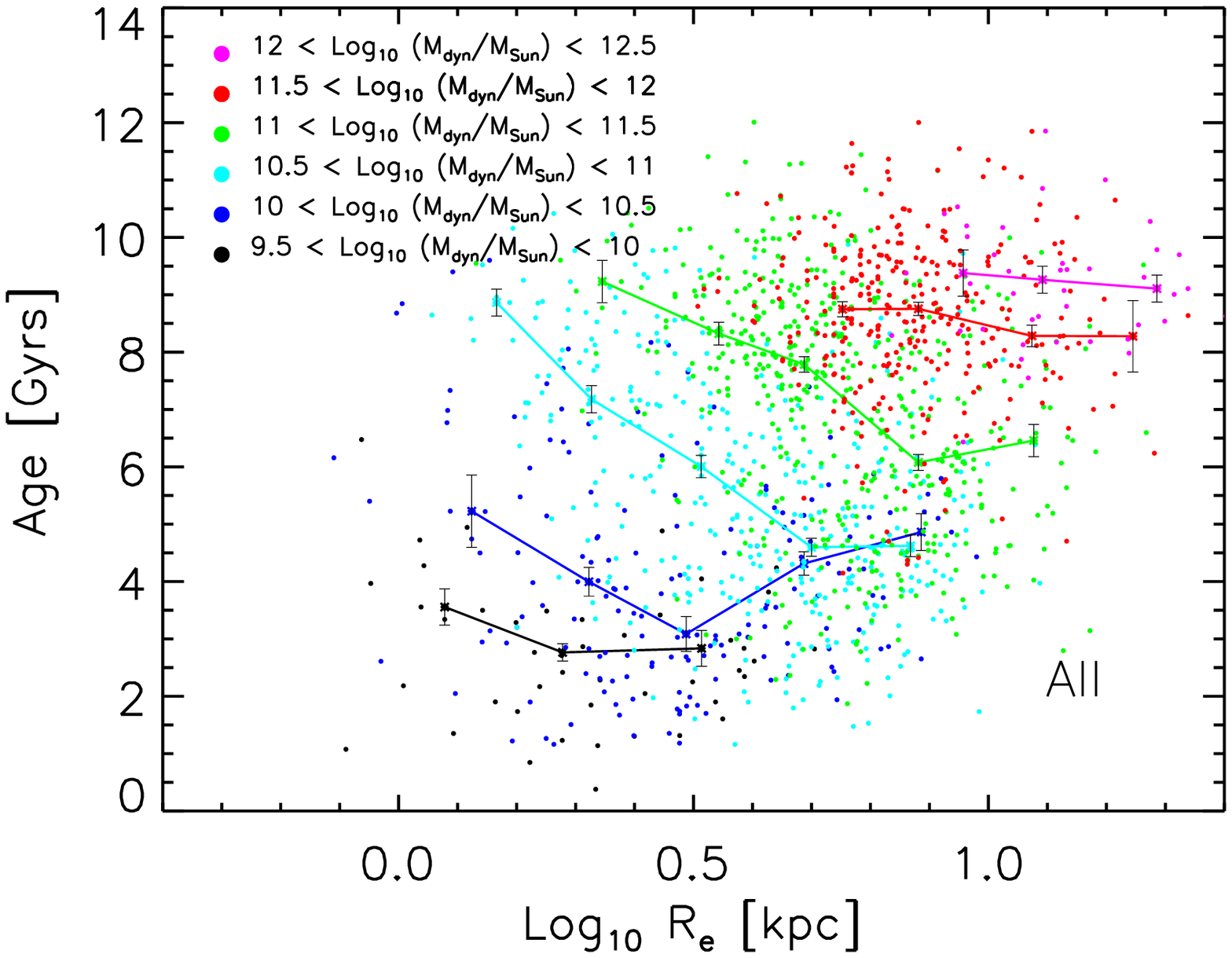}    
 \caption{Correlation between age and size at fixed dynamical mass (as labeled)
  for types E, S0, Sa, Sb, Sc-Sd, and All in the  
  Fukugita et al. (2007) sample. }
 \label{ageRMd}
\end{figure*}

\section{The age-size relation}\label{ageRe}
The previous sections studied how the distribution of 
$L$, $M_*$, $\sigma$ and $R$ depend on morphology or concentration.  
The present section shows one example of a correlation between 
observables which is particularly sensitive to morphology.  
As a result, how one chooses to select `red' sequence galaxies 
matters greatly.  

The age-size relation has been the subject of recent interest, in 
particular because, for early-type galaxies, the correlation between 
galaxy luminosity and size does not depend on age (Shankar \& Bernardi 2009).  
This is somewhat surprising, because it has been known for some 
time that early-type galaxies with large velocity dispersions tend to 
be older (e.g., Trager et al. 2000; Cattaneo \& Bernardi 2003; 
Bernardi et al. 2005; Thomas et al. 2005; Jimenez et al. 2007; 
Shankar et al. 2008), and the virial theorem implies that velocity 
dispersion and size are correlated.

To remove the effects of this correlation, Shankar et al. (2009c) and 
van der Wel et al. (2009) studied the age-size correlation at fixed 
velocity dispersion.  They find that this relation is almost flat 
(with a zero-point that depends on the velocity dispersion, of course).
At fixed dynamical mass, however, Shankar et al. still find no 
relation, whereas van der Wel et al. find a significant 
anti-correlation:  smaller galaxies are older.  What should be made 
of this discrepancy? 

Although both groups claim to be studying early-type galaxies, 
the details of how they selected their samples are different:  
Shankar et al. follow Hyde \& Bernardi (2009); the results from 
the previous sections suggest that this sample should be dominated by 
ellipticals.  van der Wel et al. use the sample of Graves et al. (2009):  
this has $0.04\le z\le 0.08$, $15 \le m_{r,{\rm Pet}} \le 18$, 
$i$-band concentration $> 2.5$; 
and likelihood of deV profile $> 1.03$ that of the exponential 
(this likelihood is output by the SDSS photometric pipeline), 
no detected emission lines (EW H$\alpha \le 0.3$\AA and OII $\le 1.7$); 
and spectra of sufficient S/N that velocity dispersions were measured
(following Bernardi et al. 2003).  
Because the requirements on the profile shape are significantly 
less stringent than those of Hyde \& Bernardi, one might expect 
this sample to include more S0s and Sas.  Moreover, recall 
that age is a strong function of morphological type (e.g. 
Figure~\ref{agemorph}), so, to see if this matters, we have studied 
how the age-size relation depends on morphological type.  

Figure~\ref{ageRsig} shows this relation, at fixed $\sigma$, for 
the full Fukugita sample (bottom right) and for the different 
morphological types (other panels).  The relation for the full sample 
is approximately flat, except at $\sigma < 100$~km~s$^{-1}$.  
However, when divided by morphological type, age (at fixed $\sigma$) 
is slightly correlated with size for ellipticals, but the relation 
is more flat for S0s and Sas.  (The mean age is a slightly increasing 
function of $\sigma$ for the later types Sb-d.)  
Thus, the flatness of the age-size relation at fixed $\sigma$ in the 
full sample hides the fact that the relation actually depends on 
morphology.  

Although this is a subtle effect for the relation at fixed $\sigma$, 
the dependence on morphology is much more pronounced when studying 
galaxies at fixed $M_{\rm dyn}\equiv 5\, R_e\sigma^2/G$.  
Figure~\ref{ageRMd} shows that, in the full sample, this relation is 
flat for $M_{\rm dyn}>10^{11.5}M_\odot$, but decreases strongly for 
lower masses, except at $M_{\rm dyn}<10^{10.5}M_\odot$, where it 
appears to be curved.  The stronger dependence here is easily understood 
from the previous figure:  since dynamical mass is $\propto R\sigma^2$, 
as one moves along lines of constant mass in the direction of increasing 
$R$, one is moving in the direction of decreasing $\sigma$.  
In Figure~\ref{ageRsig} this means that one must step downwards by 
one bin in $\log\sigma$ for every 0.4~dex to the right in $\log R$.  
For ellipticals, age is an increasing function of size at fixed $\sigma$; 
hence, the net effect of moving up and to the right (at fixed $\sigma$), 
and then stepping down to lower $\sigma$, produces an approximately 
flat age-size relation for fixed $M_{\rm dyn}$.  For Sas, on the other 
hand, keeping $M_{\rm dyn}$ fixed corresponds to shifting down and to 
the right (at fixed $\sigma$), and then stepping downwards to the 
lower $\sigma$ bin; the net result is that age decreases strongly 
as size increases.
What is remarkable is that the ellipticals show precisely the scaling 
with $M_{\rm dyn}$ reported by Shankar et al. (2009c), whereas the 
S0s and Sa's show that reported by van der Wel et al. (2009).

\section{Discussion}\label{discuss}
We compared samples selected using simple selection algorithms 
based on available photometric and spectroscopic information with 
those based on morphological information.  
Requiring concentration indices $C_r\ge 2.6$ selects a mix in which 
E+S0+Sa's account for about two-thirds of the objects; 
if $C_r\ge 2.86$ instead, then two-thirds of the sample comes from 
E+S0s; whereas Es alone account for more than two-thirds of a sample 
selected following Hyde \& Bernardi (2009) (Figures~\ref{fredC} 
and~\ref{fblueC}, and Table~\ref{purity}).  E's alone account for 
about 40\%, 50\% and 75\% of the total stellar mass in samples 
selected in these three ways.

The reddest objects at intermediate luminosities or stellar masses 
are edge-on disks (Figure~\ref{gmrmorph}).  As a result, samples 
selected on the basis of color alone, or cuts which run parallel 
to the red sequence are badly contaminated by such objects.  
However, simply adding the additional requirement that the axis ratio 
$b/a\ge 0.6$ is an easy way to remove such red edge-on disks from 
the `red' sequence; the resulting sample is similar to requiring 
$C_r\ge 2.86$.  This may provide a simple way to select relatively 
clean early-type samples in higher redshift datasets (e.g. DEEP2, $z$Cosmos).  
Our measurements provide the low redshift benchmarks against which 
such future higher redshift measurements can be compared.  

We showed how the distribution of luminosity, stellar mass, size and 
velocity dispersion in the local universe is partitioned up amongst 
different morphological types, and we compared these distributions 
with those based on simple selection algorithms based on available 
photometric and spectroscopic information 
(Figures~\ref{differentCIa}--\ref{ES0Sa}).
We described our measurements by assuming that the intrinsic 
distributions have the form given by equation~(\ref{phiX}).  
We showed how measurement errors bias the fitted parameters 
(equation~\ref{psiO}), and used this to devise a simple method which 
removes this bias.  The results, which are reported in tabular 
form in Appendix~\ref{tables}, show that ellipticals contain 
$\sim$20\% of the luminosity density and 25\% of the stellar 
mass density in the local universe, and have mean sizes of 
order 3.2~kpc.  Including S0s increases these numbers to 33\% 
and 40\%; adding Sas results in further increases, to 50\% and 60\% 
respectively.  
These numbers are in broad agreement with those from the Millennium 
Galaxy Survey of about $10^4$ objects in $37.5$~deg$^2$.  
Driver et al. (2007) report that $15\pm 5$\% of the stellar mass 
density is in ellipticals, and adding bulges increases this to 
$44\pm 9$\%.  


Our stellar mass function has more massive objects than other 
recent determinations (e.g. Cole et al. 2001; Bell et al. 2003; 
Panter et al. 2007; Li \& White 2009), similarly shifted to a 
Chabrier (2003) IMF (Figures~\ref{MsFpet} and ~\ref{MsFmod}).  
The mass scale on which the discrepancy arises is of order where 
some previous work had only a handful of objects -- our substantially 
larger volume is necessary to provide a more reliable estimate 
of these abundances. 
Using stellar masses estimated from {\tt cmodel} luminosities, 
which are more reliable than Petrosian luminosities at the large 
masses where the discrepancy in $\phi(M_*)$ is largest, gives 
stellar mass densities in objects more massive than
 $(1,2,3)\times 10^{11}\,M_\odot$ that are larger by more than 
 $\sim$~$(20,50,100)$ percent compared to Bell et al. (2003) 
(Figure~\ref{MsFcumulative}).

This analysis required that we study the sytematic differences 
between the stellar mass estimates based on $g-r$ color (our 
equation~\ref{gmrBell}, following Bell et al. 2003), 
colors in multiple bands (Blanton \& Roweis 2007), 
and on spectral features (Gallazzi et al. 2005).  
(See Gallazzi \& Bell 2009, which appeared while our work was being 
refereed, for a discussion of the pros and cons of these various 
approaches, and of the accuracy to which stellar masses can currently 
be derived.)  
The $g-r$ and Gallazzi et al. estimates are generally in good agreement 
(Figure~\ref{BG}), although the spectral based estimates suffer 
slightly from aperture effects which are complicated by the magnitude 
limit of the survey (Figures~\ref{BGz}, \ref{CompareMsz} and~\ref{MsFpet}).  
The Blanton et al. estimates are in good agreement with the other two 
provided one uses LRG-based templates to estimate masses at the 
most massive end (Figures~\ref{BG-BR}, \ref{CompareMsAll} and~\ref{MsFmod}).  
At lower masses, some combination of the LRG and other templates is 
required.  Ignoring the LRG templates altogether (e.g. Li \& White 2009) 
results in systematic underestimates of as much as 0.1~dex or more 
(Figure~\ref{BG-BR}), severely compromising estimates of the number 
of stars currently locked up in massive galaxies (Figure~\ref{MsFpet}).
If we compare our estimate of the stellar mass density in objects more 
massive than $(1,2,3)\times 10^{11}\,M_\odot$ with those from the 
Li \& White (2009) fit, then our values are $\sim$~$(140,230,400)$ 
percent larger.

Allowing more high mass objects means that major dry mergers may 
remain a viable formation mechanism at the high mass end 
(Figure~\ref{Bezanson}).  It also relieves the tension between 
estimates of the evolution of the most massive galaxies which are 
based on clustering (which predict some merging, and so some increase 
in stellar mass; Wake et al. 2008; Brown et al. 2008) and those based 
on abundances (for which comparison of high redshift measurements 
with the previous $z\sim 0$ measurements indicated little evolution; 
Wake et al. 2006; Brown et al. 2007; Cool et al. 2008).  This 
discrepancy may be related to the origin of intercluster light 
(e.g. Skibba et al. 2007; Bernardi 2009); our measurement of a larger 
local abundance in galaxies reduces the amount of stellar mass that 
must be stored in the ICL.  

It has been argued that a number of observations are better reproduced 
if one assumes a different IMF for elliptical and spiral galaxies
(e.g. Calura et al. 2009).  We showed that this acts to further increase 
the abundance of massive galaxies (Figure~\ref{MsFIMF}), 
and reduces the difference between stellar and dynamical mass, especially 
at larger masses.  At $M_*\ge 10^{11}M_\odot$, the increase due to the change 
in IMF is a factor of two with respect to models which assume a fixed IMF.  

If we sum up the observed counts to estimate the stellar mass density
in the range  $8.6 < \log_{10} M_* < 12.2$ $M_{\odot}$
($M_*$ from equation~\ref{gmrBell} using {\tt cmodel} magnitudes),
then the result is $3.05\times 10^8M_\odot$~Mpc$^{-3}$.
Using our fit to the observed distribution
(values between round brackets in Table~\ref{tabMs}) gives
a similar value ($3.06\times 10^8M_\odot$~Mpc$^{-3}$) and a slightly
smaller value if one uses the {\it intrinsic} fit 
($2.89\times 10^8M_\odot$~Mpc$^{-3}$, see Table~\ref{tabMs}). 
Our values are $\sim 15$\% and 30\% larger than those reported by 
Panter et al. (2007) and Li \& White (2009), respectively.
If we allow a type dependent IMF, the total stellar mass density 
increases by a further 30\%.  

However, although our stellar mass function has more 
$M_* > 10^{11}$ M$_\odot$ objects than other recent determinations, 
our estimate of the total stellar mass density is similar to that 
measured by Bell et al. (2003).  
It is about 20\% smaller than the value reported by Driver et al. (2007) 
(once shifted to the same IMF, for which we have chosen Chabrier 2003;
see Table~\ref{tabIMF}).
This is because differences at the mid/faint end contribute more to the
total stellar density than the difference we measured at the massive end.  

It has been suggested that direct integration of the cosmological star 
formation rate overpredicts the total local estimate of the stellar mass 
density (see, e.g., Wilkins et al. 2008, and references therein).  
However, we showed that recent determinations of the recycling factor 
(equation~\ref{eq|rhoz}) and the high-$z$ star formation rate 
(Figure~\ref{fig|SFRz}) result in better agreement (Figure~\ref{fig|rhoz}).  
This is because the former yields smaller remaining masses, and the latter 
produces fewer stars formed in the first place.  

Our measurements also show that the most luminous or most massive 
galaxies, which one might identify with BCGs, are less concentrated 
and have smaller $b/a$ ratios, than slightly less luminous or massive 
objects (Figures~\ref{cmorph}, \ref{bamorph} and~\ref{fredC}).
Their light profile is also not well represented by a pure deVaucoleur 
law. This is consistent with results in Bernardi et al. (2008)
and Bernardi (2009) who suggest that these are signatures of 
formation histories with recent radial mergers.  
In this context, note that we showed how to define seeing-corrected 
sizes, using quantities output by the SDSS pipeline, that closely 
approximate deVaucouleur bulge + Exponential disk decompositions 
(equations~\ref{Rcmodel} and~\ref{skysubr}).  Our {\tt cmodel} sizes 
represent a substantial improvement over Petrosian sizes (which are 
not seeing corrected) and pure {\tt deV} or {\tt Exp} sizes 
(Figures~\ref{cmodel} and~\ref{cmodel2}).  

And finally, our study of the age-size correlation resolves a 
discrepancy in the literature:  whereas Shankar et al. (2009c) report 
no correlation at fixed $M_{\rm dyn}$, van der Wel et al. (2009) 
report that larger galaxies tend to be younger.  We showed that 
ellipticals follow the scaling reported by Shankar et al. scaling, 
whereas S0s and Sas follow that of van der Wel et al.  
(Figure~\ref{ageRMd}), suggesting that Shankar et al. select a 
sample dominated by galaxies with elliptical morphologies, whereas 
van der Wel et al. include more S0s and Sas.  These conclusions 
about the differences between the samples are consistent with how 
the samples were actually selected.  

Since van der Wel et al. use their measurements to constrain a model 
for early-type galaxy formation, this is an instance in which having 
morphological information matters greatly for the physical 
interpretation of the data.  Our results indicate that models of 
early-type galaxy formation should distinguish between ellipticals 
and S0s because, in the projection of the age-size-mass correlation 
shown in Figure~\ref{ageRMd}, the S0s and Sas are very different 
from the other morphological types.  Whether the smaller sizes for 
older S0s are due to the gradual stripping away of a younger disk 
is an open question.  

van der Wel et al. (2009) use their observation that the age-size 
relation at fixed $\sigma$ is flat to motivate a model in which 
early-type galaxy formation requires a critical velocity dispersion 
(which they allow to be redshift dependent).  The same logic applied 
here suggests that while this may be reasonable for S0s or Sas, it 
is not well-motivated for ellipticals (Figure~\ref{ageRsig}).  
However, it might be interesting to explore a model in which 
elliptical formation requires a critical (possibly redshift 
dependent) dynamical mass rather than velocity dispersion (which 
may be redshift dependent).  This is interesting because, in 
hierarchical models, the phenomenon known as down-sizing 
(e.g., Cowie et al. 1996; Heavens et al. 2004; Sheth et al. 2006; 
Jimenez et al. 2007) is then easily understood (Sheth 2003).

\section*{Acknowledgments}
We thank Anna Gallazzi, Eric Bell, Cheng Li, Danilo Marchesini, 
Eyal Neistein and Simon White for interesting discussions.
We would also like to thank the members of the APC in Paris 7 Diderot 
and the Max-Planck Institut f\"ur Astronomie, Heidelberg, 
for their hospitality while this work was being completed. 
M.B. is grateful for support provided by NASA grant LTSA-NNG06GC19G
and NASA ADP/NNX09AD02G. FS acknowledges support from the Alexander von
Humboldt Foundation.  RS is supported in part by NSF-AST 0908241.  

Funding for the Sloan Digital Sky Survey (SDSS) and SDSS-II Archive has been
provided by the Alfred P. Sloan Foundation, the Participating Institutions, the
National Science Foundation, the U.S. Department of Energy, the National
Aeronautics and Space Administration, the Japanese Monbukagakusho, and the Max
Planck Society, and the Higher Education Funding Council for England. The
SDSS Web site is http://www.sdss.org/.

The SDSS is managed by the Astrophysical Research Consortium (ARC) for the
Participating Institutions. The Participating Institutions are the American
Museum of Natural History, Astrophysical Institute Potsdam, University of Basel,
University of Cambridge, Case Western Reserve University, The University of
Chicago, Drexel University, Fermilab, the Institute for Advanced Study, the
Japan Participation Group, The Johns Hopkins University, the Joint Institute
for Nuclear Astrophysics, the Kavli Institute for Particle Astrophysics and
Cosmology, the Korean Scientist Group, the Chinese Academy of Sciences (LAMOST),
Los Alamos National Laboratory, the Max-Planck-Institute for Astronomy (MPIA),
the Max-Planck-Institute for Astrophysics (MPA), New Mexico State University,
Ohio State University, University of Pittsburgh, University of Portsmouth,
Princeton University, the United States Naval Observatory, and the University
of Washington.




\appendix

\section{Comparison of stellar mass estimates}\label{Mscomp}

\begin{figure*}
 \centering
 \includegraphics[width=0.95\hsize]{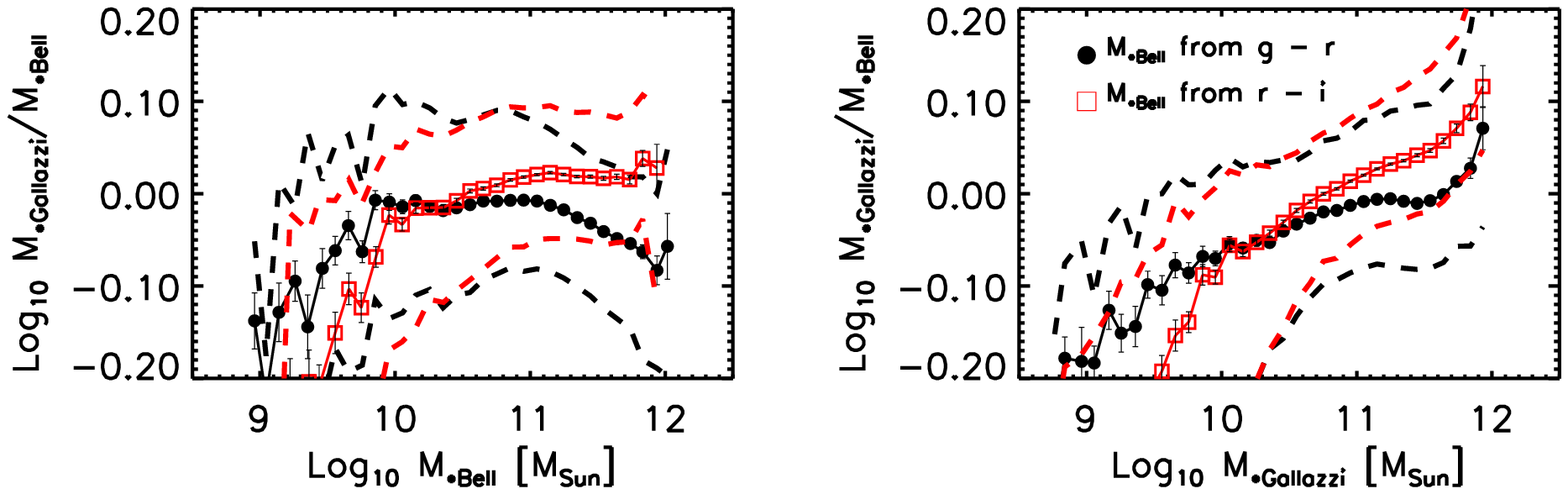}
 \caption{Comparison of stellar masses computed following 
  Bell et al. (2003) (our equations~\ref{gmrBell} (solid circles)
  and \ref{rmiBell} (open squares) with  
  {\tt Petrosian} $r$-band luminosity),   
  and Gallazzi et al. (2005), for objects with $C_r > 2.86$.}
 \label{BG}
\end{figure*}

\begin{figure*}
 \centering
 \includegraphics[width=0.45\hsize]{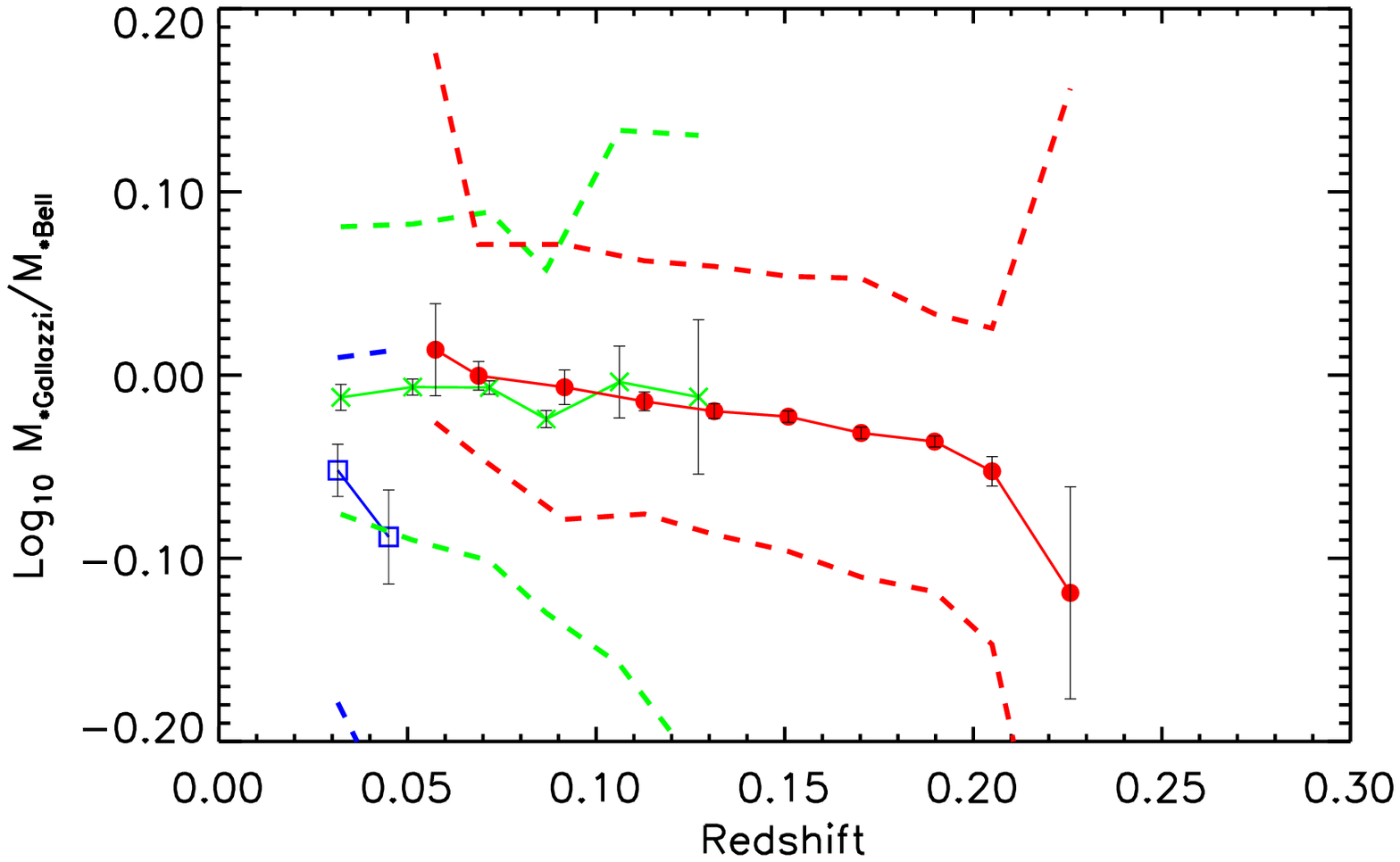}
 \includegraphics[width=0.45\hsize]{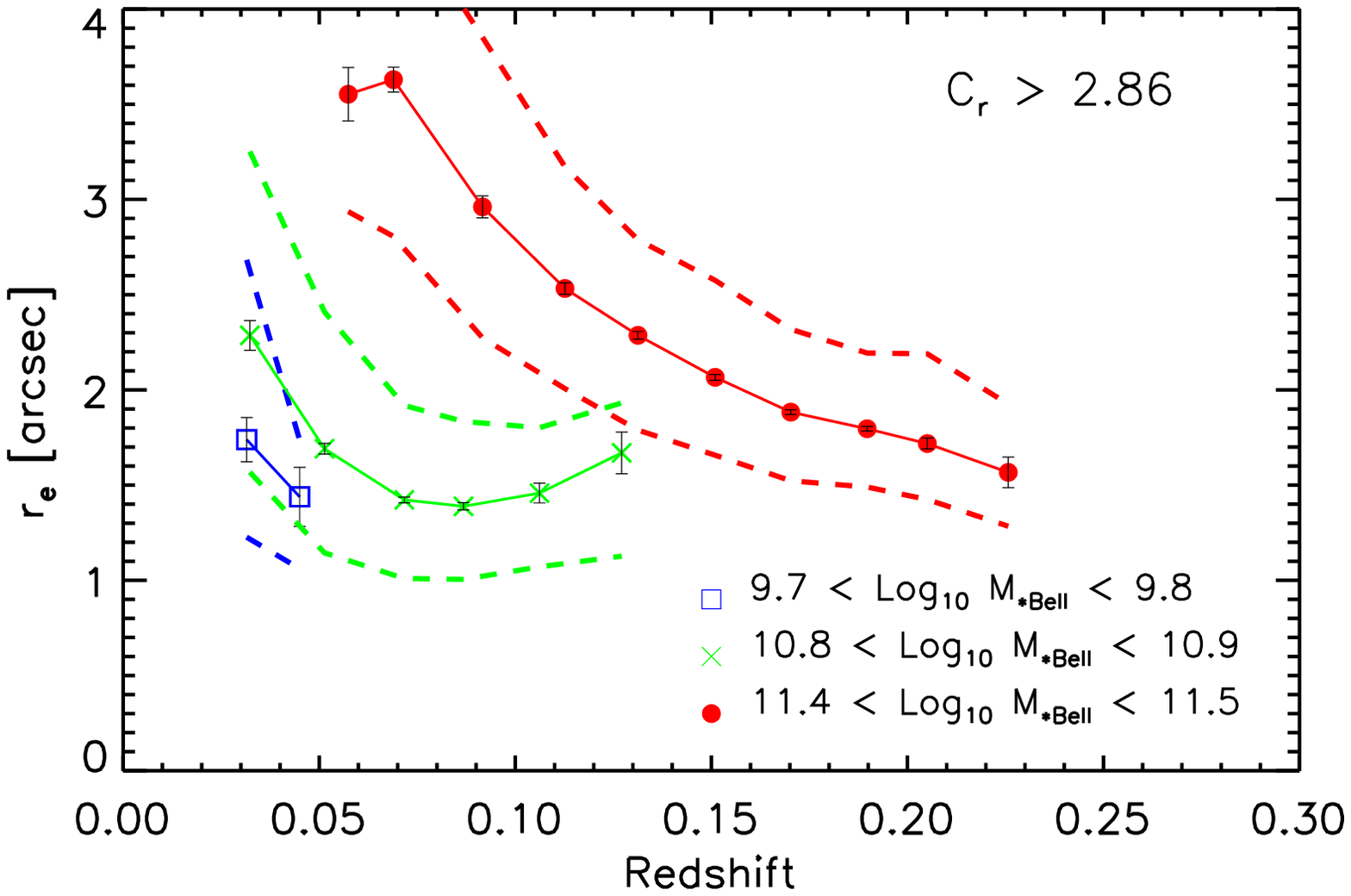}
 \includegraphics[width=0.45\hsize]{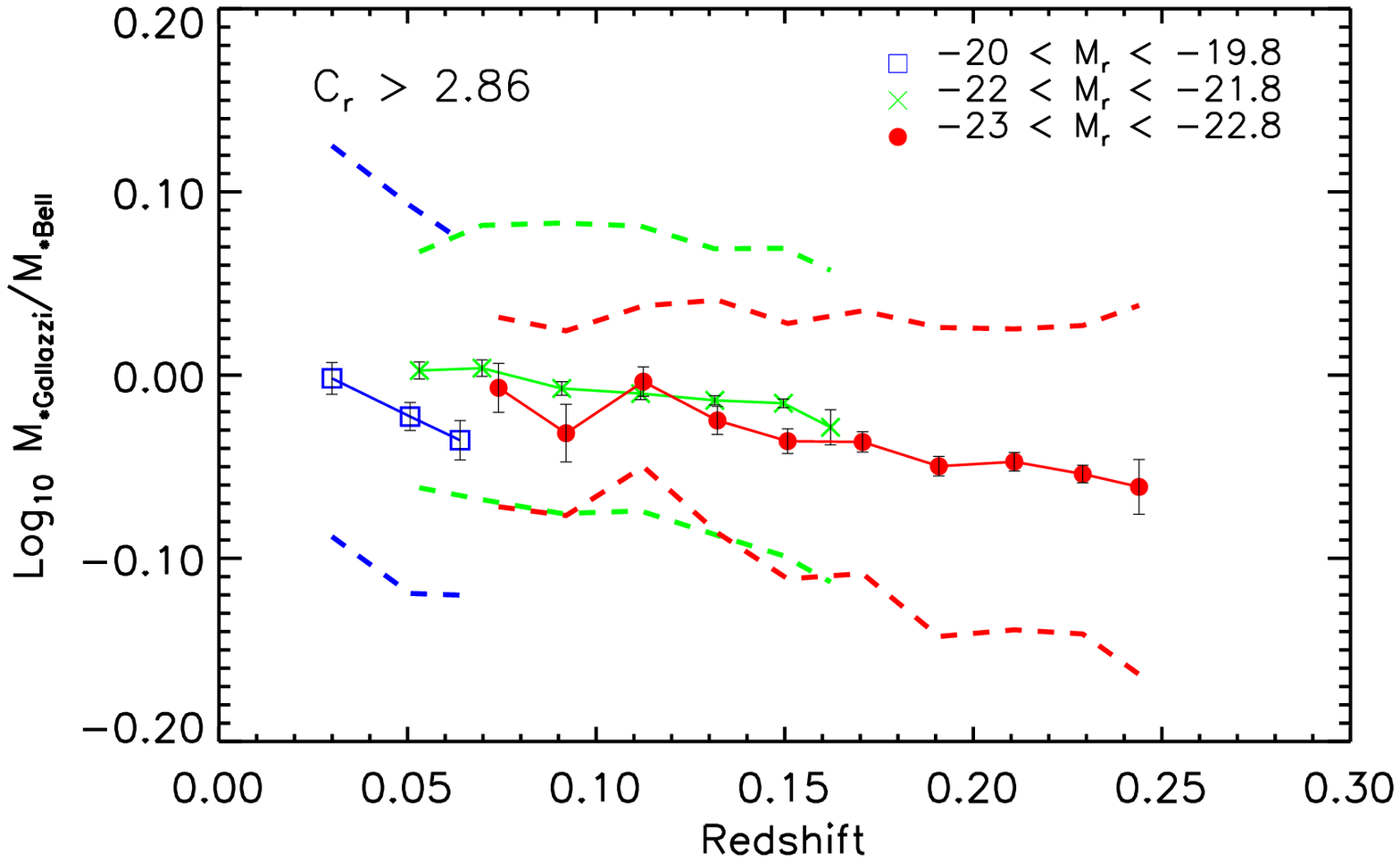}
 \includegraphics[width=0.45\hsize]{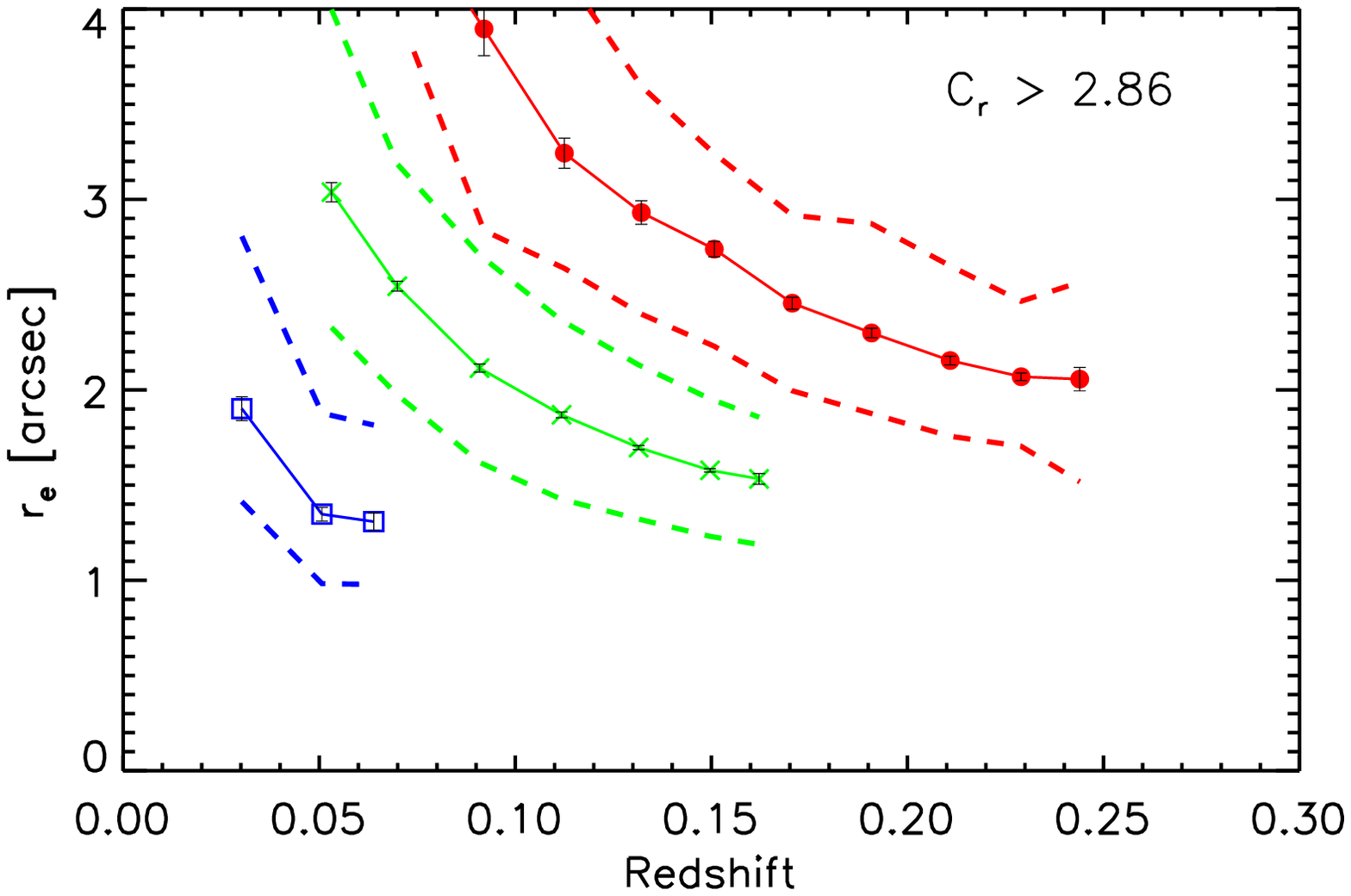}
 \caption{Aperture effects on the Gallazzi et al. (2005) 
   stellar mass estimate.  Top:  Redshift dependence of 
$M_{*{\rm Gallazzi}}/M_{*{\rm Bell}}$ 
   (left) and the angular half-light radius (right) for objects 
   with $C_r > 2.86$ in three different bins of $M_{*{\rm Bell}}$ 
   as indicated.  
   Bottom:  Similar, but now for a few narrow bins in luminosity.}
 \label{BGz}
\end{figure*}

\begin{figure*}
 \centering
 \includegraphics[width=0.95\hsize]{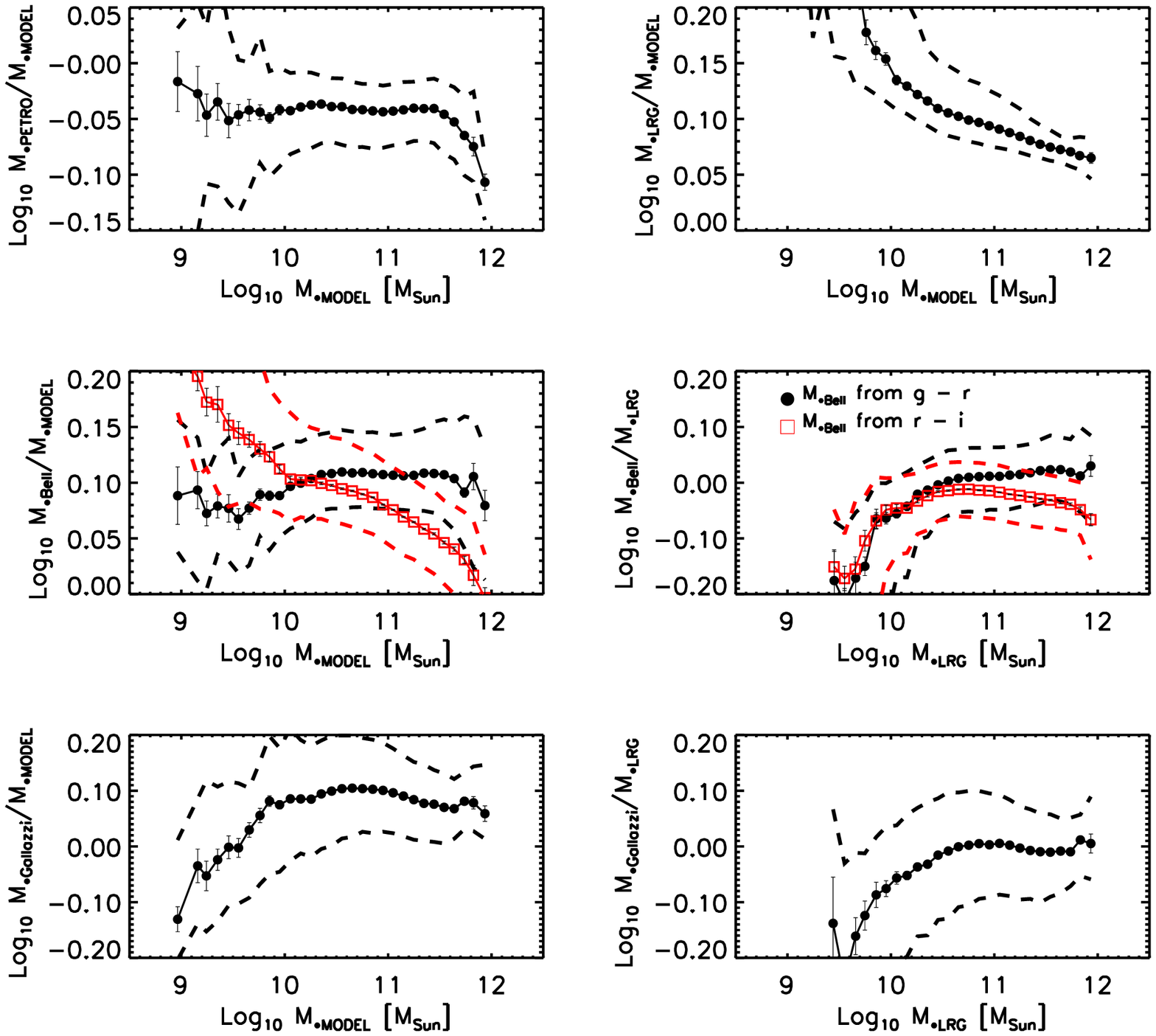}
 \caption{Comparison of various stellar mass estimates from 
  Blanton \& Roweis (2007), with $M_{*{\rm Bell}}$ and 
  $M_{*{\rm Gallazzi}}$ for objects with $C_r > 2.86$.}
 \label{BG-BR}
\end{figure*}

\begin{figure*}
 \centering
 \includegraphics[width=0.95\hsize]{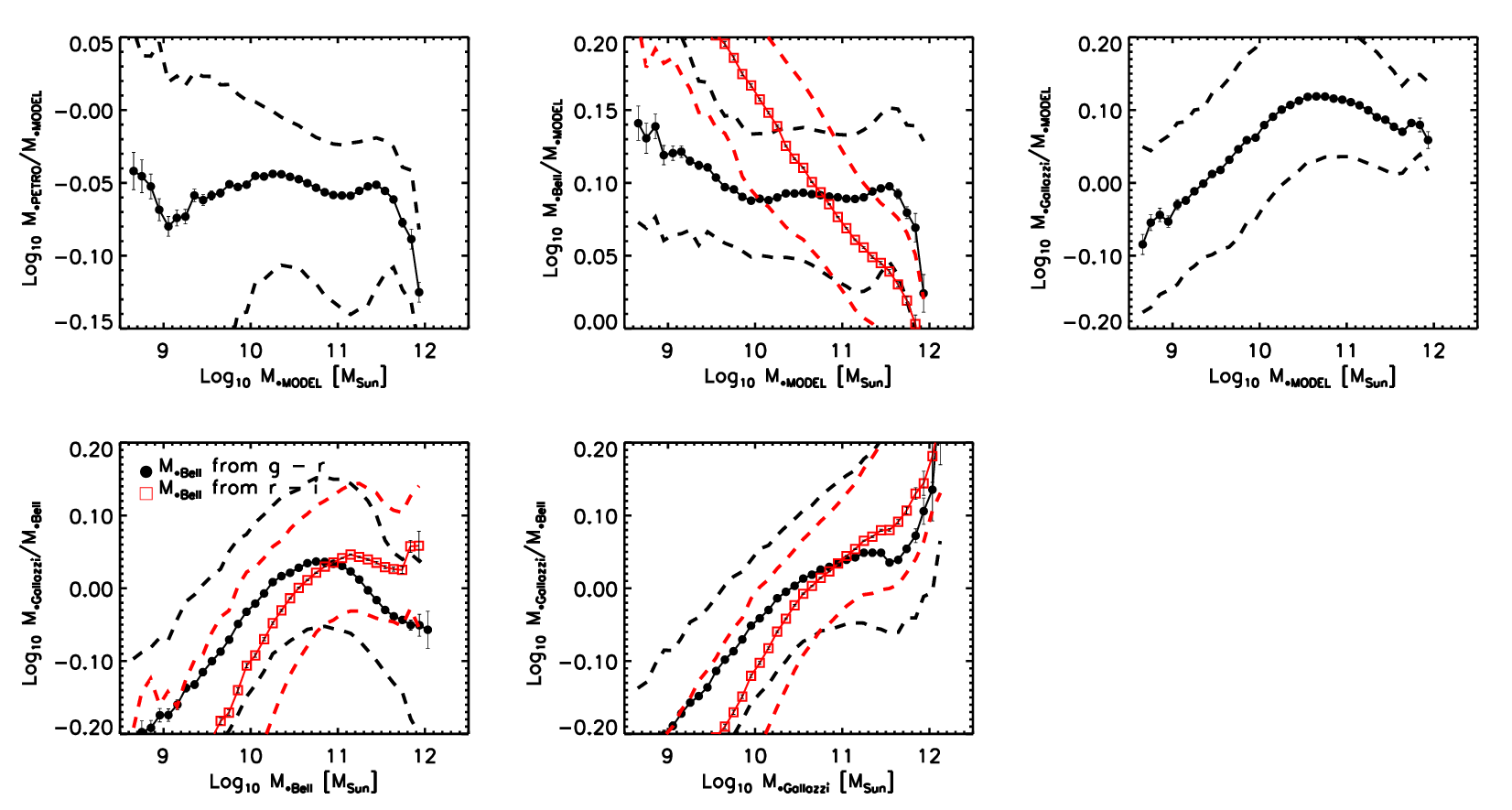}
 \caption{Comparison of our various stellar masses in the full sample 
  (i.e., no cut on concentration index).}
 \label{CompareMsAll}
\end{figure*}

\begin{figure*}
 \centering
 \includegraphics[width=0.45\hsize]{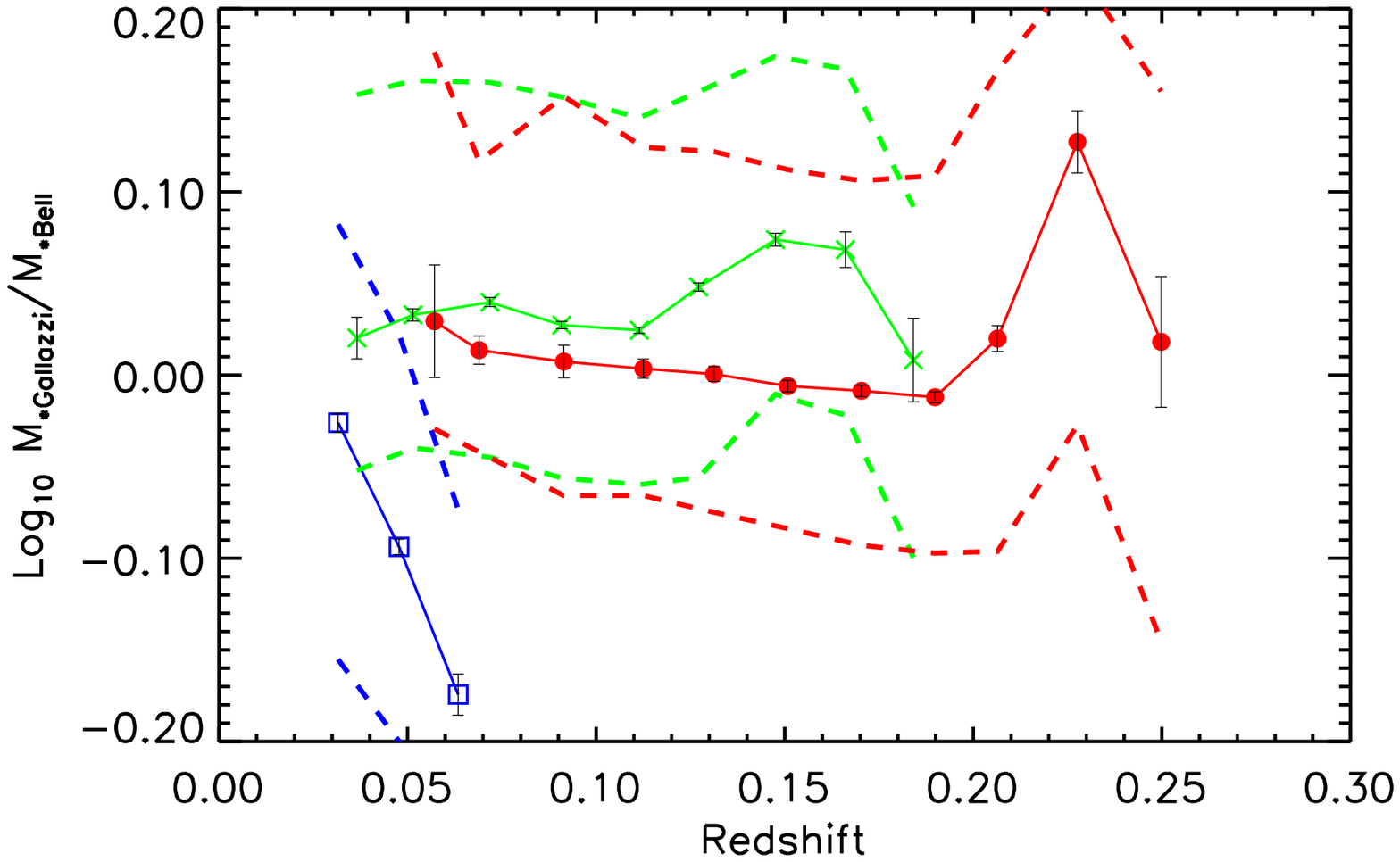}
 \includegraphics[width=0.45\hsize]{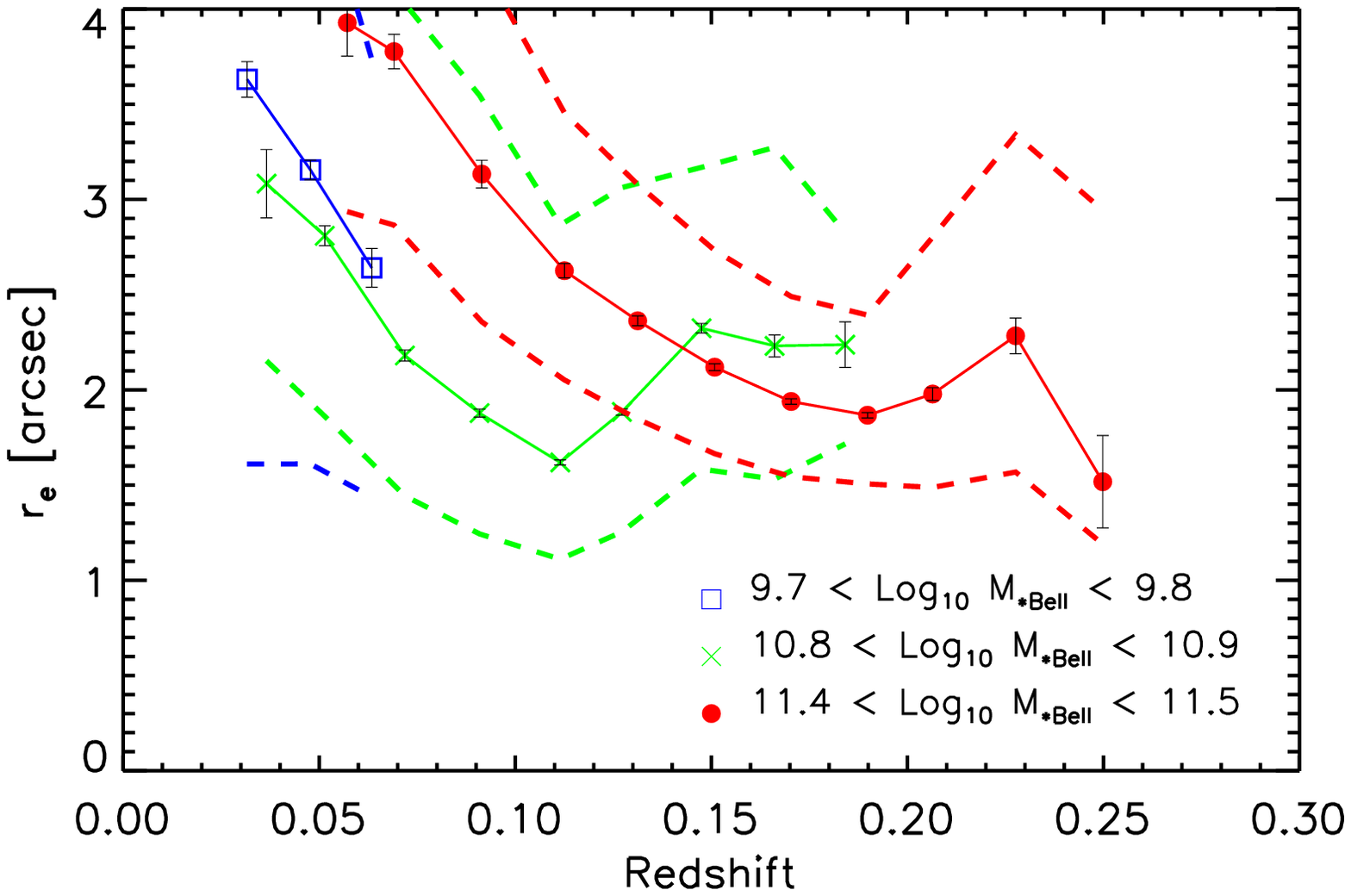}
 \includegraphics[width=0.45\hsize]{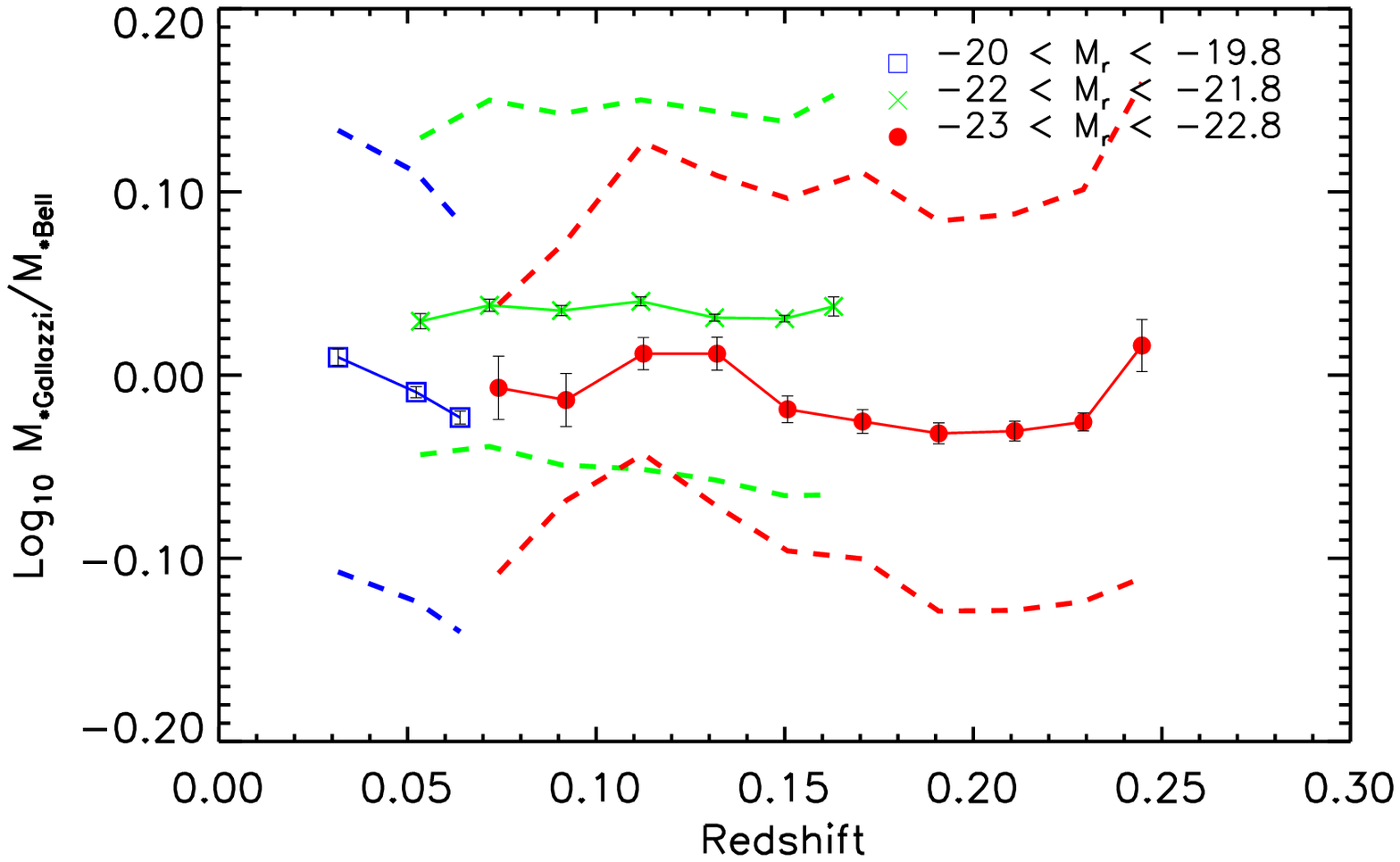}
 \includegraphics[width=0.45\hsize]{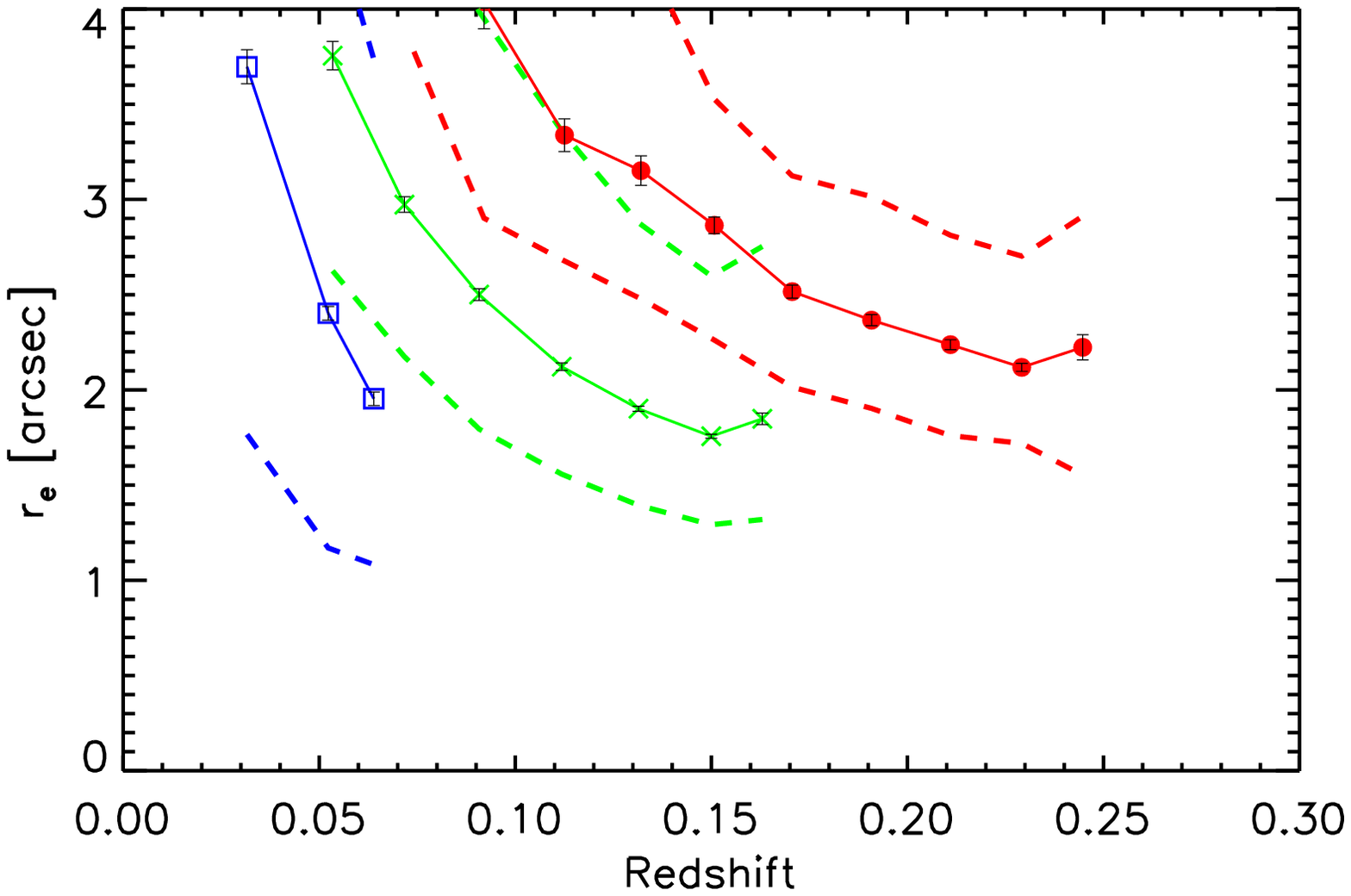}
 \caption{Same as Figure~\ref{BGz}, but now for the full sample (i.e. 
   no cut on concentration index).}
 \label{CompareMsz}
\end{figure*}

This section shows results from a study of how the different $M_*$ 
estimates compare with one another.  We first show results for 
galaxies that have $C_r\ge 2.86$ for two reasons.  First, because 
this is expected to provide a sample dominated by massive galaxies, 
and these are expected to be more homogeneous.  And second, we are 
particularly interested in studying the high mass end of the stellar 
mass function, because this is the end which is first observed in 
higher redshift datasets.  

Figure~\ref{BG} shows that the $M_{*{\rm Bell}}$ and 
$M_{*{\rm Gallazzi}}$ estimates are in good agreement, 
although $M_{*{\rm Bell}}$ tends to be larger at 
$M_{*{\rm Bell}}>10^{11} M_{\odot}$.  
The sharp downturn at small masses is due to aperture 
effects.  This is because $M_{*{\rm Bell}}$ is based on 
the {\tt model} color; for otherwise similar galaxies, this 
samples the same fraction of a galaxy's light (that within 
$R_e$) whatever its distance.  On the other hand, 
$M_{*{\rm Gallazzi}}$ is based on the light which enters the 
SDSS fiber (radius 1.5~arcsec); this samples a distance 
dependent fraction of the light of a galaxy.  For small low 
mass galaxies which have significant disks, the fiber samples 
the bulge of the nearby objects, and the disks of the more 
distant ones.  Since $M_*/L$ is larger for bulges than disks, 
the more distant objects of a given $M_{*{\rm Bell}}$ will 
have smaller $M_{*{\rm Gallazzi}}$.  
Here, of course, we have explicitly selected against disky 
galaxies, but color gradients will produce the same effect.  

Figure~\ref{BGz} shows this effect explicitly:  at fixed 
$M_{*{\rm Bell}}$, the ratio $M_{*{\rm Gallazzi}}/M_{*{\rm Bell}}$ 
decreases with increasing distance (this is a more dramatic 
effect for the lower mass systems in samples which include 
disks -- see Figure~\ref{CompareMsz}).  The panel on the right, which 
shows the angular size for objects of a fixed $M_{*{\rm Bell}}$ 
increasing at high $z$, illustrates a curious selection effect
(the upturn at higher redshift for the middle stellar mass bin, 
which is not present if when we study bins in luminosity).  
For a given $M_{*{\rm Bell}}$, the highest redshift objects in a 
magnitude limited survey will be bluer; since color is primarily 
determined by velocity dispersion, these objects will have smaller 
than average velocity dispersions, and larger sizes.  Although 
this is a small effect for the massive objects we have selected here, 
it is more dramatic in samples which include disks (as we show shortly;
see Figure~\ref{CompareMsz}).  

Figure~\ref{BG-BR} shows how these estimates compare with those 
from Blanton \& Roweis (2007).  The top left panel shows that 
$M_{*{\rm Model}}$ is $\sim 0.04$~dex larger than $M_{*{\rm Petro}}$ 
(top left panel).  
Because the two estimates are based on the same mass-to-light ratio, 
this offset is entirely due to the fact that Petrosian magnitudes are 
0.1~mags fainter than (the more realistic) {\tt model} magnitudes.  
The middle left panel shows that $M_{*{\rm Bell}}/M_{*{\rm Model}}$ is 
approximately independent of mass for $g-r$, and strongly mass dependent 
for $r-i$.  But perhaps most importantly, note that $M_{*{\rm Model}}$ 
is typically $0.1$~dex smaller than the $g-r$ based $M_{*{\rm Bell}}$.  
This offset would have been even larger if we had compared to 
$M_{*{\rm Petro}}$ -- the fairer comparison, because our 
$M_{*{\rm Bell}}$ here is based on the {\tt Petrosian} magnitude.  
We argue shortly that this offset reflects the fact that the 
templates used to estimate $M_{*{\rm Model}}$ are not appropriate 
for large masses.  In contrast, $M_{*{\rm Gallazzi}}/M_{*{\rm Model}}$ 
is strongly mass dependent at small masses -- this is because aperture 
effects matter for $M_{*{\rm Gallazzi}}$ but not for 
$M_{*{\rm Model}}$.  The behaviour at larger masses is a 
combination of aperture and template effects, but note again 
that $M_{*{\rm Model}}$ is about 0.1~dex low.  

The panels on the right show that $M_{*{\rm LRG}}$ fares much better.  
The top right panel shows that it is always substantially larger than 
$M_{*{\rm Model}}$, and the two panels below it show that 
$M_{*{\rm LRG}}$ is much more like $M_{*{\rm Bell}}$ and 
$M_{*{\rm Gallazzi}}$ than is $M_{*{\rm Model}}$.  
Comparison with the corresponding panels on the left shows that 
much of the offset is removed, strongly suggesting that neither 
$M_{*{\rm Petro}}$ nor $M_{*{\rm Model}}$ are reliable at the high
mass end.  This is not surprising -- massive galaxies are expected 
to be old, so a 10~Gyr template should provide a more accurate 
mass estimate than ones that are restricted to ages of $\sim 7$~Gyrs 
or less.  This will be important in Section~\ref{phiMass}.  At the 
low mass end, $M_{*{\rm LRG}}$ is larger than either of the other 
estimates, suggesting that it overestimates the true mass.  However, 
at these smaller masses, using $M_{*{\rm Model}}$ exclusively produces 
masses which lie below $M_{*{\rm Gallazzi}}$ and $M_{*{\rm Bell}}$.  
This suggests that the LRG template remains the better description 
for a non-negligible fraction of low mass objects.  Therefore, in 
practice, to use the Blanton \& Roweis estimates, one must devise a 
method for choosing between $M_{*{\rm Model}}$ and $M_{*{\rm LRG}}$.  

Figure~\ref{CompareMsAll} compares the different estimates of 
the stellar mass for the full sample.  This gives similar results 
to when we restricted to $C_r\ge 2.86$, although the offsets are 
more dramatic.  In this case, the selection effect associated with 
studying fixed $M_{*{\rm Bell}}$ makes the aperture effect associated 
with $M_{*{\rm Gallazzi}}$ appear more complex than it really is:  
the large bump at the highest redshift associated with a given 
$M_{*{\rm Bell}}$ is not present when one studies objects at fixed 
luminosity (compare top and bottom panels of Figure~\ref{CompareMsz}).  

\clearpage

\section{Tables}\label{tables}
This Appendix provides tables which summarize the results of fitting 
equation~(\ref{psiO}) to the measured luminosity, stellar mass, size 
and velocity dispersion distributions shown in 
Figures~\ref{differentCIa}--\ref{ES0Sa}. These measurements were
obtained using {\tt cmodel} magnitudes and sizes. 
We described our measurements by assuming that the intrinsic 
distributions have the form given by equation~(\ref{phiX}).  
The values between round brakets show the parameters 
obtained fitting equation~(\ref{phiX}) to the data 
ignoring measurement errors.

\begin{table*}
\caption[]{Best-fit parameters of equation~(\ref{psiO}) to the 
           measured $r-$band {\tt cmodel} luminosity function $\phi(L)$. 
           Values between 
           round brakets show the parameters obtained fitting 
           equation~(\ref{phiX}) to the data ignoring measurement 
           errors.}
\begin{tabular}{lccccc}
 \hline 
  Sample & $\phi_*/10^{-2}$Mpc$^{-3}$ & $L_*/10^9\,L_\odot$ & $\alpha$ & $\beta$ & $\rho_L/10^9\,L_\odot$Mpc$^{-3}$\\ 
 \hline
HB09           & ($0.095$) $ 0.095\pm  0.005$ & ($  10.69$) $  10.99\pm    4.06$ & ($   1.38$) $   1.37\pm    0.18$ & ($   0.769$) $   0.776\pm    0.071$ &  $   0.024  $  \\
CI $>$ 2.86    & ($0.174$) $ 0.174\pm  0.010$ & ($   6.96$) $   7.21\pm    2.52$ & ($   1.38$) $   1.37\pm    0.16$ & ($   0.692$) $   0.698\pm    0.051$ &  $   0.038  $  \\
CI $>$ 2.6     & ($0.382$) $ 0.382\pm  0.022$ & ($   3.16$) $   3.28\pm    1.30$ & ($   1.32$) $   1.31\pm    0.16$ & ($   0.583$) $   0.588\pm    0.039$ &  $   0.061  $  \\
All            & ($4.123$) $ 4.196\pm  1.693$ & ($  12.21$) $  12.48\pm    3.42$ & ($   0.20$) $   0.19\pm    0.10$ & ($   0.728$) $   0.734\pm    0.051$ &  $   0.135  $  \\
All ($M_r < -20$) & ($1.227$) $ 1.197\pm  0.240$ & ($   2.08$) $   1.88\pm    0.86$ & ($   1.07$) $   1.12\pm    0.24$ & ($   0.540$) $   0.533\pm    0.035$ &  $   0.125  $ \\
\hline
F07-E &   $ 0.13\pm  0.11$ & $  72.10\pm   34.38$ & $   0.320\pm    0.444$ & $   1.752\pm    0.750$ & $   0.022$ \\
F07-S0 &  $ 1.05\pm  0.65$ & $  64.05\pm   10.15$ & $   0.065\pm    0.046$ & $   1.462\pm    0.203$ & $   0.038$ \\
F07-Sa &  $ 0.80\pm  0.07$ & $  27.19\pm    1.15$ & $   0.254\pm    0.020$ & $   0.944\pm    0.023$ & $   0.058$ \\
F07-Sb &  $17.22\pm  1.46$ & $  36.01\pm    1.35$ & $   0.014\pm    0.001$ & $   1.079\pm    0.030$ & $   0.084$ \\
F07-Scd & $23.48\pm  1.93$ & $  28.37\pm    1.04$ & $   0.017\pm    0.001$ & $   0.980\pm    0.023$ & $   0.112$ \\
F07-All & $16.00\pm  8.98$ & $  25.67\pm    4.87$ & $   0.027\pm    0.016$ & $   0.934\pm    0.086$ & $   0.116$ \\
\hline
\end{tabular}
\label{tabL} 
\end{table*}

\begin{table*}
\caption[]{Best-fit parameters of equation~(\ref{psiO}) to the measured 
           stellar mass function $\phi(M_*)$.}
\begin{tabular}{lccccc}
 \hline
  Sample & $\phi_*/10^{-2}$\,Mpc$^{-3}$ & $M_*/10^9\,M_\odot$ & $\alpha$ & $\beta$ & $\rho_*/10^9\,M_\odot$~Mpc$^{-3}$\\ 
 \hline
HB09           & ($0.095$) $ 0.095\pm  0.005$ & ($  20.25$) $  25.24\pm    8.70$ & ($   1.29$) $   1.28\pm    0.14$ & ($   0.641$) $   0.696\pm    0.051$ &  $   0.066  $  \\
CI $>$ 2.86    & ($0.174$) $ 0.174\pm  0.009$ & ($  13.62$) $  17.62\pm    6.14$ & ($   1.26$) $   1.24\pm    0.13$ & ($   0.590$) $   0.640\pm    0.043$ &  $   0.105  $  \\
CI $>$ 2.6     & ($0.388$) $ 0.389\pm  0.020$ & ($   7.49$) $  10.07\pm    3.89$ & ($   1.10$) $   1.07\pm    0.12$ & ($   0.522$) $   0.563\pm    0.036$ &  $   0.166  $  \\
All            & ($5.285$) $ 5.850\pm  3.054$ & ($  34.88$) $  39.06\pm    8.81$ & ($   0.11$) $   0.09\pm    0.06$ & ($   0.650$) $   0.694\pm    0.040$ &  $   0.289  $  \\
All ($M_* > 3 \times 10^{10} M_\odot$) & ($0.761$) $ 0.672\pm  0.123$ & ($   0.14$) $   0.02\pm    0.01$ & ($   1.95$) $   2.68\pm    0.30$ & ($   0.342$) $   0.308\pm    0.010$ &  $   0.261  $ \\
\hline
F07-E &   $ 0.09\pm  0.04$ & $ 158.43\pm  115.34$ & $   0.54\pm    0.54$ & $   1.31\pm    0.53$ & $   0.062$ \\
F07-S0 &  $ 1.11\pm  0.66$ & $ 206.53\pm   35.47$ & $   0.05\pm    0.04$ & $   1.44\pm    0.25$ & $   0.104$ \\
F07-Sa &  $ 1.70\pm  2.52$ & $ 144.73\pm   45.50$ & $   0.07\pm    0.11$ & $   1.18\pm    0.23$ & $   0.151$ \\
F07-Sb &  $ 1.46\pm  1.31$ & $  90.24\pm   46.21$ & $   0.15\pm    0.18$ & $   0.96\pm    0.21$ & $   0.203$ \\
F07-Scd & $ 4.09\pm  5.46$ & $  81.34\pm   30.52$ & $   0.07\pm    0.11$ & $   0.92\pm    0.15$ & $   0.248$ \\
F07-All & $ 6.76\pm  9.92$ & $  84.20\pm   24.84$ & $   0.04\pm    0.07$ & $   0.93\pm    0.13$ & $   0.251$ \\
\hline
\end{tabular}
\label{tabMs} 
\end{table*}

\begin{table*}
\caption[]{Best-fit parameters of equation~(\ref{psiO}) to the measured 
           $r-$band {\tt cmodel} size function $\phi(R_e)$.}
\begin{tabular}{lccccc}
 \hline 
  Sample & $\phi_*/10^{-2}$Mpc$^{-3}$ & $R_*$/kpc & $\alpha$ & $\beta$ & $\langle R\rangle$/kpc\\ 
\hline 
HB09           & ($0.093$) $ 0.093\pm  0.006$ & ($  0.0001$) $  0.0003\pm   0.0002$ & ($   8.70$) $   9.29\pm    0.45$ & ($   0.293$) $   0.294\pm    0.007$ & $    3.25$ \\
CI $>$ 2.86    & ($0.173$) $ 0.173\pm  0.011$ & ($  0.0001$) $  0.0002\pm   0.0002$ & ($   9.82$) $  10.73\pm    0.53$ & ($   0.294$) $   0.290\pm    0.007$ & $    3.19$ \\
CI $>$ 2.6     & ($0.400$) $ 0.400\pm  0.024$ & ($  0.0001$) $  0.0002\pm   0.0002$ & ($   7.58$) $   8.08\pm    0.38$ & ($   0.295$) $   0.295\pm    0.006$ & $    2.74$ \\
All            & ($6.009$) $ 6.040\pm  0.766$ & ($  0.4838$) $  0.5809\pm   0.1773$ & ($   1.31$) $   1.27\pm    0.23$ & ($   0.688$) $   0.729\pm    0.045$ & $    1.41$ \\
All ($1.5 < R_e < 20$ kpc) & ($2.951$) $ 2.858\pm  0.420$ & ($   0.5523$) $   0.5612\pm    0.2033$ & ($   2.22$) $   2.39\pm    0.44$ & ($   0.765$) $   0.790\pm    0.058$ &  $   2.38  $ \\
\hline 
F07-E &   $ 0.10\pm  0.02$ & $  0.0007\pm   0.0009$ & $   6.44\pm    1.55$ & $   0.352\pm    0.032$ & $    3.13$ \\  
F07-S0 &  $ 0.19\pm  0.03$ & $  0.0008\pm   0.0008$ & $   6.07\pm    1.05$ & $   0.356\pm    0.025$ & $    2.78$ \\  
F07-Sa &  $ 0.37\pm  0.06$ & $  0.0004\pm   0.0004$ & $   6.08\pm    1.01$ & $   0.337\pm    0.021$ & $    2.46$ \\  
F07-Sb &  $ 0.83\pm  0.15$ & $  0.0123\pm   0.0104$ & $   4.24\pm    0.82$ & $   0.444\pm    0.037$ & $    2.30$ \\  
F07-Scd & $ 1.55\pm  0.21$ & $  0.0062\pm   0.0042$ & $   5.72\pm    0.76$ & $   0.431\pm    0.026$ & $    2.80$ \\  
F07-All & $ 3.92\pm  0.77$ & $  0.0049\pm   0.0033$ & $   4.02\pm    0.64$ & $   0.402\pm    0.024$ & $    1.81$ \\  
\hline 
\end{tabular}
\label{tabR} 
\end{table*}

\begin{table*}
\caption[]{Best-fit parameters of equation~(\ref{psiO}) to the measured 
           velocity dispersion function $\phi(\sigma)$.}
\label{tabS} 
\begin{tabular}{lccccc}
 \hline 
  Sample & $\phi_*/10^{-2}$Mpc$^{-3}$ & $\sigma_*$/km~s$^{-1}$ & $\alpha$ & $\beta$ & $\langle\sigma\rangle$/km~s$^{-1}$\\ 
 \hline 
HB09         &    ($0.097$) $ 0.096\pm  0.006$ & ($ 184.08$) $ 173.41\pm   16.85$ & ($   2.44$) $   2.83\pm    0.41$ & ($   2.91$) $   3.10\pm    0.35$ & $  149.15$ \\
CI $>$ 2.86    &  ($0.182$) $ 0.179\pm  0.010$ & ($ 177.34$) $ 166.50\pm   17.02$ & ($   2.17$) $   2.54\pm    0.41$ & ($   2.76$) $   2.93\pm    0.33$ & $  139.14$ \\
CI $>$ 2.6     &  ($0.663$) $ 0.590\pm  0.088$ & ($ 190.57$) $ 175.96\pm   16.88$ & ($   0.80$) $   1.06\pm    0.34$ & ($   2.86$) $   2.99\pm    0.35$ & $   92.40$ \\
All          &    ($2.099$) $ 2.099\pm  0.099$ & ($ 113.78$) $ 113.78\pm    1.06$ & ($   0.94$) $   0.94\pm    0.03$ & ($   1.85$) $   1.85\pm    0.02$ & $   63.70$ \\
All ($\sigma > 125$ km s$^{-1}$) & ($8.133$) $ 2.611\pm  0.161$ & ($ 176.99$) $ 159.57\pm    1.48$ & ($   0.11$) $   0.41\pm    0.02$ & ($   2.54$) $   2.59\pm    0.04$ &  $  44.02  $ \\
\hline 
F07-E &   $ 0.11\pm  0.04$ & $ 218.27\pm   43.06$ & $   1.53\pm    1.21$ & $   4.47\pm    1.90$ & $  131.56$ \\
F07-S0 &  $ 0.19\pm  0.05$ & $ 197.35\pm   58.97$ & $   1.66\pm    1.25$ & $   3.49\pm    1.75$ & $  128.40$ \\
F07-Sa &  $ 0.41\pm  0.13$ & $ 193.07\pm   44.05$ & $   1.06\pm    0.77$ & $   3.35\pm    1.29$ & $   99.32$ \\
F07-Sb &  $ 0.61\pm  0.25$ & $ 158.18\pm   59.10$ & $   1.22\pm    1.14$ & $   2.57\pm    1.05$ & $   93.12$ \\
F07-Scd & $ 2.47\pm  3.96$ & $ 180.45\pm   33.69$ & $   0.29\pm    0.54$ & $   2.96\pm    0.85$ & $   37.48$ \\
F07-All & $ 1.23\pm  0.31$ & $  59.04\pm   31.48$ & $   2.44\pm    1.17$ & $   1.35\pm    0.30$ & $   86.74$ \\
\hline
\end{tabular}
\end{table*}

\label{lastpage}

\end{document}